\documentclass[aps,amsmath,amssymb,superscriptaddress,showkeys,longbibliography,reprint]{revtex4-2}
\usepackage{graphicx}
\usepackage{dcolumn}
\usepackage{bm}
\usepackage{mathscinet}
\usepackage{hyperref}
\usepackage{type1cm}
\usepackage{lettrine}

\begin{document}
\title{Multidimensional political polarization in online social networks}

\author{Antonio F. Peralta}
\email{peraltaaf@ceu.edu}
\affiliation{Department of Network and Data Science, Central European University, A-1100 Vienna, Austria}
\affiliation{Helmholtz Institute for Functional Marine Biodiversity (HIFMB), 26129 Oldenburg, Germany}
\author{Pedro Ramaciotti}
\affiliation{CNRS, Complex Systems Institute of Paris Ile-de-France (ISC-PIF), Sciences Po médialab \&{} LPI, Université Paris Cité, France}
\author{J\'anos Kert\'esz}
\affiliation{Department of Network and Data Science, Central European University, A-1100 Vienna, Austria}
\affiliation{Complexity Science Hub, A-1080 Vienna, Austria}
\author{Gerardo I\~{n}iguez}
\email{iniguezg@ceu.edu}
\affiliation{Department of Network and Data Science, Central European University, A-1100 Vienna, Austria}
\affiliation{Faculty of Information Technology and Communication Sciences, Tampere University,  FI-33720 Tampere, Finland}
\affiliation{Department of Computer Science, Aalto University School of Science, FI-00076 Aalto, Finland}
\affiliation{Centro de Ciencias de la Complejidad, Universidad Nacional Auton\'{o}ma de M\'{e}xico, 04510 Ciudad de M\'{e}xico, Mexico}

\begin{abstract}
Political polarization in online social platforms is a rapidly growing phenomenon worldwide. Despite their relevance to modern-day politics,  the structure and dynamics of polarized states in digital spaces are still poorly understood. We analyze the community structure of a two-layer, interconnected network of French Twitter users, where one layer contains members of Parliament and the other one regular users. We obtain an optimal representation of the network in a four-dimensional political opinion space by combining network embedding methods and political survey data. We find structurally cohesive groups sharing common political attitudes and relate them to the political party landscape in France. The distribution of opinions of professional politicians is narrower than that of regular users, indicating the presence of more extreme attitudes in the general population. We find that politically extreme communities interact less with other groups as compared to more centrist groups. We apply an empirically tested social influence model to the two-layer network to pinpoint interaction mechanisms that can describe the political polarization seen in data, particularly for centrist groups. Our results shed light on the social behaviors that drive digital platforms towards polarization,  and uncover an informative multidimensional space to assess political attitudes online.
\end{abstract}

\date\today


\maketitle


\lettrine{U}{}nderstanding how people share information and influence each other in their political attitudes, potentially leading to ideological partisanship and political polarization \cite{mccarty2019polarization}, is a relevant yet challenging issue that has been tackled for decades using theories and methods from fields as diverse as sociology \cite{baldassarri2007dynamics}, political science \cite{fiorina2008political,prior2013media}, economics \cite{dixit2007political} and, more recently, complexity and computational social science \cite{conover2011political,bail2018exposure,flamino2023political,hohmann2023quantifying}. Mathematical modeling is a frequently applied method to elucidate the mechanisms behind social influence and ideological polarization \cite{schelling1971dynamic,granovetter1978threshold,axelrod1997dissemination,Castellano:2009,holme2015mechanistic,Baumann:2021}. Often times, however, models that are otherwise conceptually robust and even inspired by empirical data, are investigated only theoretically through analytical derivations and numerical simulations on idealized synthetic populations \cite{sobkowicz2009modelling,Peralta:2022}.  An example are models of continuous political opinions that position individuals in,  e.g.,  liberal-conservative scales, where the dimension capturing political ideology is defined {\textit{a priori}} and not as the result of data analysis in the social context of interest.  Indeed, one of the most challenging aspects of bridging opinion dynamics models and empirical observations of political attitudes in social networks is the number of dimensions determining social influence \cite{benoit2012dimensionality}. 

Ideal point estimation models \cite{Imai:2016} have been used to position large numbers of social media users in a liberal-conservative scale in several platforms \cite{barbera2015birds,bond2015quantifying}, amounting to a single-dimensional opinion analysis where users are classified from the most liberal to the most conservative.  And yet,  social scientists acknowledge that political systems in Europe \cite{bakker2012complexity} and also increasingly in the US \cite{uscinski2021american} are structured by several dimensions of opinion.  Recent advancements in multidimensional political opinion estimation methods allow to embed structural data, such as communication networks coming from social media,  into political spaces with multiple political dimensions \cite{Ramaciotti:2022}.  In these spaces, dimensions act as continuous indicators of positive or negative attitudes towards identifiable issues of political debate. As an example, the DeGroot model of opinion dynamics \cite{DeGroot:1974} has recently been used to estimate multidimensional political attitudes of users in online platforms \cite{Martin-Gutierrez:2023}.

\begin{figure}[t]
\begin{center}
\includegraphics[width=0.49\textwidth]{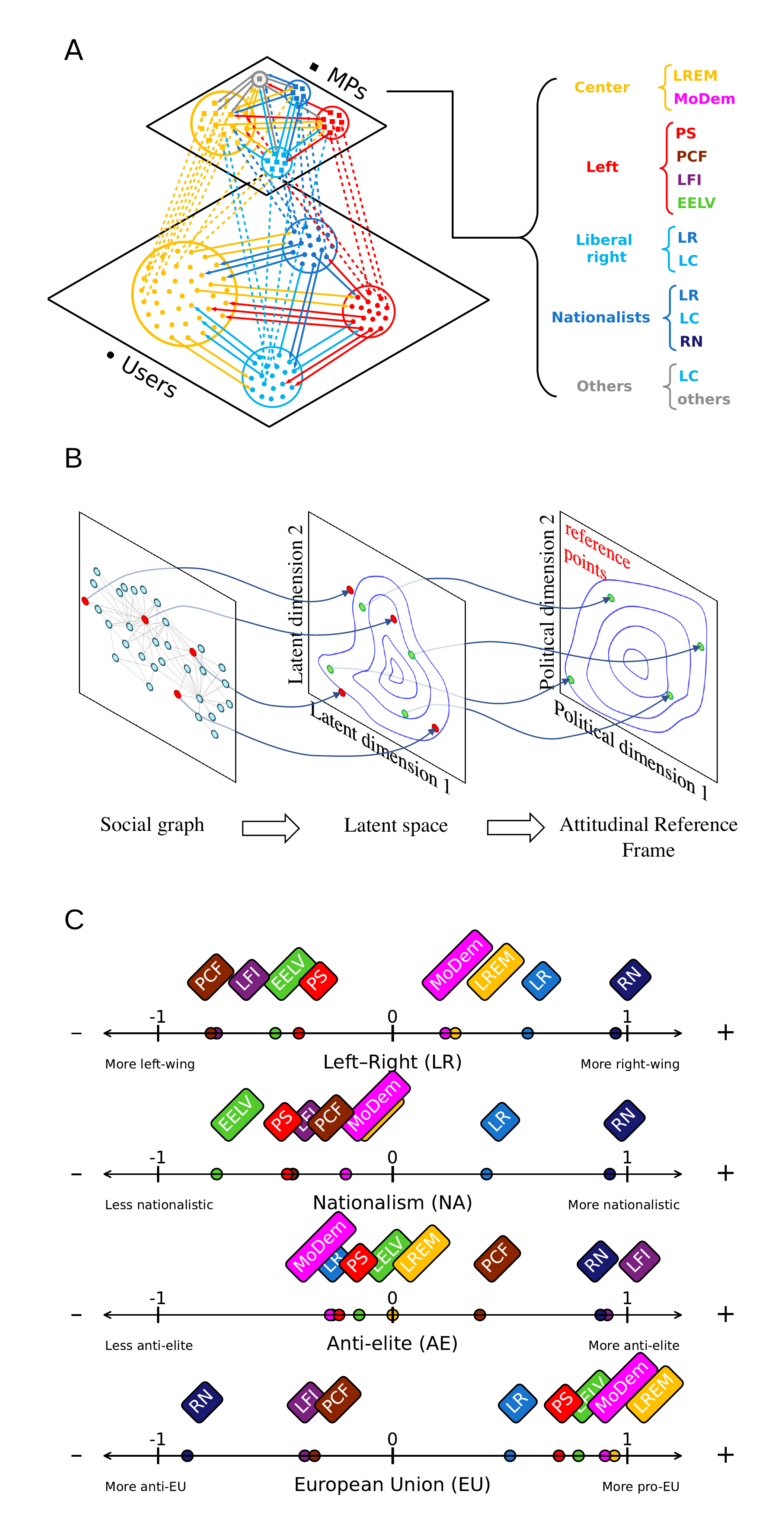}
\end{center}
\end{figure}

\begin{figure}[t]
  \caption{{\bf Uncovering political opinions via online social network data.}
 {\bf (A)} Schematic diagram of network structure in French Twitter, with Members of Parliament (MPs) on top and regular users in the bottom. Colored circles highlight communities found by the planted partition model \cite{Zhang:2020}, which divides the network by assortativity for both MPs and users. Colors follow overall political leaning. MPs belong to five communities: \emph{Center}, \emph{Left}, \emph{Liberal right}, \emph{Nationalist} and \emph{Others} (see SI, Sec. S2). Parties in each community are shown by acronyms (for details on parties see Results). Color of links corresponds to the community of the source nodes; User $\rightarrow$ User and MP $\rightarrow$ MP links are represented by solid arrows, and User $\rightarrow$ MP links by dashed arrows (we disregard MP $\rightarrow$ User links).
 {\bf (B)} Scheme of method to obtain politically relevant ideological positions of users and MPs. Twitter data is embedded in a multidimensional latent space preserving homophily: users who are close in this space have higher probability of following the same set of MPs. We compute positions of political parties in latent space as the mean position of MPs of the same party. Using these points and the corresponding party positions in political survey data, we map the network onto the opinion dimensions of the survey, forming a political Attitudinal Reference Frame (ARF, see MM and SI, Sec. S1.2.2).
 {\bf (C)} Position of French political parties used as reference points to map the position of users on latent space (resulting from homophiliy embedding) onto the ARF.  Party positions are taken from the 2019 Chapel Hill Expert Survey (CHES) data \cite{chesdata2019}, built by a panel of experts in European politics (for details see MM and SI, Sec. S1.2.2).}
 \label{fig1}
\end{figure}

Here we propose a methodology to uncover and understand patterns of online political polarization in social media by combining embedding methods with empirically grounded opinion models in a multidimensional ideological space.  Using follower networks in the online micro-blogging platform Twitter, with data from both professional politicians and regular users in France,  we estimate the ideological positions of individuals along an optimal number of four politically relevant dimensions: left-right stance and attitudes towards nationalism,  elites,  and the European Union. By means of a network community detection method based on stochastic block-modelling, we first classify politicians and regular users into groups, according to assortative patterns in the Twitter interaction structure.  We then embed the social graph in a latent space preserving homophily, where dimensions are interpreted as ideological indicators using a survey of political experts. Our four political dimensions are optimal in the sense that they capture main differences between competing parties in France. Relying on the community partition of the network and on the inferred positions along the detected four political dimensions, we propose formal measures that uncover polarized states and diverging ideologies.  Finally, we introduce an opinion dynamics model capable of reproducing the large-scale behavior of empirical data, providing a plausible explanation for the influence mechanisms underlying the structure and dynamics of political polarization in multidimensional ideological spaces.

\section*{Results}
\label{sec_results}

We gather network data from the follower $\rightarrow$ followed relations between the Twitter accounts of $M = 813$ Members of Parliament (MPs) in France and $N = 230~254$ of their followers (here denoted regular users), who follow at least 3 MPs and follow another user that follows at least 3 MPs (Fig.~\ref{fig1}\emph{A}). Based on this directed online social network, we infer politically relevant ideological positions of MPs and users via a 2-step embedding method (Fig.~\ref{fig1}\emph{B}). First, we embed nodes of the network onto a homophily-preserving latent space (users close in space follow similar sets of MPs). Then, positions in this latent space are correlated to the attitude dimensions of a standard political survey, the 2019 Chapel Hill Expert Survey (CHES) \cite{chesdata2019}. The resulting space is comprised by four real-valued variables, or political dimensions, that represent attitudes towards: (i) the political left or right (LR), intended to measure the overall ideological stance of an individual (i.e. without specifying particular political issues to survey respondents); (ii) nationalism (NA); (iii) the European Union (EU); and (iv) the establishment and elites, intended to measure anti-elite sentiment (AE). Beyond the standard left-right dimension of political cleavage, our embedding process is able to identify additional dimensions capturing relevant differences between party positions in France (Fig.~\ref{fig1}\emph{C}; for more details see Materials and Methods [MM], Ref. \cite{Ramaciotti:2022} for an implementation of the algorithm, and Supplementary Information [SI], Sec. 1).

\subsection*{Structurally cohesive groups share political attitudes}

Since MPs arguably carry the political agenda by highlighting topics of interest to their parties and the general public, we choose to focus first on the part of the network involving MPs only, i.e. the MP $\rightarrow$ MP links (Fig.~\ref{fig1}\emph{A}). We run a standard community detection algorithm on the MP layer to find the best partition into assortative groups (groups more connected to themselves than to others), by minimizing the description length of the network from an information-theoretical perspective \cite{Grunwald:2007, Zhang:2020}.

\begin{figure}[!ht]
	\begin{center}
		\includegraphics[width=0.49\textwidth]{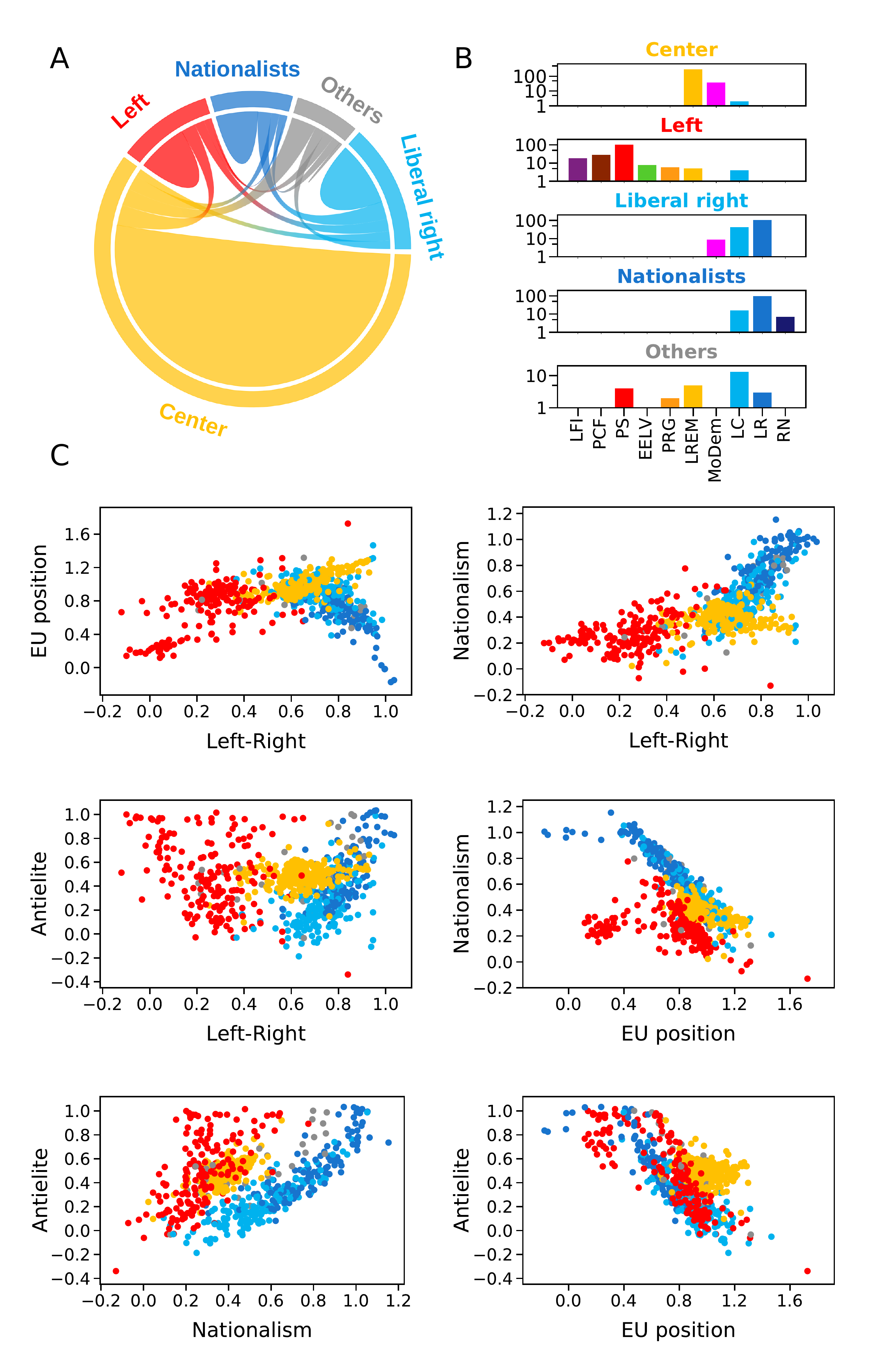}
		\caption{{\bf Communities and ideological positions of professional politicians.}
			Using the best partition (minimum description length) of the planted partition model \cite{Zhang:2020}, we find 5 MP communities: \emph{Center}, \emph{Left}, \emph{Liberal right}, \emph{Nationalists}, and \emph{Others}.
			{\bf (A)} Chord diagram indicating the connectivity (number of links) between and inside communities. The angular size of each community in the diagram is proportional to the number links that depart from it.
			{\bf (B)} Party composition of each community. The list of parties (horizontal axis) is ordered according to their positions in the LR dimension. Bars indicate the number of MPs that belong to each party in the specified community (rows). We choose convenient colors for communities and parties in order to better visualize their political attitudes (see SI, Sec. S2).
			{\bf (C)} Political positions of MPs (coloured according to their communities) in various two-dimensional projections of the 4-dimensional political space, leading to six possible pairs of opinion variables: LR-EU, LR-NA, LR-AE, EU-NA, NA-AE, and EU-AE. The positions of MPs motivate the naming of each community.}
		\label{fig2}
	\end{center}
\end{figure}

We find four assortative communities, named \emph{Center}, \emph{Left}, \emph{Liberal right}, and \emph{Nationalists} by following traditional distinctions in French politics \cite{remond2014droites}, plus a non-assortative group denoted \emph{Others} (Fig. \ref{fig2}\emph{A}; see also MM and SI, Sec. S2). These names are determined by the positions of the corresponding MPs along the identified political dimensions (Fig.~\ref{fig1}\emph{C} and Fig.~\ref{fig2}\emph{B--C}). The \emph{Center} is composed mainly of members of the French parties LREM (Macron's party Republic on the Move), 
and MoDem (Moderate Democrats), displaying centrist positions in all dimensions except EU (where it is the most pro-Europe community). The \emph{Left} has a markedly left-leaning distribution, assembling most MPs from known left-wing parties (LFI, PCF, PS, and EELV, standing respectively for Indomitable France, the Communist, Socialist, and Ecologists parties). The rightmost communities in the LR dimension are named \emph{Liberal right} (including some MPs from MoDem, but mostly from the LC and LR parties, standing for the Centrist and Republican parties) and \emph{Nationalists} (including MPs from LC and LR, and notably all MPs from Le Pen's National Rally party, RN) to account for their differences along the NA and AE dimensions.

The \emph{Others} group is composed of several parties across the whole political spectrum, and the structural patterns of its MPs do not fit any of the other groups. We observe that the dimensions of the latent space properly capture the attitudes of politicians expected by their party allegiance, and MPs of the same group are clustered together in opinion space, except for \emph{Others}. We also identify interesting features in the opinion overlaps between groups: the \emph{Liberal right} and \emph{Nationalists} exhibit some overlap in their LR attitudes, but occupy different regions in the NA and AE dimensions (Fig. \ref{fig2}\emph{C}).

These results further cement the need and real-world relevance of the political attitude dimensions comprising our multidimensional latent space. Beyond the traditional left-right cleavage, we find a dimension of attitudes towards institutions and elites (previously identified as relevant in French politics and political Twitter in general \cite{ramaciotti2021unfolding}), an ideological position towards nationalism that differentiates between two right-wing tendencies (\emph{Liberal right} and \emph{Nationalists}), and a variable encapsulating opinions with respect to the EU, also deemed significant in French politics \cite{hooghe2018cleavage} (see MM for a detailed discussion on the selection of these dimensions).

\subsection*{Professional politicians have less extreme attitudes than regular users}

\begin{figure*}[!ht]
\begin{center}
\includegraphics[width=0.95\textwidth]{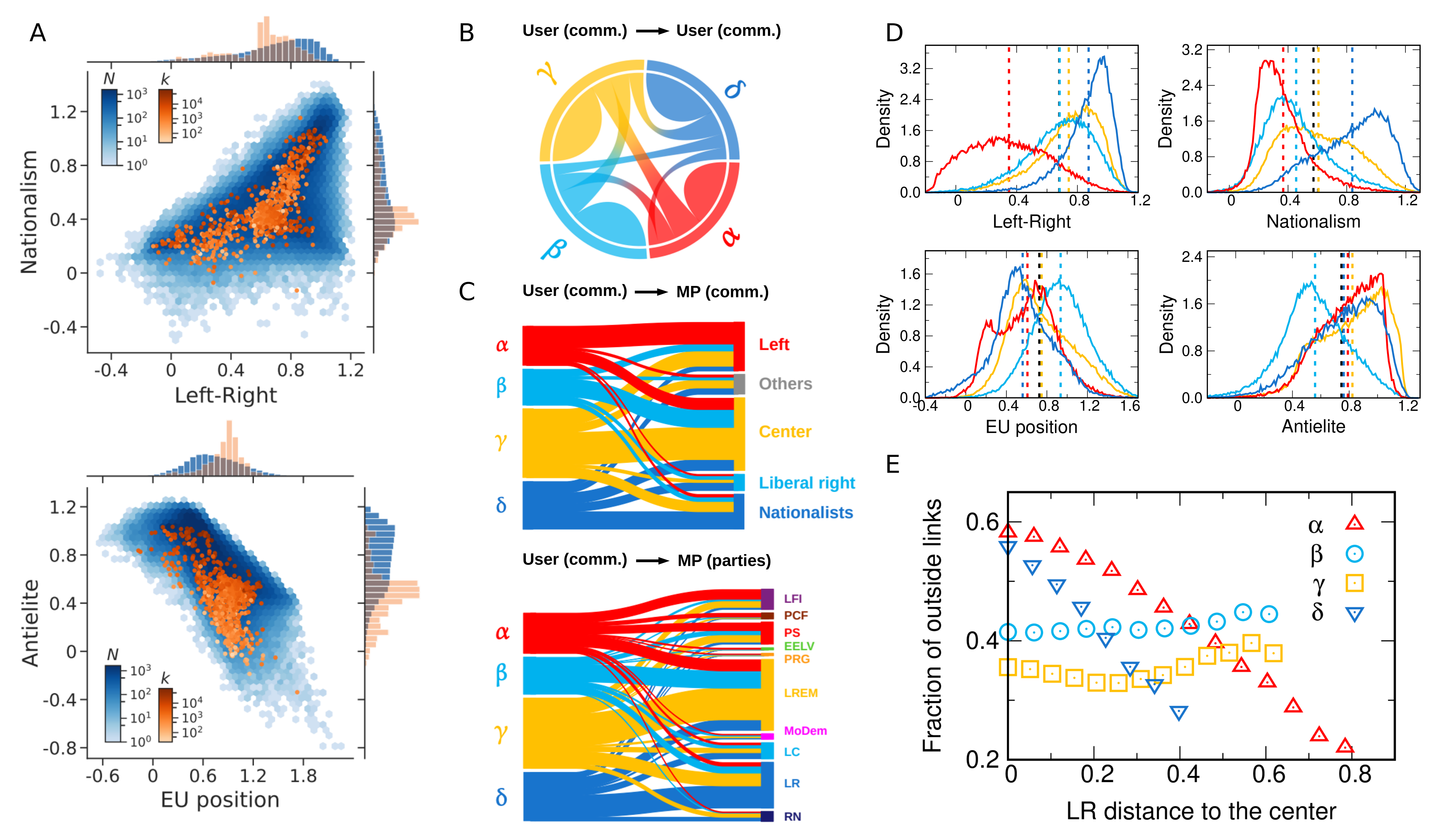}
\caption{{\bf Communities and ideological positions of regular users and their relation to professional politicians.}
{\bf (A)} Number distribution of ideological positions of users (MPs), represented by blue (orange) dots in a two-dimensional opinion space for pairs of opinion variables: LR-NA (top) and EU-AE (bottom) (other pairs in SI, Sec. S1.3). Color shading for MPs is proportional to the number of followers $k$ (users) of each MP in logarithmic scale, i.e. in-degree in the User $\rightarrow$ MP network, $k \equiv k^{\text{in/um}}_{m}$ (see SI, Sec. S1.1.1). Corresponding marginal probability densities of users (blue) and MPs (orange) are plotted in linear scale. 
{\bf (B)} Communities in the user layer correspond to the best partition (minimum description length) of the planted partition model \cite{Zhang:2020} with a fixed number of communities equal to four. The color of each community is chosen according to its characteristic political attitude in relation to the MP layer. The chord diagram indicates the connectivity (number of links) between and inside communities of users. 
{\bf (C)} Sankey diagrams indicating the connectivity between user groups and both MP communities (top) and their parties (bottom). Size of flows is proportional to the number of links in the User $\rightarrow$ MP network, whose source nodes belong to a particular community of users (indicated by colors). Link colors are chosen according to user communities.
{\bf (D)} Probability densities of opinion variables (LR, NA, EU and AE) of users in each community (as indicated by line color). Colored dashed lines represent the average opinion of each community, and the black dashed line is the global average.
{\bf (E)} Fraction of links of users (in the user layer) pointing outside of their community as a function of the distance of their opinion from the average opinion of all users. Plot corresponds to the LR dimension (for others see SI, Sec. S3.3.4). While members of $\beta $ and $\gamma$ connect freely to other communities despite of political differences, this function rapidly decreases with ideological distance for members of $\alpha$ and $\delta$.}
\label{fig3}
\end{center}
\end{figure*}

We analyze the positions of regular users in latent space and compare them to the ideological positions of the MPs they follow (Fig. \ref{fig3}). Political positions of MPs lie exclusively within the limits of the distribution of values for users. This means that, in our sample, the most extreme attitudes in French Twitter are held by the regular audience of the platform (Fig. \ref{fig3}\emph{A}). The difference in ideological extremism between MPs and users is most salient for the positions along the AE dimension, with MPs having a noticeable less anti-elite leaning than users (see SI, Sec. S1.3). From a political science strategic standpoint, this is to be expected, as politicians seek to position themselves as appealing to the largest possible number of users \cite{riker1968theory} (see SI, Sec. S1.3). Users also tend to align with the most popular MPs based on their number of followers, which ultimately produces a high concentration of opinions around popular MPs (see SI, Sec. S1.3). This is a first indication of the social influence mechanisms potentially driving the dynamics of political Twitter, such as opinion imitation or assimilation \cite{Dandekar:2013,Peralta:2022}, which we explore further below.

We compare the structural patterns of connectivity of users and MPs by running the same community detection algorithm in the user layer, but now under a constraint of four groups, the same number of assortative communities found for MPs (Fig.~\ref{fig3}\emph{B}). In other words, we focus on the particular level of the hierarchy of community structure among the general audience of Twitter that corresponds to the group divisions imposed by professional politicians (for details see SI, Sec. S2). We denote the resulting four communities by $\alpha$, $\beta$, $\gamma$ and $\delta$. At this level of granularity, the assortative communities of users also exhibit opinion coherence, both in terms of the distribution of political attitudes and their relation to the communities and parties of MPs (Fig.~\ref{fig3}\emph{C}).

The predominant political positions of user groups are strikingly informative (Fig.~\ref{fig3}\emph{D}). The $\alpha$ community is the only one having a marked left-wing stance over the LR dimension. The $\delta$ group, the rightmost in the LR dimension, is the only one skewed towards the nationalist side of the NA spectrum. The more centrist communities $\beta$ and $\gamma$ lie in between $\alpha$ and $\delta$ in the LR dimension and show a low nationalist stance. Yet they differ in their attitudes towards elites and the EU, with the $\gamma$ community showing a marked antielite sentiment. Three of these communities ($\alpha$, $\beta$ and $\delta$, respectively) somewhat correspond to identifiable political ideologies in France and their associated MP groups: the traditional left and two types of center to right-wing stances, the liberal and the nationalist right \cite{hooghe2018cleavage}. On the other extreme, community $\gamma$ does not correspond to a single group in the MP layer; its users mainly follow \emph{Center} politicians (see Fig.~\ref{fig3}\emph{C}) and show a strong antielite sentiment.

\subsection*{Groups with extreme political attitudes are more segregated}

Assortative communities of users in political Twitter also differ in the way they connect to each other despite their ideological disagreement, as captured by our attitudinal latent space (Fig.~\ref{fig3}\emph{E}). For each user community, we partition one of the dimensions of latent space (say, LR) into chunks, and compute the fraction of links (of users in that community and opinion interval) that lead to one of the other groups. This is a measure of community segregation, or political polarization, as a function of attitudinal positions in LR space (for details and other dimensions see SI, Sec. S3.3.4). The $\beta$ and $\gamma$ communities, roughly corresponding to the \emph{Liberal right} and \emph{Center}, show a flat trend, meaning that individuals identifying with these political ideologies interact with other groups despite their differences. Notably, the more right-wing $\beta$ community is slightly better connected to others than the more centrist $\gamma$ group (i.e. the fraction of outside links is larger on average). On the other hand, the politically extreme $\alpha$ and $\delta$ communities, mostly associated to the \emph{Left} and \emph{Nationalists}, have a decreasing trend in their connectivity to users with diverging ideologies, highlighting their segregation both in terms of structural connectivity and attitudinal positions in latent space. The $\alpha$ and $\delta$ groups are in this sense more heterogeneous yet assortative, since the political stances of their users are a strong indication of their degree of homophily with peers, echoing recent findings of increasing political polarization in US Twitter \cite{flamino2023political,hohmann2023quantifying}.

\begin{figure*}[!ht]
\begin{center}
\includegraphics[width=0.95\textwidth]{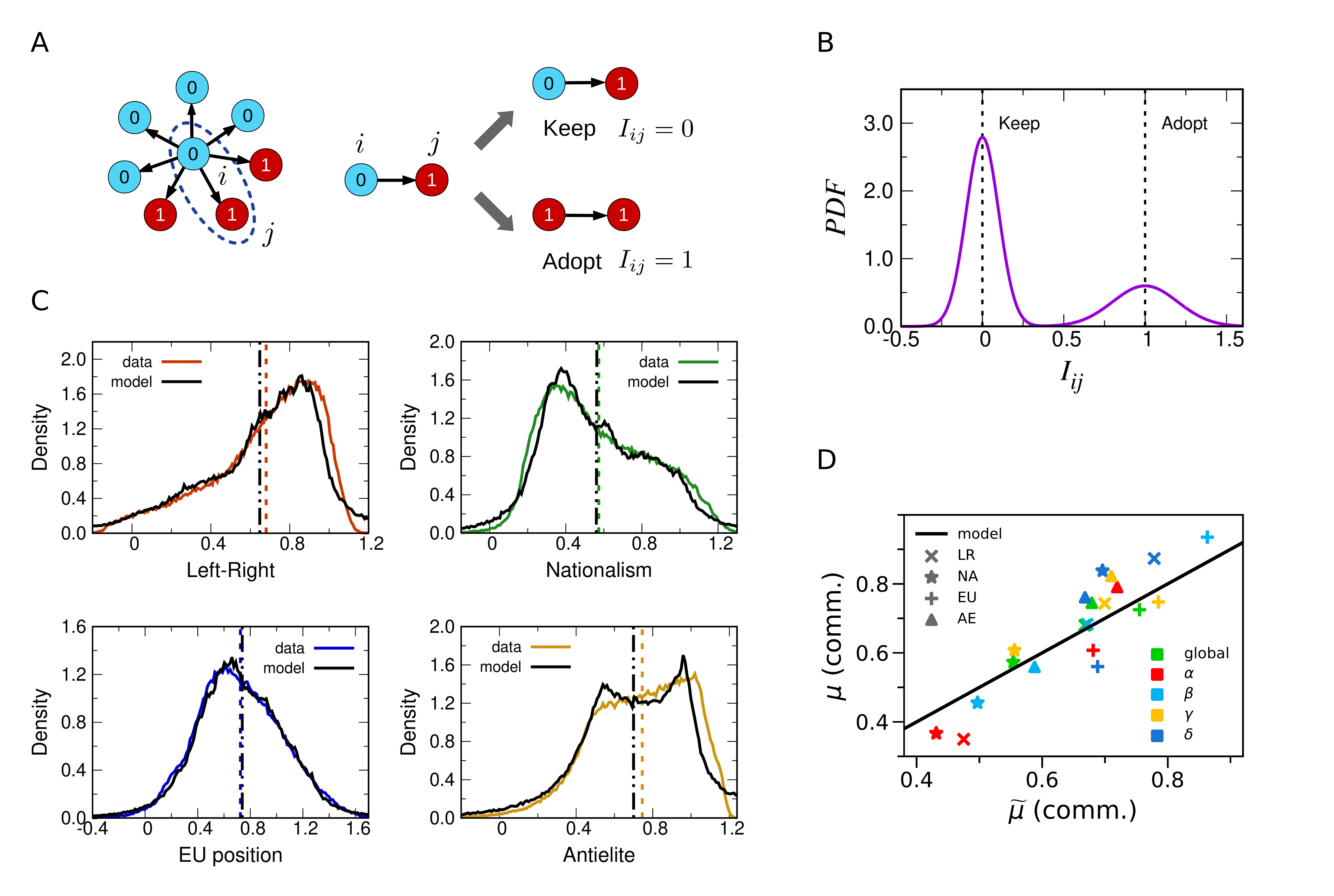}
\caption{{\bf Modelling multidimensional political polarization online.} 
{\bf (A)} In our minimal social influence model, user $i$ interacts with its neighbor, another user $j$ or MP $m$, and decides to either keep its own opinion or incrementally adopt the neighbor's position according to influence factor $I_{ij}$ (see Eqs.~(\ref{copy_user})--(\ref{copy_mps})).
{\bf (B)} We model the influence factor $I_{ij}$ as a sum of two Gaussians peaked around $I_{ij}=0$ and $I_{ij}=1$. The height and width of the peaks are parametrized by $(p, \sigma_{K}, \sigma_{A})$, which we determine by fitting (see MM).
{\bf (C)} Probability density function of attitudinal positions of users in latent space (LR, NA, EU and AE) in both empirical data (colored solid lines) and best fit of stationary state of the model (black solid lines). Colored (black) dashed lines represent the average opinion of data (model) in a given political dimension.
{\bf (D)} Average opinion $\mu$ of users as a function of the weighted average opinion $\widetilde{\mu}$ of the MPs they follow (see Eq. (\ref{mean_opinions})). Weights are the in-degrees of the MPs coming from the considered set of users. Each point corresponds to either the whole network (global) or certain community of users (indicated by colors; see Fig.~\ref{fig3}) and for a given opinion variable (LR, NA, EU, AE). The straight line is the degree-based mean field approximation $\mu = \widetilde{\mu}$ (see SI, Sec. S3.2).}
\label{fig4}
\end{center}
\end{figure*}

\subsection*{Modeling multidimensional political polarization online}

The positioning of professional politicians and regular users of French Twitter in a multidimensional attitudinal space indicates that people form structurally cohesive groups that become more segregated as their political ideologies diverge. Regular users also tend to be more extreme towards topics of political debate, while concentrating their attention on popular MPs. In order to identify potential idealized mechanisms that might explain this behavior, we explore an opinion dynamics model with social influence processes based on the results of controlled psychological experiments \cite{Moussaid:2013,Chacoma:2015}.

In the model, each user $i$ holds a vector opinion $\vec{v}_{i}(t)=(x_{i}(t), y_{i}(t), z_{i}(t), w_{i}(t))$ at time $t$ that determines its position in attitudinal latent space (the LR, NA, EU, and AE dimensions, respectively). Each MP $m$ has a static vector opinion $\vec{V}_m=(X_{m}, Y_{m}, Z_{m}, W_{m})$ that we extract from data. This assumption can be understood as a separation of time scales: regular users of Twitter change their minds and who they follow faster than MPs, who pursue the political agenda of their parties at a slower pace. Explicitly, the fast time scale, the opinion dynamics of users, is coupled to the slow one, the opinion dynamics of MPs, following its changes closely; which simplifies the analysis within the context of the slower dynamics.
	
In reality, both user and MP opinions evolve over time and influence each other. However, due to the visibility of the political position of MPs in the public sphere, we can argue that their opinion is in some sense more influential than that of regular users. Factors supporting this premise include the presence of MPs in the news across various media beyond Twitter, such as newspapers, television, radio, other social platforms, and websites. Another factor that bolsters this assumption is that MPs, due to the organization of political parties and the nature of the political agenda, are less prone to idiosyncratic changes. From a political theoretical standpoint, this follows a strategic logic in which politicians position themselves by displaying ideological and issue positions as the supply, which the public (the demand side) can chose via voting, or, in the case of our study, via following \cite{riker1968theory}. This distinction in the nature of opinions can be summarized as a public of users holding ideologies and attitudes that determine their choices, while MPs choose to display a set of ideologies and opinions not subject to change due to the positions of other MPs. In summary, we assume that users are influenced by other users and MPs, but not the other way around (meaning we ignore links from MPs to users; see Fig.~\ref{fig1}\emph{A}).

The dynamics of the model is as follows (Fig. \ref{fig4}\emph{A}). In a time step $\Delta t = 1/N$, a randomly selected user $i$ interacts with one of its neighbors, either another user $j$ or an MP $m$, who influences the opinion of $i$ according to
\begin{equation}
\label{copy_user}  
\vec{v}_{i}(t+\Delta t)=\vec{v}_{i}(t)+I_{ij} [\vec{v}_{j}(t)-\vec{v}_{i}(t)],
\end{equation}
or
\begin{equation}
\label{copy_mps}  
\vec{v}_{i}(t+\Delta t)=\vec{v}_{i}(t)+I_{im} [\vec{V}_{m}-\vec{v}_{i}(t)],
\end{equation}
where the influence factors $I_{ij}$ and $I_{im}$ are drawn at each time step from a predefined probability density function $f(I)$ (Fig.~\ref{fig4}\emph{B}) ( note that, in our model, opinion dimensions do not significantly interact; for related modeling approaches see \cite{Wang:2020, Baumann:2021}). We follow the dynamics until the system is stationary, that is, until the distribution of opinion values in all dimensions is stable. The distribution $f(I)$ captures a spectrum between prototypical influence processes, here denoted \emph{Keep} ($I=0$) and \emph{Adopt} ($I=1$), i.e. not being influenced by a neighbor or fully imitating its behavior [for details on the choice of $f(I)$ see MM].

In Ref. \cite{Chacoma:2015}, the authors quantify the change in the opinion of subjects under the influence of others, Eqs. (\ref{copy_user})--(\ref{copy_mps}), and obtain the probability kernel $f(I)$. This exhibits two pronounced peaks at $I=0$ and $I=1$, with some dispersion for intermediate values. Inspired by these experimental results, we propose a parametrization of the kernel consisting of two Gaussians (Fig.~\ref{fig4}\emph{B}). The free parameters of the kernel are the probability $p$ of having an influence factor around $I=0$ ($1-p$ around $I=1$), and the standard deviations $\sigma_{K}$ and $\sigma_{A}$ of the Gaussians peaked around $I=0$ and $I=1$, respectively. We also assume that the two influence factors $I_{ij}$ and $I_{im}$ are drawn from the same distribution. Crucially, we introduce a parameter $\lambda$ controlling the ratio of rates of interactions with either users [Eq. (\ref{copy_user})] or MPs [Eq. (\ref{copy_mps})] (see SI, Sec. S3.1).

We fit the model by estimating $f(I)$ (i.e., $p$, $\sigma_{K}$, and $\sigma_{A}$) and $\lambda$ such that the difference in the marginal distributions of all opinion components between data and model are minimized (for details see SI, Sec. S3.3.1). The fitted parameters take reasonable values, and the shape of $f(I)$ is comparable with that of experiments \cite{Chacoma:2015}. Notably, we obtain a high value of $p$, which indicates that keeping your opinion after an interaction is more probable than adopting another one. We also obtain a high value of $\lambda$, showing that MPs influence users more often than other users do (further details on fitting, its accuracy, and a table of parameter values in SI, Sec. S3.3.1). Despite its simplicity, numerical simulations of the stationary state of the fitted model recover the attitudinal positions of most users across the entire latent space (Fig.~\ref{fig4}\emph{C}), with some deviations at the extremes of the political spectrum, particularly in the LR, NA, and AE dimensions. Our results imply that the collective decisions of users to either keep their own opinions or incrementally get influenced by others are compatible with the amount of political polarization seen in data.

The levels of political polarization across communities in this latent space are further clarified by a degree-based mean field analysis \cite{Porter:2016} of our model (Fig.~\ref{fig4}\emph{D}). Since the attitudes of MPs are static, the average opinion $\mu$ of users along a given dimension is approximately equal to the degree-weighted average opinion $\widetilde{\mu}$ of MPs they follow. In terms of, say, the stationary opinions $x^{(\text{st})}_{i}$ in the LR dimension, we have
\begin{equation}
\label{mean_opinions}
\frac{1}{N}\sum_{i=1}^{N} x^{(\text{st})}_{i} =\mu \approx \widetilde{\mu}=\dfrac{\sum_{m=1}^{M} k^{\text{in/um}}_m X_{m}}{\sum_{m=1}^{M} k^{\text{in/um}}_m},
\end{equation}
where $k^{\text{in/um}}_m$ is the number of users following MP $m$ (see SI, Sec. S3.2). Attitudinal positions in French Twitter roughly follow the mean-field trend $\mu \approx \widetilde{\mu}$ in all dimensions of the latent space, both at the global level and when separating users by their political communities (see Fig.~\ref{fig3}). In data, however, this approximately linear relation has a slope higher than 1, implying that users have even more radical attitudes than the MPs they follow, especially at the extremes of the multidimensional political spectrum (see Fig. \ref{fig3}\emph{A}).

\section*{Discussion}
\label{sec_discussion}

Our results show that political polarization in online social networks cannot be reduced to a single dimension. This contrasts with a stream of recent research leveraging ideological scaling in social media data, which focuses in uni-dimensional left-right models. Using embedding methods based on large-scale Twitter and political survey data, we uncover at least four dimensions that capture relevant attitudinal differences across political groups in France. We observe that both professional politicians and regular users of Twitter create cohesive communities of similarly-minded people, but users are more extreme in their attitudes and may distance themselves from groups with dissimilar political leanings, further polarizing the online platform. Indeed, there is a clear but nuanced relationship between the group segregation in this multidimensional latent space and the political party structure in France, highlighting how real-world political cleavages are reflected in online activity.

Political polarization is intrinsically multidimensional and thus depends on particular topics of public debate. In France, the political left-right and the nationalism issue segregate online communities the most,  while attitudes towards the European Union and against socio-economic elites are less polarizing.  We observe a strong relationship between the intra- and inter-connectivity of communities and the political opinions of their members.  The centrist communities $\beta$ and $\gamma$ interact quite uniformly with other groups, while the more extreme communities $\alpha$ and $\delta$ (in the left and right of the political spectrum) connect less with other groups as the political disagreement between them increases.

Identifying and understanding the characteristics of individuals in distinct regions of multidimensional political spaces is of importance to several lines of research, with broad implications for policy making, political campaigning, grassroots movements, and collective social phenomena in democratic spaces.  By virtue of their engagement with professional politicians, the inferred attitudinal positions of a sample set of citizens could be harnessed in, e.g., the run up to elections. From a political space competition perspective \cite{downs1957economic}, eligible candidates might take positions appealing to voters in certain regions of a previously identified latent space.  Identifying the users and political spatial regions under-served by existing candidates is a potential benefit of our methodology,  which, together with text analysis of opinions in social media, may have relevant implications for political strategies. Other applications include the study of online social movements \cite{cointet2021colours}, discovery of political biases in algorithms \cite{bozdag2013bias,Peralta:2021a,huszar2022algorithmic}, and polarization in online news media consumption \cite{watts2021measuring,Falkenberg2022,flamino2023political}.

To complement our statistical analysis, we have explored a model that replicates the positions of professional politicians and Twitter users in opinion space and pinpoints the basic social mechanisms, such as imitation, that might drive the levels of political polarization seen online. The fitted parameter $\lambda$ (a ratio of the frequency of interactions with politicians) indicates, in accordance with empirical observations, that MPs lead the dynamics. The relation between the opinions of users and MPs predicted by the model shows a good fit with data. Notably, the global opinion average of users is independent of model parameters and might be thought of as a fundamental property of the proposed imitation mechanism. The model recovers this fundamental property at the global level, but there are some deviations for individual communities. The discrepancies are mostly at the extremes: the average opinion of communities with extremist individuals is more extreme than predicted by the model. This indicates that, in addition to imitation, further mechanisms are potentially at play in the dynamics of polarization in French Twitter.

Taking into account influence mechanisms based on similarity or other socio-cognitive biases is a further step to investigate in the future. Indications that the introduction of biases would enhance the accuracy of the model are suggested by our results (see Fig. 4D). The linear trend that results from the mean field approach matches the data, but the slope seems to be higher than predicted by the model. The introduction of biases in the interaction mechanisms, like bias assimilation as proposed in Ref. \cite{Dandekar:2013}, may increase this slope. Additionally, the deviations observed between model and data for extremist users might be corrected by introducing biases.

The emergence of political cleavages as indicated by interactions in online social media is an inherently temporal and cultural phenomenon. As the political agenda evolves and the topics of national debate transition from one administration to the next, the dimensions of our ideological space relevant to political polarization will also change. The results of the embedding process might also depend on the selected online platform and the country for which data is gathered. How does this opinion space vary across countries and time? And, perhaps more crucially, what characteristic dimensions of political polarization are common around the world, despite cultural differences? Our results offer a flexible framework to further explore these tantalizing questions.

\section*{Materials and Methods}

\subsection*{Twitter network data}

The network is obtained via the Twitter accounts of Members of Parliament (MPs) in France \footnote{Obtained from: \url{http://www2.assemblee-nationale.fr/deputes/liste/reseaux-sociaux} for deputies, and \url{http://www.senat.fr/espace\_presse/actualites/201402/les senateurs\_sur\_twitter.html} for senators.}. We have data on 813 MPs (out of 925), including 348 senators and 577 deputies, each one belonging to one of ten political parties: LREM (\textit{La République en Marche}), LR (\textit{Les Républicains}), PS (\textit{Le Parti Socialiste}), LFI (\textit{La France Insoumise}), LC (\textit{Les Centristes}), RN (\textit{Rassemblement National}), PCF (\textit{Parti Communiste Français}), MoDem (\textit{Mouvement démocrate}), PRG (\textit{Parti Radical de Gauche}), and EELV (\textit{Europe Écologie -- Les Verts}). Followers of the MPs were collected in May 2019, from which we keep only the $230~254$ users with sufficiently high number of political interactions on Twitter (see SI, Sec. S1.1 for details on how we filter data).

Considering these two types of nodes, MPs and users, we categorize their links (follower $\rightarrow$ followed interactions collected for the period August--December 2020) as: User $\rightarrow$ User (63~625~921), User $\rightarrow$ MP (3~351~359), MP $\rightarrow$ MP (113~596) and MP $\rightarrow$ User (515~882). The average number of followers of the MPs (4122) is higher than that of users (276) (see SI, Sec. S1.1 for additional statistics on data collection).

\subsection*{Latent space embedding}

For the political positions of MPs and users we rely on the computation of Ref. \cite{Ramaciotti:2022} (see acknowledgements for details). From the described data, political positions of individuals in a four-dimensional space are computed in two steps as follows (see Fig.~\ref{fig1}\emph{B}). In the first step, we consider the bipartite network of MPs and users (User $\rightarrow$ MP links) and create an embedding in a multidimensional latent space preserving homophily: Users closer in space have higher chances of following the same MPs, and MPs closer in space have higher chances of being followed by the same users. To produce this embedding, a generative homophilic process is considered for the bipartite network of MPs and their follower users \cite{barbera2015birds,ramaciotti2021unfolding}:
\begin{equation}
P\left(\text{User}_i \rightarrow \text{MP}_j \right) = \text{logit}^{-1}\left(\alpha_i + \beta_j - \gamma \|\phi_i - \phi_j  \|^2 \right),
\label{eq:homophilic_model}
\end{equation}
where $P\left(\text{User}_i \rightarrow \text{MP}_j \right)$ is the probability of observing $\text{User}_i$ following $\text{MP}_j$, $\alpha_i$ is the level of activity of $\text{User}_i$ (number of followed friends), $\beta_j$ is the popularity of $\text{MP}_j$ (number of followers), $\gamma$ is a sensitivity parameter, and $\phi_i$ and $\phi_j$ are unobservable positions of $\text{User}_i$ and $\text{MP}_j$ in latent space.

The first step takes the bipartite graph of MPs and users as observations to compute Bayesian inference of $\phi$ values for them (see Fig.~\ref{fig1}\emph{B}). This is done by performing a correspondence analysis \cite{greenacre2017correspondence} of the adjacency matrix of the bipartite graph as an approximation of the unobservable positions of MPs and users in Eq.~(\ref{eq:homophilic_model}) \cite{lowe2008understanding}. Correspondence analysis, being a factor analysis method, preserves global properties such as distance; i.e. up to affine transformations. This latent space embedding for the bipartite graph assures that the relation of relative distances are preserved, in contrast to other network embedding methods \cite{chari2023specious} (see Ref. \cite{barbera2015tweeting} for an evaluation of this approximation, and SI, Sec. S1.2.1 for the first step leading to the latent space embedding).

\subsection*{Political survey data}

The second step of the embedding process uses political survey data to map latent space positions onto a second space where dimensions do have explicit meaning, as they stand for attitudes towards identifiable issues of political debate (see Fig.~\ref{fig1}\emph{B}). The 2019 Chapel Hill Expert Survey (CHES) data \cite{chesdata2019} contains positions of political parties in France (and across Europe) in 51 policy and ideological dimensions. We call this space the Attitudinal Reference Frame (ARF; SI, Sec. S1.2.2). To map positions from the latent space onto this ARF, we use positions of political parties to compute an affine transformation. For each party, we compute the position in latent space as the centroid or mean of the positions of MPs that belong to that party. Knowing party positions on both the latent space and the ARF, we compute an affine transformation mapping positions of the former onto the latter by choosing the number of latent dimensions that fully determine the parameters of the affine transformation (see SI, Sec. S1.2.2 for more details on this transformation).

The positions of French political parties, as captured by the 51 CHES dimensions, can be described almost completely with only 4 dimensions, as shown by principal component analysis of the CHES dimensions (see Sec. IV in Ref. \cite{ramaciotti2022embedding}). The four dimensions deemed relevant for our analysis are: left-right (LR, variable \textit{lrgen} in CHES), antielite salience (AE, variable \textit{antielite\_salience} in CHES), attitudes towards European integration (EU, variable \textit{eu\_position} in CHES), and nationalism (NA, variable \textit{nationalism} in CHES). The ARF is built with explicit spatial reference points: e.g., the question that experts answer to position parties on the left-right scale is ``Where do you position the party in terms of its overall ideological stance, 0 being extreme left, 5 being centrist, and 10 being extreme right?'' (for the questions defining all four dimensions in CHES data see SI, Sec. S1.2.2). We further normalize the scales so that bounds of each dimension of the survey match the $[0, 1]$ interval, making them comparable.

\subsection*{Validation of embedding and robustness}

To test the validity of positions in ARF and their robustness, we use text written by users on their Twitter profiles. We select subsets of users by keywords that reveal their political leaning in their bio profiles, and check that they are correctly positioned in, e.g., the left-right scale (see Ref. \cite{ramaciotti2022measuring} for a detailed presentation of this text-based validation approach). We focus on a limited set of keywords that must be correctly positioned: ``left'' (``gauche'') and ``right'' (``droite'') on the LR dimension; ``Europe'' and ``European'' (``européen'') on the EU dimension; ``people'' (``peuple'') and ``elites'' on the AE dimension; and ``patriot'' (``patriote'') on the NA dimension. When computing metrics for a correct positioning of users that use these keywords along our four political dimensions, we further filter out users that have written a bio profile with negative sentiment (as computed via a sentiment analysis model), to minimize the probability of a user uttering criticism rather than support for an issue.

We evaluate two qualities for the density of users that express support for these issues. First, we evaluate positioning; users that express support should be concentrated in the corresponding region of ideological space. For example, most users with the keyword ``right'' with positive sentiment are positioned to the right of value 0.5 on the LR dimension. Second, we evaluate the monotonicity of the density of users that use a keyword over the various dimensions. For example, the proportion of users with the keyword ``right'' on their profile with positive sentiment grows with higher values along the LR dimension (see Ref. \cite{ramaciotti2023geometry} for a bootstrap robustness analysis of the positions of users, and SI, Sec. S1.2.2 for more details on this dataset and the quality metrics for positioning individuals along the four dimensions).

\subsection*{Network community detection}

The community analysis of the network data is performed with the Python library ``graph-tool''~\cite{peixoto_graph-tool_2014}. We use the minimum description length of communities as a measure of goodness of fit, which is optimal in the sense that it avoids under- and over-fitting and minimizes the occurrence of spurious communities. For the community detection model, we choose a version of the stochastic block model known as planted partition model \cite{Zhang:2020}, which constrains the community search to find structural patterns based on assortativity properties. Assortative groups are characterized by nodes that are connected mostly to other nodes of the same group (see SI, Sec. S2). First we apply the method to find communities in the MP layer and find 5 groups, 4 of which are assortative and one non-assortative. Then we identify the modular structure in the user layer by constraining the search to 4 communities (for more details see SI, Sec. S2).

\subsection*{Opinion dynamics model}

The agent-based model defined by Eqs.~(\ref{copy_user})--(\ref{copy_mps}) establishes the dynamics of opinions and evolves in discrete time steps. It considers two possible outcomes for each user at every  time step: one represents the adoption or imitation of the opinion of a neighbor (``Adopt''), and the other the preservation of its current opinion (``Keep''). We consider a bimodal distribution for the influence factor $I$, determining whether a user keeps its opinion or adopts a new one (see Fig.~\ref{fig4}\emph{B}). We parametrize this bimodal distribution as a mixture model: $f(I)=p \mathcal{N}(I;0,\sigma_{K}^2) + (1-p) \mathcal{N}(I;1,\sigma_{A}^2)$, where $\mathcal{N}(x; \mu,\sigma^2)=\frac{1}{\sigma \sqrt{2 \pi}} \exp \left[-{(x-\mu)^2}/{2\sigma^2}\right]$ is a normal distribution with mean $\mu$ and variance $\sigma^2$. We use the same distribution for both $I_{ij}$ and $I_{im}$ in Eqs.~(\ref{copy_user})--(\ref{copy_mps}).

The parameter input of the model is as follows: (i) $(p,\sigma_{K},\sigma_{A})$ for the influence distribution $f(I)$; (ii) $\lambda$ as a measure of the frequency at which users interact with MPs compared to other users (both (i) and (ii) are fitting parameters); and (iii) the network of interactions and the opinions of MPs, $\lbrace X_{m} \rbrace_{m=1, \dots, M}$, which we assume to be constant and extract from the data. The dynamics and stationarity of the model can be obtained by means of numerical (Monte Carlo) simulations, for which we can optionally apply boundary conditions in opinion space (see SI, Sec. S3.3.2). At the degree-based, mean-field level (SI, Sec. S3.2), the average opinion of users in Eq. (\ref{mean_opinions}) depends only on input (iii) through the degree-weighted average of MP opinions.  The variance of user opinions depends also on (i) and (ii).

\section*{Acknowledgments}

This work has been funded by the ``European Polarisation Observatory'' (EPO) of the CIVICA Consortium.  P.R.  acknowledges support by the Data Intelligence Institute of Paris (diiP) through the French National Agency for Research (ANR) grant ANR-18-IDEX-0001 ``IdEx Université de Paris'' and SoMe4Dem (Grant No. 101094752) Horizon Europe project.  G.I. and J.K. acknowledge support from AFOSR (Grant No. FA8655-20-1-7020), project EU H2020 Humane AI-net (Grant No. 952026), and CHIST-ERA project SAI (Grant No. FWF I 5205-N). J.K. acknowledges support from European Union’s Horizon 2020 research and innovation programme under grant agreement ERC No 810115 - DYNASNET.  Data declared in 19 March 2020 and 15 July 2021 at the registry of data processing of \textit{Fondation Nationale de Sciences Politiques} (Sciences Po) in accordance with General Data Protection Regulation 2016/679 (GDPR) and Twitter policy. For further details and the respective legal notice, please visit \url{https://medialab.sciencespo.fr/en/activities/epo/}.

\section*{Author Contribution}
A.F.P., P.R., J.K., and G.I. conceived and designed the study. P.R. collected Twitter data and performed the embedding process. A.F.P. analyzed the data and explored the opinion dynamics model. J.K. and G.I. contributed to the network analysis. A.F.P., P.R., J.K., and G.I. interpreted the results and wrote the paper.

\bibliography{references}

\begin{thebibliography}{54}%
\makeatletter
\providecommand \@ifxundefined [1]{%
 \@ifx{#1\undefined}
}%
\providecommand \@ifnum [1]{%
 \ifnum #1\expandafter \@firstoftwo
 \else \expandafter \@secondoftwo
 \fi
}%
\providecommand \@ifx [1]{%
 \ifx #1\expandafter \@firstoftwo
 \else \expandafter \@secondoftwo
 \fi
}%
\providecommand \natexlab [1]{#1}%
\providecommand \enquote  [1]{``#1''}%
\providecommand \bibnamefont  [1]{#1}%
\providecommand \bibfnamefont [1]{#1}%
\providecommand \citenamefont [1]{#1}%
\providecommand \href@noop [0]{\@secondoftwo}%
\providecommand \href [0]{\begingroup \@sanitize@url \@href}%
\providecommand \@href[1]{\@@startlink{#1}\@@href}%
\providecommand \@@href[1]{\endgroup#1\@@endlink}%
\providecommand \@sanitize@url [0]{\catcode `\\12\catcode `\$12\catcode `\&12\catcode `\#12\catcode `\^12\catcode `\_12\catcode `\%12\relax}%
\providecommand \@@startlink[1]{}%
\providecommand \@@endlink[0]{}%
\providecommand \url  [0]{\begingroup\@sanitize@url \@url }%
\providecommand \@url [1]{\endgroup\@href {#1}{\urlprefix }}%
\providecommand \urlprefix  [0]{URL }%
\providecommand \Eprint [0]{\href }%
\providecommand \doibase [0]{https://doi.org/}%
\providecommand \selectlanguage [0]{\@gobble}%
\providecommand \bibinfo  [0]{\@secondoftwo}%
\providecommand \bibfield  [0]{\@secondoftwo}%
\providecommand \translation [1]{[#1]}%
\providecommand \BibitemOpen [0]{}%
\providecommand \bibitemStop [0]{}%
\providecommand \bibitemNoStop [0]{.\EOS\space}%
\providecommand \EOS [0]{\spacefactor3000\relax}%
\providecommand \BibitemShut  [1]{\csname bibitem#1\endcsname}%
\let\auto@bib@innerbib\@empty
\bibitem [{\citenamefont {McCarty}(2019)}]{mccarty2019polarization}%
  \BibitemOpen
  \bibfield  {author} {\bibinfo {author} {\bibfnamefont {N.}~\bibnamefont {McCarty}},\ }\href@noop {} {\emph {\bibinfo {title} {Polarization: What everyone needs to know}}}\ (\bibinfo  {publisher} {Oxford University Press},\ \bibinfo {year} {2019})\BibitemShut {NoStop}%
\bibitem [{\citenamefont {Baldassarri}\ and\ \citenamefont {Bearman}(2007)}]{baldassarri2007dynamics}%
  \BibitemOpen
  \bibfield  {author} {\bibinfo {author} {\bibfnamefont {D.}~\bibnamefont {Baldassarri}}\ and\ \bibinfo {author} {\bibfnamefont {P.}~\bibnamefont {Bearman}},\ }\bibfield  {title} {\bibinfo {title} {Dynamics of political polarization},\ }\href@noop {} {\bibfield  {journal} {\bibinfo  {journal} {Am. Sociol. Rev.}\ }\textbf {\bibinfo {volume} {72}},\ \bibinfo {pages} {784} (\bibinfo {year} {2007})}\BibitemShut {NoStop}%
\bibitem [{\citenamefont {Fiorina}\ and\ \citenamefont {Abrams}(2008)}]{fiorina2008political}%
  \BibitemOpen
  \bibfield  {author} {\bibinfo {author} {\bibfnamefont {M.~P.}\ \bibnamefont {Fiorina}}\ and\ \bibinfo {author} {\bibfnamefont {S.~J.}\ \bibnamefont {Abrams}},\ }\bibfield  {title} {\bibinfo {title} {Political polarization in the american public},\ }\href@noop {} {\bibfield  {journal} {\bibinfo  {journal} {Annu. Rev. Polit. Sci.}\ }\textbf {\bibinfo {volume} {11}},\ \bibinfo {pages} {563} (\bibinfo {year} {2008})}\BibitemShut {NoStop}%
\bibitem [{\citenamefont {Prior}(2013)}]{prior2013media}%
  \BibitemOpen
  \bibfield  {author} {\bibinfo {author} {\bibfnamefont {M.}~\bibnamefont {Prior}},\ }\bibfield  {title} {\bibinfo {title} {Media and political polarization},\ }\href@noop {} {\bibfield  {journal} {\bibinfo  {journal} {Annu. Rev. Polit. Sci.}\ }\textbf {\bibinfo {volume} {16}},\ \bibinfo {pages} {101} (\bibinfo {year} {2013})}\BibitemShut {NoStop}%
\bibitem [{\citenamefont {Dixit}\ and\ \citenamefont {Weibull}(2007)}]{dixit2007political}%
  \BibitemOpen
  \bibfield  {author} {\bibinfo {author} {\bibfnamefont {A.~K.}\ \bibnamefont {Dixit}}\ and\ \bibinfo {author} {\bibfnamefont {J.~W.}\ \bibnamefont {Weibull}},\ }\bibfield  {title} {\bibinfo {title} {Political polarization},\ }\href@noop {} {\bibfield  {journal} {\bibinfo  {journal} {Proc. Nat. Acad. Sci. USA}\ }\textbf {\bibinfo {volume} {104}},\ \bibinfo {pages} {7351} (\bibinfo {year} {2007})}\BibitemShut {NoStop}%
\bibitem [{\citenamefont {Conover}\ \emph {et~al.}(2011)\citenamefont {Conover}, \citenamefont {Ratkiewicz}, \citenamefont {Francisco}, \citenamefont {Gon{\c{c}}alves}, \citenamefont {Menczer},\ and\ \citenamefont {Flammini}}]{conover2011political}%
  \BibitemOpen
  \bibfield  {author} {\bibinfo {author} {\bibfnamefont {M.}~\bibnamefont {Conover}}, \bibinfo {author} {\bibfnamefont {J.}~\bibnamefont {Ratkiewicz}}, \bibinfo {author} {\bibfnamefont {M.}~\bibnamefont {Francisco}}, \bibinfo {author} {\bibfnamefont {B.}~\bibnamefont {Gon{\c{c}}alves}}, \bibinfo {author} {\bibfnamefont {F.}~\bibnamefont {Menczer}},\ and\ \bibinfo {author} {\bibfnamefont {A.}~\bibnamefont {Flammini}},\ }\bibfield  {title} {\bibinfo {title} {{Political polarization on Twitter}},\ }in\ \href@noop {} {\emph {\bibinfo {booktitle} {Proceedings of the International AAAI Conference on Web and Social Media}}},\ Vol.~\bibinfo {volume} {5}\ (\bibinfo {year} {2011})\ pp.\ \bibinfo {pages} {89--96}\BibitemShut {NoStop}%
\bibitem [{\citenamefont {Bail}\ \emph {et~al.}(2018)\citenamefont {Bail}, \citenamefont {Argyle}, \citenamefont {Brown}, \citenamefont {Bumpus}, \citenamefont {Chen}, \citenamefont {Hunzaker}, \citenamefont {Lee}, \citenamefont {Mann}, \citenamefont {Merhout},\ and\ \citenamefont {Volfovsky}}]{bail2018exposure}%
  \BibitemOpen
  \bibfield  {author} {\bibinfo {author} {\bibfnamefont {C.~A.}\ \bibnamefont {Bail}}, \bibinfo {author} {\bibfnamefont {L.~P.}\ \bibnamefont {Argyle}}, \bibinfo {author} {\bibfnamefont {T.~W.}\ \bibnamefont {Brown}}, \bibinfo {author} {\bibfnamefont {J.~P.}\ \bibnamefont {Bumpus}}, \bibinfo {author} {\bibfnamefont {H.}~\bibnamefont {Chen}}, \bibinfo {author} {\bibfnamefont {M.~F.}\ \bibnamefont {Hunzaker}}, \bibinfo {author} {\bibfnamefont {J.}~\bibnamefont {Lee}}, \bibinfo {author} {\bibfnamefont {M.}~\bibnamefont {Mann}}, \bibinfo {author} {\bibfnamefont {F.}~\bibnamefont {Merhout}},\ and\ \bibinfo {author} {\bibfnamefont {A.}~\bibnamefont {Volfovsky}},\ }\bibfield  {title} {\bibinfo {title} {Exposure to opposing views on social media can increase political polarization},\ }\href@noop {} {\bibfield  {journal} {\bibinfo  {journal} {Proc. Nat. Acad. Sci. USA}\ }\textbf {\bibinfo {volume} {115}},\ \bibinfo {pages} {9216} (\bibinfo {year} {2018})}\BibitemShut {NoStop}%
\bibitem [{\citenamefont {Flamino}\ \emph {et~al.}(2023)\citenamefont {Flamino}, \citenamefont {Galeazzi}, \citenamefont {Feldman}, \citenamefont {Macy}, \citenamefont {Cross}, \citenamefont {Zhou}, \citenamefont {Serafino}, \citenamefont {Bovet}, \citenamefont {Makse},\ and\ \citenamefont {Szymanski}}]{flamino2023political}%
  \BibitemOpen
  \bibfield  {author} {\bibinfo {author} {\bibfnamefont {J.}~\bibnamefont {Flamino}}, \bibinfo {author} {\bibfnamefont {A.}~\bibnamefont {Galeazzi}}, \bibinfo {author} {\bibfnamefont {S.}~\bibnamefont {Feldman}}, \bibinfo {author} {\bibfnamefont {M.~W.}\ \bibnamefont {Macy}}, \bibinfo {author} {\bibfnamefont {B.}~\bibnamefont {Cross}}, \bibinfo {author} {\bibfnamefont {Z.}~\bibnamefont {Zhou}}, \bibinfo {author} {\bibfnamefont {M.}~\bibnamefont {Serafino}}, \bibinfo {author} {\bibfnamefont {A.}~\bibnamefont {Bovet}}, \bibinfo {author} {\bibfnamefont {H.~A.}\ \bibnamefont {Makse}},\ and\ \bibinfo {author} {\bibfnamefont {B.~K.}\ \bibnamefont {Szymanski}},\ }\bibfield  {title} {\bibinfo {title} {{Political polarization of news media and influencers on Twitter in the 2016 and 2020 US presidential elections}},\ }\href@noop {} {\bibfield  {journal} {\bibinfo  {journal} {Nat. Human Behav.}\ }\textbf {\bibinfo {volume} {7}},\ \bibinfo {pages} {904} (\bibinfo {year} {2023})}\BibitemShut {NoStop}%
\bibitem [{\citenamefont {Hohmann}\ \emph {et~al.}(2023)\citenamefont {Hohmann}, \citenamefont {Devriendt},\ and\ \citenamefont {Coscia}}]{hohmann2023quantifying}%
  \BibitemOpen
  \bibfield  {author} {\bibinfo {author} {\bibfnamefont {M.}~\bibnamefont {Hohmann}}, \bibinfo {author} {\bibfnamefont {K.}~\bibnamefont {Devriendt}},\ and\ \bibinfo {author} {\bibfnamefont {M.}~\bibnamefont {Coscia}},\ }\bibfield  {title} {\bibinfo {title} {{Quantifying ideological polarization on a network using generalized Euclidean distance}},\ }\href@noop {} {\bibfield  {journal} {\bibinfo  {journal} {Sci. Adv.}\ }\textbf {\bibinfo {volume} {9}},\ \bibinfo {pages} {eabq2044} (\bibinfo {year} {2023})}\BibitemShut {NoStop}%
\bibitem [{\citenamefont {Schelling}(1971)}]{schelling1971dynamic}%
  \BibitemOpen
  \bibfield  {author} {\bibinfo {author} {\bibfnamefont {T.~C.}\ \bibnamefont {Schelling}},\ }\bibfield  {title} {\bibinfo {title} {Dynamic models of segregation},\ }\href@noop {} {\bibfield  {journal} {\bibinfo  {journal} {J. Math. Sociol.}\ }\textbf {\bibinfo {volume} {1}},\ \bibinfo {pages} {143} (\bibinfo {year} {1971})}\BibitemShut {NoStop}%
\bibitem [{\citenamefont {Granovetter}(1978)}]{granovetter1978threshold}%
  \BibitemOpen
  \bibfield  {author} {\bibinfo {author} {\bibfnamefont {M.}~\bibnamefont {Granovetter}},\ }\bibfield  {title} {\bibinfo {title} {Threshold models of collective behavior},\ }\href@noop {} {\bibfield  {journal} {\bibinfo  {journal} {Am. J. Sociol.}\ }\textbf {\bibinfo {volume} {83}},\ \bibinfo {pages} {1420} (\bibinfo {year} {1978})}\BibitemShut {NoStop}%
\bibitem [{\citenamefont {Axelrod}(1997)}]{axelrod1997dissemination}%
  \BibitemOpen
  \bibfield  {author} {\bibinfo {author} {\bibfnamefont {R.}~\bibnamefont {Axelrod}},\ }\bibfield  {title} {\bibinfo {title} {The dissemination of culture: A model with local convergence and global polarization},\ }\href@noop {} {\bibfield  {journal} {\bibinfo  {journal} {J. Conf. Resolut.}\ }\textbf {\bibinfo {volume} {41}},\ \bibinfo {pages} {203} (\bibinfo {year} {1997})}\BibitemShut {NoStop}%
\bibitem [{\citenamefont {Castellano}\ \emph {et~al.}(2009)\citenamefont {Castellano}, \citenamefont {Fortunato},\ and\ \citenamefont {Loreto}}]{Castellano:2009}%
  \BibitemOpen
  \bibfield  {author} {\bibinfo {author} {\bibfnamefont {C.}~\bibnamefont {Castellano}}, \bibinfo {author} {\bibfnamefont {S.}~\bibnamefont {Fortunato}},\ and\ \bibinfo {author} {\bibfnamefont {V.}~\bibnamefont {Loreto}},\ }\bibfield  {title} {\bibinfo {title} {Statistical physics of social dynamics},\ }\href@noop {} {\bibfield  {journal} {\bibinfo  {journal} {Rev. Mod. Phys.}\ }\textbf {\bibinfo {volume} {81}},\ \bibinfo {pages} {591} (\bibinfo {year} {2009})}\BibitemShut {NoStop}%
\bibitem [{\citenamefont {Holme}\ and\ \citenamefont {Liljeros}(2015)}]{holme2015mechanistic}%
  \BibitemOpen
  \bibfield  {author} {\bibinfo {author} {\bibfnamefont {P.}~\bibnamefont {Holme}}\ and\ \bibinfo {author} {\bibfnamefont {F.}~\bibnamefont {Liljeros}},\ }\bibfield  {title} {\bibinfo {title} {Mechanistic models in computational social science},\ }\href@noop {} {\bibfield  {journal} {\bibinfo  {journal} {Front. Phys.}\ }\textbf {\bibinfo {volume} {3}},\ \bibinfo {pages} {78} (\bibinfo {year} {2015})}\BibitemShut {NoStop}%
\bibitem [{\citenamefont {Baumann}\ \emph {et~al.}(2021)\citenamefont {Baumann}, \citenamefont {Lorenz-Spreen}, \citenamefont {Sokolov},\ and\ \citenamefont {Starnini}}]{Baumann:2021}%
  \BibitemOpen
  \bibfield  {author} {\bibinfo {author} {\bibfnamefont {F.}~\bibnamefont {Baumann}}, \bibinfo {author} {\bibfnamefont {P.}~\bibnamefont {Lorenz-Spreen}}, \bibinfo {author} {\bibfnamefont {I.~M.}\ \bibnamefont {Sokolov}},\ and\ \bibinfo {author} {\bibfnamefont {M.}~\bibnamefont {Starnini}},\ }\bibfield  {title} {\bibinfo {title} {Emergence of polarized ideological opinions in multidimensional topic spaces},\ }\href@noop {} {\bibfield  {journal} {\bibinfo  {journal} {Phys. Rev. X}\ }\textbf {\bibinfo {volume} {11}},\ \bibinfo {pages} {011012} (\bibinfo {year} {2021})}\BibitemShut {NoStop}%
\bibitem [{\citenamefont {Sobkowicz}(2009)}]{sobkowicz2009modelling}%
  \BibitemOpen
  \bibfield  {author} {\bibinfo {author} {\bibfnamefont {P.}~\bibnamefont {Sobkowicz}},\ }\bibfield  {title} {\bibinfo {title} {Modelling opinion formation with physics tools: Call for closer link with reality},\ }\href@noop {} {\bibfield  {journal} {\bibinfo  {journal} {JASSS}\ }\textbf {\bibinfo {volume} {12}},\ \bibinfo {pages} {11} (\bibinfo {year} {2009})}\BibitemShut {NoStop}%
\bibitem [{\citenamefont {Peralta}\ \emph {et~al.}(2022)\citenamefont {Peralta}, \citenamefont {Kertész},\ and\ \citenamefont {Iñiguez}}]{Peralta:2022}%
  \BibitemOpen
  \bibfield  {author} {\bibinfo {author} {\bibfnamefont {A.~F.}\ \bibnamefont {Peralta}}, \bibinfo {author} {\bibfnamefont {J.}~\bibnamefont {Kertész}},\ and\ \bibinfo {author} {\bibfnamefont {G.}~\bibnamefont {Iñiguez}},\ }\bibfield  {title} {\bibinfo {title} {Opinion dynamics in social networks: From models to data},\ }\href@noop {} {\bibfield  {journal} {\bibinfo  {journal} {arXiv e-print}\ } (\bibinfo {year} {2022})},\ \bibinfo {note} {arXiv:2201.01322}\BibitemShut {NoStop}%
\bibitem [{\citenamefont {Benoit}\ and\ \citenamefont {Laver}(2012)}]{benoit2012dimensionality}%
  \BibitemOpen
  \bibfield  {author} {\bibinfo {author} {\bibfnamefont {K.}~\bibnamefont {Benoit}}\ and\ \bibinfo {author} {\bibfnamefont {M.}~\bibnamefont {Laver}},\ }\bibfield  {title} {\bibinfo {title} {The dimensionality of political space: Epistemological and methodological considerations},\ }\href@noop {} {\bibfield  {journal} {\bibinfo  {journal} {Eur. Union Polit.}\ }\textbf {\bibinfo {volume} {13}},\ \bibinfo {pages} {194} (\bibinfo {year} {2012})}\BibitemShut {NoStop}%
\bibitem [{\citenamefont {Imai}\ \emph {et~al.}(2016)\citenamefont {Imai}, \citenamefont {Lo},\ and\ \citenamefont {Olmsted}}]{Imai:2016}%
  \BibitemOpen
  \bibfield  {author} {\bibinfo {author} {\bibfnamefont {K.}~\bibnamefont {Imai}}, \bibinfo {author} {\bibfnamefont {J.}~\bibnamefont {Lo}},\ and\ \bibinfo {author} {\bibfnamefont {J.}~\bibnamefont {Olmsted}},\ }\bibfield  {title} {\bibinfo {title} {Fast estimation of ideal points with massive data},\ }\href@noop {} {\bibfield  {journal} {\bibinfo  {journal} {Am. Polit. Sci. Rev.}\ }\textbf {\bibinfo {volume} {110}},\ \bibinfo {pages} {631} (\bibinfo {year} {2016})}\BibitemShut {NoStop}%
\bibitem [{\citenamefont {Barber{\'a}}(2015)}]{barbera2015birds}%
  \BibitemOpen
  \bibfield  {author} {\bibinfo {author} {\bibfnamefont {P.}~\bibnamefont {Barber{\'a}}},\ }\bibfield  {title} {\bibinfo {title} {Birds of the same feather tweet together: Bayesian ideal point estimation using twitter data},\ }\href@noop {} {\bibfield  {journal} {\bibinfo  {journal} {Polit. Anal.}\ }\textbf {\bibinfo {volume} {23}},\ \bibinfo {pages} {76} (\bibinfo {year} {2015})}\BibitemShut {NoStop}%
\bibitem [{\citenamefont {Bond}\ and\ \citenamefont {Messing}(2015)}]{bond2015quantifying}%
  \BibitemOpen
  \bibfield  {author} {\bibinfo {author} {\bibfnamefont {R.}~\bibnamefont {Bond}}\ and\ \bibinfo {author} {\bibfnamefont {S.}~\bibnamefont {Messing}},\ }\bibfield  {title} {\bibinfo {title} {{Quantifying social media’s political space: Estimating ideology from publicly revealed preferences on Facebook}},\ }\href@noop {} {\bibfield  {journal} {\bibinfo  {journal} {Am. Polit. Sci. Rev.}\ }\textbf {\bibinfo {volume} {109}},\ \bibinfo {pages} {62} (\bibinfo {year} {2015})}\BibitemShut {NoStop}%
\bibitem [{\citenamefont {Bakker}\ \emph {et~al.}(2012)\citenamefont {Bakker}, \citenamefont {Jolly},\ and\ \citenamefont {Polk}}]{bakker2012complexity}%
  \BibitemOpen
  \bibfield  {author} {\bibinfo {author} {\bibfnamefont {R.}~\bibnamefont {Bakker}}, \bibinfo {author} {\bibfnamefont {S.}~\bibnamefont {Jolly}},\ and\ \bibinfo {author} {\bibfnamefont {J.}~\bibnamefont {Polk}},\ }\bibfield  {title} {\bibinfo {title} {Complexity in the european party space: Exploring dimensionality with experts},\ }\href@noop {} {\bibfield  {journal} {\bibinfo  {journal} {Eur. Union Polit.}\ }\textbf {\bibinfo {volume} {13}},\ \bibinfo {pages} {219} (\bibinfo {year} {2012})}\BibitemShut {NoStop}%
\bibitem [{\citenamefont {Uscinski}\ \emph {et~al.}(2021)\citenamefont {Uscinski}, \citenamefont {Enders}, \citenamefont {Seelig}, \citenamefont {Klofstad}, \citenamefont {Funchion}, \citenamefont {Everett}, \citenamefont {Wuchty}, \citenamefont {Premaratne},\ and\ \citenamefont {Murthi}}]{uscinski2021american}%
  \BibitemOpen
  \bibfield  {author} {\bibinfo {author} {\bibfnamefont {J.~E.}\ \bibnamefont {Uscinski}}, \bibinfo {author} {\bibfnamefont {A.~M.}\ \bibnamefont {Enders}}, \bibinfo {author} {\bibfnamefont {M.~I.}\ \bibnamefont {Seelig}}, \bibinfo {author} {\bibfnamefont {C.~A.}\ \bibnamefont {Klofstad}}, \bibinfo {author} {\bibfnamefont {J.~R.}\ \bibnamefont {Funchion}}, \bibinfo {author} {\bibfnamefont {C.}~\bibnamefont {Everett}}, \bibinfo {author} {\bibfnamefont {S.}~\bibnamefont {Wuchty}}, \bibinfo {author} {\bibfnamefont {K.}~\bibnamefont {Premaratne}},\ and\ \bibinfo {author} {\bibfnamefont {M.~N.}\ \bibnamefont {Murthi}},\ }\bibfield  {title} {\bibinfo {title} {American politics in two dimensions: Partisan and ideological identities versus anti-establishment orientations},\ }\href@noop {} {\bibfield  {journal} {\bibinfo  {journal} {Am. J. Polit. Sci.}\ }\textbf {\bibinfo {volume} {65}},\ \bibinfo {pages} {877} (\bibinfo {year} {2021})}\BibitemShut {NoStop}%
\bibitem [{\citenamefont {Ramaciotti~Morales}\ \emph {et~al.}(2023{\natexlab{a}})\citenamefont {Ramaciotti~Morales}, \citenamefont {Cointet}, \citenamefont {Mu{\~n}oz~Zolotoochin}, \citenamefont {Fern{\'a}ndez~Peralta}, \citenamefont {I{\~n}iguez},\ and\ \citenamefont {Pournaki}}]{Ramaciotti:2022}%
  \BibitemOpen
  \bibfield  {author} {\bibinfo {author} {\bibfnamefont {P.}~\bibnamefont {Ramaciotti~Morales}}, \bibinfo {author} {\bibfnamefont {J.-P.}\ \bibnamefont {Cointet}}, \bibinfo {author} {\bibfnamefont {G.}~\bibnamefont {Mu{\~n}oz~Zolotoochin}}, \bibinfo {author} {\bibfnamefont {A.}~\bibnamefont {Fern{\'a}ndez~Peralta}}, \bibinfo {author} {\bibfnamefont {G.}~\bibnamefont {I{\~n}iguez}},\ and\ \bibinfo {author} {\bibfnamefont {A.}~\bibnamefont {Pournaki}},\ }\bibfield  {title} {\bibinfo {title} {Inferring attitudinal spaces in social networks},\ }\href@noop {} {\bibfield  {journal} {\bibinfo  {journal} {Social Network Analysis and Mining}\ }\textbf {\bibinfo {volume} {13}},\ \bibinfo {pages} {14} (\bibinfo {year} {2023}{\natexlab{a}})}\BibitemShut {NoStop}%
\bibitem [{\citenamefont {DeGroot}(1974)}]{DeGroot:1974}%
  \BibitemOpen
  \bibfield  {author} {\bibinfo {author} {\bibfnamefont {M.~H.}\ \bibnamefont {DeGroot}},\ }\bibfield  {title} {\bibinfo {title} {Reaching a consensus},\ }\href@noop {} {\bibfield  {journal} {\bibinfo  {journal} {J. Am. Stat. Assoc.}\ }\textbf {\bibinfo {volume} {69}},\ \bibinfo {pages} {118} (\bibinfo {year} {1974})}\BibitemShut {NoStop}%
\bibitem [{\citenamefont {Martin-Gutierrez}\ \emph {et~al.}(2023)\citenamefont {Martin-Gutierrez}, \citenamefont {Losada},\ and\ \citenamefont {Benito}}]{Martin-Gutierrez:2023}%
  \BibitemOpen
  \bibfield  {author} {\bibinfo {author} {\bibfnamefont {S.}~\bibnamefont {Martin-Gutierrez}}, \bibinfo {author} {\bibfnamefont {J.~C.}\ \bibnamefont {Losada}},\ and\ \bibinfo {author} {\bibfnamefont {R.~M.}\ \bibnamefont {Benito}},\ }\bibfield  {title} {\bibinfo {title} {Multipolar social systems: Measuring polarization beyond dichotomous contexts},\ }\href@noop {} {\bibfield  {journal} {\bibinfo  {journal} {Chaos, Solitons \& Fractals}\ }\textbf {\bibinfo {volume} {169}},\ \bibinfo {pages} {113244} (\bibinfo {year} {2023})}\BibitemShut {NoStop}%
\bibitem [{\citenamefont {Zhang}\ and\ \citenamefont {Peixoto}(2020)}]{Zhang:2020}%
  \BibitemOpen
  \bibfield  {author} {\bibinfo {author} {\bibfnamefont {L.}~\bibnamefont {Zhang}}\ and\ \bibinfo {author} {\bibfnamefont {T.~P.}\ \bibnamefont {Peixoto}},\ }\bibfield  {title} {\bibinfo {title} {Statistical inference of assortative community structures},\ }\href@noop {} {\bibfield  {journal} {\bibinfo  {journal} {Phys. Rev. Research}\ }\textbf {\bibinfo {volume} {2}},\ \bibinfo {pages} {043271} (\bibinfo {year} {2020})}\BibitemShut {NoStop}%
\bibitem [{\citenamefont {Jolly}\ \emph {et~al.}(2022)\citenamefont {Jolly}, \citenamefont {Bakker}, \citenamefont {Hooghe}, \citenamefont {Marks}, \citenamefont {Polk}, \citenamefont {Rovny}, \citenamefont {Steenbergen},\ and\ \citenamefont {Vachudova}}]{chesdata2019}%
  \BibitemOpen
  \bibfield  {author} {\bibinfo {author} {\bibfnamefont {S.}~\bibnamefont {Jolly}}, \bibinfo {author} {\bibfnamefont {R.}~\bibnamefont {Bakker}}, \bibinfo {author} {\bibfnamefont {L.}~\bibnamefont {Hooghe}}, \bibinfo {author} {\bibfnamefont {G.}~\bibnamefont {Marks}}, \bibinfo {author} {\bibfnamefont {J.}~\bibnamefont {Polk}}, \bibinfo {author} {\bibfnamefont {J.}~\bibnamefont {Rovny}}, \bibinfo {author} {\bibfnamefont {M.}~\bibnamefont {Steenbergen}},\ and\ \bibinfo {author} {\bibfnamefont {M.~A.}\ \bibnamefont {Vachudova}},\ }\bibfield  {title} {\bibinfo {title} {Chapel hill expert survey trend file, 1999–2019},\ }\href@noop {} {\bibfield  {journal} {\bibinfo  {journal} {Electoral Studies}\ }\textbf {\bibinfo {volume} {75}},\ \bibinfo {pages} {102420} (\bibinfo {year} {2022})}\BibitemShut {NoStop}%
\bibitem [{\citenamefont {Grunwald}(2007)}]{Grunwald:2007}%
  \BibitemOpen
  \bibfield  {author} {\bibinfo {author} {\bibfnamefont {P.~D.}\ \bibnamefont {Grunwald}},\ }\href@noop {} {\emph {\bibinfo {title} {The Minimum Description Length Principle}}}\ (\bibinfo  {publisher} {The MIT Press, Cambridge, MA},\ \bibinfo {year} {2007})\BibitemShut {NoStop}%
\bibitem [{\citenamefont {R{\'e}mond}(2014)}]{remond2014droites}%
  \BibitemOpen
  \bibfield  {author} {\bibinfo {author} {\bibfnamefont {R.}~\bibnamefont {R{\'e}mond}},\ }\href@noop {} {\emph {\bibinfo {title} {Les droites en France}}}\ (\bibinfo  {publisher} {Aubier},\ \bibinfo {year} {2014})\BibitemShut {NoStop}%
\bibitem [{\citenamefont {Ramaciotti~Morales}\ \emph {et~al.}(2021)\citenamefont {Ramaciotti~Morales}, \citenamefont {Cointet},\ and\ \citenamefont {Mu\~{n}oz Zolotoochin}}]{ramaciotti2021unfolding}%
  \BibitemOpen
  \bibfield  {author} {\bibinfo {author} {\bibfnamefont {P.}~\bibnamefont {Ramaciotti~Morales}}, \bibinfo {author} {\bibfnamefont {J.-P.}\ \bibnamefont {Cointet}},\ and\ \bibinfo {author} {\bibfnamefont {G.}~\bibnamefont {Mu\~{n}oz Zolotoochin}},\ }\bibfield  {title} {\bibinfo {title} {Unfolding the dimensionality structure of social networks in ideological embeddings},\ }in\ \href@noop {} {\emph {\bibinfo {booktitle} {Proceedings of the 2021 IEEE/ACM International Conference on Advances in Social Networks Analysis and Mining}}}\ (\bibinfo {year} {2021})\ pp.\ \bibinfo {pages} {333--338}\BibitemShut {NoStop}%
\bibitem [{\citenamefont {Hooghe}\ and\ \citenamefont {Marks}(2018)}]{hooghe2018cleavage}%
  \BibitemOpen
  \bibfield  {author} {\bibinfo {author} {\bibfnamefont {L.}~\bibnamefont {Hooghe}}\ and\ \bibinfo {author} {\bibfnamefont {G.}~\bibnamefont {Marks}},\ }\bibfield  {title} {\bibinfo {title} {Cleavage theory meets europe’s crises: Lipset, rokkan, and the transnational cleavage},\ }\href@noop {} {\bibfield  {journal} {\bibinfo  {journal} {J. Eur. Public Policy}\ }\textbf {\bibinfo {volume} {25}},\ \bibinfo {pages} {109} (\bibinfo {year} {2018})}\BibitemShut {NoStop}%
\bibitem [{\citenamefont {Riker}\ and\ \citenamefont {Ordeshook}(1968)}]{riker1968theory}%
  \BibitemOpen
  \bibfield  {author} {\bibinfo {author} {\bibfnamefont {W.~H.}\ \bibnamefont {Riker}}\ and\ \bibinfo {author} {\bibfnamefont {P.~C.}\ \bibnamefont {Ordeshook}},\ }\bibfield  {title} {\bibinfo {title} {A theory of the calculus of voting},\ }\href@noop {} {\bibfield  {journal} {\bibinfo  {journal} {Am. Pol. Sci. Rev.}\ }\textbf {\bibinfo {volume} {62}},\ \bibinfo {pages} {25} (\bibinfo {year} {1968})}\BibitemShut {NoStop}%
\bibitem [{\citenamefont {Dandekar}\ \emph {et~al.}(2013)\citenamefont {Dandekar}, \citenamefont {Goel},\ and\ \citenamefont {Lee}}]{Dandekar:2013}%
  \BibitemOpen
  \bibfield  {author} {\bibinfo {author} {\bibfnamefont {P.}~\bibnamefont {Dandekar}}, \bibinfo {author} {\bibfnamefont {A.}~\bibnamefont {Goel}},\ and\ \bibinfo {author} {\bibfnamefont {D.~T.}\ \bibnamefont {Lee}},\ }\bibfield  {title} {\bibinfo {title} {Biased assimilation, homophily, and the dynamics of polarization},\ }\href@noop {} {\bibfield  {journal} {\bibinfo  {journal} {Proc. Nat. Acad. Sci. USA}\ }\textbf {\bibinfo {volume} {110}},\ \bibinfo {pages} {5791} (\bibinfo {year} {2013})}\BibitemShut {NoStop}%
\bibitem [{\citenamefont {Moussaïd}\ \emph {et~al.}(2013)\citenamefont {Moussaïd}, \citenamefont {Kämmer}, \citenamefont {Analytis},\ and\ \citenamefont {Neth}}]{Moussaid:2013}%
  \BibitemOpen
  \bibfield  {author} {\bibinfo {author} {\bibfnamefont {M.}~\bibnamefont {Moussaïd}}, \bibinfo {author} {\bibfnamefont {J.~E.}\ \bibnamefont {Kämmer}}, \bibinfo {author} {\bibfnamefont {P.~P.}\ \bibnamefont {Analytis}},\ and\ \bibinfo {author} {\bibfnamefont {H.}~\bibnamefont {Neth}},\ }\bibfield  {title} {\bibinfo {title} {Social influence and the collective dynamics of opinion formation},\ }\href@noop {} {\bibfield  {journal} {\bibinfo  {journal} {PLoS ONE}\ }\textbf {\bibinfo {volume} {8}},\ \bibinfo {pages} {e78433} (\bibinfo {year} {2013})}\BibitemShut {NoStop}%
\bibitem [{\citenamefont {Chacoma}\ and\ \citenamefont {Zanette}(2015)}]{Chacoma:2015}%
  \BibitemOpen
  \bibfield  {author} {\bibinfo {author} {\bibfnamefont {A.}~\bibnamefont {Chacoma}}\ and\ \bibinfo {author} {\bibfnamefont {D.~H.}\ \bibnamefont {Zanette}},\ }\bibfield  {title} {\bibinfo {title} {Opinion formation by social influence: From experiments to modeling},\ }\href@noop {} {\bibfield  {journal} {\bibinfo  {journal} {PLOS ONE}\ }\textbf {\bibinfo {volume} {10}},\ \bibinfo {pages} {e0140406} (\bibinfo {year} {2015})}\BibitemShut {NoStop}%
\bibitem [{\citenamefont {Wang}\ \emph {et~al.}(2020)\citenamefont {Wang}, \citenamefont {Sirianni}, \citenamefont {Tang}, \citenamefont {Zheng},\ and\ \citenamefont {Fu}}]{Wang:2020}%
  \BibitemOpen
  \bibfield  {author} {\bibinfo {author} {\bibfnamefont {X.}~\bibnamefont {Wang}}, \bibinfo {author} {\bibfnamefont {A.~D.}\ \bibnamefont {Sirianni}}, \bibinfo {author} {\bibfnamefont {S.}~\bibnamefont {Tang}}, \bibinfo {author} {\bibfnamefont {Z.}~\bibnamefont {Zheng}},\ and\ \bibinfo {author} {\bibfnamefont {F.}~\bibnamefont {Fu}},\ }\bibfield  {title} {\bibinfo {title} {Public discourse and social network echo chambers driven by socio-cognitive biases},\ }\href@noop {} {\bibfield  {journal} {\bibinfo  {journal} {Phys. Rev. X}\ }\textbf {\bibinfo {volume} {10}},\ \bibinfo {pages} {041042} (\bibinfo {year} {2020})}\BibitemShut {NoStop}%
\bibitem [{\citenamefont {Porter}\ and\ \citenamefont {Gleeson}(2016)}]{Porter:2016}%
  \BibitemOpen
  \bibfield  {author} {\bibinfo {author} {\bibfnamefont {M.~A.}\ \bibnamefont {Porter}}\ and\ \bibinfo {author} {\bibfnamefont {J.~P.}\ \bibnamefont {Gleeson}},\ }\bibfield  {title} {\bibinfo {title} {Dynamical systems on networks},\ }\href@noop {} {\bibfield  {journal} {\bibinfo  {journal} {Front. Appl. Dyn. Syst. Rev. Tutor.}\ }\textbf {\bibinfo {volume} {4}} (\bibinfo {year} {2016})}\BibitemShut {NoStop}%
\bibitem [{\citenamefont {Downs}(1957)}]{downs1957economic}%
  \BibitemOpen
  \bibfield  {author} {\bibinfo {author} {\bibfnamefont {A.}~\bibnamefont {Downs}},\ }\bibfield  {title} {\bibinfo {title} {An economic theory of political action in a democracy},\ }\href@noop {} {\bibfield  {journal} {\bibinfo  {journal} {J. Polit. Econ.}\ }\textbf {\bibinfo {volume} {65}},\ \bibinfo {pages} {135} (\bibinfo {year} {1957})}\BibitemShut {NoStop}%
\bibitem [{\citenamefont {Cointet}\ \emph {et~al.}(2021)\citenamefont {Cointet}, \citenamefont {Morales}, \citenamefont {Cardon}, \citenamefont {Froio}, \citenamefont {Mogoutov}, \citenamefont {Ooghe},\ and\ \citenamefont {Plique}}]{cointet2021colours}%
  \BibitemOpen
  \bibfield  {author} {\bibinfo {author} {\bibfnamefont {J.-P.}\ \bibnamefont {Cointet}}, \bibinfo {author} {\bibfnamefont {P.~R.}\ \bibnamefont {Morales}}, \bibinfo {author} {\bibfnamefont {D.}~\bibnamefont {Cardon}}, \bibinfo {author} {\bibfnamefont {C.}~\bibnamefont {Froio}}, \bibinfo {author} {\bibfnamefont {A.}~\bibnamefont {Mogoutov}}, \bibinfo {author} {\bibfnamefont {B.}~\bibnamefont {Ooghe}},\ and\ \bibinfo {author} {\bibfnamefont {G.}~\bibnamefont {Plique}},\ }\bibfield  {title} {\bibinfo {title} {{What colours are the yellow vests? An ideological scaling of Facebook groups}},\ }\href@noop {} {\bibfield  {journal} {\bibinfo  {journal} {Statistique et Soci{\'e}t{\'e}}\ }\textbf {\bibinfo {volume} {9}},\ \bibinfo {pages} {79} (\bibinfo {year} {2021})}\BibitemShut {NoStop}%
\bibitem [{\citenamefont {Bozdag}(2013)}]{bozdag2013bias}%
  \BibitemOpen
  \bibfield  {author} {\bibinfo {author} {\bibfnamefont {E.}~\bibnamefont {Bozdag}},\ }\bibfield  {title} {\bibinfo {title} {Bias in algorithmic filtering and personalization},\ }\href@noop {} {\bibfield  {journal} {\bibinfo  {journal} {Ethics Inf. Technol}\ }\textbf {\bibinfo {volume} {15}},\ \bibinfo {pages} {209} (\bibinfo {year} {2013})}\BibitemShut {NoStop}%
\bibitem [{\citenamefont {Peralta}\ \emph {et~al.}(2021)\citenamefont {Peralta}, \citenamefont {Neri}, \citenamefont {Kert\'esz},\ and\ \citenamefont {I\~niguez}}]{Peralta:2021a}%
  \BibitemOpen
  \bibfield  {author} {\bibinfo {author} {\bibfnamefont {A.~F.}\ \bibnamefont {Peralta}}, \bibinfo {author} {\bibfnamefont {M.}~\bibnamefont {Neri}}, \bibinfo {author} {\bibfnamefont {J.}~\bibnamefont {Kert\'esz}},\ and\ \bibinfo {author} {\bibfnamefont {G.}~\bibnamefont {I\~niguez}},\ }\bibfield  {title} {\bibinfo {title} {Effect of algorithmic bias and network structure on coexistence, consensus, and polarization of opinions},\ }\href@noop {} {\bibfield  {journal} {\bibinfo  {journal} {Phys. Rev. E}\ }\textbf {\bibinfo {volume} {104}},\ \bibinfo {pages} {044312} (\bibinfo {year} {2021})}\BibitemShut {NoStop}%
\bibitem [{\citenamefont {Husz{\'a}r}\ \emph {et~al.}(2022)\citenamefont {Husz{\'a}r}, \citenamefont {Ktena}, \citenamefont {O’Brien}, \citenamefont {Belli}, \citenamefont {Schlaikjer},\ and\ \citenamefont {Hardt}}]{huszar2022algorithmic}%
  \BibitemOpen
  \bibfield  {author} {\bibinfo {author} {\bibfnamefont {F.}~\bibnamefont {Husz{\'a}r}}, \bibinfo {author} {\bibfnamefont {S.~I.}\ \bibnamefont {Ktena}}, \bibinfo {author} {\bibfnamefont {C.}~\bibnamefont {O’Brien}}, \bibinfo {author} {\bibfnamefont {L.}~\bibnamefont {Belli}}, \bibinfo {author} {\bibfnamefont {A.}~\bibnamefont {Schlaikjer}},\ and\ \bibinfo {author} {\bibfnamefont {M.}~\bibnamefont {Hardt}},\ }\bibfield  {title} {\bibinfo {title} {{Algorithmic amplification of politics on Twitter}},\ }\href@noop {} {\bibfield  {journal} {\bibinfo  {journal} {Proc. Nat. Acad. Sci. USA}\ }\textbf {\bibinfo {volume} {119}},\ \bibinfo {pages} {e2025334119} (\bibinfo {year} {2022})}\BibitemShut {NoStop}%
\bibitem [{\citenamefont {Watts}\ \emph {et~al.}(2021)\citenamefont {Watts}, \citenamefont {Rothschild},\ and\ \citenamefont {Mobius}}]{watts2021measuring}%
  \BibitemOpen
  \bibfield  {author} {\bibinfo {author} {\bibfnamefont {D.~J.}\ \bibnamefont {Watts}}, \bibinfo {author} {\bibfnamefont {D.~M.}\ \bibnamefont {Rothschild}},\ and\ \bibinfo {author} {\bibfnamefont {M.}~\bibnamefont {Mobius}},\ }\bibfield  {title} {\bibinfo {title} {Measuring the news and its impact on democracy},\ }\href@noop {} {\bibfield  {journal} {\bibinfo  {journal} {Proc. Nat. Acad. Sci. USA}\ }\textbf {\bibinfo {volume} {118}},\ \bibinfo {pages} {e1912443118} (\bibinfo {year} {2021})}\BibitemShut {NoStop}%
\bibitem [{\citenamefont {Falkenberg}\ \emph {et~al.}(2022)\citenamefont {Falkenberg}, \citenamefont {Galeazzi}, \citenamefont {Torricelli}, \citenamefont {Marco}, \citenamefont {Larosa}, \citenamefont {Sas}, \citenamefont {Mekacher}, \citenamefont {Pearce}, \citenamefont {Zollo}, \citenamefont {Quattrociocchi},\ and\ \citenamefont {Baronchelli}}]{Falkenberg2022}%
  \BibitemOpen
  \bibfield  {author} {\bibinfo {author} {\bibfnamefont {M.}~\bibnamefont {Falkenberg}}, \bibinfo {author} {\bibfnamefont {A.}~\bibnamefont {Galeazzi}}, \bibinfo {author} {\bibfnamefont {M.}~\bibnamefont {Torricelli}}, \bibinfo {author} {\bibfnamefont {N.~D.}\ \bibnamefont {Marco}}, \bibinfo {author} {\bibfnamefont {F.}~\bibnamefont {Larosa}}, \bibinfo {author} {\bibfnamefont {M.}~\bibnamefont {Sas}}, \bibinfo {author} {\bibfnamefont {A.}~\bibnamefont {Mekacher}}, \bibinfo {author} {\bibfnamefont {W.}~\bibnamefont {Pearce}}, \bibinfo {author} {\bibfnamefont {F.}~\bibnamefont {Zollo}}, \bibinfo {author} {\bibfnamefont {W.}~\bibnamefont {Quattrociocchi}},\ and\ \bibinfo {author} {\bibfnamefont {A.}~\bibnamefont {Baronchelli}},\ }\bibfield  {title} {\bibinfo {title} {Growing polarization around climate change on social media},\ }\href@noop {} {\bibfield  {journal} {\bibinfo  {journal} {Nature Climate Change}\ }\textbf {\bibinfo {volume} {12}},\ \bibinfo {pages} {1114} (\bibinfo {year} {2022})}\BibitemShut
  {NoStop}%
\bibitem [{Note1()}]{Note1}%
  \BibitemOpen
  \bibinfo {note} {Obtained from: \protect \url {http://www2.assemblee-nationale.fr/deputes/liste/reseaux-sociaux} for deputies, and \protect \url {http://www.senat.fr/espace\protect \_presse/actualites/201402/les senateurs\protect \_sur\protect \_twitter.html} for senators.}\BibitemShut {Stop}%
\bibitem [{\citenamefont {Greenacre}(2017)}]{greenacre2017correspondence}%
  \BibitemOpen
  \bibfield  {author} {\bibinfo {author} {\bibfnamefont {M.}~\bibnamefont {Greenacre}},\ }\href@noop {} {\emph {\bibinfo {title} {Correspondence analysis in practice}}}\ (\bibinfo  {publisher} {Chapman and Hall},\ \bibinfo {year} {2017})\BibitemShut {NoStop}%
\bibitem [{\citenamefont {Lowe}(2008)}]{lowe2008understanding}%
  \BibitemOpen
  \bibfield  {author} {\bibinfo {author} {\bibfnamefont {W.}~\bibnamefont {Lowe}},\ }\bibfield  {title} {\bibinfo {title} {Understanding wordscores},\ }\href@noop {} {\bibfield  {journal} {\bibinfo  {journal} {Polit. Anal.}\ }\textbf {\bibinfo {volume} {16}},\ \bibinfo {pages} {356} (\bibinfo {year} {2008})}\BibitemShut {NoStop}%
\bibitem [{\citenamefont {Chari}\ and\ \citenamefont {Pachter}(2023)}]{chari2023specious}%
  \BibitemOpen
  \bibfield  {author} {\bibinfo {author} {\bibfnamefont {T.}~\bibnamefont {Chari}}\ and\ \bibinfo {author} {\bibfnamefont {L.}~\bibnamefont {Pachter}},\ }\bibfield  {title} {\bibinfo {title} {The specious art of single-cell genomics},\ }\href@noop {} {\bibfield  {journal} {\bibinfo  {journal} {PLOS Computational Biology}\ }\textbf {\bibinfo {volume} {19}},\ \bibinfo {pages} {e1011288} (\bibinfo {year} {2023})}\BibitemShut {NoStop}%
\bibitem [{\citenamefont {Barber{\'a}}\ \emph {et~al.}(2015)\citenamefont {Barber{\'a}}, \citenamefont {Jost}, \citenamefont {Nagler}, \citenamefont {Tucker},\ and\ \citenamefont {Bonneau}}]{barbera2015tweeting}%
  \BibitemOpen
  \bibfield  {author} {\bibinfo {author} {\bibfnamefont {P.}~\bibnamefont {Barber{\'a}}}, \bibinfo {author} {\bibfnamefont {J.~T.}\ \bibnamefont {Jost}}, \bibinfo {author} {\bibfnamefont {J.}~\bibnamefont {Nagler}}, \bibinfo {author} {\bibfnamefont {J.~A.}\ \bibnamefont {Tucker}},\ and\ \bibinfo {author} {\bibfnamefont {R.}~\bibnamefont {Bonneau}},\ }\bibfield  {title} {\bibinfo {title} {Tweeting from left to right: Is online political communication more than an echo chamber?},\ }\href@noop {} {\bibfield  {journal} {\bibinfo  {journal} {Psychol. Sci.}\ }\textbf {\bibinfo {volume} {26}},\ \bibinfo {pages} {1531} (\bibinfo {year} {2015})}\BibitemShut {NoStop}%
\bibitem [{\citenamefont {Ramaciotti~Morales}\ and\ \citenamefont {Vagena}(2022)}]{ramaciotti2022embedding}%
  \BibitemOpen
  \bibfield  {author} {\bibinfo {author} {\bibfnamefont {P.}~\bibnamefont {Ramaciotti~Morales}}\ and\ \bibinfo {author} {\bibfnamefont {Z.}~\bibnamefont {Vagena}},\ }\bibfield  {title} {\bibinfo {title} {Embedding social graphs from multiple national settings in common empirical opinion spaces},\ }in\ \href@noop {} {\emph {\bibinfo {booktitle} {Proceedings of the 2022 IEEE/ACM International Conference on Advances in Social Networks Analysis and Mining}}}\ (\bibinfo {year} {2022})\BibitemShut {NoStop}%
\bibitem [{\citenamefont {Ramaciotti~Morales}\ and\ \citenamefont {Mu{\~n}oz~Zolotoochin}(2022)}]{ramaciotti2022measuring}%
  \BibitemOpen
  \bibfield  {author} {\bibinfo {author} {\bibfnamefont {P.}~\bibnamefont {Ramaciotti~Morales}}\ and\ \bibinfo {author} {\bibfnamefont {G.}~\bibnamefont {Mu{\~n}oz~Zolotoochin}},\ }\bibfield  {title} {\bibinfo {title} {Measuring the accuracy of social network ideological embeddings using language models},\ }in\ \href@noop {} {\emph {\bibinfo {booktitle} {Information Technology and Systems: Proceedings of ICITS 2022}}}\ (\bibinfo  {publisher} {Springer},\ \bibinfo {year} {2022})\ pp.\ \bibinfo {pages} {267--276}\BibitemShut {NoStop}%
\bibitem [{\citenamefont {Ramaciotti~Morales}\ \emph {et~al.}(2023{\natexlab{b}})\citenamefont {Ramaciotti~Morales}, \citenamefont {Berriche},\ and\ \citenamefont {Cointet}}]{ramaciotti2023geometry}%
  \BibitemOpen
  \bibfield  {author} {\bibinfo {author} {\bibfnamefont {P.}~\bibnamefont {Ramaciotti~Morales}}, \bibinfo {author} {\bibfnamefont {M.}~\bibnamefont {Berriche}},\ and\ \bibinfo {author} {\bibfnamefont {J.-P.}\ \bibnamefont {Cointet}},\ }\bibfield  {title} {\bibinfo {title} {The geometry of misinformation: embedding twitter networks of users who spread fake news in geometrical opinion spaces},\ }in\ \href@noop {} {\emph {\bibinfo {booktitle} {Proceedings of the International AAAI Conference on Web and Social Media}}},\ Vol.~\bibinfo {volume} {17}\ (\bibinfo {year} {2023})\ pp.\ \bibinfo {pages} {730--741}\BibitemShut {NoStop}%
\bibitem [{\citenamefont {Peixoto}(2014)}]{peixoto_graph-tool_2014}%
  \BibitemOpen
  \bibfield  {author} {\bibinfo {author} {\bibfnamefont {T.~P.}\ \bibnamefont {Peixoto}},\ }\bibfield  {title} {\bibinfo {title} {The graph-tool python library},\ }\bibfield  {journal} {\bibinfo  {journal} {Figshare}\ }\href {https://doi.org/10.6084/m9.figshare.1164194} {10.6084/m9.figshare.1164194} (\bibinfo {year} {2014})\BibitemShut {NoStop}%
\end{thebibliography}%


\begin{thebibliography}{10}

\bibitem{Ramaciotti:2022}
Pedro Ramaciotti~Morales, Jean-Philippe Cointet, Gabriel Mu{\~n}oz~Zolotoochin,
  Antonio Fern{\'a}ndez~Peralta, Gerardo I{\~n}iguez, and Armin Pournaki.
\newblock Inferring attitudinal spaces in social networks.
\newblock {\em Social Network Analysis and Mining}, 13(1):1--18, 2023.

\bibitem{barbera2015birds}
Pablo Barber{\'a}.
\newblock Birds of the same feather tweet together: Bayesian ideal point
  estimation using twitter data.
\newblock {\em Polit. Anal.}, 23(1):76--91, 2015.

\bibitem{barbera2015tweeting}
Pablo Barber{\'a}, John~T Jost, Jonathan Nagler, Joshua~A Tucker, and Richard
  Bonneau.
\newblock Tweeting from left to right: Is online political communication more
  than an echo chamber?
\newblock {\em Psychol. Sci.}, 26(10):1531--1542, 2015.

\bibitem{metin2022tweeting}
Omer~Faruk Metin and Pedro Ramaciotti~Morales.
\newblock Tweeting apart: Democratic backsliding, new party cleavage and
  changing media ownership in turkey.
\newblock {\em Party Politics}, 2022.

\bibitem{ramaciotti2021unfolding}
Pedro Ramaciotti~Morales, Jean-Philippe Cointet, and Gabriel Mu\~{n}oz
  Zolotoochin.
\newblock Unfolding the dimensionality structure of social networks in
  ideological embeddings.
\newblock In {\em Proceedings of the 2021 IEEE/ACM International Conference on
  Advances in Social Networks Analysis and Mining}, pages 333--338, 2021.

\bibitem{mcpherson2001birds}
Miller McPherson, Lynn Smith-Lovin, and James~M Cook.
\newblock Birds of a feather: Homophily in social networks.
\newblock {\em Annu. Rev. Sociol.}, 27(1):415--444, 2001.

\bibitem{greenacre2017correspondence}
Michael Greenacre.
\newblock {\em Correspondence analysis in practice}.
\newblock Chapman and Hall, 2017.

\bibitem{lowe2008understanding}
Will Lowe.
\newblock Understanding wordscores.
\newblock {\em Polit. Anal.}, 16(4):356--371, 2008.

\bibitem{chari2023specious}
Tara Chari and Lior Pachter.
\newblock The specious art of single-cell genomics.
\newblock {\em PLOS Computational Biology}, 19(8):e1011288, 2023.

\bibitem{chesdata2019}
R.~Bakker, L.~Hooghe, S.~Jolly, G.~Marks, J.~Polk, J.~Rovny, M.~Steenbergen,
  and M.~A. Vachudova.
\newblock {2019 Chapel Hill Expert Survey}.
\newblock {\em Chapel Hill}, 2020.
\newblock www.chesdata.eu.

\bibitem{barbera2015understanding}
Pablo Barber{\'a} and Gonzalo Rivero.
\newblock {Understanding the political representativeness of Twitter users}.
\newblock {\em Soc. Sci. Comput. Rev.}, 33(6):712--729, 2015.

\bibitem{hooghe2018cleavage}
Liesbet Hooghe and Gary Marks.
\newblock Cleavage theory meets europe’s crises: Lipset, rokkan, and the
  transnational cleavage.
\newblock {\em J. Eur. Public Policy}, 25(1):109--135, 2018.

\bibitem{grossman2019economic}
Emiliano Grossman and Nicolas Sauger.
\newblock Economic internationalization and the decline of the left--right
  dimension.
\newblock {\em Party Polit.}, 25(1):36--49, 2019.

\bibitem{ramaciotti2022embedding}
Pedro Ramaciotti~Morales and Zografoula Vagena.
\newblock Embedding social graphs from multiple national settings in common
  empirical opinion spaces.
\newblock In {\em Proceedings of the 2022 IEEE/ACM International Conference on
  Advances in Social Networks Analysis and Mining}, 2022.

\bibitem{Baumann:2021}
Fabian Baumann, Philipp Lorenz-Spreen, Igor~M. Sokolov, and Michele Starnini.
\newblock Emergence of polarized ideological opinions in multidimensional topic
  spaces.
\newblock {\em Phys. Rev. X}, 11:011012, 2021.

\bibitem{Zhang:2020}
Lizhi Zhang and Tiago~P. Peixoto.
\newblock Statistical inference of assortative community structures.
\newblock {\em Phys. Rev. Research}, 2:043271, Nov 2020.

\bibitem{Porter:2016}
M.~A. Porter and J.~P. Gleeson.
\newblock Dynamical systems on networks.
\newblock {\em Front. Appl. Dynam. Syst.}, 4, 2016.

\bibitem{Sood:2008}
V.~Sood, Tibor Antal, and S.~Redner.
\newblock Voter models on heterogeneous networks.
\newblock {\em Phys. Rev. E}, 77:041121, 2008.

\bibitem{Carro:2016}
Adrian Carro, Ra{\'u}l Toral, and Maxi San~Miguel.
\newblock The noisy voter model on complex networks.
\newblock {\em Scientific reports}, 6:24775, 2016.

\bibitem{Chacoma:2015}
Andrés Chacoma and Damián~H. Zanette.
\newblock Opinion formation by social influence: From experiments to modeling.
\newblock {\em PLOS ONE}, 10(10):e0140406, 10 2015.

\bibitem{Moussaid:2013}
Mehdi Moussaïd, Juliane~E. Kämmer, Pantelis~P. Analytis, and Hansjörg Neth.
\newblock Social influence and the collective dynamics of opinion formation.
\newblock {\em PLOS ONE}, 8(11):1--8, 11 2013.

\end{thebibliography}
\end{document}


\begin{center}
{\LARGE Supplemental Information for}\\[0.7cm]
{\Large \textbf{Multidimensional political polarization in online social networks}}\\[0.5cm]
{\large A. F. Peralta, P. Ramaciotti, J. Kertész, and G. Iñiguez}\\[0.7cm]
\end{center}

\addtocontents{toc}{\protect\setstretch{0.1}}
\tableofcontents

\section{Data details}

\subsection{Network}\label{sec_network_data}

For establishing a French political Twitter network we take as starting point the dataset collected in \cite{Ramaciotti:2022}. The network is seeded using the Twitter accounts of the members of the parliament (MPs) in France, these are 831 out of a total of 925 (577 in the National Assembly and 348 in the Senate). The followers of the MPs were collected in May 2019, resulting in $4~487~430$ users. To filter inactive or bot accounts, as well as accounts without enough strong ideological interactional signals, the dataset only contains users that have at least 25 followers and that follow at least 3 MPs (see Section 3.1 in \cite{Ramaciotti:2022}). These criteria have also been used by similar ideological scaling methods applied to Twitter friendship networks \cite{barbera2015birds,barbera2015tweeting,metin2022tweeting}. As a next step, we identified the edges in the friendship network (the follower $\rightarrow$ followed relations are the links of the network) for the period August to December 2020 and kept only the $231~067$ total number of users who are connected to at least one other user. This restriction is necessary to disregard from the study the nodes that do not play an active role in the political interaction on Twitter. We neglect the MPs that do not have followers which results in a total number of MPs (m) equal to $M = 813$, and we also neglect the users that are not connected to other users and only follow a few MPs which results in a number of users (u) equal to $N = 230~254$ (the total number is $N+M=231~067$). Taking into account these two node categories, i.e., MPs and users, we have four possible types of links: (i) User $\rightarrow$ User (uu), (ii) User $\rightarrow$ MP (um), (iii) MP $\rightarrow$ MP (mm),  and (iv) MP $\rightarrow$ User (mu). The link count of each link category is: {63~625~921} for (uu), 3~351~359 for (um), 113~596 for (mm), and 515~882 for (mu), and the corresponding out/in average degrees are: $\langle k^{\text{out/uu}} \rangle = \langle k^{\text{in/uu}} \rangle = 276.33$; $\langle k^{\text{out/um}} \rangle = 14.56$, $\langle k^{\text{in/um}} \rangle = 4~122.21$;  $\langle k^{\text{out/mm}} \rangle = \langle k^{\text{in/mm}} \rangle = 139.72$; and $\langle k^{\text{out/mu}} \rangle = 634.54$, $\langle k^{\text{in/mu}} \rangle = 2.24$. The average degree of the User $\rightarrow$ User subgraph is high indicating that in the network considered the users, as well as the MPs, are highly active in politically related issues in Twitter. The difference in the number of links in the User $\rightarrow$ MP and MP $\rightarrow$ User subgraphs demonstrates the lack of reciprocity in the interactions between users and MPs. For this reason, we will not consider the MP $\rightarrow$ User links as they will not be as important in the analysis. This assumption is fundamental both in the study of the structure of the network and in the modelling part.

\subsubsection{Degree distributions} 
\begin{figure}[h!]
\begin{center}
\subfloat[]{\label{degree_dist:a}\includegraphics[width=0.49\textwidth]{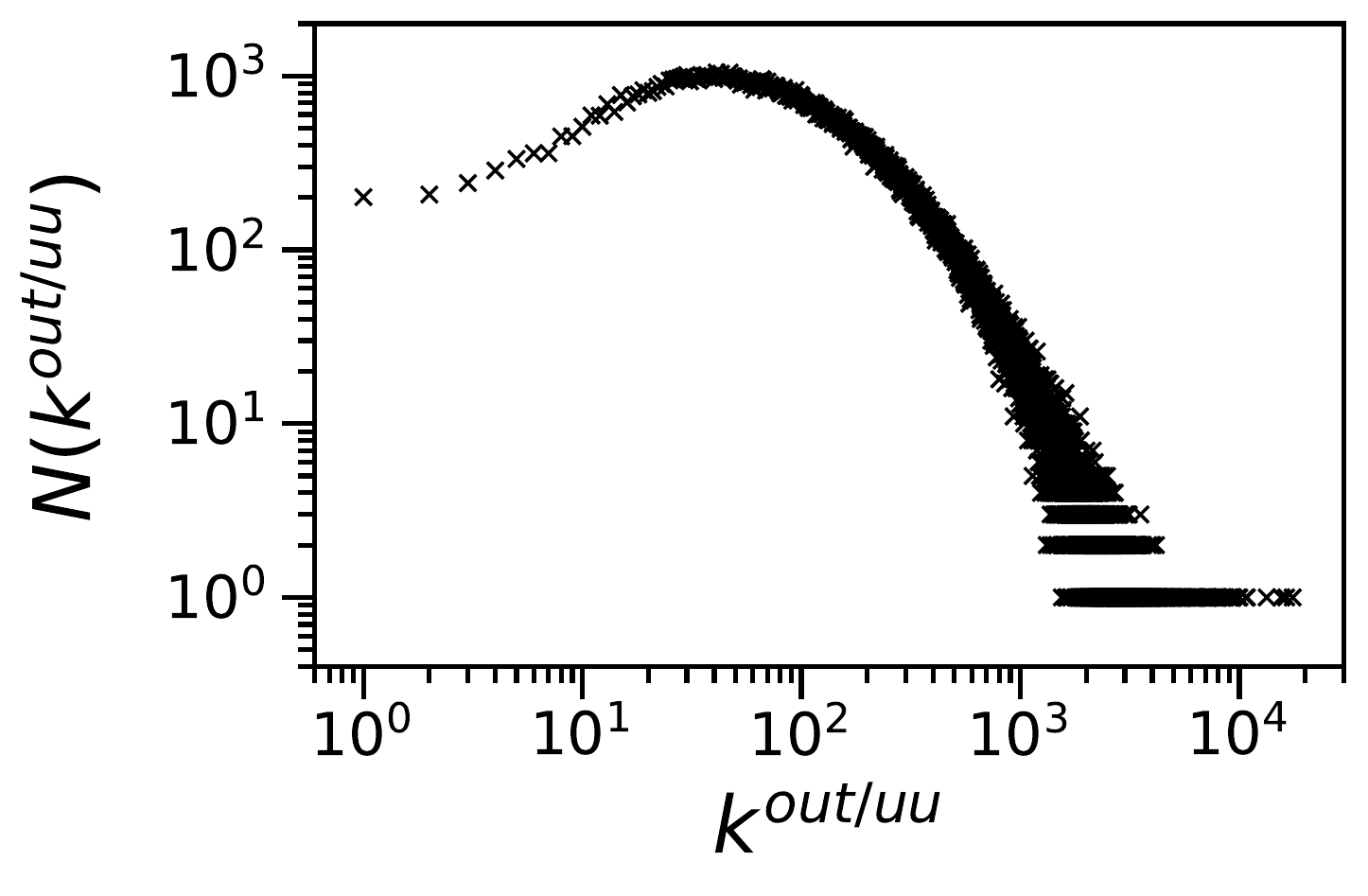}}
\subfloat[]{\label{degree_dist:b}\includegraphics[width=0.49\textwidth]{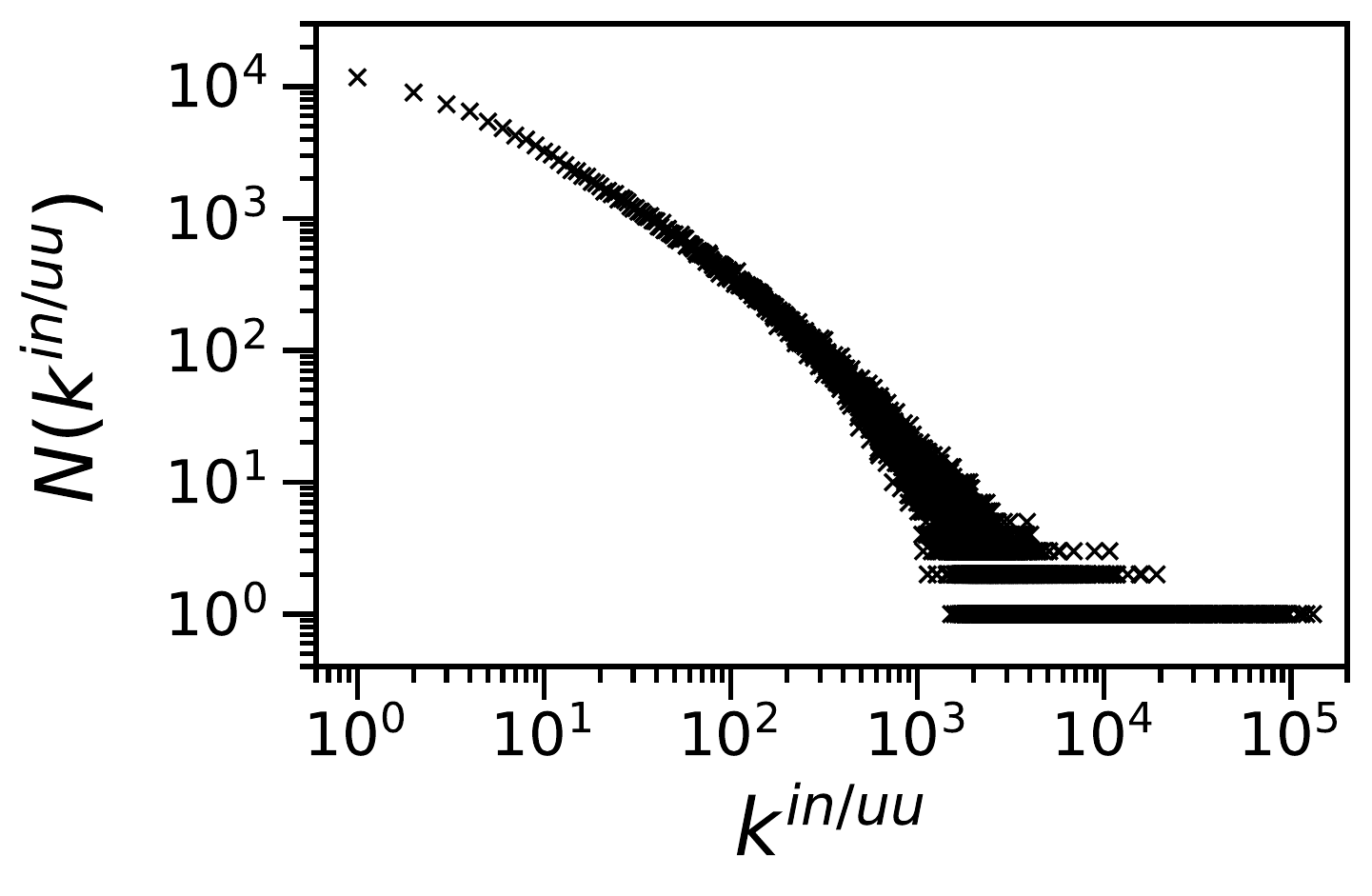}}
 
\subfloat[]{\label{degree_dist:c}\includegraphics[width=0.49\textwidth]{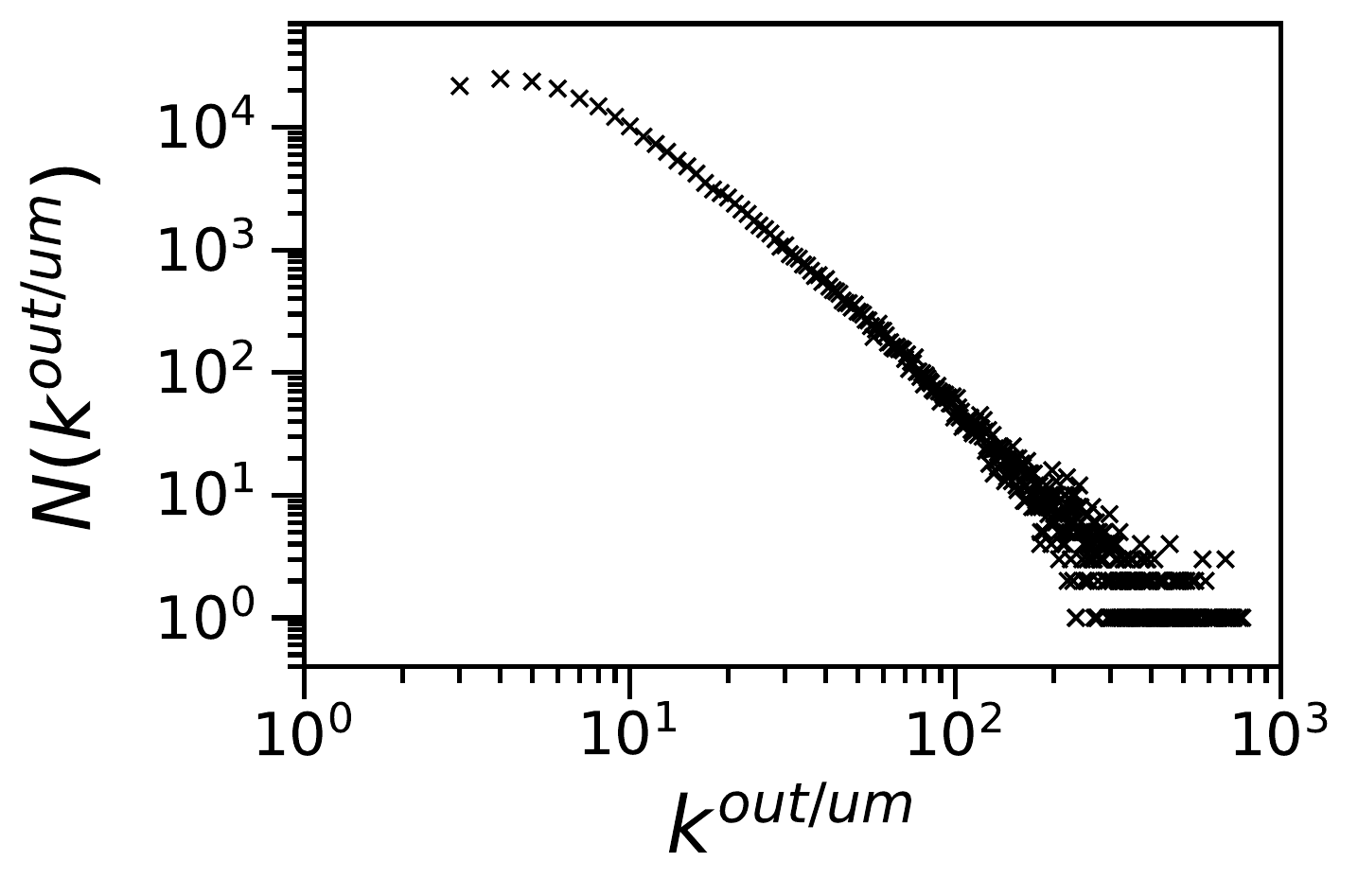}}
\subfloat[]{\label{degree_dist:d}\includegraphics[width=0.49\textwidth]{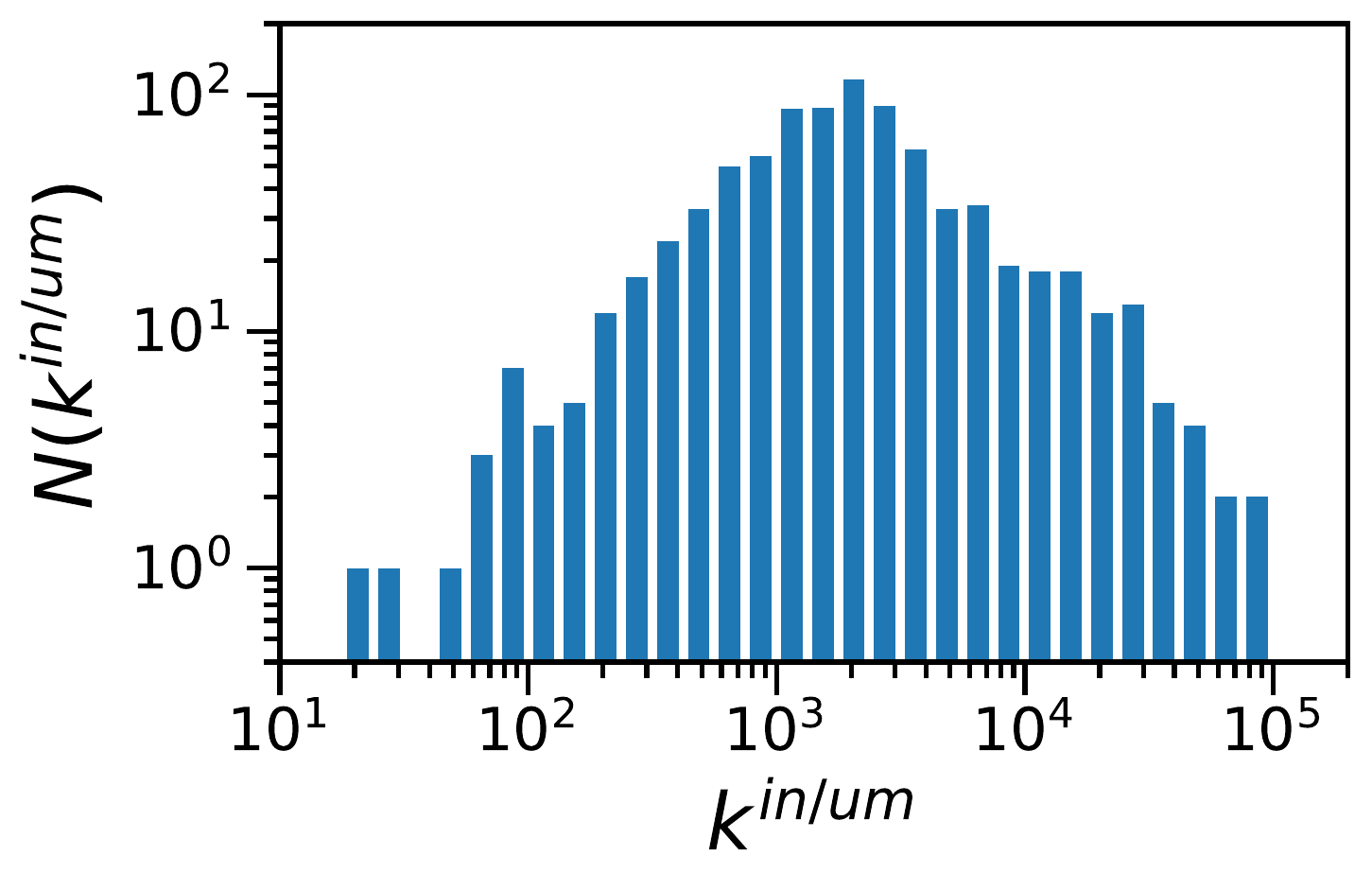}}
\caption{Degree distributions of the User $\rightarrow$ User and User $\rightarrow$ MP subgraphs in log-log scale, i.e., $N(k) \equiv$ number of nodes with degree $k$ as a function of $k$, where the degree is for (a) $k^{\text{out/uu}}$, (b) $k^{\text{in/uu}}$, (c) $k^{\text{out/um}}$, and (d) $k^{\text{in/um}}$. In panels (a), (b) and (c) the numbers $N(k)$ of all possible values of $k$ are shown, while for panel (d) we use logarithmic binning for visualization purposes as the number of MPs is not large enough.}
\label{degree_dist}
\end{center}
\end{figure}

We present now a more detailed analysis of the degrees of the nodes in the network for the User $\rightarrow$ User (uu) and User $\rightarrow$ MP (um) subgraphs. We define the adjacency matrix of these subgraphs as follows: for the User $\rightarrow$ User subgraph it is $A_{ij}=1$ if there is a directed link between user $i$ (source) and user $j$ (target) with $i, j = 1, \dots, N$ and $A_{ij}=0$ otherwise; for the User $\rightarrow$ MP subgraph it is $P_{im}=1$ if there is a directed link between user $i$ (source) and MP $m$ (target) with $i = 1, \dots, N$, $m= 1, \dots, M$, and $P_{im}=0$ otherwise. There are four possibles degrees in these subgraphs, i.e.: $k^{\text{out/uu}}_{i} = \sum_{j=1}^{N} A_{ij}$ and $k^{\text{in/uu}}_{i} = \sum_{j=1}^{N} A_{ji}$ with $i=1, \dots, N$, $k^{\text{out/um}}_{i} = \sum_{m=1}^{M} P_{im}$ with $i=1, \dots, N$ and $k^{\text{in/um}}_{m} = \sum_{i=1}^{M} P_{im}$ with $m=1, \dots, M$. The degree distributions of the sequences $\lbrace k^{\text{in/uu}}_{i}, k^{\text{out/uu}}_{i}, k^{\text{out/um}}_{i} \rbrace_{i=1, \dots, N}$ and $\lbrace k^{\text{in/um}}_{m} \rbrace_{m=1, \dots, M}$ are shown in Fig. S\ref{degree_dist}. The distributions of $k^{\text{out/uu}}$, $k^{\text{out/um}}$, and $k^{\text{in/um}}$ have maximum (mode) around 40, 4 and 2000 respectively, while for $k^{\text{in/uu}}$ it is a decreasing function, and all the distributions are fat tailed. Other properties that we observe is that all users follow at least three MPs and that all MPs have at least 20 followers. Comparing the in-degree distribution of the users $k^{\text{in/uu}}$ and MPs $k^{\text{in/um}}$ we see that the MPs have a huge number of followers and that only a few users are comparable to them in popularity, which confirms our previous consideration that the MPs play the central role and that the network revolves around politically related issues.

\begin{figure}[h!]
\begin{center}
\subfloat[]{\label{degree_corr:a}\includegraphics[width=0.49\textwidth]{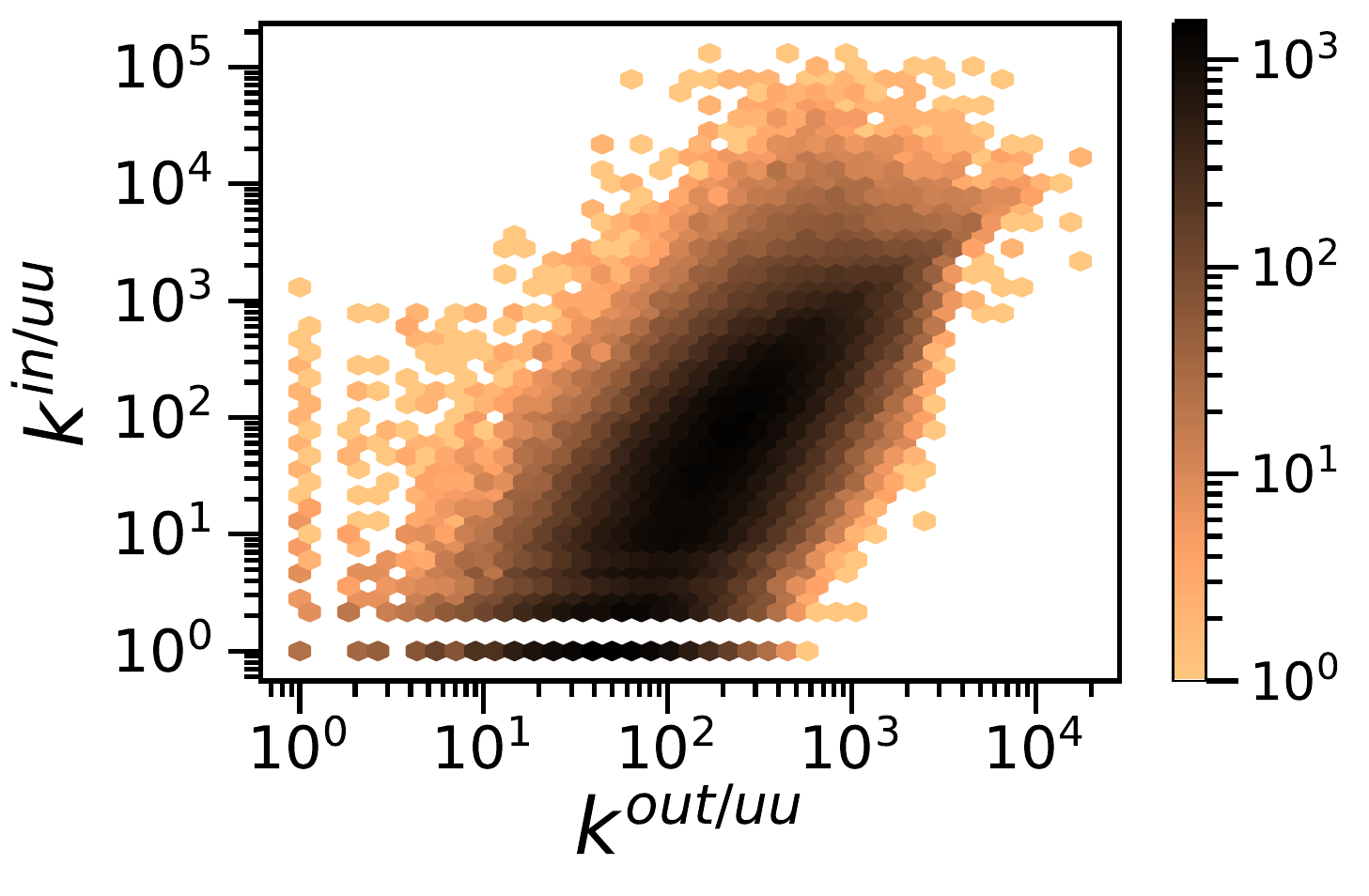}}
\subfloat[]{\label{degree_corr:b}\includegraphics[width=0.49\textwidth]{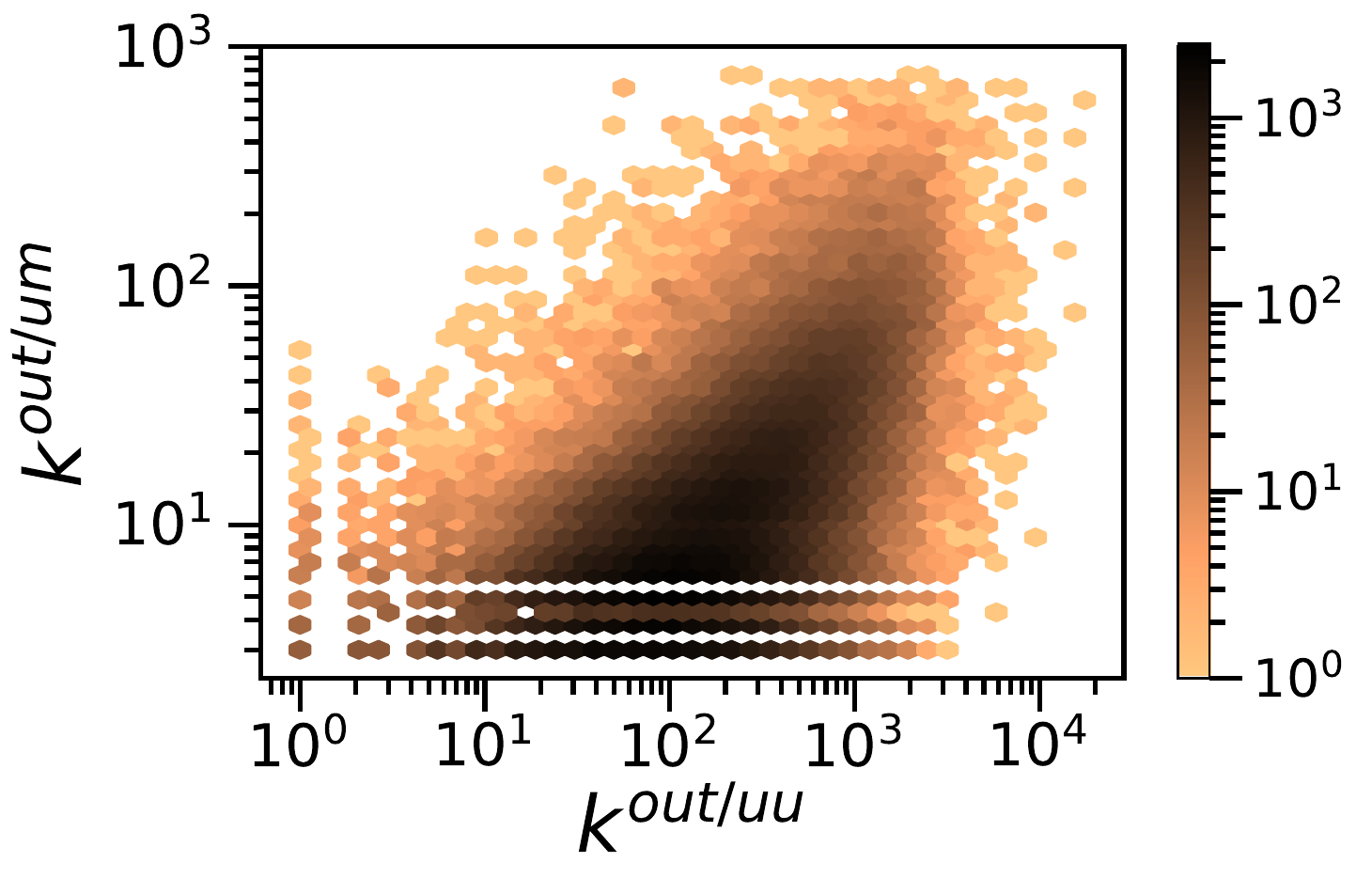}}
\caption{Number of nodes with degrees $k^{\text{in/uu}}$ and $k^{\text{out/uu}}$ in panel (a), and with degrees $k^{\text{out/um}}$ and $k^{\text{out/uu}}$ in panel (b), in log-log scale and with hexagonal logarithmic binning.}
\label{degree_corr}
\end{center}
\end{figure}

In Fig. S\ref{degree_corr} we show the degree correlations $k^{\text{in/uu}}_{i}$ vs $k^{\text{out/uu}}_{i}$ and $k^{\text{out/um}}_{i}$ vs $k^{\text{out/uu}}_{i}$ to examine if there is a connection between the different degrees of users. We observe that some correlations are present in both cases, although we also observe a very high level of noise. We compute the Pearson correlation coefficient ($r$) and Kendall rank correlation coefficient ($\tau$), which measure the linear correlation and ordinal association (rank correlation or similarity in the orderings) between two variables. For $k^{\text{in/uu}}_{i}$ vs $k^{\text{out/uu}}_{i}$ we have $r=0.29$ and $\tau=0.56$ which shows some degree of correlation, the linear coefficient is not very high but the rank coefficient is higher indicating that the two variables are associated but the relation has some non-linearity, as it can be seen in Fig. S\ref{degree_corr}. For $k^{\text{out/um}}_{i}$ vs $k^{\text{out/uu}}_{i}$ we have $r=0.49$ and $\tau=0.40$ which shows a higher level of linear correlation but still with a high level of noise. 

\begin{figure}[h!]
\begin{center}
\subfloat[]{\label{degree_assorta:a}\includegraphics[width=0.49\textwidth]{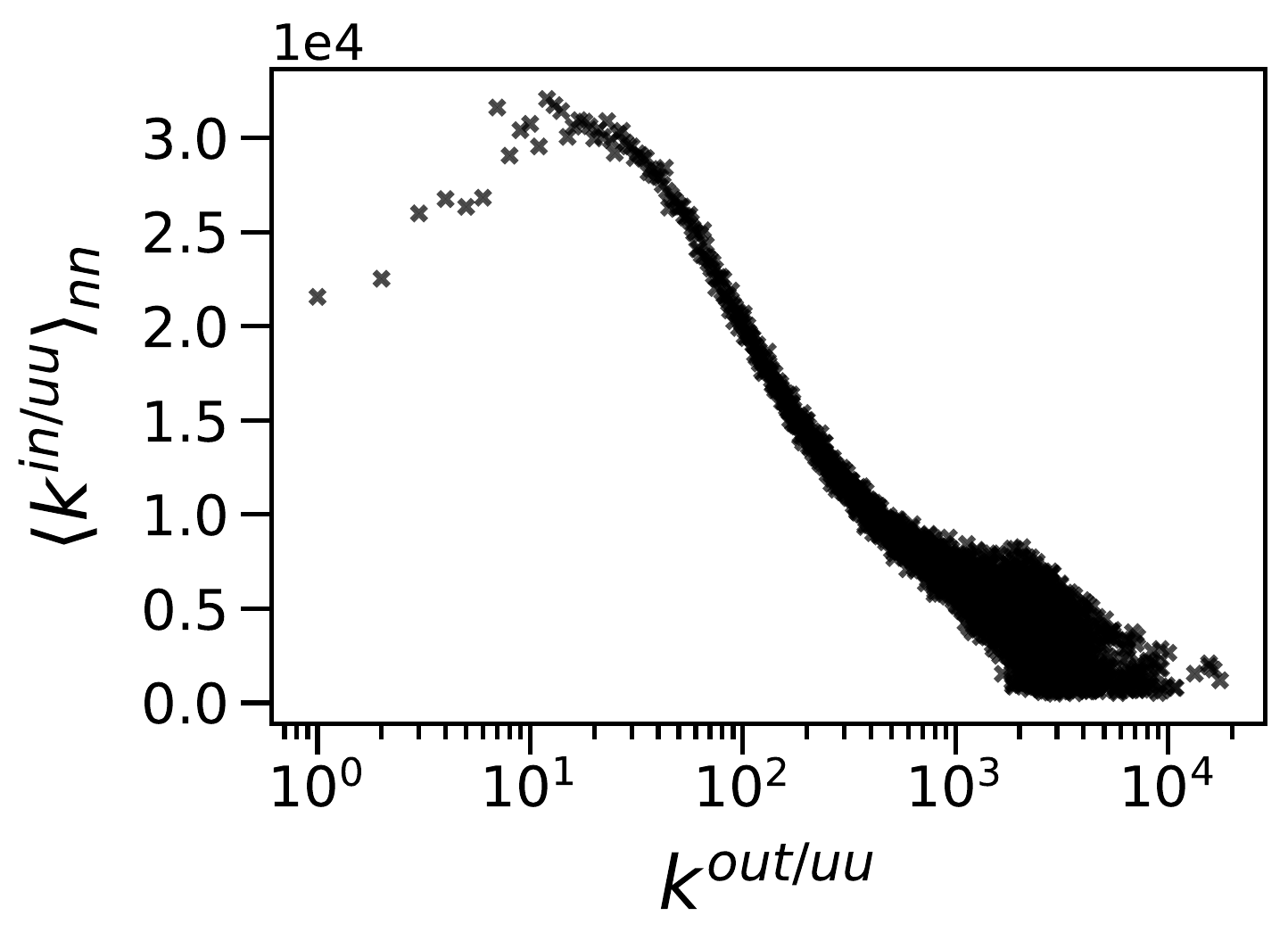}}
\subfloat[]{\label{degree_assorta:b}\includegraphics[width=0.49\textwidth]{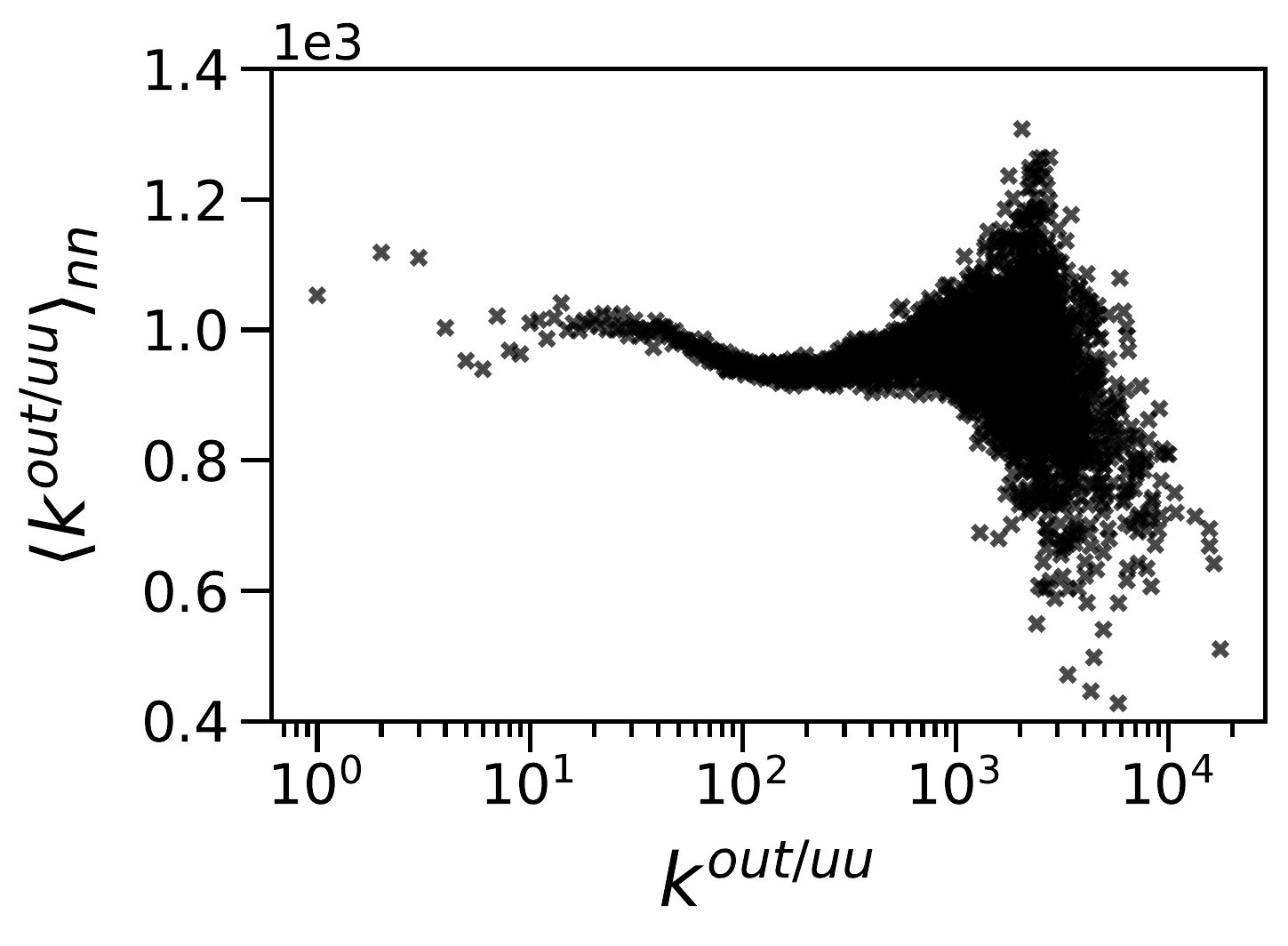}}

\subfloat[]{\label{degree_assorta:c}\includegraphics[width=0.49\textwidth]{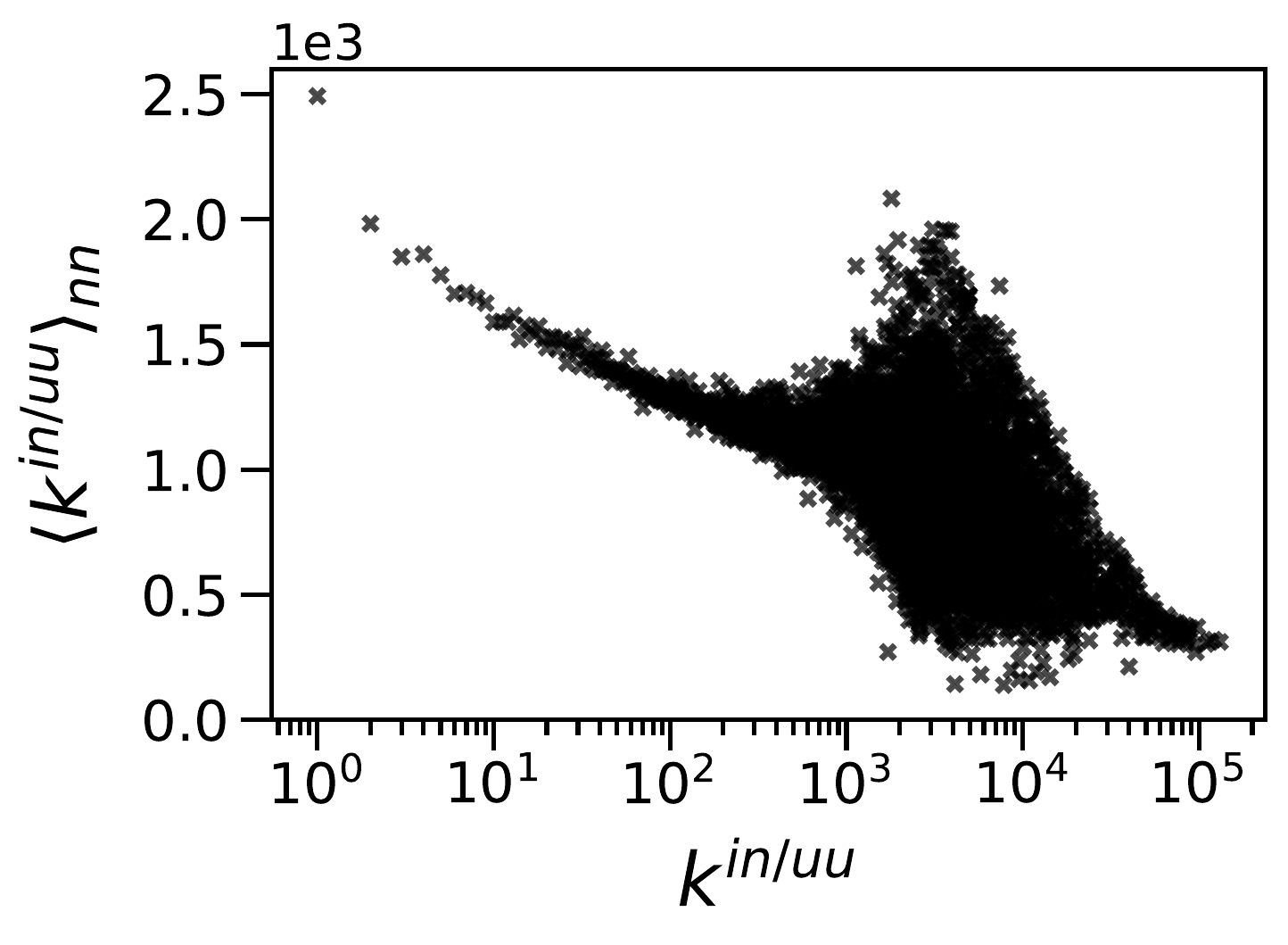}}
\subfloat[]{\label{degree_assorta:d}\includegraphics[width=0.49\textwidth]{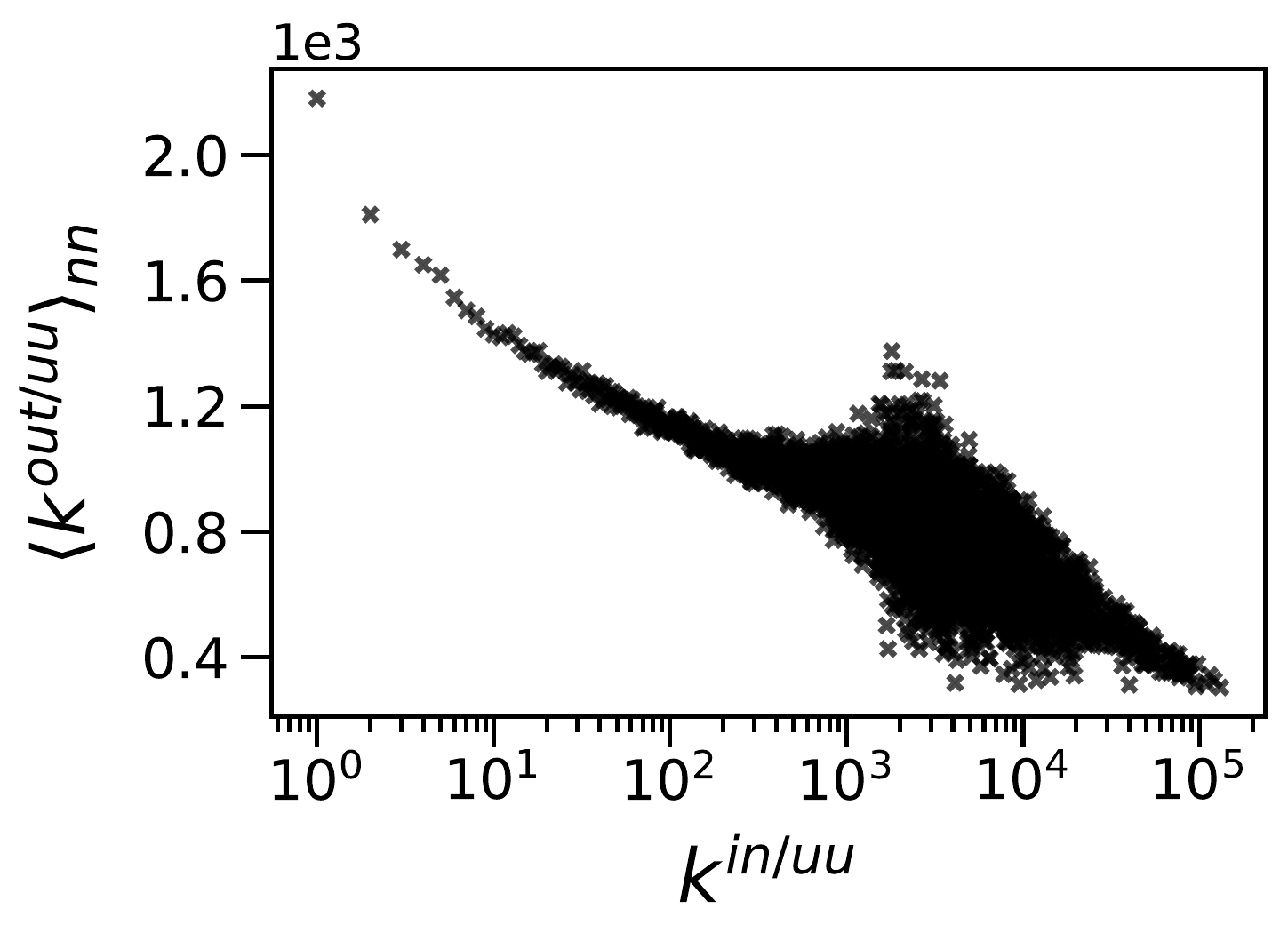}}
\caption{Assortativity of the User $\rightarrow$ User network, i.e., average degree $\langle k^{\text{in/uu}} \rangle_{\text{nn}}$ and $\langle k^{\text{out/uu}} \rangle_{\text{nn}}$ of the (nearest) neighbors of nodes with degrees $k^{\text{out/uu}}$ and  $k^{\text{in/uu}}$: (a) $\langle k^{\text{in/uu}} \rangle_{\text{nn}}$ vs $k^{\text{out/uu}}$, (b) $\langle k^{\text{out/uu}} \rangle_{\text{nn}}$ vs $k^{\text{out/uu}}$, (c) $\langle k^{\text{in/uu}} \rangle_{\text{nn}}$ vs $k^{\text{in/uu}}$, and (d) $\langle k^{\text{out/uu}} \rangle_{\text{nn}}$ vs $k^{\text{in/uu}}$ in log-lin scale.}
\label{degree_assorta}
\end{center}
\end{figure}

In Fig. S\ref{degree_assorta} we show the assortativity of the User $\rightarrow$ User subgraph, which is calculated as the average degree $k_{j}$ of the neighbors $j$ of $i$ ($A_{ij}$=1), averaged over all nodes $i$ that have the same degree $k_{i}$, depending if the degrees are in or out we have four possibilities. If we interpret the out-degree as a measure of activity and the in-degree as a measure of popularity we draw the following conclusions: (i) for $\langle k^{\text{in/uu}} \rangle_{\text{nn}}$ vs $k^{\text{out/uu}}$ we have that the average popularity of your neighbors decreases as a function of your activity, (ii) for $\langle k^{\text{out/uu}} \rangle_{\text{nn}}$ vs $k^{\text{out/uu}}$ the average activity of your neighbors does not depend on your activity, (iii) for $\langle k^{\text{in/uu}} \rangle_{\text{nn}}$ vs $k^{\text{in/uu}}$ the average popularity of your neighbors decreases as a function of your popularity, and (iv) for $\langle k^{\text{out/uu}} \rangle_{\text{nn}}$ vs $k^{\text{in/uu}}$ the average activity of your neighbors decreases as a function of your popularity. These properties agree with what one can expect from directed dissasortative networks (where low degree nodes are more probable to be connected to high degree nodes). Note also that noise increases in the tails of the distributions Fig. S\ref{degree_dist}, i.e., $k^{\text{out/uu}} \sim 10^3-10^4$ and $k^{\text{in/uu}} \sim 10^3-10^5$.

\subsection{Network embedding}

We rely on the method exposed in \cite{Ramaciotti:2022} for the estimation of the position of Users and MPs along our four selected political dimensions. This method consist in two steps illustrated in Fig. 1B of the article: 1) latent space embedding, and 2) attitudinal embedding using political survey data. The whole method can be summarized as follows, and will be detailed in the following subsections. In the first step, we rely on a possible generative probabilistic model that might be underlying observed parts of the Twitter social graph. The model in which we rely is the multidimensional ideal point estimation \cite{barbera2015tweeting, ramaciotti2021unfolding}. This models posits an underlying homophilic mechanism for the bipartite part of the social graph linking MPs with their followers: users that follow a similar set of MPs should be positioned in proximity in a latent space representing political preferences. Using the observed bipartite graph we infer latent space positions. In the second step we tackle the problem of inferring semantical meaning for the dimensions of the space. For this, we map positions of the latent space onto the dimensions of an instrument having explicit meaning and references points. In this case, we use the Chapel Hill Expert Survey (CHES) data. To compute the map between latent and the dimension of the instrument we use guiding points existing in both: estimated positions of political parties.

\subsubsection{Latent space embedding}

We first rely on our seed nodes, the 831 MPs, to collect their followers. We then collected (between August and December 2020) the neighbors of these followers whenever they followed at least 3 MPs and were followed by at least 25 other users. This resulted in a social graph of $230~254$ users that followed at least 3 MPs. We now consider the adjacency matrix $P$ of the bipartite graph linking MPs and their followers: $P_{im}=1$ if user $i$ follows MP $m$, and $P_{im}=0$ if not. We consider an homophilic \cite{mcpherson2001birds} process hinging on unobservable multidimensional quantities $\phi_i$ and $\phi_m$ for users $i$ and MP $m$ \cite{barbera2015birds}:

\begin{equation}
    \text{Prob.}\left(P_{im}=1 | \alpha_i,\beta_m,\gamma, \phi_i,\phi_m \right) = \text{logistic}\left( \alpha_i + \alpha_m -\gamma \|\phi_i-\phi_m\|^2 \right),
\end{equation}
where $\alpha_i$ is the level of activity of User $i$ in number followed users, $\beta_m$ is the popularity of MP $m$ in number of followers, and $\phi_i$ and $\phi_m$ are unobservable positions in a multidimensional space that might be explaining the observed bipartite network. As in \cite{barbera2015tweeting,ramaciotti2021unfolding,Ramaciotti:2022}, we rely on the fact that Correspondence Analysis (CA) \cite{greenacre2017correspondence} principal components of $P$ approximate ideal point estimation \cite{lowe2008understanding}. We compute the Correspondence Analysis of $P$ using the Language-Independent Network Attitudinal Embedding (LINATE) package \footnote{\url{http:www.github.com/pedroramaciotti/linate}}.
Producing estimates of positions $\phi$ via an embedding with CA has the advantage introducing only linear transformations, in contrast with non-linear embedding methods such as UMAP of t-SNE \cite{chari2023specious}. 
This linearity means that if position $\phi_i$ is twice as far from position $\phi_j$ than position $\phi_{i'}$ (i.e., $\|\phi_{i}-\phi_j\| = 2 \|\phi_{i'}-\phi_j\|$), this relation will be preserved in the latent space achieved via CA.
The preservation of relative is distances up to linear transformations, excluding non-linear transformation, is needed in the next step of the two-step embedding procedure (see Fig. 1B in the main article), described in the next subsection.

\subsubsection{Attitudinal embedding}

To provide a spatialization with explicit spatial semantics, we map positions in this latent space to a second space in which dimensions stand as indicators of continuous positive or negative attitudes towards specified issues of political debate. We consider the 51 dimensional space of the CHES data as the Attitudinal Reference Frame (AFR) in which we will analyze the collected social network. This ARF was constructed using the assessment of the position of 277 parties from 32 countries (including 8 French parties also included in our Twitter dataset) in this 51-dimensional space. See \cite{chesdata2019} for further details. To map positions from the latent space to this ARF, we use positions of parties to compute an affine transformation. For each party, we compute the position on the latent space as the centroid or mean of the positions of MPs that belong to that party. Knowing party positions on both the latent space and the ARF, we compute an affine transformation mapping position of the former onto the latter choosing the number of latent dimensions that fully determine the parameters of the affine transformation. For 8 political parties (and thus reference points en both spaces) the affine transformation is fully determined taking the first 7 dimensions of the latent space. The question of the number of CHES dimensions needed to represent the latent space data has been treated for this dataset in \cite{ramaciotti2021unfolding}, yielding that at least three dimensions of the latent space are related to CHES dimensions in how MPs are positioned.


To select the dimension for our study, we begin by considering the traditional left-right (LR) dimension leveraged in most social media studies \cite{barbera2015birds,barbera2015understanding}. We then include additional political dimensions to cover the dimensional complexity of our sample population (i.e., of at least three dimensions) choosing issue dimensions that have been identified in the literature as relevant for the French case. This is needed to address possible distinctions identified by political science literature within the left- and right-wing politicians and individuals. We identified three types of distinctions in the literature that need to addressed. First, and particularly in the case of France, recent works have shown that attitudes towards elites (AE) -- from the CHES data -- are the most structuring political element in social media, and in particular in structuring political Twitter networks \cite{ramaciotti2021unfolding}. Second, beyond LR and AE dimensions, we need to include a dimension distinguishing traditional liberal right from the nationalist right. This distinction has been identified for several years now and in particular for France \cite{hooghe2018cleavage}, and motivates the inclusion of the CHES dimension of attitudes towards nationalism (NA). Finally, we include a fourth dimension, addressing a third distinction identified as important and independent from left-right dimension. Recent studies have identified the how internationalization of legal frameworks and economies relates to decline in importance of the left-right dimension in politics \cite{grossman2019economic}. In particular in France this internationalization is first and foremost related to the European integration process. This motivates the inclusion of the CHES dimension of attitudes towards the European Union (EU). Moreover, the positions of French political parties, as captured by the 51 CHES dimensions, can be described almost completely with only 4 dimensions, as shown by a Principal Component Analysis of the CHES dimensions (see Section IV of \cite{ramaciotti2022embedding}). 

In the survey, experts are asked to put parties in scales with respect to the following reference positions, and according to the following descriptions:

\begin{itemize}
\item \textbf{Left-right} (LR): \textit{``position of the party in terms of its overall ideological stance''}, with 0 being \textit{``Extreme left''}, 5 being \textit{``Center''} and 10 being \textit{``Extreme right''}.
\item \textbf{Anti-elite salience} (AE): \textit{``salience of anti-establishment and anti-elite rhetoric''}, with 0 being \textit{``Not important at all''} and 10 being \textit{``Extremely important''}.
\item \textbf{EU position:} (EU): \textit{``overall orientation of the party leadership towards European integration''}, with 1 being \textit{``Strongly opposed''} and 7 being \textit{``Strongly in favor''}
\item \textbf{Nationalism} (NA): \textit{``position towards cosmopolitanism vs. nationalism''}, with 0 being \textit{``Strongly promotes cosmopolitan conceptions of society''} and 10 being \textit{``Strongly promotes nationalist conceptions of society''}.
\end{itemize}

While the positions for political parties are bounded by construction of the CHES instrument, the positions of individual MPs and their followers are not bounded to these intervals. The computed affine transformation will map party position on the latent space to the corresponding party positions in the CHES ARF. Individual MPs and followers inside the convex hull formed by party positions are mapped into the convex hull of party positions in the ARF, and thus are placed within the limit values specified on the CHES for each dimension. However, because party positions in the latent space are computed as centroid of MPs of that party, a fraction of MPs of any given party, are bound to have more extreme positions that those of the party, and can thus lay outside the CHES bounds in the ARF. Because of these four dimensions have difference reference points, we further map them onto the $0-1$ interval. More precisely, we rescale each dimensions such that:

\begin{itemize}
\item \textbf{Left-right} (LR): position 0 is mapped to 0, and position 10 to 1;
\item \textbf{Anti-elite salience} (AE): position 0 is mapped to 0, and position 10 to 1;
\item \textbf{EU position} (EU): position 1 is mapped to 0, and position 7 to 1;
\item \textbf{Nationalism} (NA): position 0 is mapped to 0, and position 10 to 1;
\end{itemize}
Fig.~S\ref{opinions_blues} in the next section shows the distributions of our Twitter sample in several 2D projections of this 4-dimensional political opinion space.

\begin{figure}[ht]
    \centering
    \subfloat[]{\includegraphics[width=0.49\textwidth]{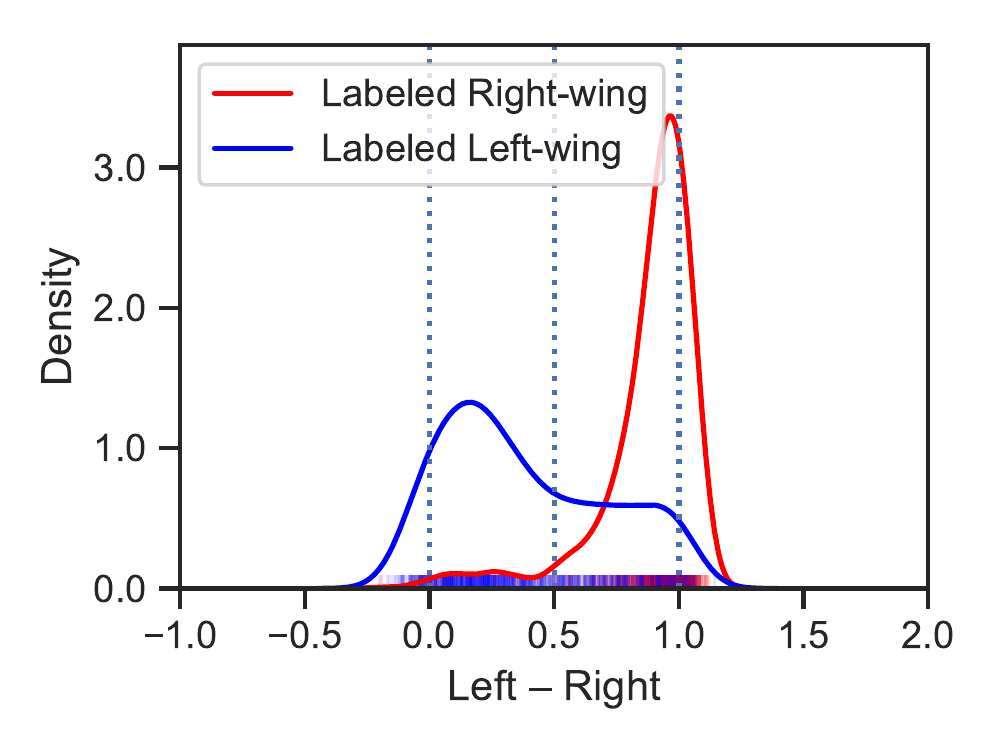}}
    \subfloat[]{\includegraphics[width=0.49\textwidth]{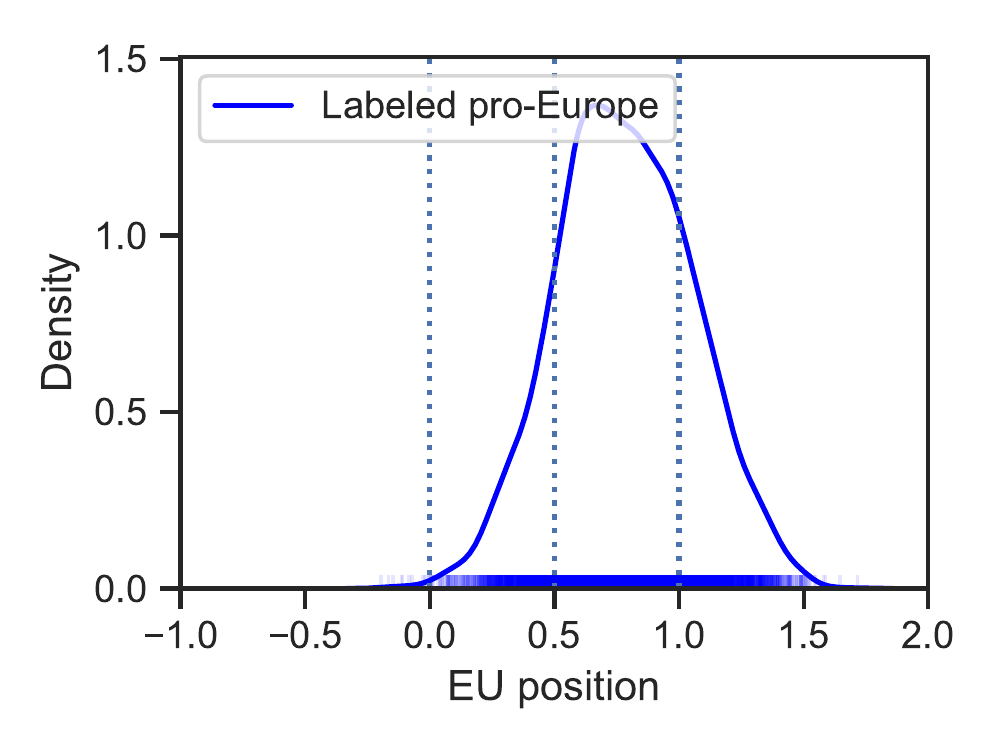}}
    
    \subfloat[]{\includegraphics[width=0.49\textwidth]{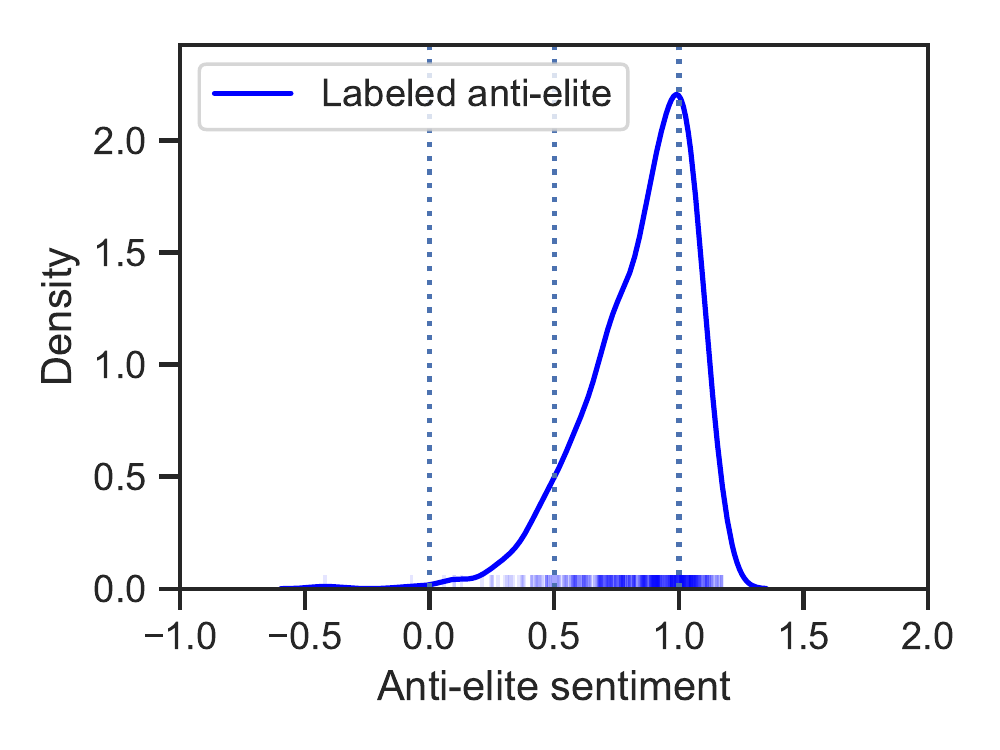}}
    \subfloat[]{\includegraphics[width=0.49\textwidth]{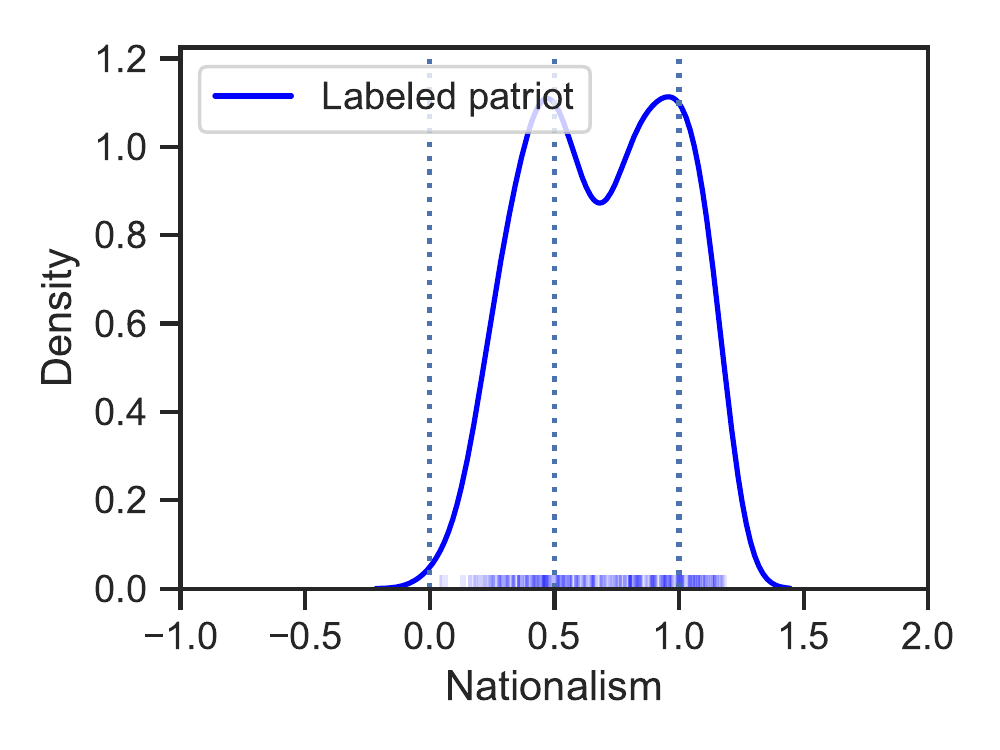}}
    \caption{Density of five groups of users defined by use of keywords used in their Twitter bio profile text, according to their positions along the dimensions on which we expect them to be skewed to show that, they are positioned mostly in the corresponding half of the space. For example, users labeled as Right-wingers, are mostly concentrated to the right of position 0.5, marking the explicitly defined center of the LR dimension. The same is true for the other groups in their corresponding dimensions.}
    \label{fig:distributions_flags_dims}
\end{figure}

Next, we tackled the question of the validation of such space. The construction of this ARF embedding rests in three hypotheses: 1) users are aware of the existence of MPs but choose to follow those with whom they feel political affinity, such that the User $\rightarrow$ MP edges are homophilic to some degree, 2) latent space preserves this homophily, in that users positioned in proximity in space follow similar sets of MPs, and 3) that party positions perceived by users in the way they decide to follow MPs from parties resembles that of experts (in particular in ordering along our 4 different dimensions). In order to test the validity of the embedding, we use an independent set of data: the text produced by users in the bio profile self-description. Indeed, users sometimes display their political positions voluntarily in their profiles: e.g., ``\textit{si vous êtes fier d'être de droite, dites le!}'' (``if you'r proud to be a right-winger, say it!''), or ``\textit{créons une meilleurs Europe pour nos enfants}'' (``Let's create a better Europe for our children''). In order to test the positions of users along our four dimensions, we leverage test data to identify subset of users that have a clear stance on the issues related to these four dimensions with minimal place for ambiguity. In doing so, we will also measure the \textit{sentiment} of each profile text using NLPTown's BERT multilingual uncased sentiment detector\footnote{\url{https://huggingface.co/nlptown/bert-base-multilingual-uncased-sentiment}}. With these elements, we identify the following subset of users to test our dimensions:

\begin{itemize}
    \item \textit{Labeled Right-wing} (for testing dimension LR): users that use the keywork ``\textit{droite}'' (``right'') without negative sentiment (to minimize possible utterances of criticism towards the political right);
    \item \textit{Labeled Left-wing} (for testing dimension LR): users that use the keyword ``\textit{gauche}'' (``left'') without negative sentiment (to minimize possible utterances of criticism towards the political left);
    \item \textit{Labeled pro-Europe} (for testing dimension EU): users that use the keyword ``\textit{Europe}'' or ``\textit{European}'' (in both genders and numbers) without negative sentiment (to minimize possible utterances of criticism towards Europe);
    \item \textit{Labeled anti-elite} (for testing dimension AE): users that use the keyword ``\textit{elite}'' or ``\textit{peuple}'' (``people''). We argue that users that use these keywords subscribe a worldview that opposes two supposedly homogeneous and antagonistic groups, “the elites” and “the people”, one of the most common definitions of populism;
    \item \textit{Labeled patriot} (for testing dimension NA):  users that use the keyword ``\textit{patriote}'' (``patriot'') without negative sentiment (to minimize possible utterances of criticism towards the patriotism). While patriotism and nationalism are not the same conceptually, we argue that people that declare themselves publicly to be patriots in France are also nationalist, solving the issue to finding a keyword to capture the more concept notion of nationalism.
\end{itemize}
Of course, our keywords are not exhaustive: the keyword ``left'' does not account for all possible ways with which left-wing users can express that they are affiliated or sympathizers to the political left. Rather, each our five labeled groups represent a subpopulation that, we argue, has unequivocal skewed position in one of our four political dimensions. In Fig.~S\ref{fig:distributions_flags_dims} we plot the density of these five groups along the dimensions on which we expect them to be skewed to show that, in fact, they are positioned mostly in the corresponding half of the space: e.g., users labeled as Right-wingers, are mostly concentrated to the right of position 0.5, marking the explicitly defined center of the LR dimension. The same is true for the other groups in their corresponding dimensions.

\begin{figure}[ht]
    \centering
    \subfloat[]{\includegraphics[width=0.33\textwidth]{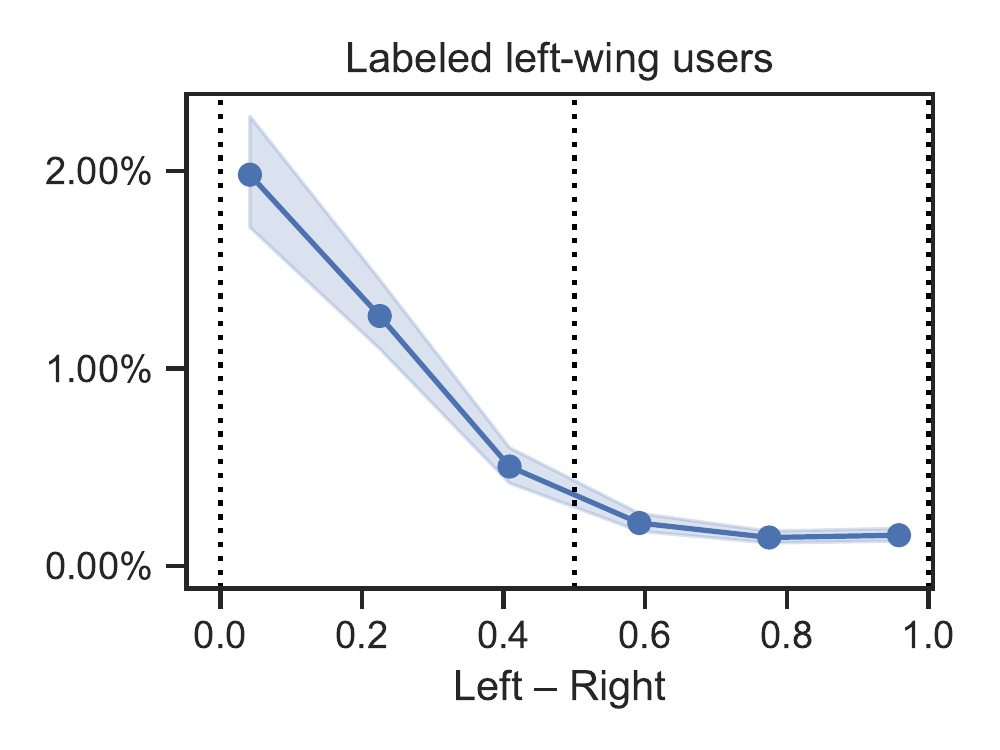}}
    \subfloat[]{\includegraphics[width=0.33\textwidth]{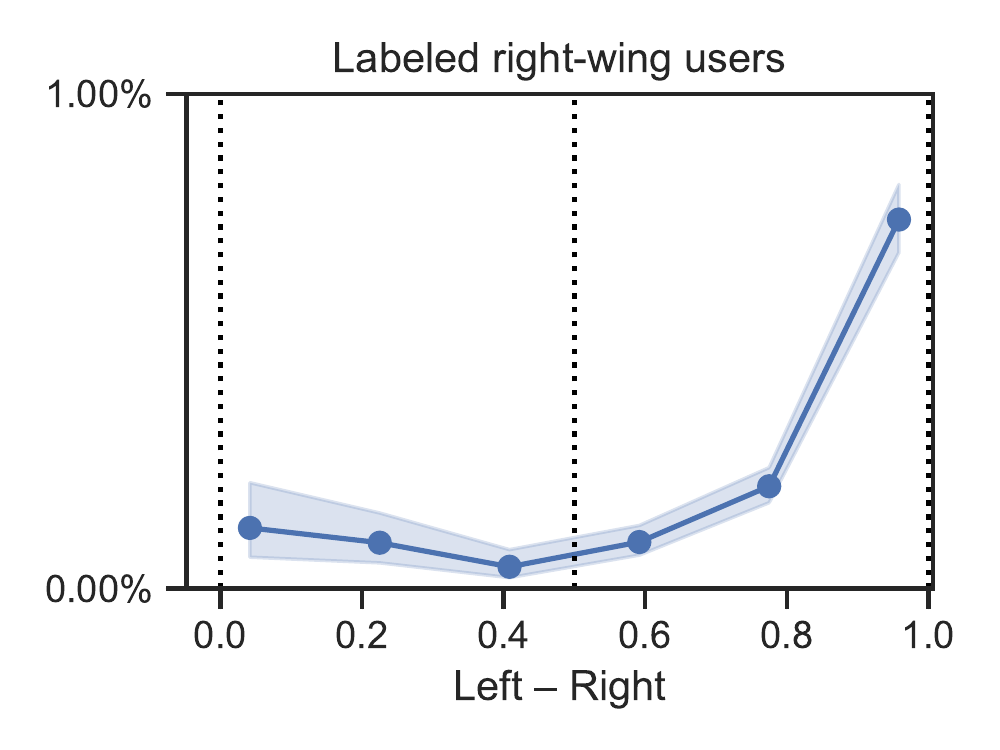}}
    \subfloat[]{\includegraphics[width=0.33\textwidth]{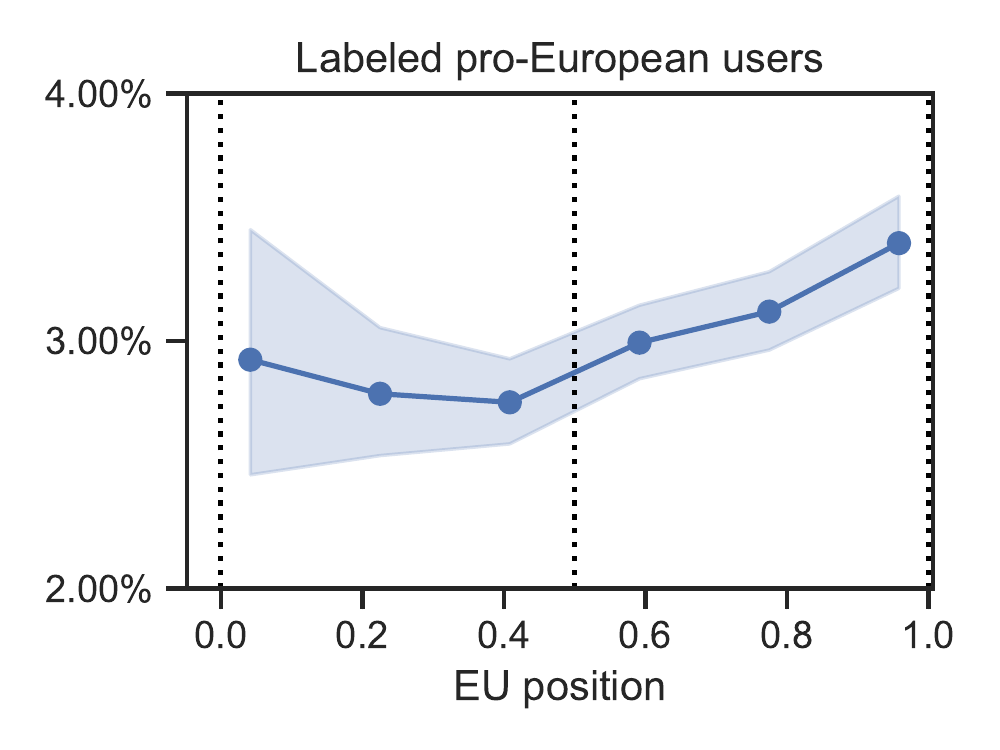}}
    
    \subfloat[]{\includegraphics[width=0.33\textwidth]{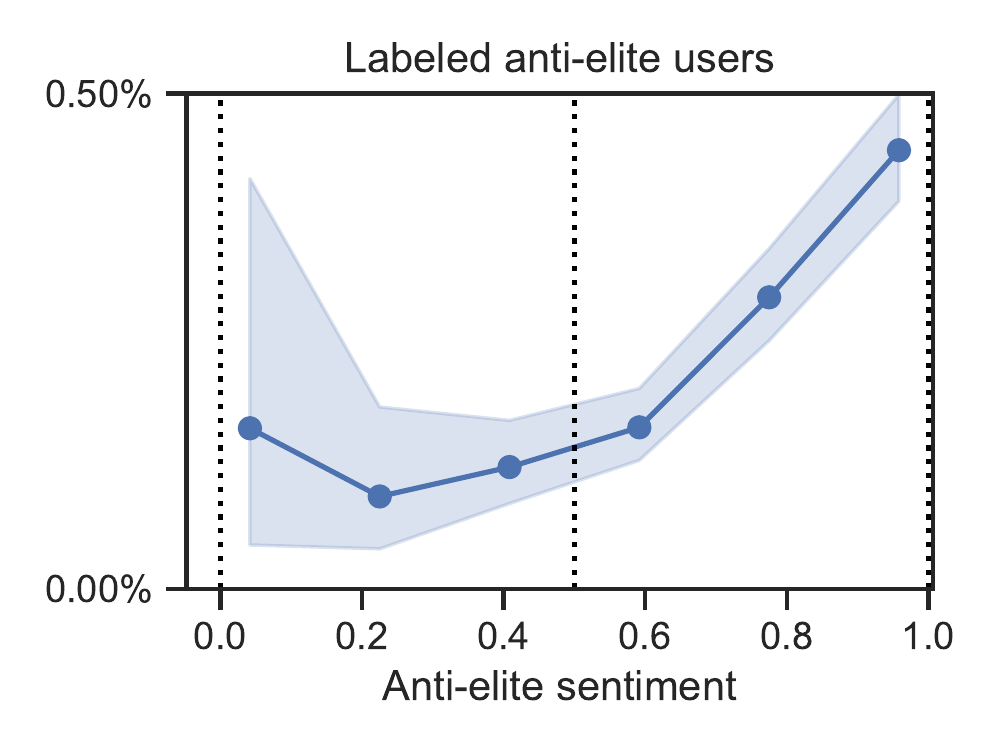}}
    \subfloat[]{\includegraphics[width=0.33\textwidth]{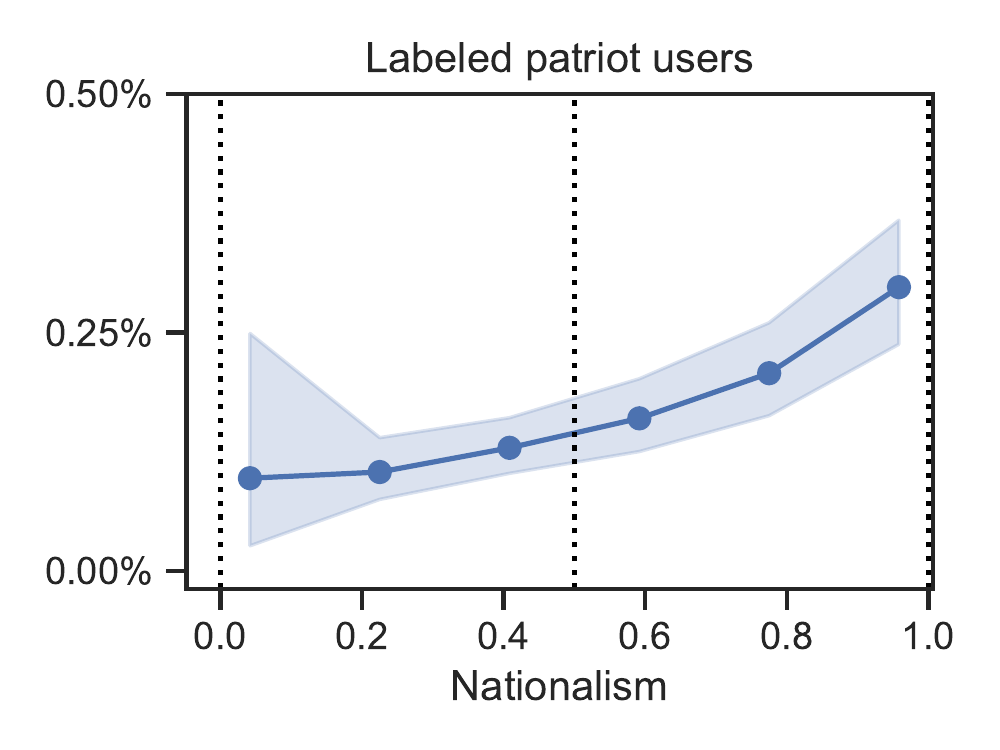}}
    \caption{The 0-1 interval (where most of the users are, and thus where we can obtain a reliable proportion) along each one of the 4 political dimensions divided into 6 bins. For each bin, we compute the proportion of labeled users with respect to the total number of users in the bin. Confidence intervals ($\alpha=0.05$) are computed  using the Clopper-Pearson estimates. The observed density of labeled users grows monotonically with the direction on the corresponding dimension.}
    \label{fig:densities_flags_dims}
\end{figure}

To further assess the quality of the embedded positions we measure the density of our groups of labeled users at different positions along our four different dimensions. To do so, we divide the $0-1$ interval (where most of the users are, and thus where we can obtain a reliable proportion) in 6 bins. On each bin, we compute the proportion of labeled users with respect to the total number of users in the bin. We also compute the confidence intervals using the Clopper-Pearson estimates for $\alpha=0.05$ confidence. As shown in Fig.~S\ref{fig:densities_flags_dims}, the density of labeled users grows with the direction on the corresponding dimension. For example, the density of users labeled as Left-wingers, grows monotonically to the left, i.e., negative LR direction. The expected corresponding trends verify for all 5 groups in our 4 dimensions. 

\subsection{Variable statistics}

\begin{figure}[h!]
\begin{center}
\subfloat[]{\label{opinions_blues:a}\includegraphics[width=0.33\textwidth]{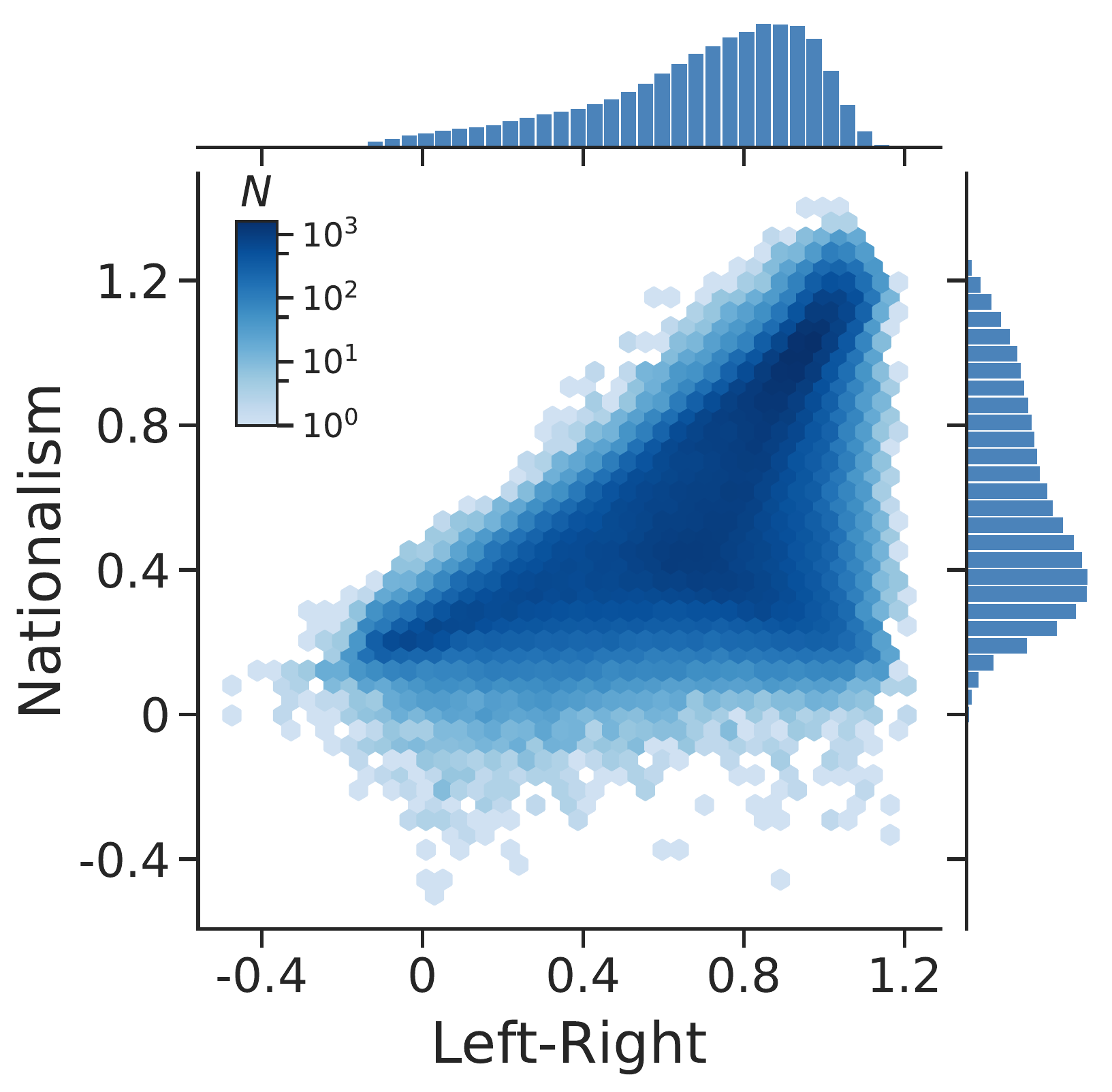}}
\subfloat[]{\label{opinions_blues:b}\includegraphics[width=0.33\textwidth]{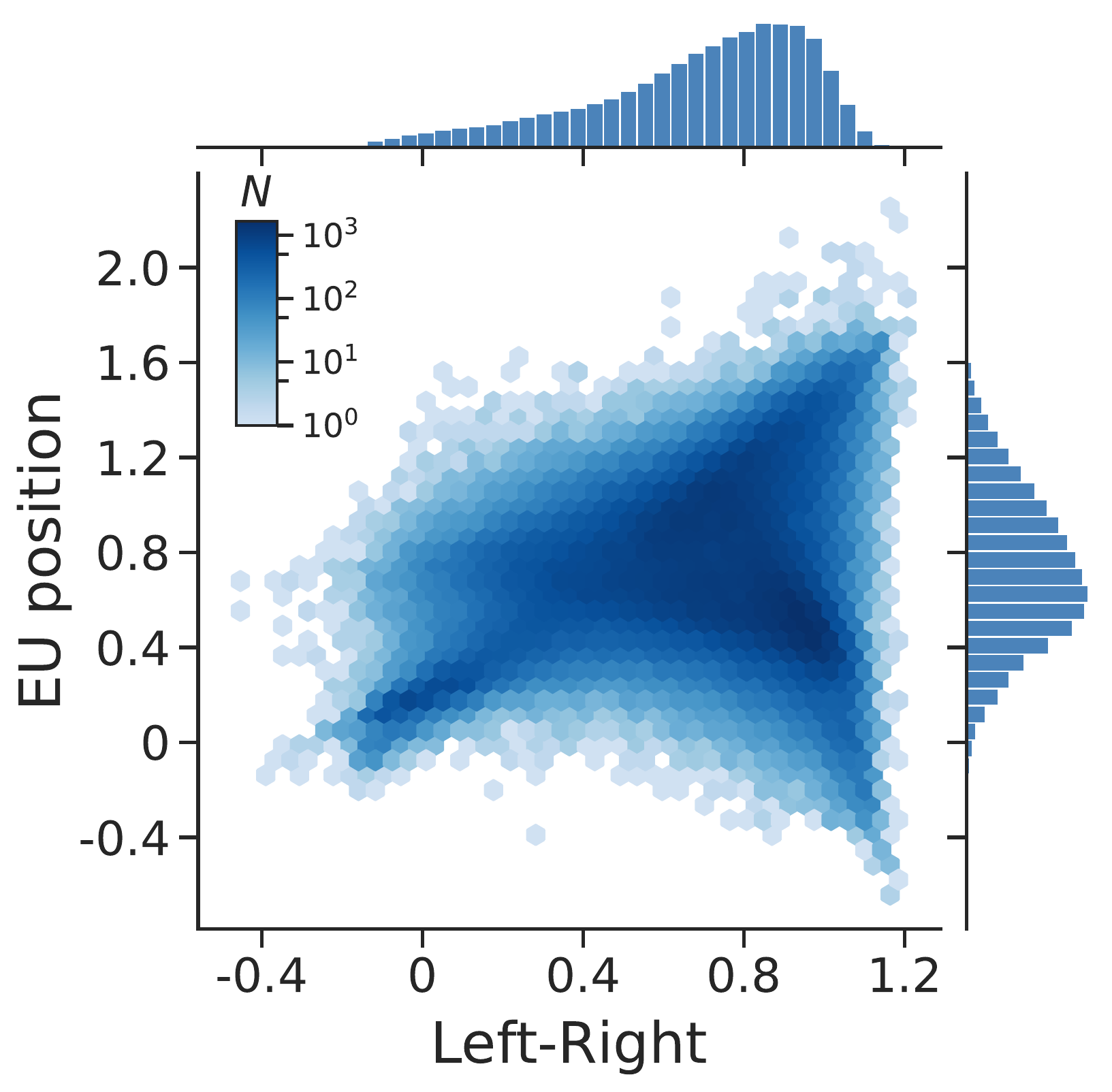}}
\subfloat[]{\label{opinions_blues:c}\includegraphics[width=0.33\textwidth]{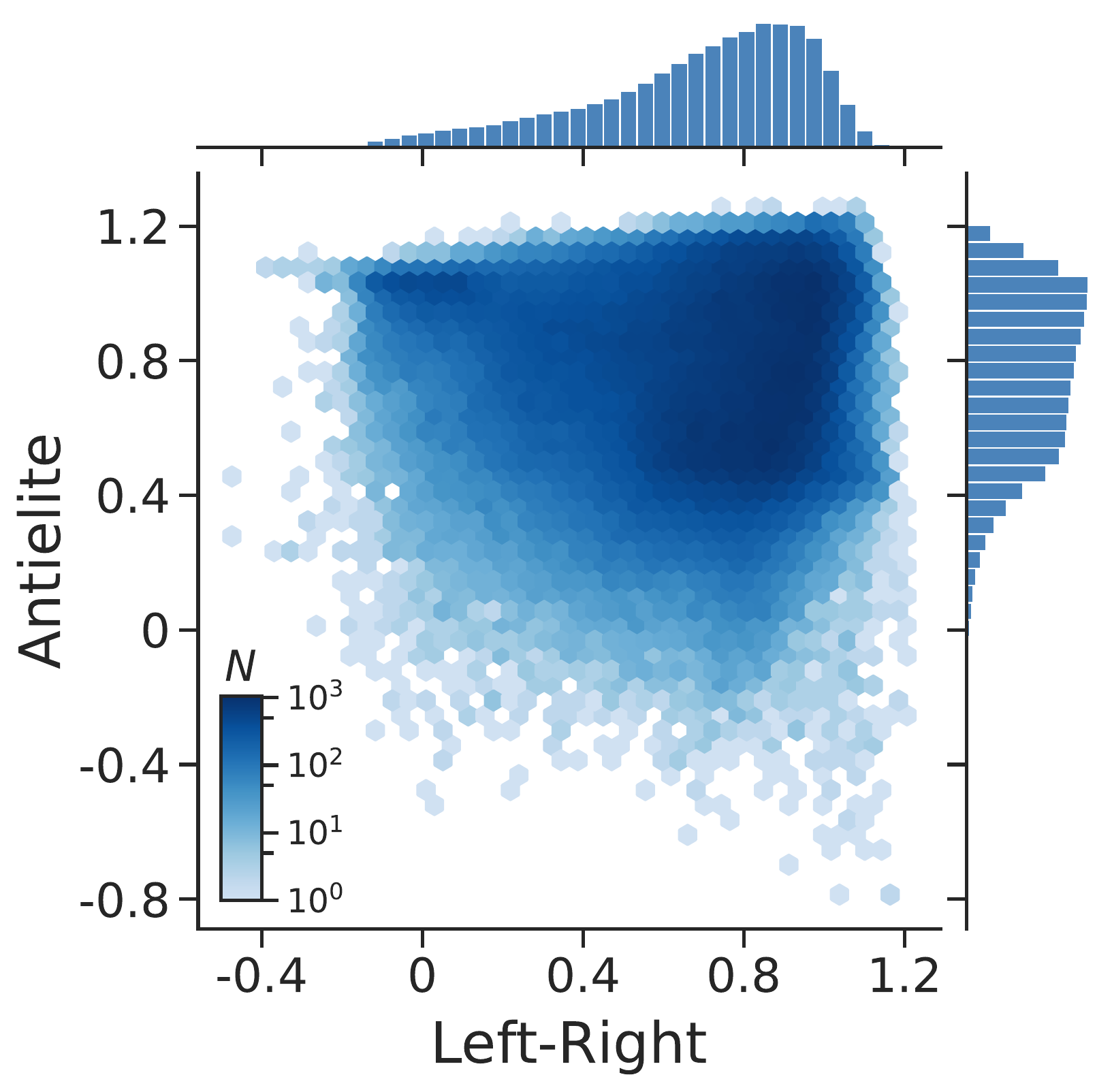}}

\subfloat[]{\label{opinions_blues:d}\includegraphics[width=0.33\textwidth]{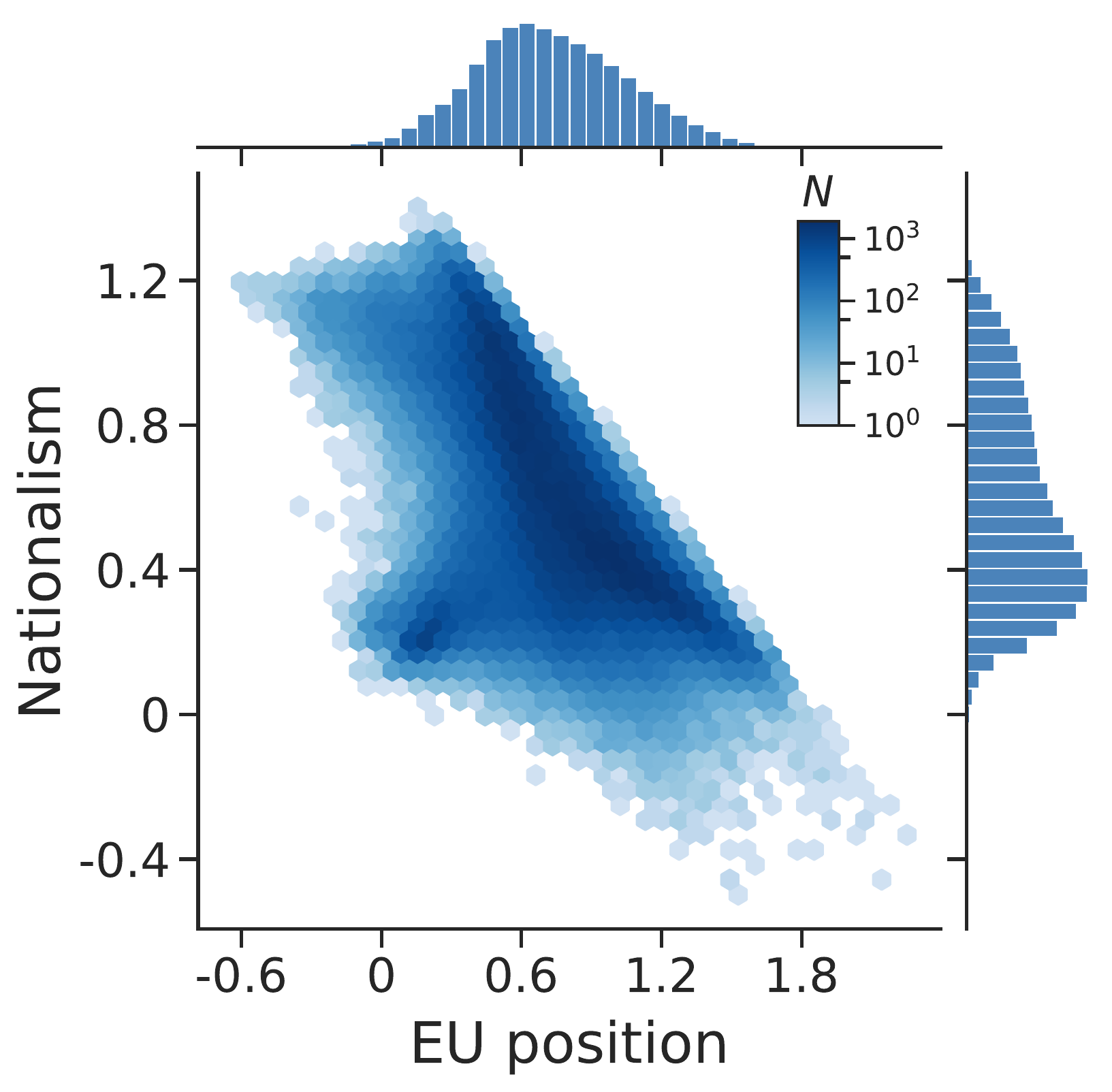}}
\subfloat[]{\label{opinions_blues:e}\includegraphics[width=0.33\textwidth]{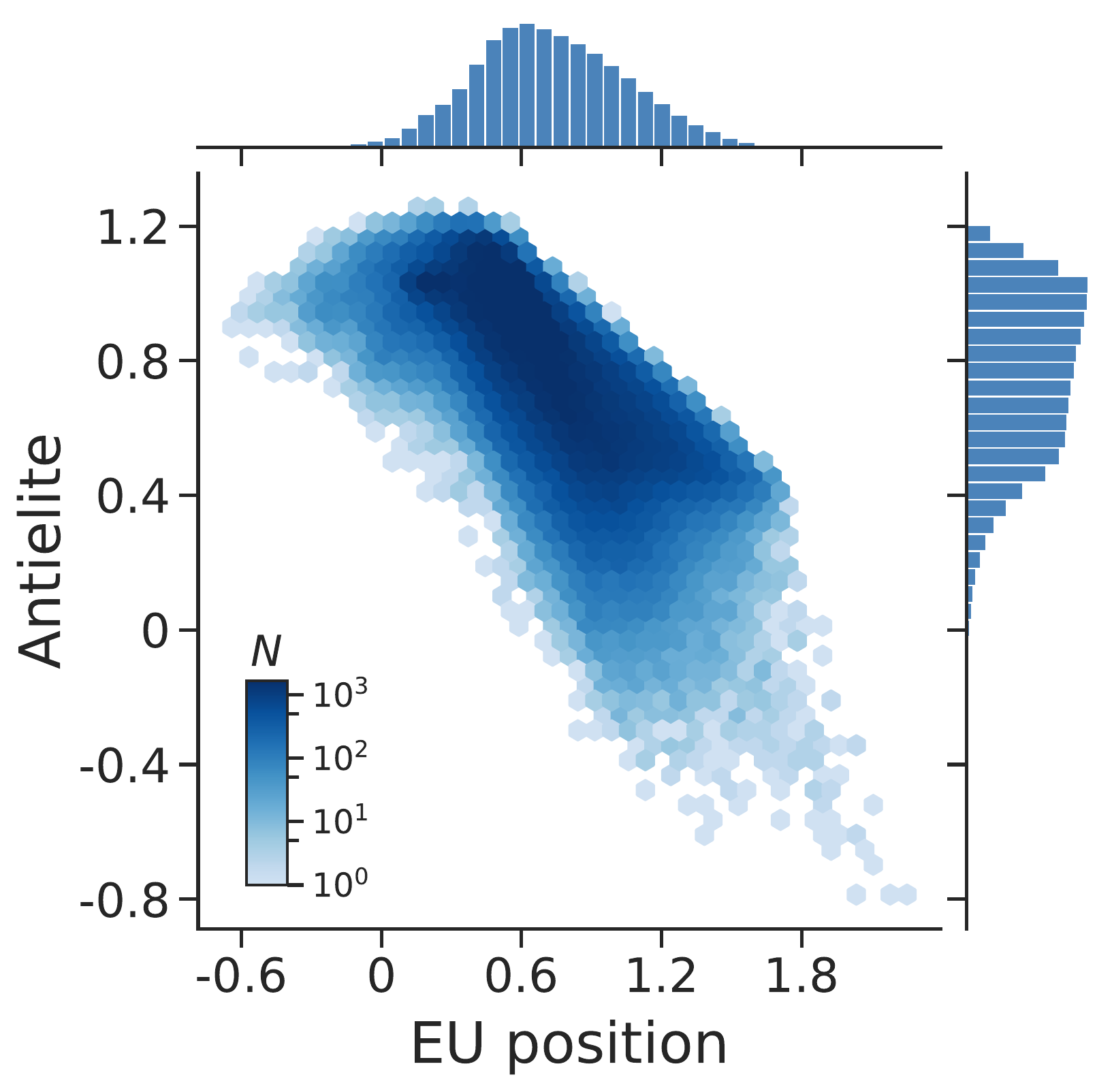}}
\subfloat[]{\label{opinions_blues:f}\includegraphics[width=0.33\textwidth]{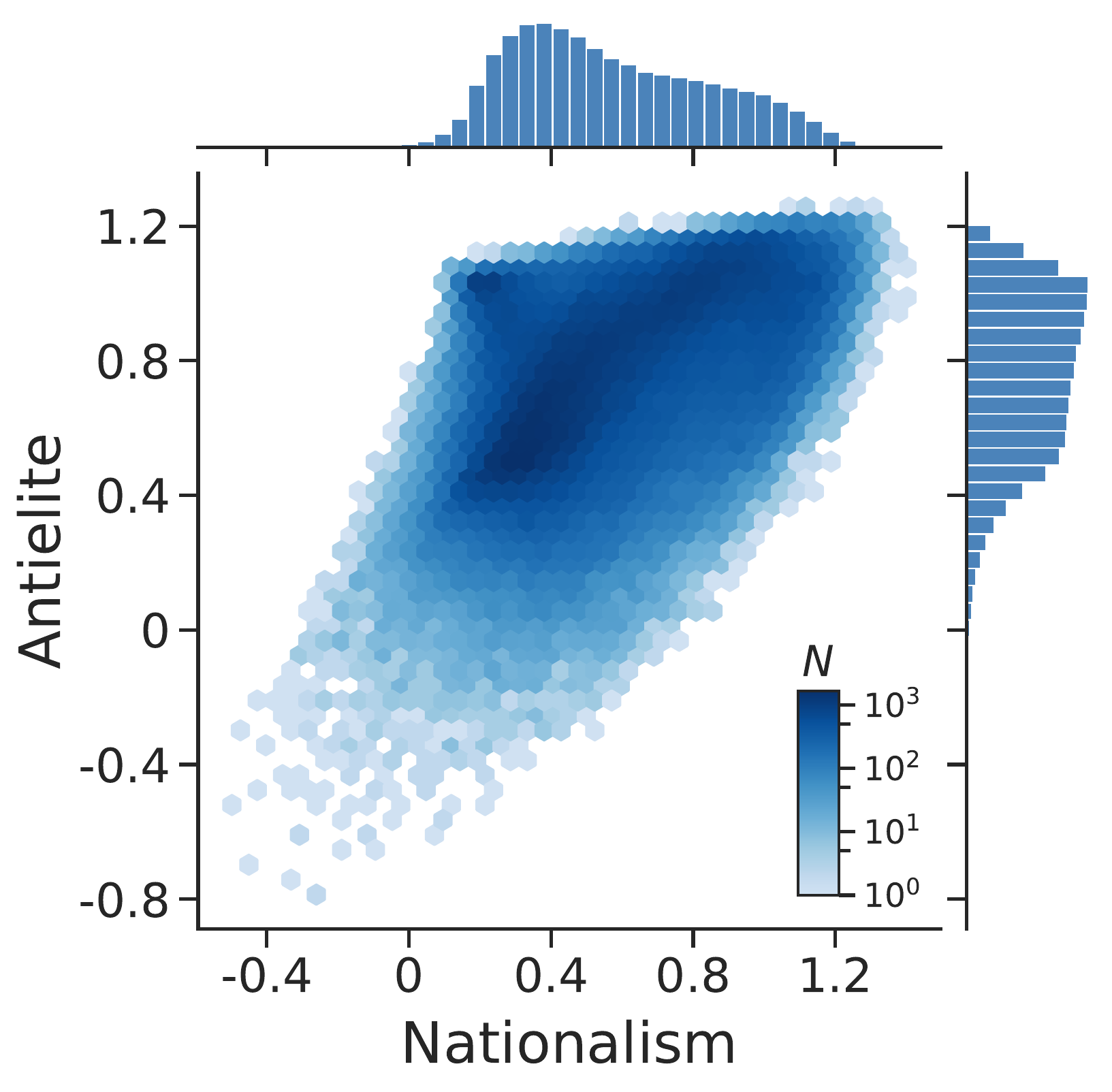}}
\caption{Number of users (N) with opinion variables at a particular position (hexbins) in a two dimensional state space for all possible pair of variables: (a) NA vs LR, (b) EU vs LR, (c) AE vs LR, (d) NA vs EU, (e) AE vs EU, (f) AE vs NA, hexagonal logarithmic binning is used for visualization purposes with gridsize = 40. The marginal distributions of each separate variable are shown in the corresponding margin of each subfigure.}
\label{opinions_blues}
\end{center}
\end{figure}

\begin{figure}[h!]
\begin{center}
\subfloat[]{\label{opinions_blues_mps:a}\includegraphics[width=0.33\textwidth]{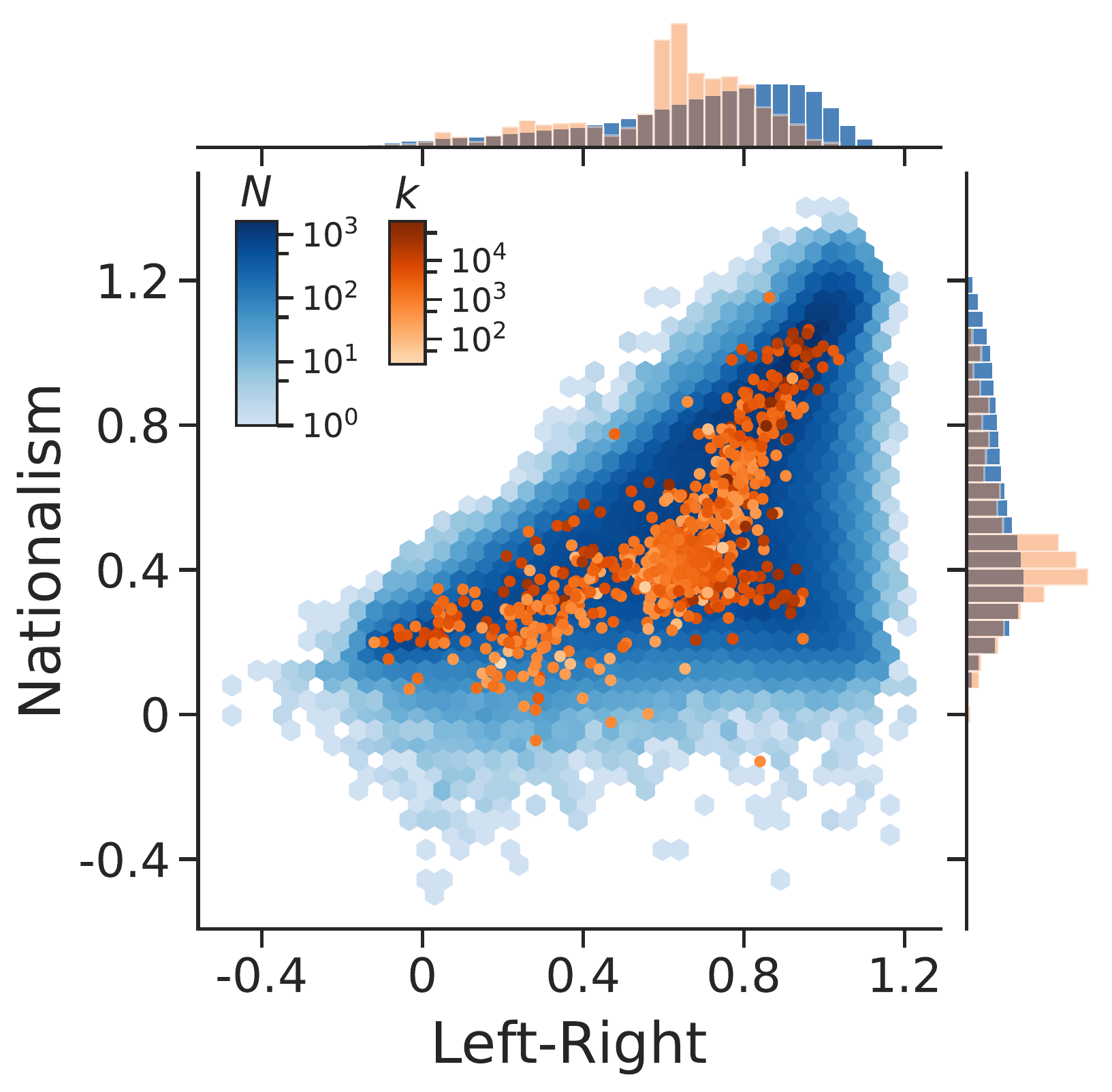}}
\subfloat[]{\label{opinions_blues_mps:b}\includegraphics[width=0.33\textwidth]{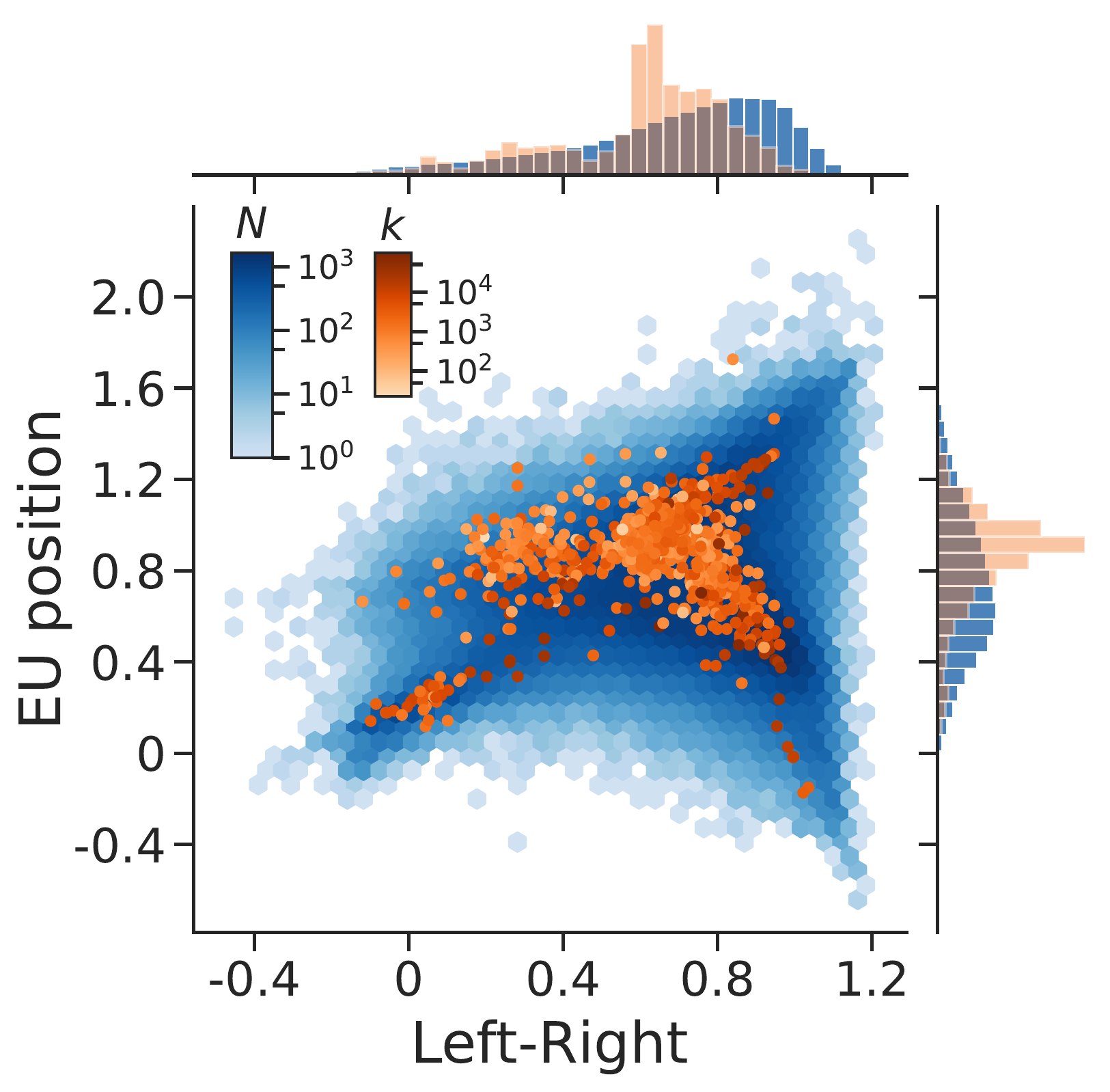}}
\subfloat[]{\label{opinions_blues_mps:c}\includegraphics[width=0.33\textwidth]{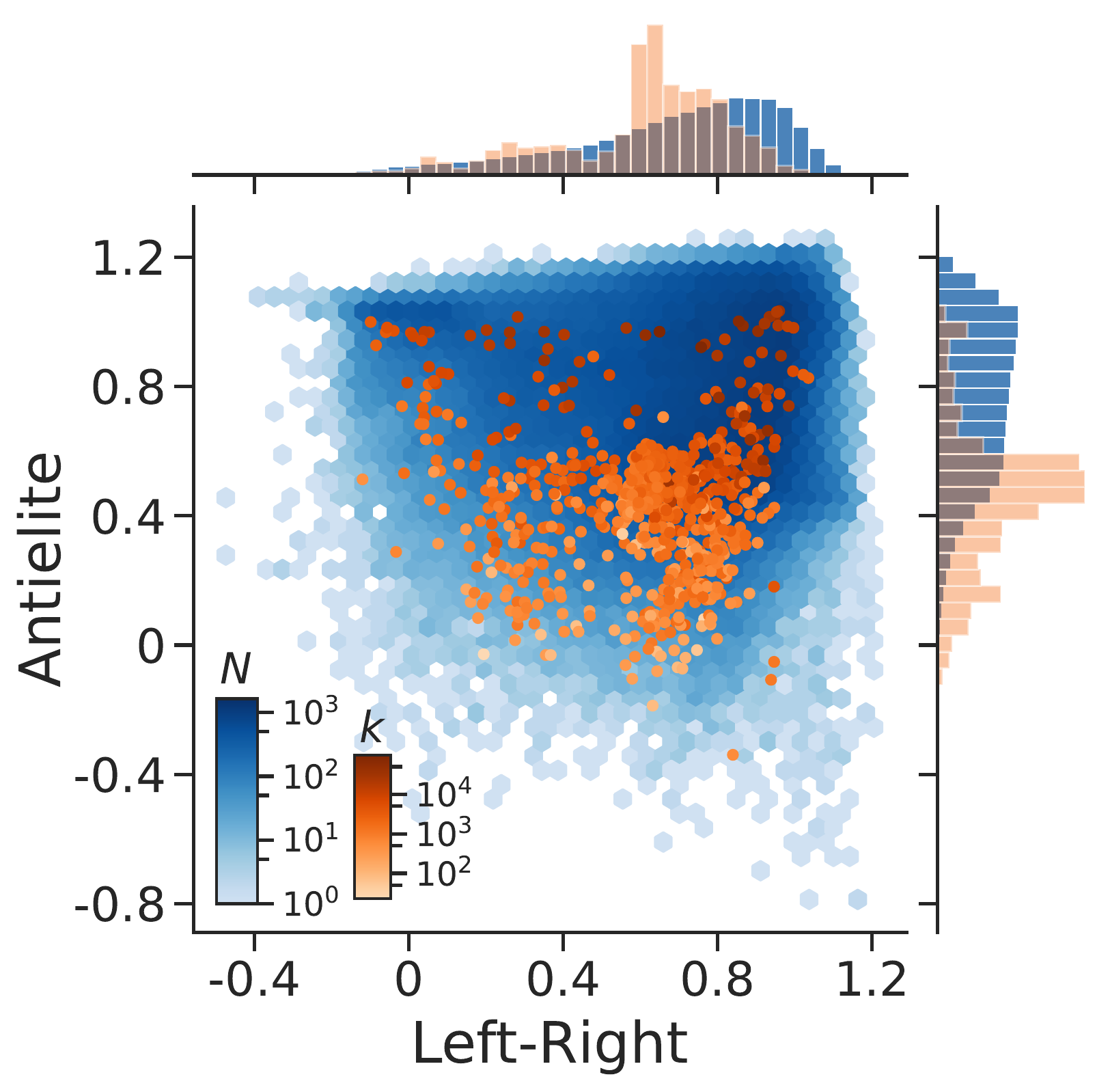}}

\subfloat[]{\label{opinions_blues_mps:d}\includegraphics[width=0.33\textwidth]{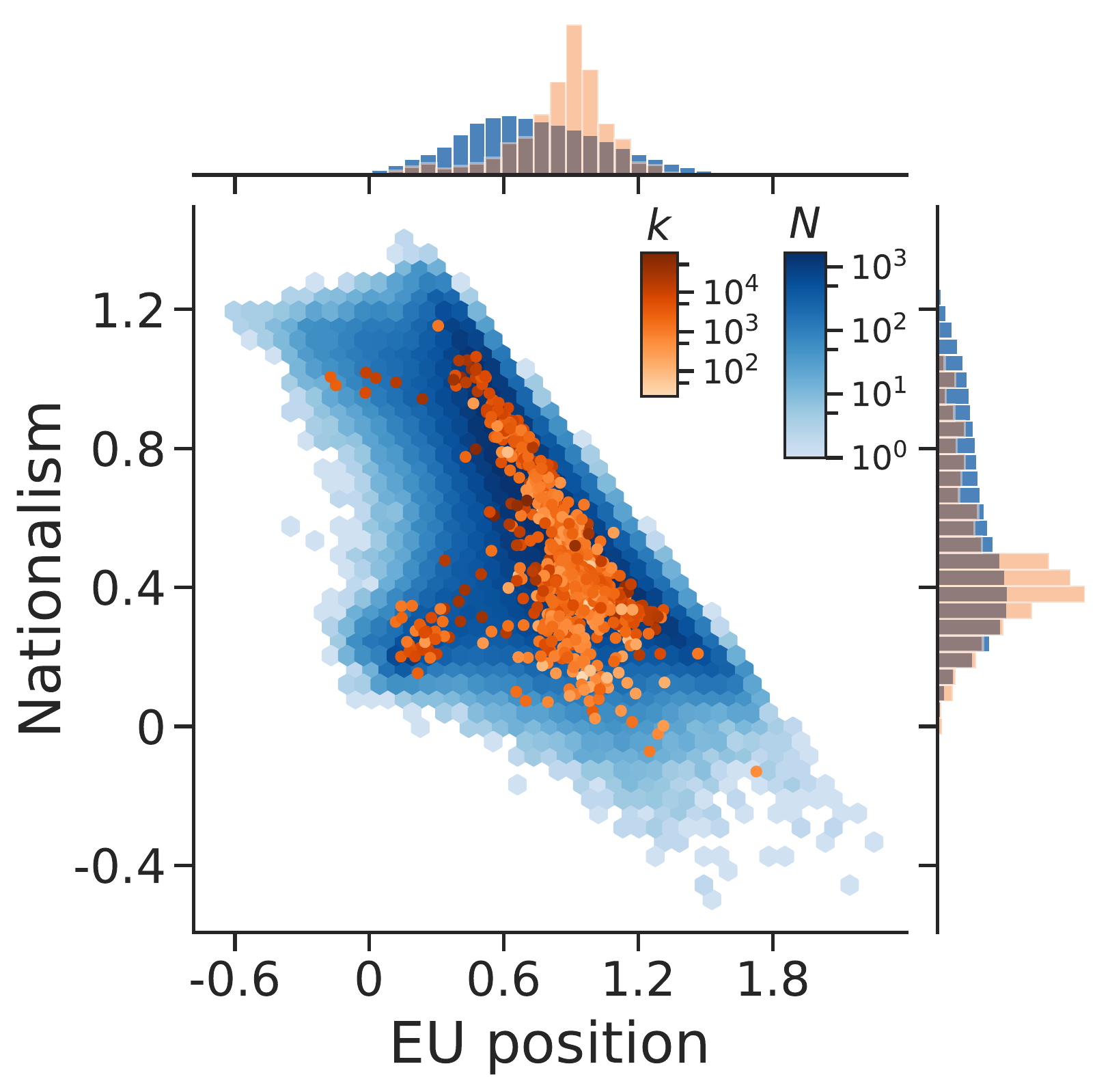}}
\subfloat[]{\label{opinions_blues_mps:e}\includegraphics[width=0.33\textwidth]{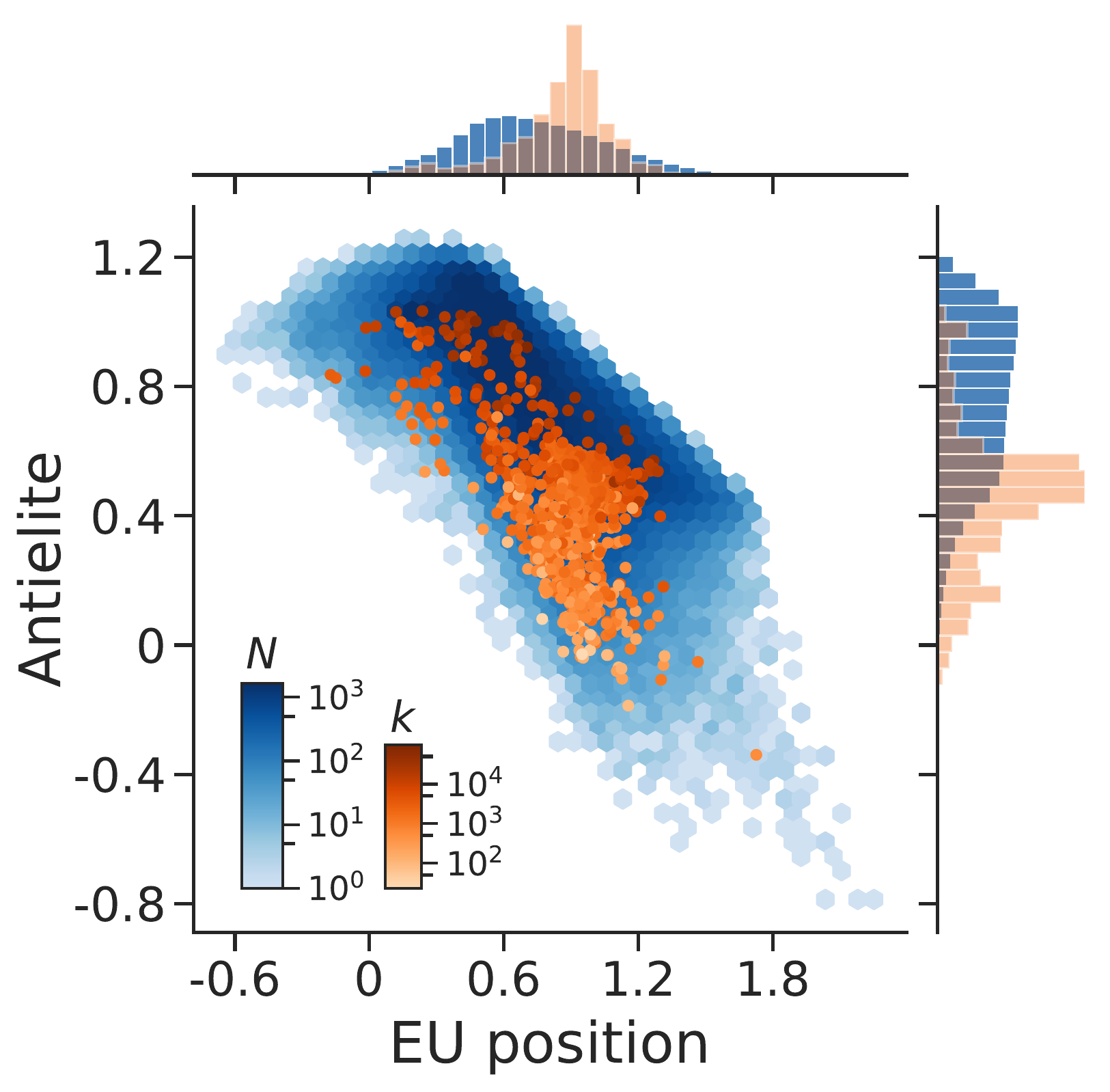}}
\subfloat[]{\label{opinions_blues_mps:f}\includegraphics[width=0.33\textwidth]{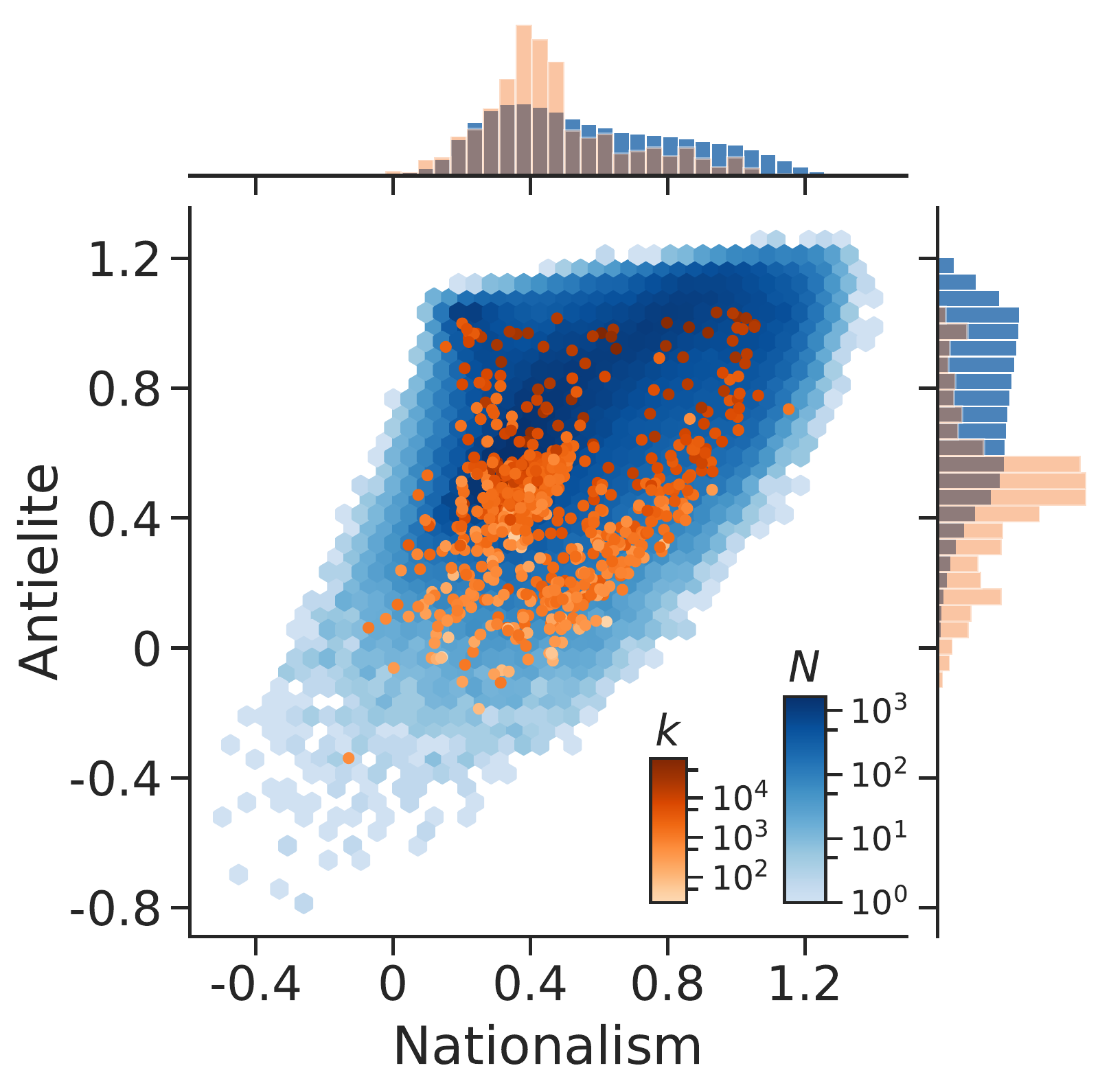}}
\caption{Number of users ($N$) with state variables in a two dimensional space in blue using hexagonal logarithmic binning, and state variables of MPs in orange for all possible pairs of variables: (a) NA vs LR, (b) EU vs LR, (c) AE vs LR, (d) NA vs EU, (e) AE vs EU, (f) AE vs NA. The color (orange) shading of the orange points is associated to the in-degree $k^{\text{in/um}}$ ($k$) in the User $\rightarrow$ MP subgraph in logarithmic scale. The marginal distributions of each separate dimension are shown in the corresponding margin of each subfigure, in blue for the users and in orange for the MPs.}
\label{opinions_blues_mps}
\end{center}
\end{figure}

In this section we show some basic statistical properties of the state (opinion) of users and MPs, captured by our four real variables (LR, NA, EU, AE). In Fig. S\ref{opinions_blues} we show the distribution of the opinion variables of users, i.e., for each variable separately with the marginal distributions and also for the correlations of all possible pairs of variables. In Fig. S\ref{opinions_blues_mps} we reproduce the same figure (Fig. S\ref{opinions_blues}) but now we include, as a matter of comparison, the opinion variables of MPs organised by ``popularity'' (orange color), i.e., number of followers $k^{\text{in/um}}$ of the MPs. These figures confirm in more detail (for all pairs of variables) the line of reasoning given in the main text, i.e., the explanation of the relation between the opinion of the users, that of the MPs, and the network structure, as well as the explanation of the differences in the opinion distributions of MPs and users.

In Table S\ref{tabcorrelations} we computed some measures of the correlations between variables for users and MPs by means of the Pearson correlation coefficient ($r$) and the Kendall rank correlation coefficient ($\tau$). These results indicate that some correlations exist between variables, which may be explained by the presence of ``ideologies'' \cite{Baumann:2021} that synthesize the opinions of individuals in different, in principle, uncorrelated topics in a single unified group of ideas. The correlations between variables are also a reflect of the structural correlations in the interactions between individuals. Note the parallelism between the correlation of variables of the MPs and users, that result to be very similar and which indicate that the same ideologies are shared between MPs and users.

\begin{table}[ht]
\centering
\caption{Correlations (Pearson, $r$, and Kendall, $\tau$) of all possible pairs of opinion variables (LR vs NA, LR vs EU, LR vs AE, EU vs NA, EU vs AE, and NA vs AE) of users and MPs.}
\begin{tabular}{l|c|c|c|c|c|c}
  & LR vs NA & LR vs EU & LR vs AE & EU vs NA & EU vs AE & NA vs AE \\
\midrule
 $r$ (Users) & $0.61$ & $0.21$ & $-0.01$ & $-0.57$ & $-0.73$ & $0.52$   \\
 $r$ (MPs) & $0.69$ & $0.22$ & $-0.07$ & $-0.47$ & $-0.60$ & $0.31$   \\
 \midrule
 $\tau$ (Users) & $0.44$ & $0.07$ & $0.02$ & $-0.43$ & $-0.57$ & $0.36$   \\
 $\tau$ (MPs) & $0.53$ & $0.05$ & $0.00$ & $-0.40$ & $-0.32$ & $0.16$   \\
\bottomrule
\end{tabular}
\label{tabcorrelations}
\end{table}

\section{Communities}

In order to find communities in the network and identify the best partition in terms of assortative groups we use the nonparametric Bayesian formulation of the planted partition model (PP) \cite{Zhang:2020}, a version of the more general stochastic block model (SBM). It is not our intention here to describe the details of this model, neither to give an extensive explanation of why it is more convenient than other possibilities in the literature. We will just clarify the most important features and advantages of this partition algorithm, how we apply it to our network and understand the results, for more details about the partition algorithm and model the reader should refer to the original reference \cite{Zhang:2020}. We decided to use the (PP) because it considers assortativity, i.e., the tendency of the nodes of a group to be connected to other nodes of the same group (homophily), as the main mixing pattern in the network. The (PP) simplifies the analysis as compared to the more general (SBM), where other possible mixing patterns are also included, and also allow us to focus on the relation between assortative structural patterns and the political positions of users and MPs. The (PP) is more convenient than other algorithms such as modularity maximization because avoids overfitting and underfitting problems while, at the same time, it can find an arbitrarily large number of communities as long as there is enough statistical evidence of them, without a prior knowledge of the optimum number.

\begin{figure}[h!]
\begin{center}
\subfloat[]{\label{communities_mps:a}\includegraphics[width=0.40\textwidth]{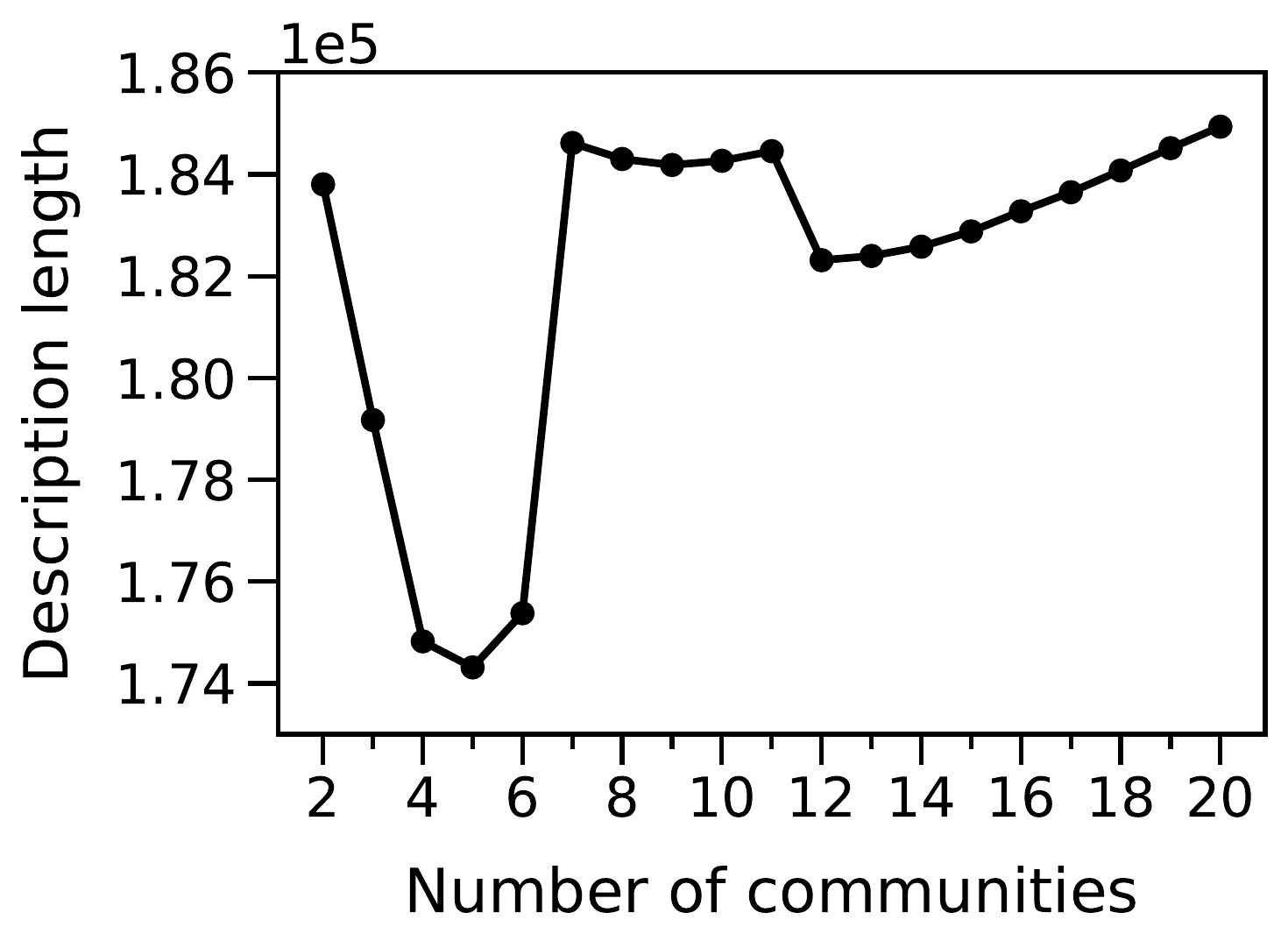}}
\hspace{0.5cm}
\subfloat[]{\label{communities_mps:b}\includegraphics[width=0.23\textwidth]{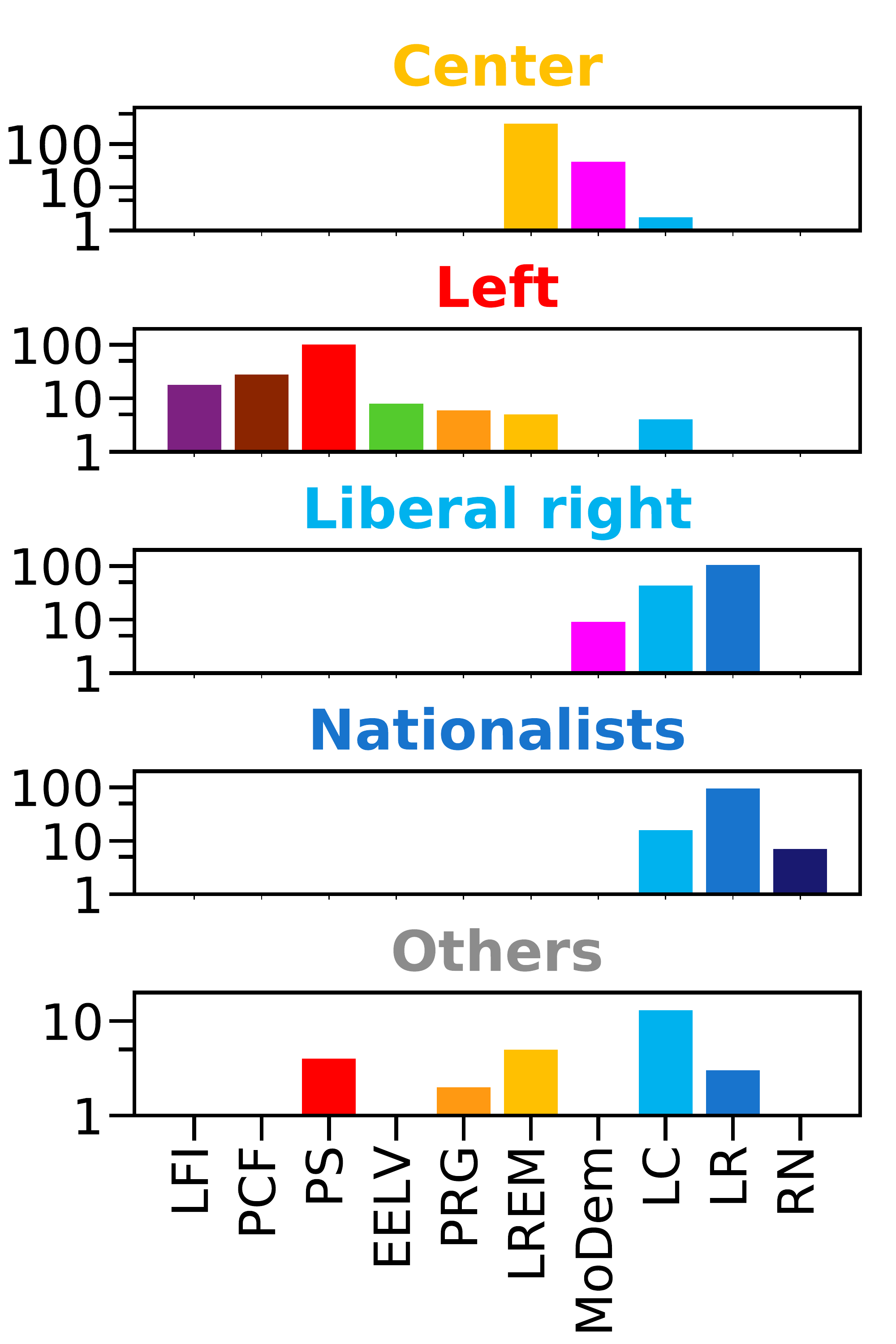}}

\subfloat[]{\label{communities_mps:c}\includegraphics[width=0.35\textwidth]{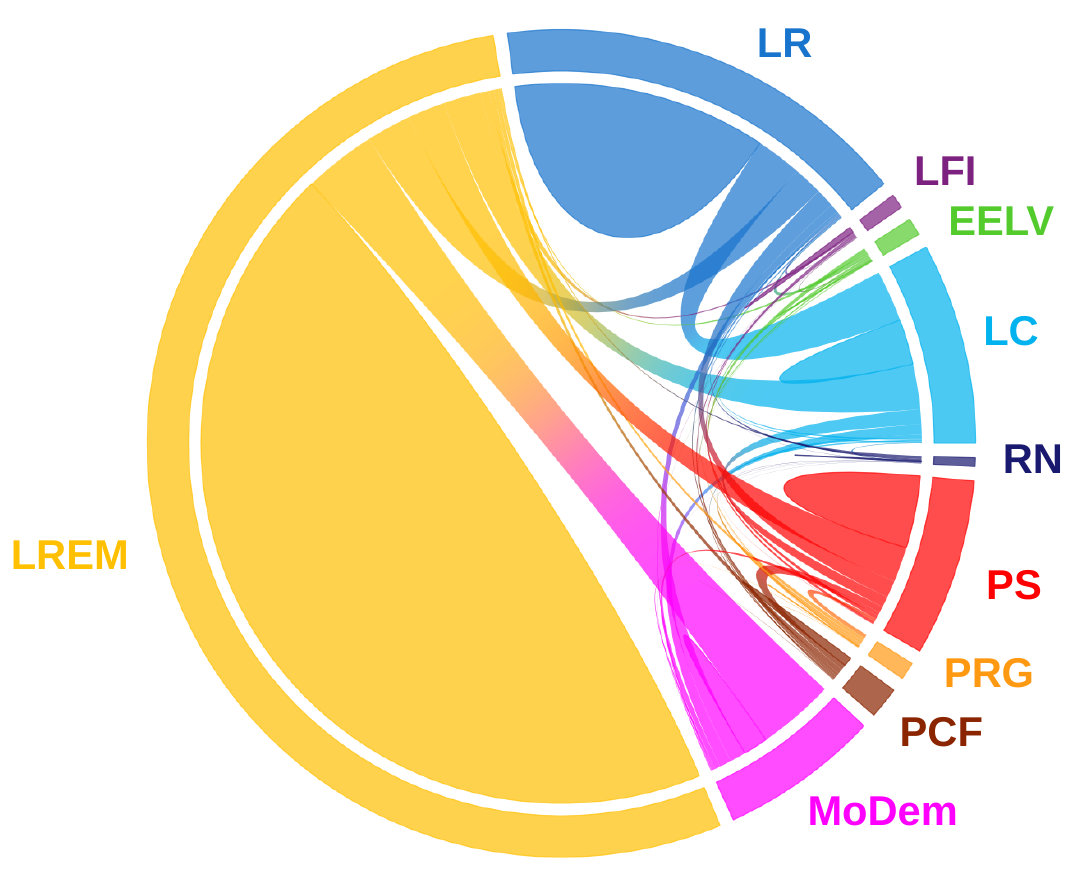}}
\subfloat[]{\label{communities_mps:d}\includegraphics[width=0.35\textwidth]{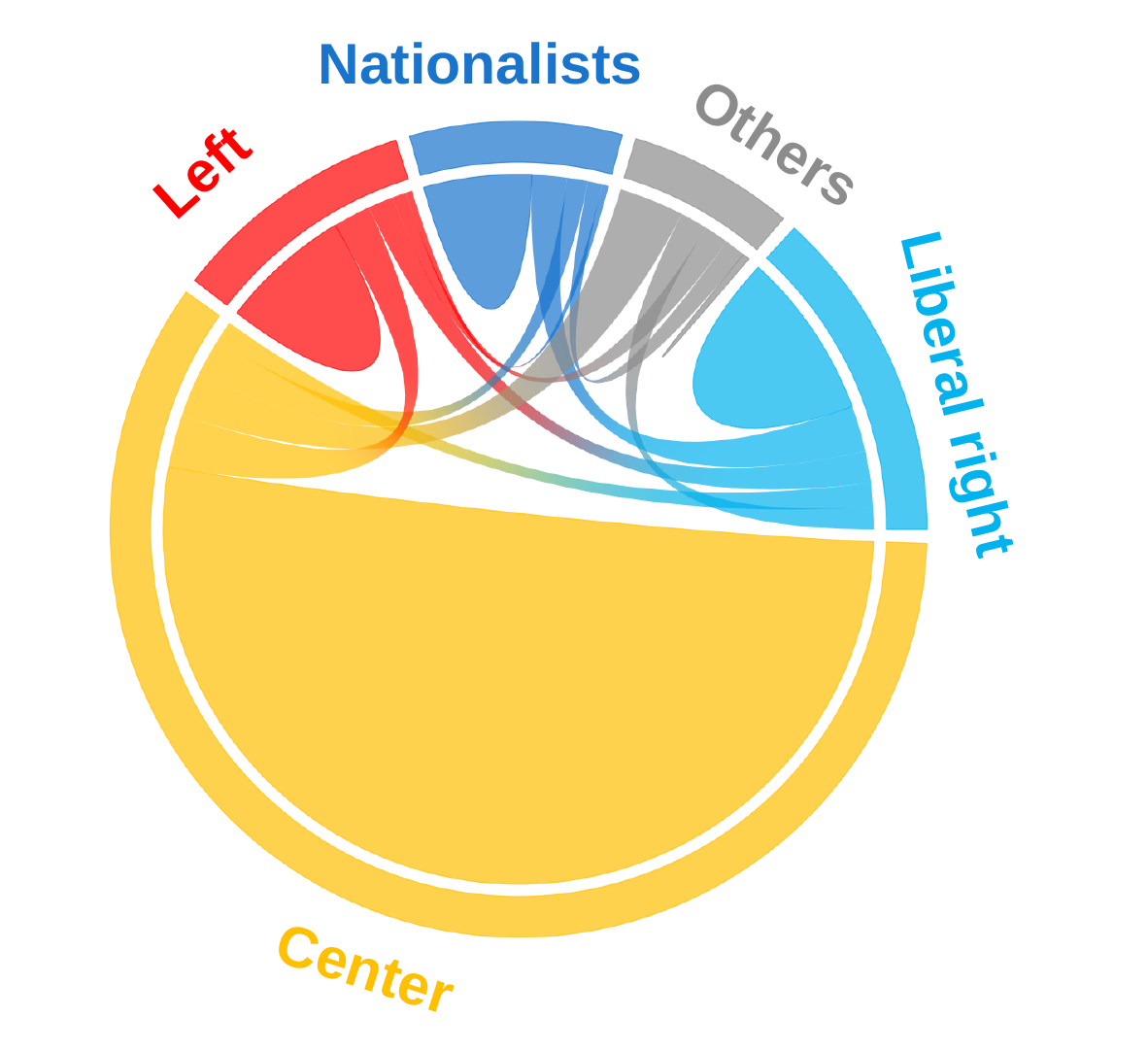}}
\caption{Panel (a) description length of the planted partition community detection model as a function of the number of communities in the MP $\rightarrow$ MP subgraph. Panel (b) party composition of the minimum description length result for communities, i.e., number of MPs (vertical axis) belonging to each party (horizontal axis) for each community separately (rows). Panel (c,d) chord diagrams indicating the connectivity (number of links) between/inside parties (c) and communities (d) for the minimum description length partition. The angular size of each party/community in the diagram is proportional to the number links that depart from that party/community, the party/community of destination of the links is indicated in the chord.}
\label{communities_mps}
\end{center}
\end{figure}

Given the distinction between the two types of nodes that we use as a basis, i.e., MPs and users, we develop our analysis of the community structure of the network in two steps in a specific order: (i) we apply the (PP) on the MP $\rightarrow$ MP subgraph without restricting the number of communities and infer the optimum number, and (ii) we apply the (PP) on the User $\rightarrow$ User subgraph restricting the number of communities according to step (i). The reason for applying the algorithm in this way is to simplify the analysis by first understating the structure of the politicians network, comparing altogether the communities obtained with party membership of their constituents, and then establish a parallelism (if possible) between the MPs and users communities. Note that the optimum number of communities in the User $\rightarrow$ User subgraph, without restricting it, is much larger than for the MP $\rightarrow$ MP. However, our criterion of restricting the number of communities in the User $\rightarrow$ User subgraph does not prevent us from studying the heterogeneities in the connectivity patterns of the users of a same group, as we carry out in section \ref{sec_results_comm}. In Fig. S\ref{communities_mps} we show the results of step (i), i.e, the community detection of the MP $\rightarrow$ MP network. The description length, i.e., a measure of the goodness of the community detection model \cite{Zhang:2020}, as a function of the number of communities has a minimum at 5, which corresponds to the best partition. This result is the one that we show in the main text, where we also discuss in detail the name and composition of each community. The two second best results are 4 and 6 communities which are very similar to 5, the differences being: for 4 communities the ``Liberal right'' and ``Nationalist'' groups are merged together, and for 6 communities we find two different non-assortative groups instead of one (``Others'').

\begin{figure}[h!]
\begin{center}
\subfloat[]{\label{opinion_mps:a}\includegraphics[width=0.33\textwidth]{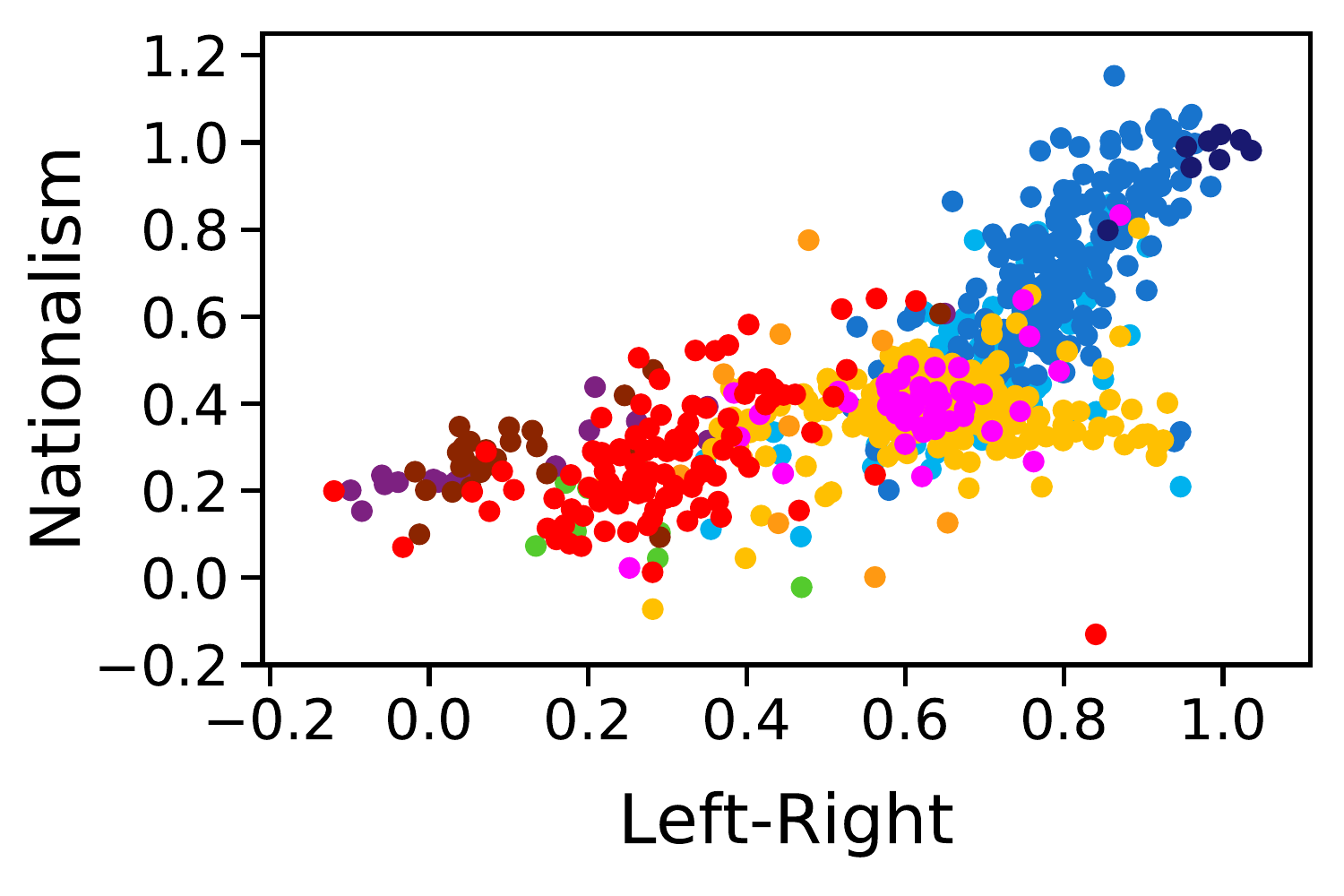}}
\subfloat[]{\label{opinion_mps:b}\includegraphics[width=0.33\textwidth]{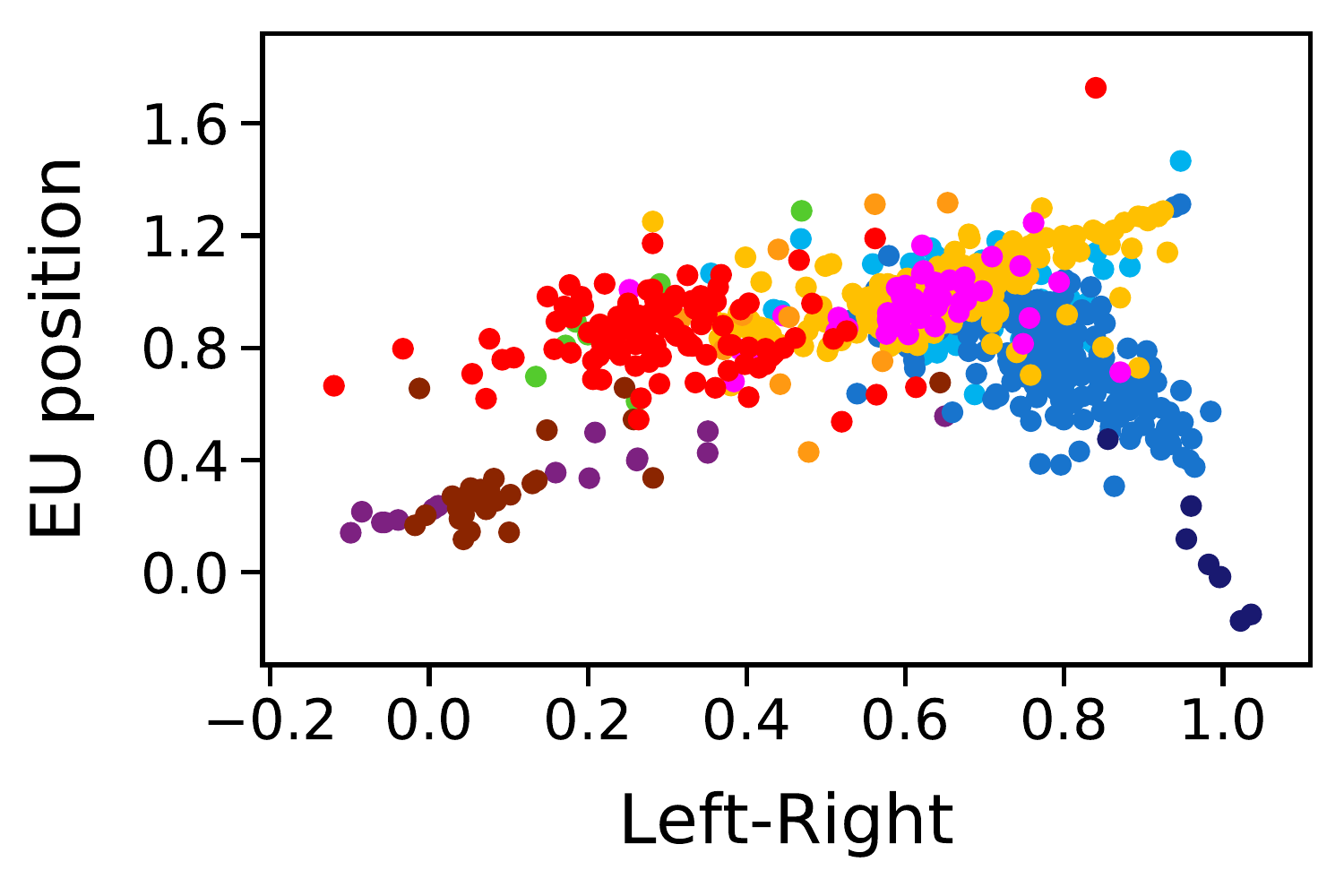}}
\subfloat[]{\label{opinion_mps:c}\includegraphics[width=0.33\textwidth]{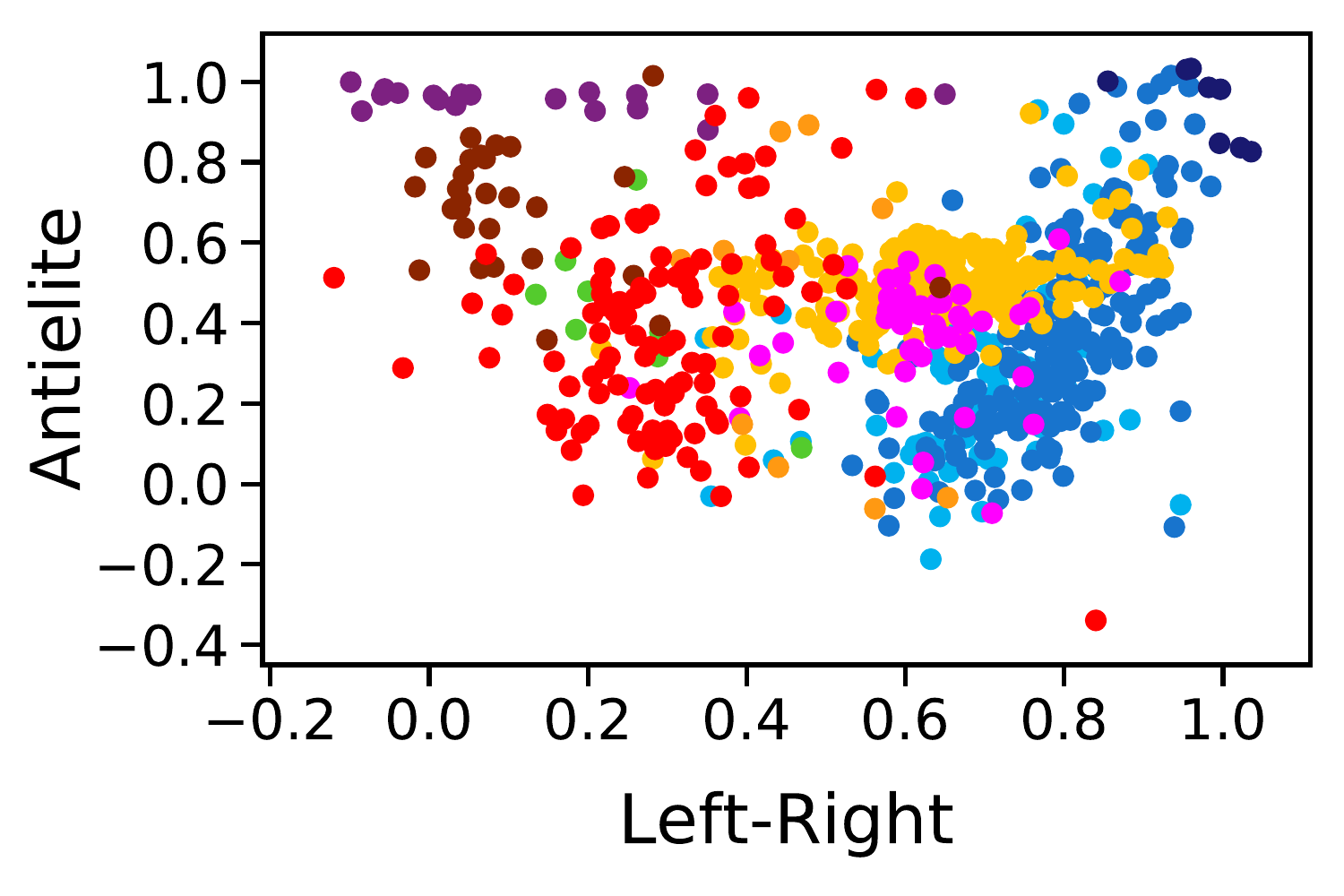}}

\vspace*{0.10cm}
\subfloat[]{\label{opinion_mps:d}\includegraphics[width=0.33\textwidth]{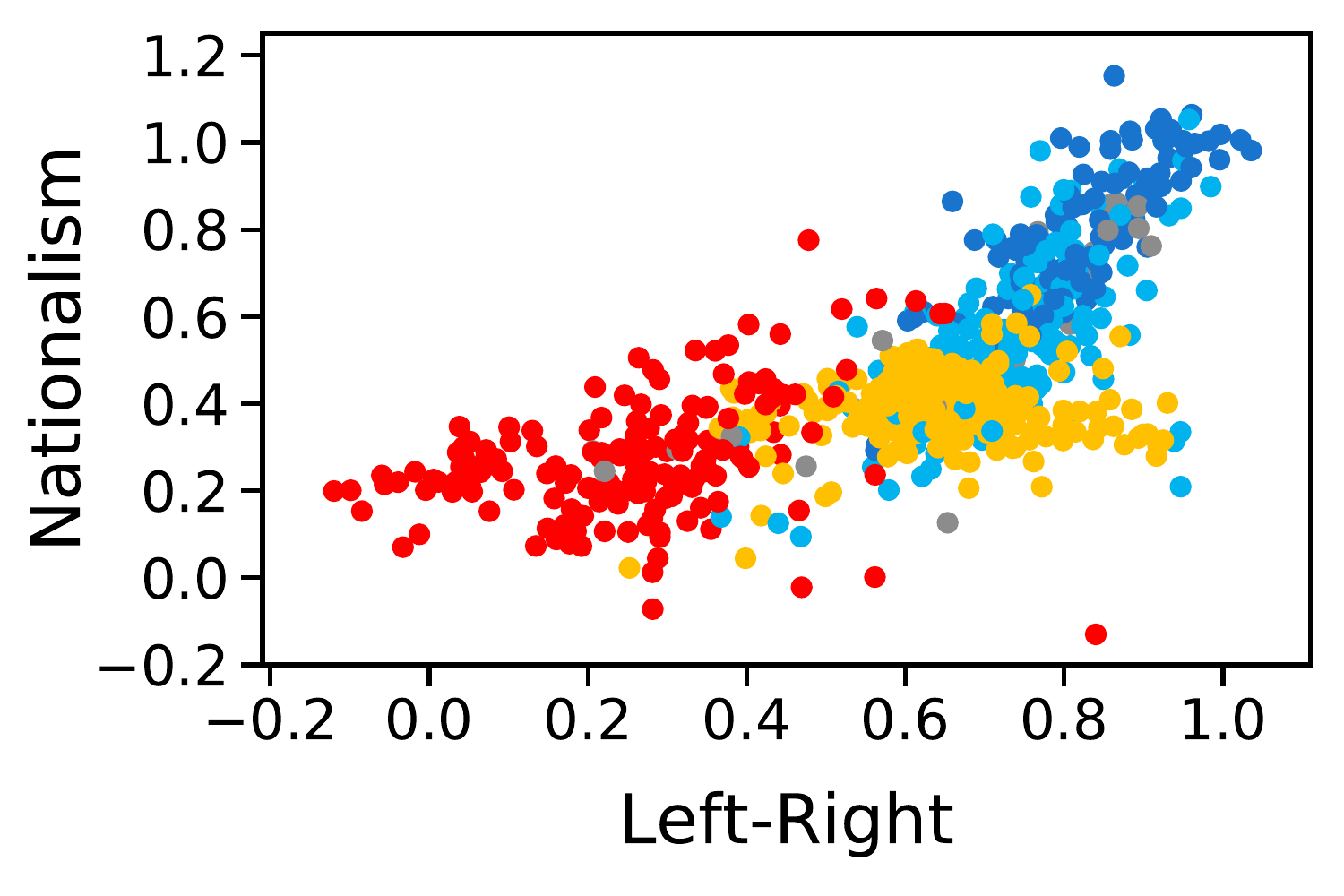}}
\subfloat[]{\label{opinion_mps:e}\includegraphics[width=0.33\textwidth]{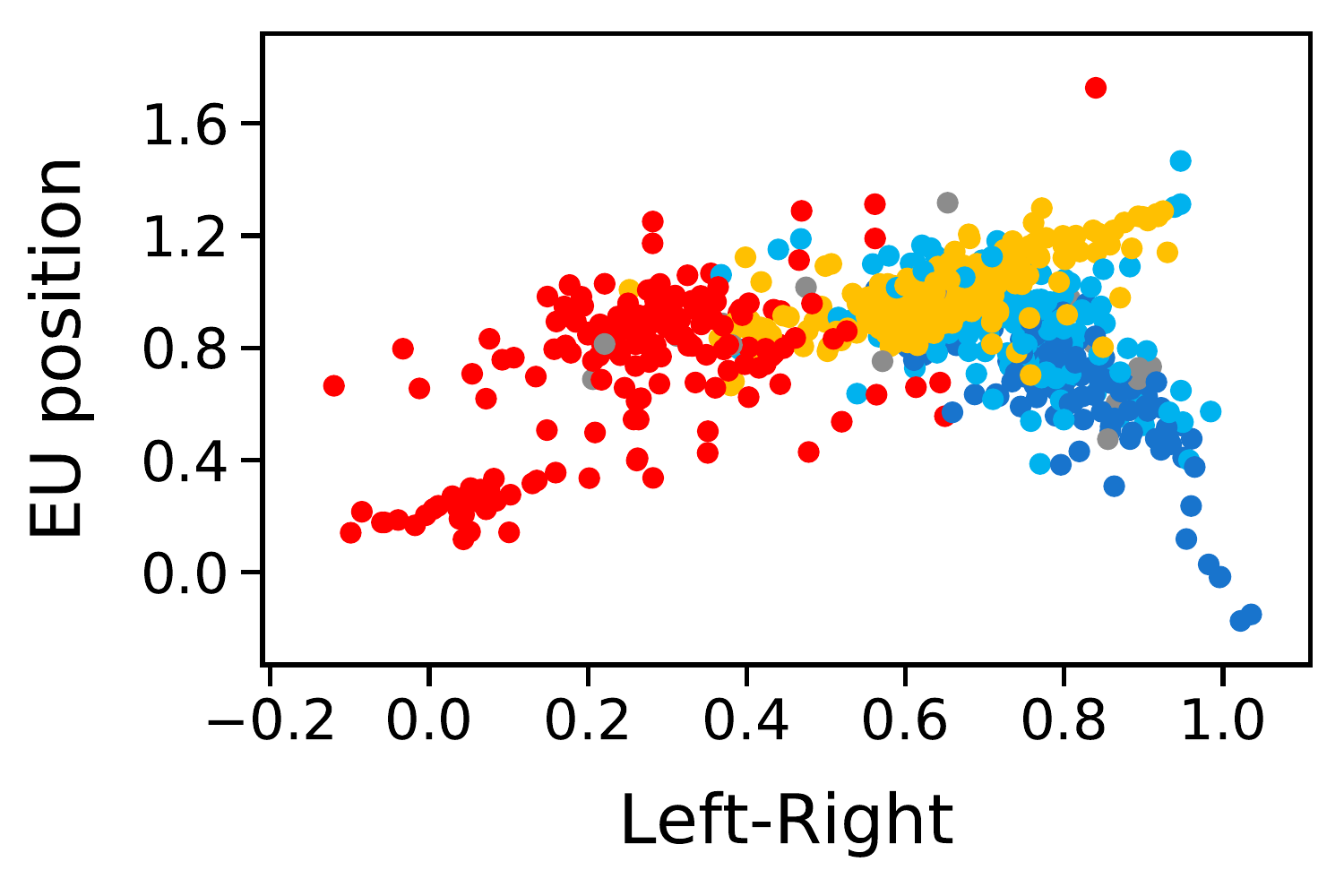}}
\subfloat[]{\label{opinion_mps:f}\includegraphics[width=0.33\textwidth]{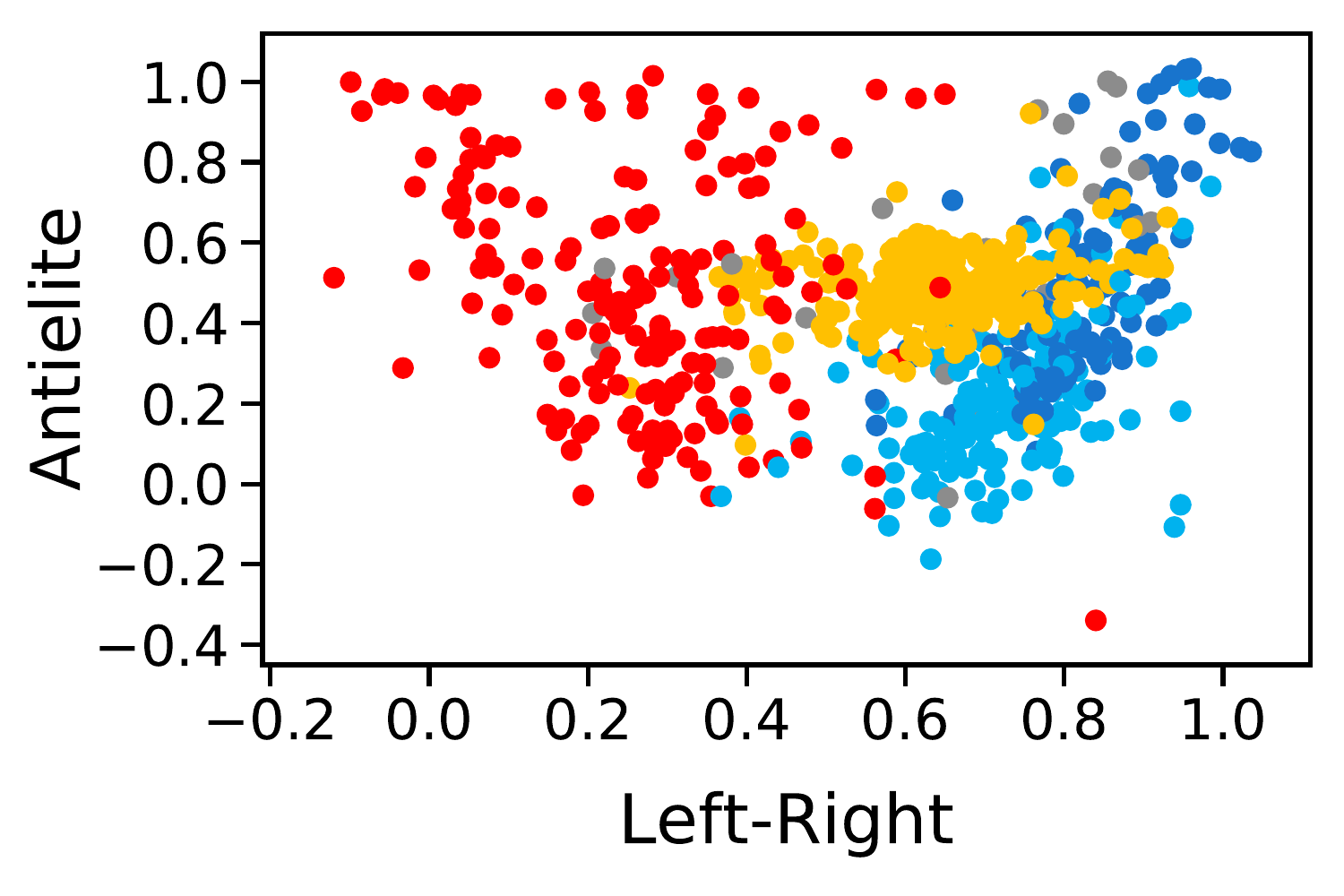}}

\vspace*{+0.20cm}

\subfloat[]{\label{opinion_mps:g}\includegraphics[width=0.33\textwidth]{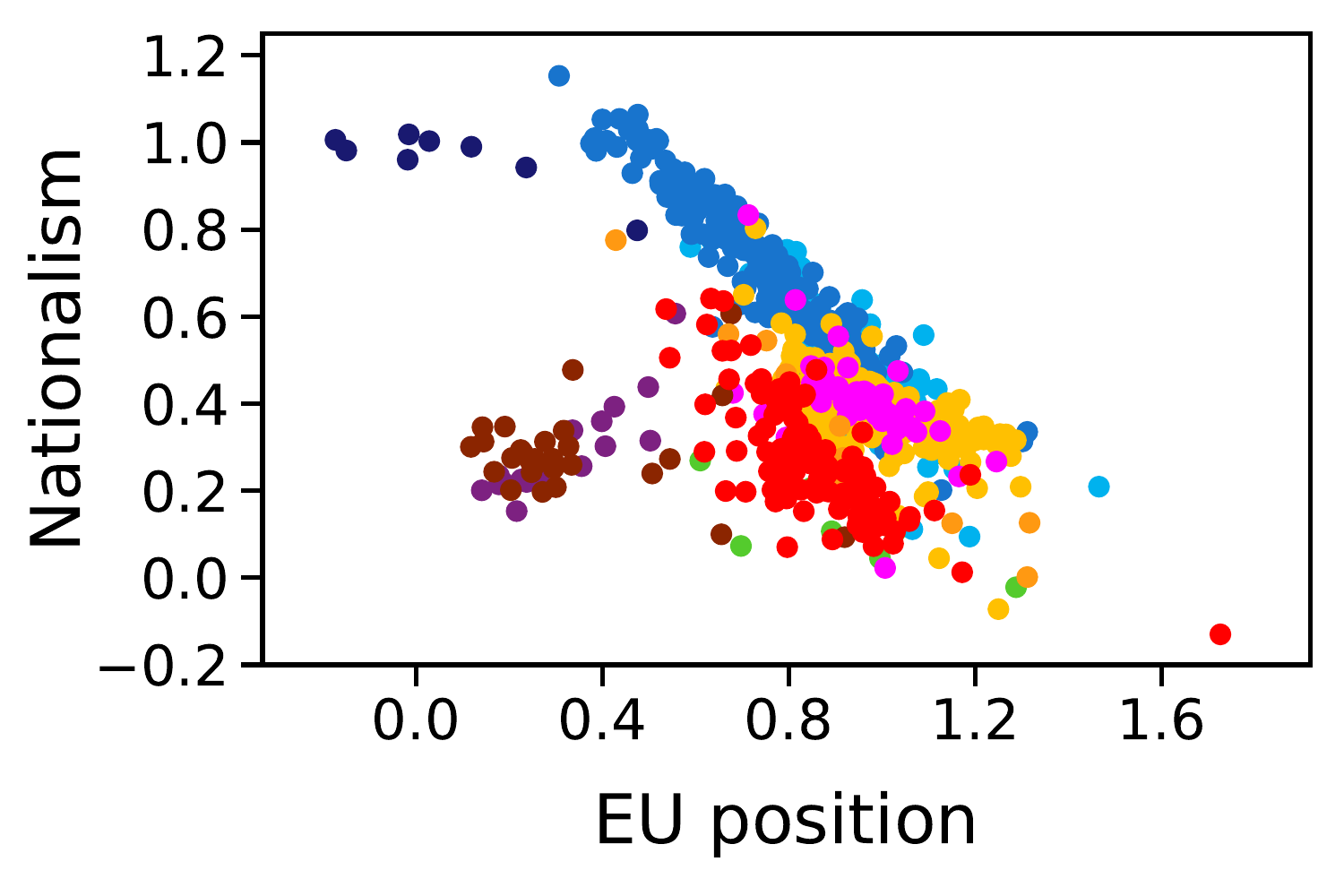}}
\subfloat[]{\label{opinion_mps:h}\includegraphics[width=0.33\textwidth]{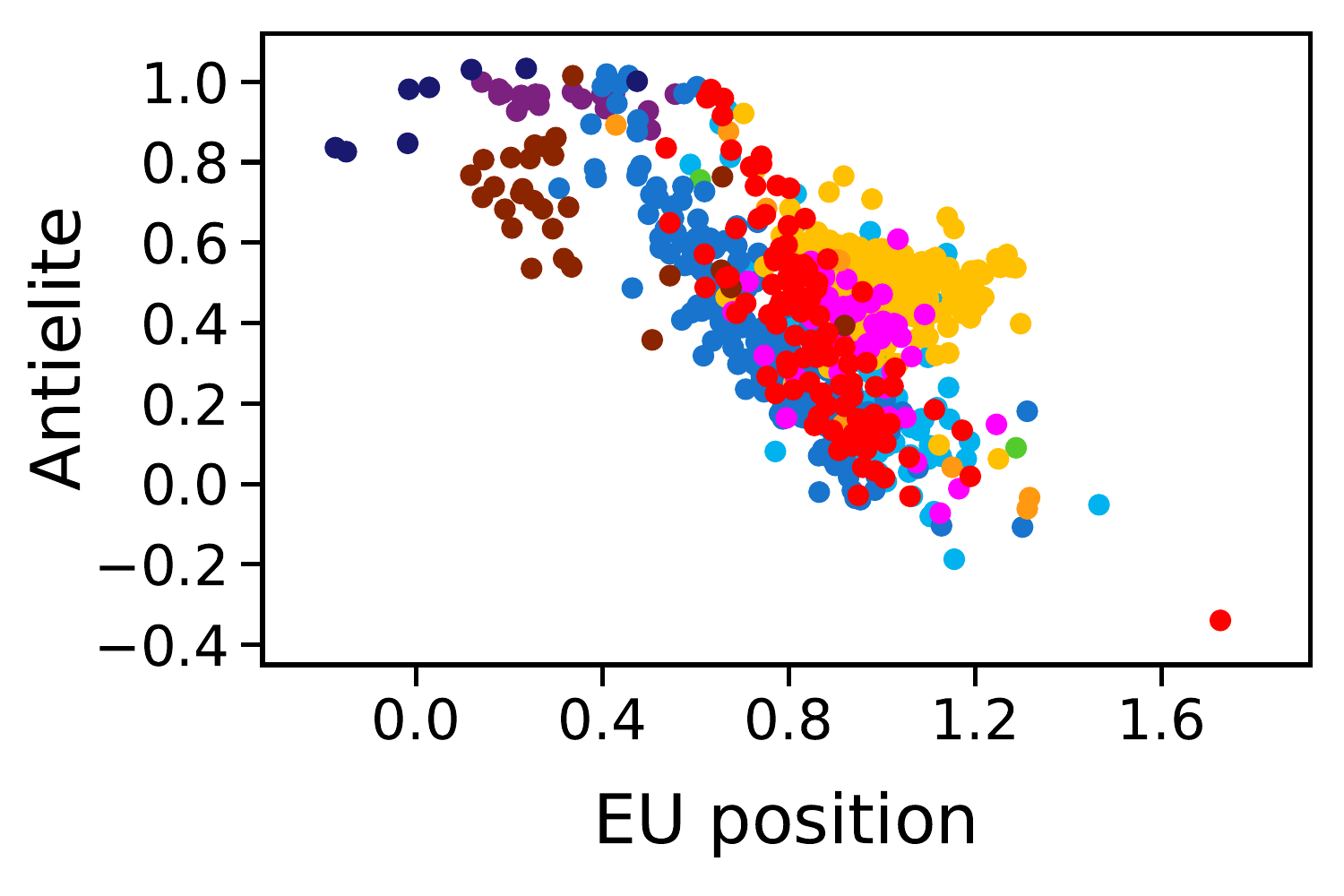}}
\subfloat[]{\label{opinion_mps:i}\includegraphics[width=0.33\textwidth]{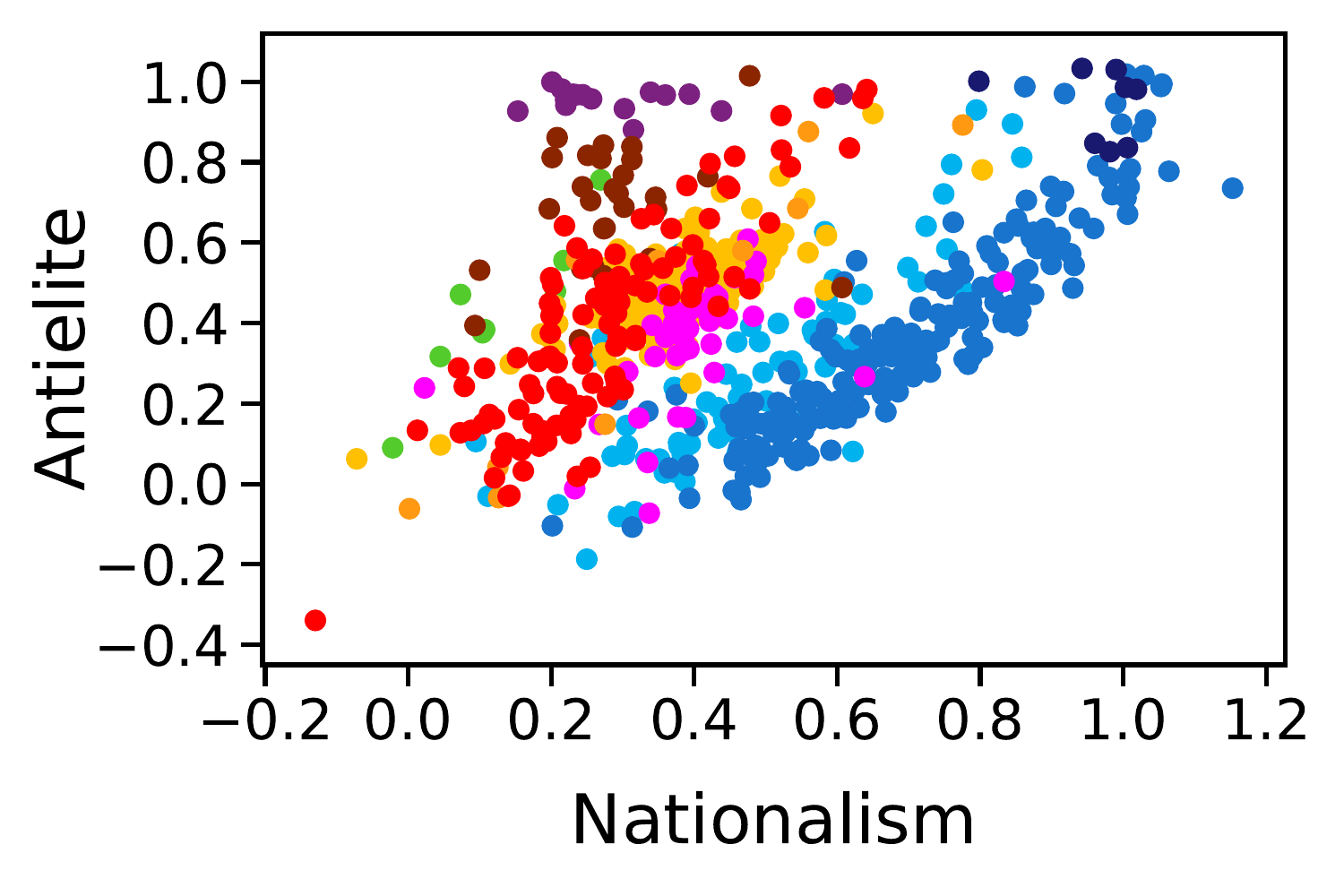}}

\vspace*{0.10cm}
\subfloat[]{\label{opinion_mps:j}\includegraphics[width=0.33\textwidth]{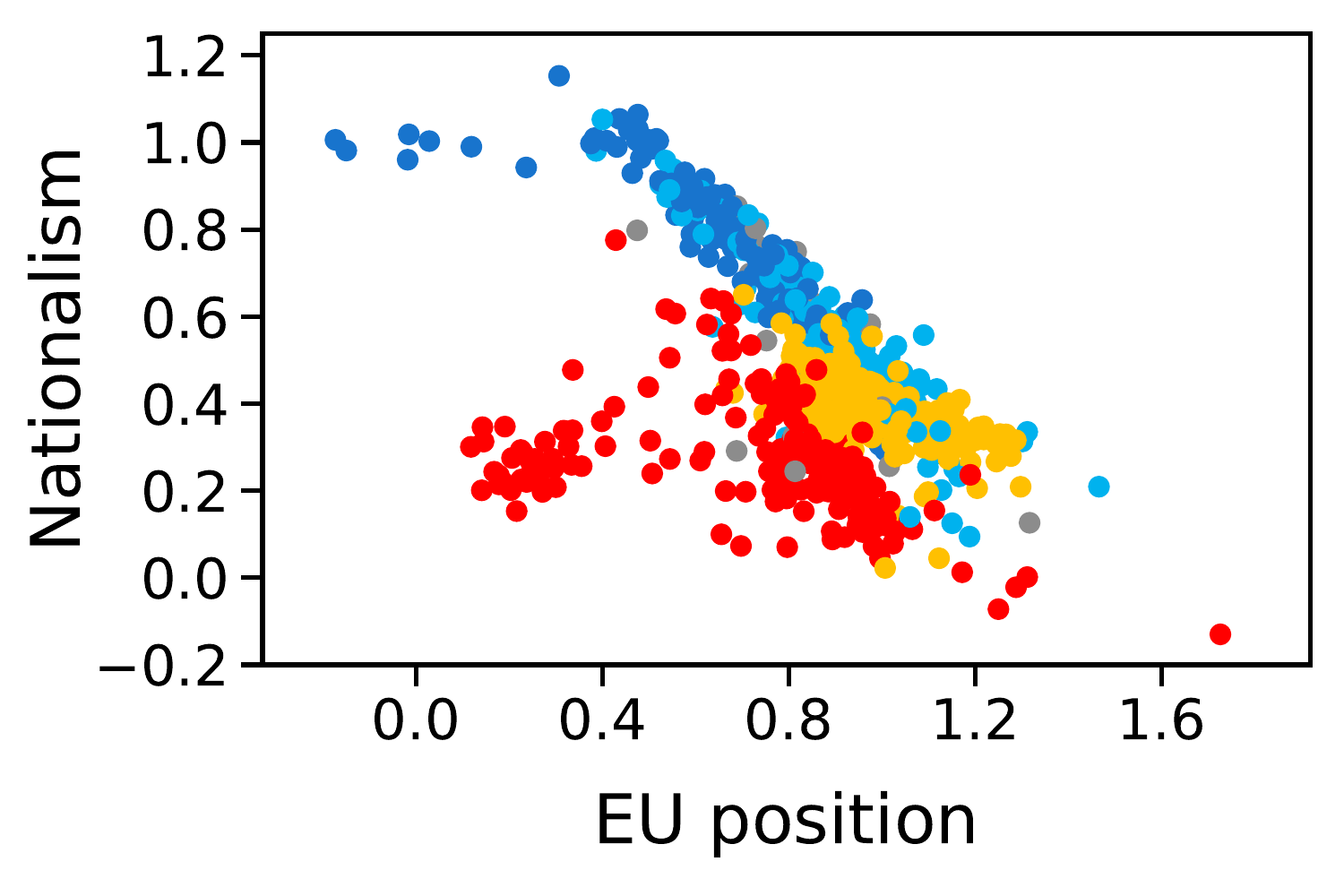}}
\subfloat[]{\label{opinion_mps:k}\includegraphics[width=0.33\textwidth]{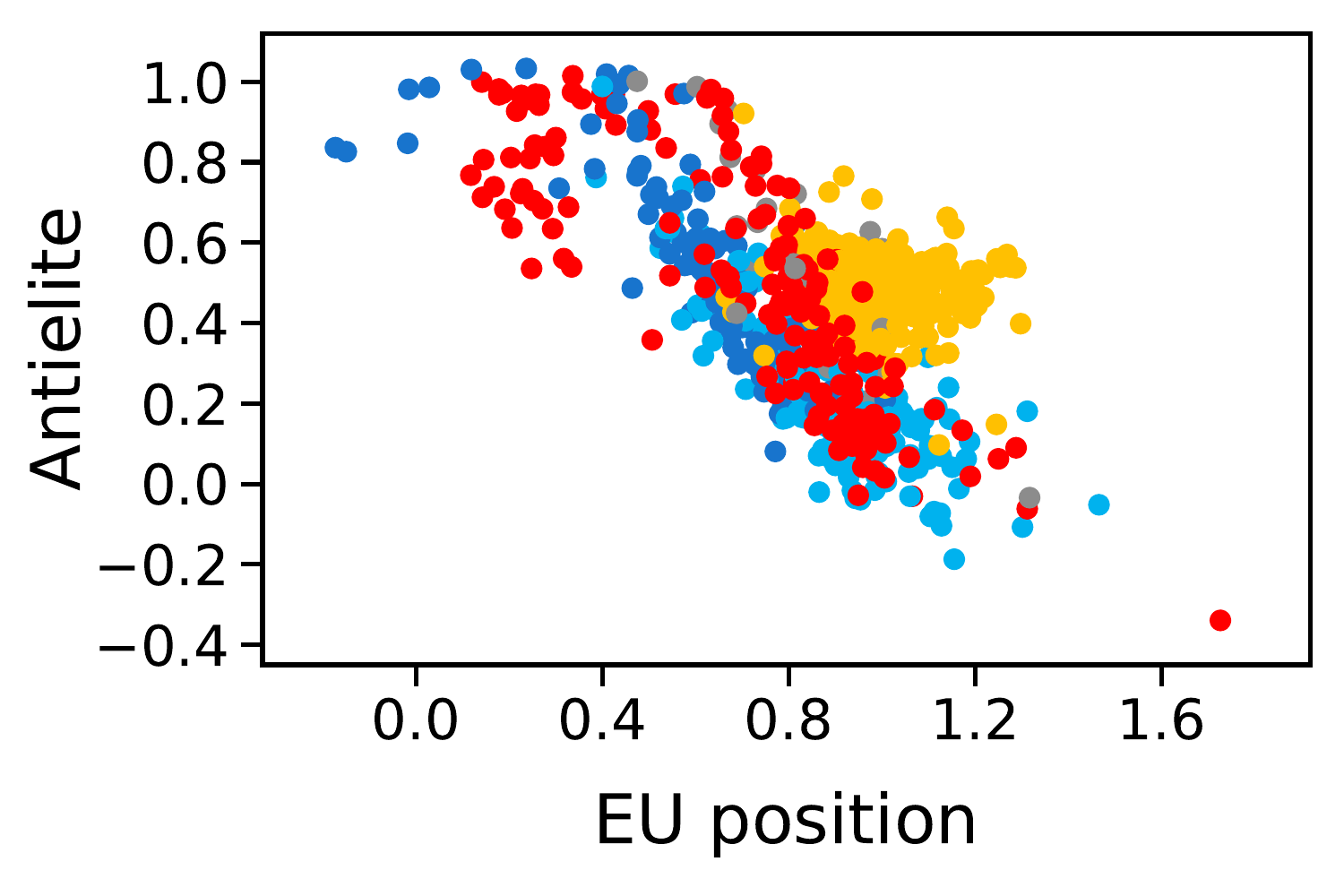}}
\subfloat[]{\label{opinion_mps:l}\includegraphics[width=0.33\textwidth]{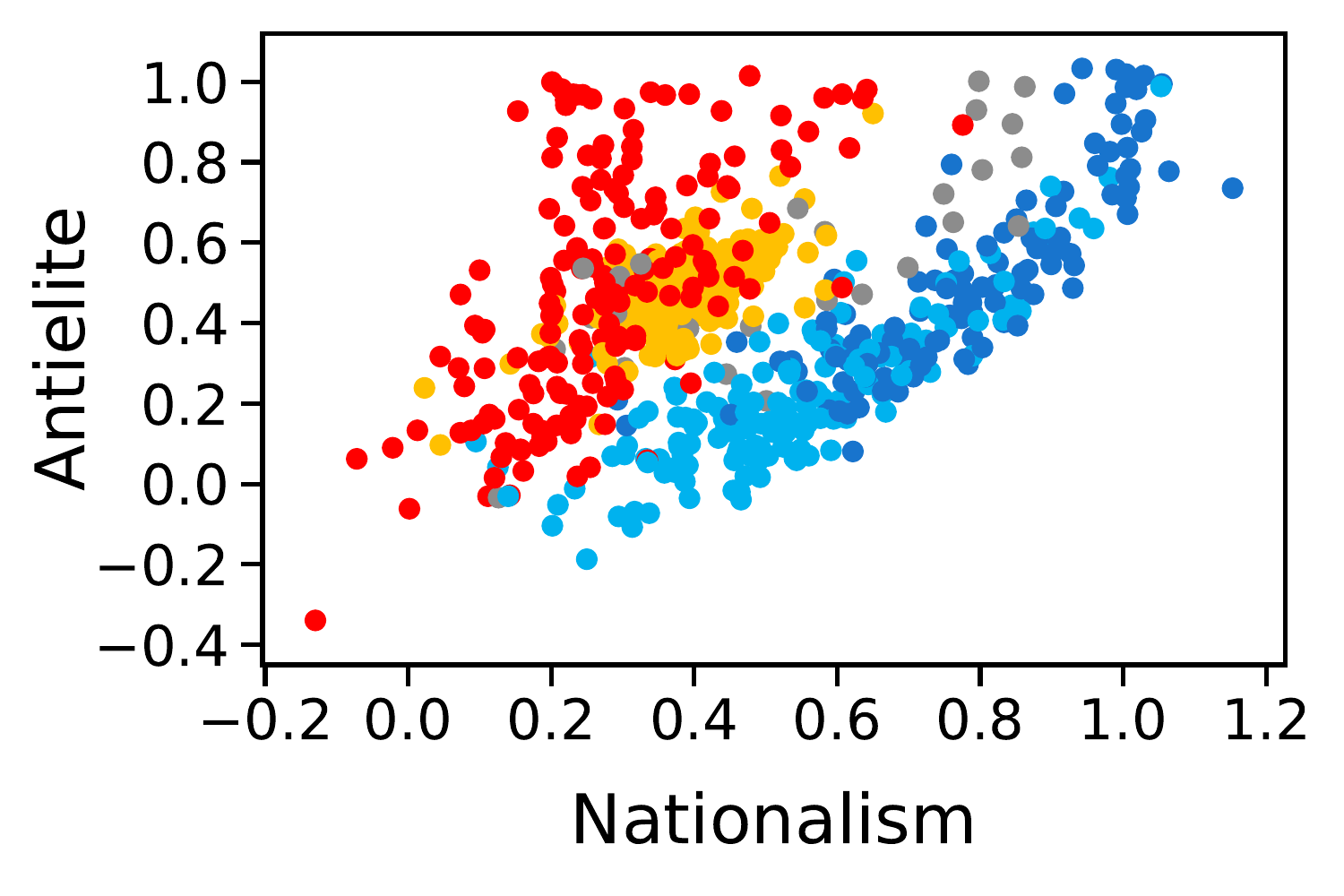}}

\caption{States of the MPs in a two-dimensional opinion space for the six possible pairs of variables (a/d) LR-EU, (b/e) LR-NA, (c/f) LR-AE, (g/j) EU-NA, (h/k) NA-AE and (i/l) EU-AE. The color of the points indicate the party, for panels (a-c) and (g-i), and the community, for panels (d-f) and (j-i), to which the corresponding MP belongs.}
\label{opinion_mps}
\end{center}
\end{figure}

A key question that we have to address is why the number of communities is smaller than the number of parties. In Fig. S\ref{communities_mps} we show the party composition and connectivities of the communities in comparison to the parties. The community $\rightarrow$ community and party $\rightarrow$ party connectivity shows that some parties are non-assortative and consequently are grouped together in different communities. The ``Left'' and ``Others'' are the two main communities that aggregate these non-assortative parties. Thus, we conclude that the assortative structural patterns of the MP $\rightarrow$ MP subgraph do not identify some minority parties as a separated unit. Despite this we remark that some difference can be observed in the political positions of these parties, see Fig. S\ref{opinion_mps}, which would allow us to differentiate between them. These difference, however, can be inferred from the (assortative) structural patterns of the User $\rightarrow$ MP subgraph but not from the MP $\rightarrow$ MP.

\section{Model}

\subsection{Definition}

\sloppy
We consider a population of $N$ individuals (users) that hold a time-dependent vector variable $\vec{v}_{i}(t)=(x_{i}(t), y_{i}(t), z_{i}(t), w_{i}(t)) \in \mathbb{R}^4$ defining the state of individual $i=1, \dots, N$ in the different (four) dimensions at time $t$, and a set of zealots, i.e., parliamentarians (MPs) in this case, $m=1, \dots, M$ that hold a vector variable $\vec{V}_m=(X_{m}, Y_{m}, Z_{m}, W_{m}) \in \mathbb{R}^4$ that we assume to be fixed in time. Individuals interact between them through a (follower $\rightarrow$ following) network structure, that we map onto the usual (symmetric) adjacency matrix $A_{ij}$ for $i, j =1, \dots, N$. Additionally, individuals interact with zealots through a bipartite network with adjacency matrix $P_{im}$ for $i =1, \dots, N$ and $m =1, \dots, M$. In the model we consider that these two types of interactions, i.e., individuals with other individuals or individuals with zealots, occur at a different rate and for this reason we introduce a set of parameters $\lbrace \lambda_{m} \rbrace_{m = 1, \dots, M}$ for all zealots\footnote{Note that in the main text we only introduce one parameter $\lambda \equiv \lambda_{m}$ for all zealots $m$, but here we relax this assumption for convenience in the calculations.}. This parameter $\lambda_{m}$ measures the ratio between the frequency at which a neighbor zealot $m$ of an individual $i$, $\lbrace m \rbrace$ with $P_{im} =1$, is chosen for interaction as compared to a neighbor individual $j$ of $i$, $\lbrace j \rbrace$ with $A_{ij} =1$. If individual $i$ follows $k^{\text{out/uu}}_{i}=\sum_{j=1}^{N} A_{ij}$ other individuals and $k^{\text{out/um}}_{i}=\sum_{m=1}^{M} P_{im}$ zealots, then the probabilities of interacting with a neighbor individual $\text{prob}^{uu}_{i \rightarrow j}$ or a zealot $\text{prob}^{um}_{i \rightarrow m}$ are:

\begin{align}
\label{prob_interact}
&\text{prob}^{uu}_{i \rightarrow j} = A_{ij} \mathcal{C}_{i}, \hspace{15pt} \text{prob}^{um}_{i \rightarrow m} = P_{im} \lambda_{m} \mathcal{C}_{i},\notag\\
&\mathcal{C}_{i}^{-1} = k^{\text{out/uu}}_{i} + \sum_{m=1}^{M} P_{im} \lambda_{m},
\end{align}
with normalization $\sum_{j=1}^{N} \text{prob}^{uu}_{i \rightarrow j}+\sum_{m=1}^{M} \text{prob}^{um}_{i \rightarrow m} \equiv \text{prob}^{uu}_{i}+ \text{prob}^{um}_{i} =1$. Thus, the total probabilities of $i$ interacting with any other individual $j=1, \dots, N$ or zealot $m=1, \dots, M$ are:
\begin{align}
\label{prob_interact_tot}
&\text{prob}^{uu}_{i} = \frac{k^{\text{out/uu}}_{i}}{k^{\text{out/uu}}_{i} + \sum_{m=1}^{M} P_{im} \lambda_{m}}, \hspace{15pt} \text{prob}^{um}_{i} = \frac{\sum_{m=1}^{M} P_{im} \lambda_{m}}{k^{\text{out/uu}}_{i} + \sum_{m=1}^{M} P_{im} \lambda_{m}}.
\end{align}

The dynamical rules of the model come defined by the following steps:

\begin{enumerate}
    \item An individual $i$ is selected at random from the $N$ possibilities.
    \item An interaction process takes place between $i$ and one of its neighbors in the network selected at random with probabilities Eq. (\ref{prob_interact}), which can be another individual $j$ or a zealot $m$.
    \item The selected neighbor influences the state of $i$ as follows
    \begin{equation}
    \label{copy_individual}  
    \vec{v}_{i}(t+\Delta t)=\vec{v}_{i}(t)+I_{ij} \cdot (\vec{v}_{j}(t)-\vec{v}_{i}(t)),
    \end{equation}
    when the selected neighbor $j$ is another individual, and
    \begin{equation}
    \label{copy_zealot}  
    \vec{v}_{i}(t+\Delta t)=\vec{v}_{i}(t)+I_{im} \cdot (\vec{V}_{m}-\vec{v}_{i}(t)),
    \end{equation}
    when the selected neighbor $m$ is a zealot, where the influence factors $I_{ij}$ and $I_{im}$ are real numbers drawn independently at each interaction step from a predefined probability density function $f(I)$. 
    \item Time is increased $t \rightarrow t + \Delta t$ with $\Delta t = 1/N$ and after that another influence process takes place, i.e., return to step 1., until the desired time limit is reached.
\end{enumerate}

The parameters of the model are $\lbrace \lambda_{m} \rbrace_{m = 1, \dots, M}$ and the probability density $f(I)$ which can take any value/shape, we will specify later our choice for these parameters. The influence process defined by Eqs. (\ref{copy_individual}, \ref{copy_zealot}) can be understood as a situation where an individual approaches the state of a neighbor, where the influence factor $I$ determines the magnitude of this approach, e.g., for $I=0$ the individual does not change state while for $I=1$ it copies the state of the neighbor, where $I$ can take any real value depending on the behavior. The steps 1.--4. define the model and they are the rules that we follow in the simulations to obtain numerical results of the dynamics. In order to have some analytical results of the model too, and thus improve our understating of its macroscopic behavior, we will make different assumptions in the following sections with respect to the ingredients that define the dynamics, especially those related to the structure of the network of interactions.

\subsection{Mean Field approaches}

In the so-called mean field approach we make assumptions about the structure of the network, depending on the level of approximation and on the structural details it can be a good approximation or not. For our case, see Section \ref{sec_network_data}, a degree based approach \cite{Porter:2016} is a good estimation of the dynamics on top of the real network.  The simplest and coarser approach is to assume that everyone can interact with everyone (fully connected network), which is equivalent to neglecting all structural features. Although simple, it can be a helpful approximation to improve our comprehension of the dynamics and the impact of the parameters of the model. A better approach is the degree-based approximation (or heterogeneous mean field) in which the distribution of the degree (number of contacts) of the network is taken into account. These mean field approaches work well for highly dense networks (with a high average degree). In the following sections we show the most important results for both mean field approaches.
\subsubsection*{Fully connected}

In a fully connected network we have that $A_{ij}=1$ for all $i, j = 1, \dots, N$ and $P_{im}=1$ for all $i = 1, \dots, N$ and $j = 1, \dots, M$. In this case the interaction probabilities Eqs. (\ref{copy_individual}, \ref{copy_zealot}) reduce to
\begin{align}
\label{prob_interact_fc}
&\text{prob}^{uu}_{i \rightarrow j} = \mathcal{C}, \hspace{15pt} \text{prob}^{um}_{i \rightarrow m} = \lambda_{m} \mathcal{C},\notag\\
&\mathcal{C}^{-1} = N-1 + \sum_{m=1}^{M} \lambda_{m},
\end{align}
and the total probability that $i$ interacts with any individual or zealot Eq. (\ref{prob_interact_tot}) reduces to
\begin{align}
\label{prob_interact_fc_tot}
&\text{prob}^{uu} = \frac{N-1}{N-1 + \sum_{m=1}^{M} \lambda_{m}}, \hspace{15pt} \text{prob}^{um} = \frac{\sum_{m=1}^{M} \lambda_{m}}{N-1 + \sum_{m=1}^{M} \lambda_{m}}.
\end{align}
As every individual $i$ has the same interaction probabilities, we will be able to obtain a time-evolution equation for the density $P(\vec{v},t)$, defined as the probability that the state of an individual $i$ is in between $\vec{v} < \vec{v}_{i} < \vec{v}+d \vec{v}$ at time $t$, with the normalization condition $\int P(\vec{v},t) d \vec{v}=1$. This analysis is carried out in detail in the next sections.

\subsubsection*{Degree-based approach} \label{sec_degree_approach}
Taking the degree sequence of the individuals $\lbrace k^{\text{out/uu}}_{i}, k^{\text{in/uu}}_{i}, k^{\text{out/um}}_{i} \rbrace_{i=1, \dots, N}$ and the zealots $\lbrace k^{\text{in/um}}_{m} \rbrace_{m=1, \dots, M}$, we replace the actual value of the adjacency matrix $A_{ij}$ and $P_{im}$ by the ensemble average over all networks that have the same degree sequence \cite{Sood:2008, Carro:2016}, i.e.,

\begin{equation}
\label{annealed}
A_{ij} \approx \frac{k^{\text{out/uu}}_{i} k^{\text{in/uu}}_{j}}{ N \langle k^{\text{out/uu}} \rangle}, \hspace{15pt} P_{im} \approx \frac{k^{\text{out/um}}_{i} k^{\text{in/um}}_{m}}{ N \langle k^{\text{out/um}} \rangle}.
\end{equation}
Note that Eq. (\ref{annealed}) fulfills the necessary conditions: $k^{\text{out/uu}}_{i}=\sum_{j=1}^{N} A_{ij}$, $k^{\text{in/uu}}_{i}=\sum_{i=1}^{N} A_{ij}$, $\langle k^{\text{out/uu}} \rangle = \langle k^{\text{in/uu}} \rangle$, $k^{\text{out/um}}_{i}=\sum_{m=1}^{M} P_{im}$, $k^{\text{in/um}}_{m}=\sum_{i=1}^{N} P_{im}$, and $N \langle k^{\text{out/uu}} \rangle = M \langle k^{\text{in/uu}} \rangle$. Introducing this in the interaction probabilities Eq. (\ref{prob_interact}) we have
\begin{align}
\label{prob_interact_degree}
&\text{prob}^{uu}_{i \rightarrow j} = \frac{k^{\text{out/uu}}_{i} k^{\text{in/uu}}_{j}}{ N \langle k^{\text{out/uu}} \rangle} \mathcal{C}_{i}, \hspace{15pt} \text{prob}^{um}_{i \rightarrow m} = \frac{k^{\text{out/um}}_{i} k^{\text{in/um}}_{m}}{ M \langle k^{\text{in/um}} \rangle} \lambda_{m} \mathcal{C}_{i},\notag\\
&\mathcal{C}_{i}^{-1} = k^{\text{out/uu}}_{i} + k^{\text{out/um}}_{i} \frac{\sum_{m=1}^{M} k^{\text{in/um}}_{m} \lambda_{m}}{M \langle k^{\text{in/um}} \rangle},
\end{align}
and the total probability that $i$ interacts with any individual or zealot Eq. (\ref{prob_interact_tot}) reduces to
\begin{align}
\label{prob_interact_degree_tot}
&\text{prob}^{uu}_{i} = \frac{k^{\text{out/uu}}_{i}}{k^{\text{out/uu}}_{i} + k^{\text{out/um}}_{i} \dfrac{\sum_{m=1}^{M} k^{\text{in/um}}_{m} \lambda_{m}}{M \langle k^{\text{in/um}} \rangle}}, \hspace{15pt} \text{prob}^{um}_{i} = \frac{k^{\text{out/um}}_{i}\dfrac{\sum_{m=1}^{M} k^{\text{in/um}}_{m} \lambda_{m}}{M \langle k^{\text{in/um}} \rangle}}{k^{\text{out/uu}}_{i} + k^{\text{out/um}}_{i} \dfrac{\sum_{m=1}^{M} k^{\text{in/um}}_{m} \lambda_{m}}{M \langle k^{\text{in/um}} \rangle}}.
\end{align}
In contrast to the fully connected result Eq. (\ref{prob_interact_fc_tot}), the probabilities Eq. (\ref{prob_interact_degree_tot}) depend on the individual $i$, we note, however, that if $k^{\text{out/um}}_{i} = \mathcal{A} k^{\text{out/uu}}_{i}$ the degrees cancel out and this dependency is lost. If this assumption is true, see Fig. S\ref{degree_corr:a}, then the degree based and fully connected approaches coincide after an adequate re-scaling of parameters. The constant $\mathcal{A}$ can be determined from consistency conditions $\mathcal{A} = \dfrac{\langle k^{\text{out/um}} \rangle}{\langle k^{\text{out/uu}} \rangle}$, which leads to a reduced expression for the interaction probabilities 
\begin{align}
\label{prob_interact_degree_tot_reduced}
&\text{prob}^{uu} = \frac{1}{1 + \dfrac{\langle k^{\text{out/um}} \rangle}{\langle k^{\text{out/uu}} \rangle} \dfrac{\sum_{m=1}^{M} k^{\text{in/um}}_{m} \lambda_{m}}{M \langle k^{\text{in/um}} \rangle}}, \hspace{15pt} \text{prob}^{um} = \dfrac{\dfrac{\langle k^{\text{out/um}} \rangle}{\langle k^{\text{out/uu}} \rangle} \dfrac{\sum_{m=1}^{M} k^{\text{in/um}}_{m} \lambda_{m}}{M \langle k^{\text{in/um}} \rangle}}{1 + \dfrac{\langle k^{\text{out/um}} \rangle}{\langle k^{\text{out/uu}} \rangle} \dfrac{\sum_{m=1}^{M} k^{\text{in/um}}_{m} \lambda_{m}}{M \langle k^{\text{in/um}} \rangle}},
\end{align}
that do not depend on $i$.

\subsubsection*{Dynamical equation for the density $\pmb{P(\vec{v},t)}$ of users with opinion $\vec{v}$ at time $t$}\label{sec_eq_prob}

Note that the rules Eqs. (\ref{copy_individual}, \ref{copy_zealot}) are applied for each component separately, thus we only need to find the dynamical equation of a single component say $x_{i}(t)$. In the mean field description we consider a fully connected network and the rules of the model can be written as:
\begin{equation}
 x_{i}(t+1/N) =
    \begin{cases}
      x_{i}(t)+I_{ij} \cdot (x_{j}(t)-x_{i}(t)), \hspace {8pt} \text{prob}^{uu} \equiv \alpha_{0},\\
      x_{i}(t)+I_{im} \cdot (X_{m}-x_{i}(t)), \hspace {10pt} \text{prob}^{um}_{i \rightarrow m} \equiv \alpha_{m},
    \end{cases}    
\end{equation}
with $j\neq i = 1, \dots, N$, any of them chosen at random, $m=1,\dots,M$, and the probabilities $\alpha_{0}$, $\lbrace \alpha_{m} \rbrace$ are
\begin{eqnarray}
\alpha_{0}= \frac{N-1}{N-1 + \sum_{m=1}^{M} \lambda_{m}}, \hspace{15pt} \alpha_{m}=\frac{\lambda_{m}}{N-1 + \sum_{m=1}^{M} \lambda_{m}},
\end{eqnarray} 
with normalization $\alpha_{0}+\sum_{m=1}^{M} \alpha_{m}=1$. This formulation of the model is equivalent to the one given in the original reference \cite{Chacoma:2015}. 

It is possible to write an evolution equation for the probability density $P(x,t)$ of finding an individual with opinion between $x$ and $x+dx$ at time $t$
\begin{align}
& \frac{P(x,t+\Delta t) - P(x,t)}{\Delta t} = - P(x,t) \notag\\
& + \alpha_{0} \int dI \int dy \int dz P(y,t) P(z,t) f(I) \delta (x-y-I(z-y)) \notag\\
& + \sum_{m=1}^{M} \alpha_{m} \int dI \int dy P(y,t) f(I) \delta (x-y-I(X_{m}-y)),
\end{align}
which can be read as the change in the density of individuals with state $x$ per unit time is equal to minus the probability of selecting an individual with state $x$ that changes to any other state, plus the probability of selecting an individual with state $y$ that changes to $x$. For continuous time $\Delta t \rightarrow 0$ it is
\begin{align}\label{dynamical_eq_prob_delta}
& \frac{\partial P(x,t)}{\partial t} + P(x,t)=  \notag\\
& + \alpha_{0} \int dI \int dy \int dz P(y,t) P(z,t) f(I) \delta (x-y-I(z-y)) \notag\\
& + \sum_{m=1}^{M} \alpha_{m} \int dI \int dy P(y,t) f(I) \delta (x-y-I(X_{m}-y)),
\end{align}
and simplifying the Dirac deltas
\begin{equation}\label{dynamical_eq_prob}
\frac{\partial P(x,t)}{\partial t} + P(x,t) = \alpha_{0} \int \frac{dI f(I)}{1-I} \int dz P(z,t) P \left(\frac{x-I z}{1-I},t \right) + \sum_{m=1}^{M} \alpha_{m} \int \frac{dI f(I)}{1-I}  P \left(\frac{x-I X_{m}}{1-I},t \right).
\end{equation}
Equation [\ref{dynamical_eq_prob}] is an integro-differential equation for the probability density $P(x,t)$ that we cannot solve analytically. However, as we show in next section, it is possible to obtain some closed expressions for the moments of the distribution.
\subsubsection*{Average opinion and variance of users, time evolution and stationary}\label{sec_eq_moments}

First we consider the evolution equation of the average state $\langle x \rangle = \int x P(x,t) dx$, integrating both sides of Eq. (\ref{dynamical_eq_prob}) $\int x (\dots) dx$ we find
\begin{align}
\frac{d \langle x \rangle}{d t} &= \langle I\rangle \sum_{m=1}^{M} \alpha_{m} (X_{m}-\langle x \rangle),
\end{align}
with $\langle I\rangle = \int f(I) I dI$.
The solution is
\begin{equation}\label{dynamic_average}
\langle x \rangle (t) = \langle x \rangle_{\text{st}} + \big(\langle x \rangle (0) - \langle x \rangle_{\text{st}} \big) e^{-\langle I\rangle (1-\alpha_{0}) t},
\end{equation}
with stationary value
\begin{equation}\label{stationary_average}
\langle x \rangle_{\text{st}}=\dfrac{\sum_{m=1}^{M} \alpha_{m} X_{m}}{\sum_{m=1}^{M} \alpha_{m}} \equiv \widetilde{\mu}.
\end{equation}
According to the solution Eq. (\ref{stationary_average}), the stationary average state of individuals is equal to a weighted average of the zealots. It is also possible to derive evolution equations for higher order moments, but for the sake of concreteness we will only show the results for the variance $\sigma^2=\left\langle \left(x-\langle x \rangle \right)^{2} \right\rangle = \int dx \left(x-\langle x \rangle \right)^{2} P(x,t)$. Integrating both sides of Eq. (\ref{dynamical_eq_prob}) $\int x^2 (\dots) dx$ we find

\begin{align}
 \frac{d \sigma^2}{dt} &= \left(2 \langle I \rangle - (1+\alpha_{0}) \langle I^{2} \rangle\right) \left(\sigma_{\text{st}}^2 -\sigma^2 \right),
 \end{align}
 whose solution is
 \begin{equation}
 \label{dynamic_variance}
 \sigma^{2} (t) = \sigma^{2}_{\text{st}} + \left(\sigma^{2} (0) - \sigma^{2}_{\text{st}}\right) e^{- \left(2 \langle I \rangle - (1+\alpha_{0}) \langle I^{2} \rangle\right) t},
 \end{equation}
  with stationary value
 \begin{equation}
 \label{stationary_variance}
 \sigma^2_{\text{st}} = \frac{(1-\alpha_{0}) \langle I^{2} \rangle}{2 \langle I \rangle - (1+\alpha_{0}) \langle I^{2} \rangle} \widetilde{\sigma}^2, \hspace{0.7cm} \widetilde{\sigma}^2 \equiv \dfrac{\sum_{m=1}^{M} \alpha_{m} \left( X_{m} - \widetilde{\mu} \right)^2}{\sum_{m=1}^{M} \alpha_{m}},
 \end{equation}
 where $\langle I^{n} \rangle \equiv \int f(I) I^{n} dI$, and $\langle I \rangle$, $\langle I^2 \rangle$  are the first  moments of the influence factor distribution.
 According to the stationary solution Eq. (\ref{stationary_variance}) the state variance of individuals is equal to a weighted variance of zealots times a scaling factor. The time dependent solutions Eqs. (\ref{dynamic_average}, \ref{dynamic_variance}) are exponential decays that reach a steady state (see Section \ref{model_dynamics}) whose properties, time scales and stationary values $\langle x \rangle_{\text{st}}$, $\sigma^2_{\text{st}}$, depend on the parameters of the model and the network. Equivalent evolution equations for $\langle x \rangle(t)$ and $\sigma^2(t)$ are obtained in Ref. \cite{Chacoma:2015} for general $\left\lbrace \alpha_{m} \right\rbrace_{m=0}^{M}$, we additionally provide here the expressions for heterogeneous networks. In the degree-based approach and assuming, as we do in the main text, that $\lambda_{m} \equiv \lambda$ for $m=1, \dots, M$ the probabilities $\lbrace \alpha_{m} \rbrace_{m=0}^{M}$ can be written as a function of $\lambda$ and the degree sequences as

\begin{equation}
\label{alpha0}
\alpha_{0} = \left(1 + \dfrac{\langle k^{\text{out/um}} \rangle}{\langle k^{\text{out/uu}} \rangle} \lambda \right)^{-1}, \hspace{0.7cm} \alpha_{m} = \alpha_{0} \dfrac{\langle k^{\text{out/um}} \rangle}{\langle k^{\text{out/uu}} \rangle} \dfrac{ \lambda k^{\text{in/um}}_{m}}{M \langle k^{\text{in/um}} \rangle}.  
\end{equation}
Using the proportionality relation of $\alpha_{m} \propto k^{\text{in/um}}_{m}$ we have that the weighted average $\widetilde{\mu}$ and variance $\widetilde{\sigma}^2$ can be written as
\begin{eqnarray}\label{weighted_average_simplify}
\widetilde{\mu}&=&\dfrac{\sum_{m=1}^{M} k^{\text{in/um}}_m X_{m}}{\sum_{m=1}^{M} k^{\text{in/um}}_m}, \\
\label{weighted_variance_simplify}
\widetilde{\sigma}^2 &\equiv& \dfrac{\sum_{m=1}^{M} k^{\text{in/um}}_m \left( X_{m} - \widetilde{\mu} \right)^2}{\sum_{m=1}^{M} k^{\text{in/um}}_m},
\end{eqnarray}
which means that the weights can be replaced simply by the degrees $k^{\text{in/um}}_{m}$, i.e., the number of user followers of MP $m$.

Summarizing, we found that the average and variance opinion of users evolve exponentially in time and reach a steady state value. The time scale of the time evolution Eqs. (\ref{dynamic_average}, \ref{dynamic_variance}) depend on the parameters of the model: $\langle I \rangle$ for $\langle x \rangle(t)$, $\langle I \rangle$ and $\langle I^2 \rangle$ for $\sigma^2(t)$, in addition to $\alpha_{0}$ which depends on $\lambda$ and the average out degree of users $\langle k^{\text{out/um}} \rangle$ and $\langle k^{\text{out/uu}} \rangle$. For the stationary average value $\langle x \rangle_{\text{st}}$ we found that it does not depend on the parameters of the model, but it is simply equal to the degree weighted average of the MPs ($\widetilde{\mu}$) Eq. (\ref{weighted_average_simplify}). The stationary variance $\sigma^2_{\text{st}}$ is equal to the degree weighted variance of the MPs ($\widetilde{\sigma}^2$) Eq. (\ref{weighted_variance_simplify}) times a scaling factor that depends on the parameters of the model $\langle I \rangle$, $\langle I^2 \rangle$ and $\alpha_{0}$.

\subsubsection*{General properties of the mean field solution}\label{sec_properties} The relations Eqs. (\ref{stationary_average}, \ref{stationary_variance}) relate the basic statistical properties of the states of the individuals (the stationary average $\langle x \rangle_{\text{st}}$ and variance $\sigma_{\text{st}}^2$) with the parameters of the model ($\alpha_{0}$, $\lbrace \alpha_{m} \rbrace_{m=1}^{M}$ and $f(I)$) and the statistics of the states of the zealots (the weighted moments $\widetilde{\mu}$, $\widetilde{\sigma}^2$). Note that in the degree-based mean field the details of the network and the parameter $\lambda$ are absorbed in the probabilities $\alpha_{0}$, $\lbrace \alpha_{m} \rbrace_{m=1}^{M}$, Eq. (\ref{alpha0}). From these expressions we obtain valuable information about the behavior of the users and what we can expect from the results of simulation on top of the real network. For example, we can identify some general properties that the influence distribution $f(I)$ has to fulfill in order to get a well behaved solution. Using the expression for the stationary variance Eq. (\ref{stationary_variance}) we find that the variance is finite (which is equivalent to state that the distribution $P(x,t)$ converges to a well defined function) when the condition
\begin{equation}
\label{convergence_condition}
\frac{1+\alpha_{0}}{2} \langle I^{2} \rangle < \langle I \rangle,
\end{equation}
is fulfilled.
Another condition for the variance that we obtain from observation of the opinion statistics of the data is that $\sigma^2/\widetilde{\sigma}^2>1$ which leads to 
\begin{equation}
\label{condition_influence}
\langle I^{2} \rangle > \langle I \rangle.
\end{equation}
This condition implies that the influence factor $I$ has to take values beyond the interval $I \in [0, 1]$, which can be interpreted by noting that in the data the opinions of the individuals exceed the limits of the opinions of the zealots, this corresponds in the model to an influence process with $I<0$ or $I>1$. Note also that the conditions Eq. (\ref{convergence_condition}) and Eq. (\ref{condition_influence}) are contradictory for $\alpha_{0}=1$, thus the two conditions are fulfilled at the same time only for some values of the parameter $\alpha_{0}$. The dependence of the variance $\sigma_{\text{st}}^2/\widetilde{\sigma}^2$ as a function of $\alpha_{0}$ strongly depends on the condition Eq. (\ref{condition_influence}). For $\langle I^{2} \rangle < \langle I \rangle$ we have that $\sigma^2$ is a monotonically decreasing function of $\alpha_{0}$, while for $\langle I^{2} \rangle > \langle I \rangle$ it is monotonically increasing and diverging for $\alpha_{0}(\text{max.}) \equiv 2 \dfrac{\langle I \rangle}{\langle I^2 \rangle}-1$. Thus if the condition Eq. (\ref{condition_influence}) is fulfilled by the model then $\alpha_{0} < \alpha_{0}(\text{max.})$, i.e., $\alpha_{0}$ is bounded.

\subsection{Results}
\subsubsection{Fitting the parameters of the model and errors}\label{sec_fitting}

The free parameters of the model are $\lambda$ and the influence distribution $f(I)$. For simplicity, and in analogy with the experimental results in Ref. \cite{Chacoma:2015}, we parameterize the influence distribution as
$f(I)=p \mathcal{N}(I;0,\sigma_{K}^2) + (1-p) \mathcal{N}(I;1,\sigma_{A}^2)$, where $\mathcal{N}(x; \mu,\sigma^2)=\frac{1}{\sigma \sqrt{2 \pi}} \exp \left[-\frac{(x-\mu)^2}{2\sigma^2}\right]$ is a normal distribution with mean $\mu$ and variance $\sigma^2$. Thus we take as $f(I)$ two Gaussians, peaked around two prototypical behaviors: \emph{Keep} $I=0$, when the individual does not change its opinion, and \emph{Adopt} $I=1$, when it copies the opinion of its neighbor. Thus, the free parameters of the model are $(p,\sigma_{K},\sigma_{A})$ which cannot be inferred or measured directly from the data. In order to find an appropriate value for these parameters, we carry out a fitting process using the results of the opinions of users $\vec{v}_{i}(t)=(x_{i}(t), y_{i}(t), z_{i}(t), w_{i}(t)) \in \mathbb{R}^4$ coming from numerical simulations, and compare them with the data. For this purpose we maximize the ``Similarity'', defined between two distributions $1$ and $2$ as
\begin{equation}
\label{sim_def1}
\text{Similarity} = 1- \frac{1}{2} \int_{x_{\text{min}}}^{x_{\text{max}}} \big\vert \rho_{1}(x) - \rho_{2}(x) \big\vert dx \in [0, 1],
\end{equation}
where $\rho_{1,2}(x)$ are the corresponding densities and $x_{\text{min}}$, $x_{\text{max}}$ their limits, note that the similarity is equal to $0$ when $\rho_{1}(x)$, $\rho_{2}(x)$ are non-overlapping and equal to $1$ when they are identical. The function of the parameters that we choose for maximizing is the average similarity between data and model
\begin{equation}
\label{sim_def2}
\big\langle \text{Similarity} \big\rangle [\lambda, p, \sigma_{K},\sigma_{A}] = 1- \frac{1}{2}  \int_{t_{0}}^{T} \frac{dt}{T-t_{0}} \int_{x_{\text{min}}}^{x_{\text{max}}} \big\vert \rho_{\text{data}}(x) - \rho_{\text{model}}(x,t) \big\vert dx,
\end{equation}
over a sample of time steps of the model in the interval $t_{0} \le t \le T$, where $t_{0}$ is the number of thermalizing time steps after which the dynamics is assumed to have reached the stationary state. The density $\rho_{\text{data}}(x)$ corresponds to the marginal distribution of the corresponding opinion dimension taking the sample $\lbrace x_{i} \rbrace_{i=1}^{N}$ from the data, while $\rho_{\text{model}}(x,t)$ is the density coming from the sample $\lbrace x_{i}(t) \rbrace_{i=1}^{N}$ of the model, i.e., a snapshot of the opinions of the users from the simulations at time $t$. Due to the uncoupling of the dynamics of the model between different opinion dimensions, we will carry out the fitting process for each dimension separately, i.e., we will obtain the maximum of Eq. (\ref{sim_def2}) for (LR, NA, EU, AE). Another important ingredient of the fitting is the limits $(x_{\text{min}}, x_{\text{max}})$ used to compute Eq. (\ref{sim_def2}) that we extract from the data and that correspond to: LR $(-0.48, 1.20)$, NA $(-0.50, 1.40)$, EU $(-0.65, 2.24)$, and AE $(-0.79, 1.25)$. For convenience of the fitting process we will impose an extra condition in the model, that is a boundary condition such that: if during an influence process Eqs. (\ref{copy_individual}, \ref{copy_zealot}) the opinion of an individual would exceed the limits $x_{i}(t+\Delta t)<x_{\text{min}}$ or $x_{i}(t+\Delta t)>x_{\text{max}}$ that process is simply discarded (does not occur). More information about the implications of this boundary conditions can be found in the next Section \ref{sec_boundary}.

\begin{figure}[h!]
\begin{center}
\includegraphics[width=0.26\textwidth]{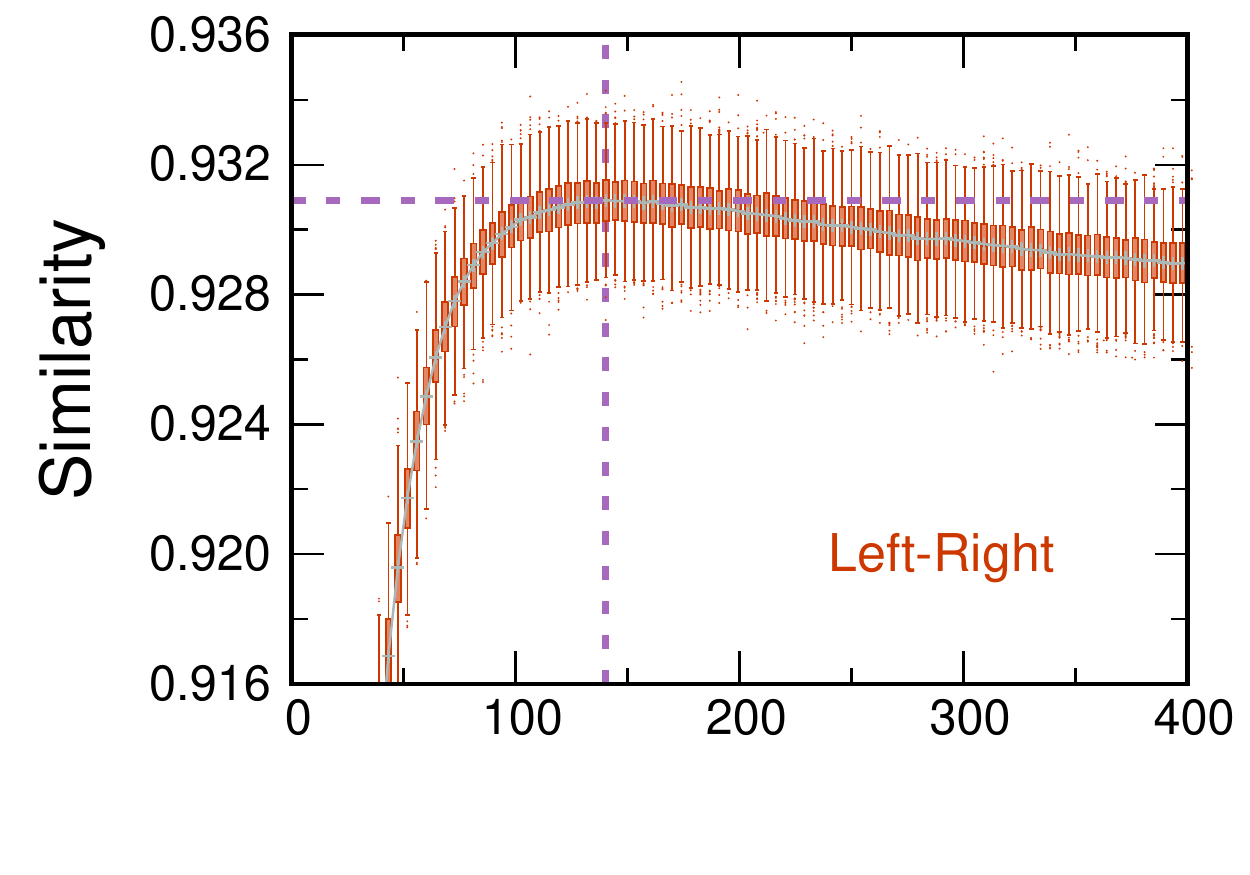}\hspace*{-0.4cm}
\includegraphics[width=0.26\textwidth]{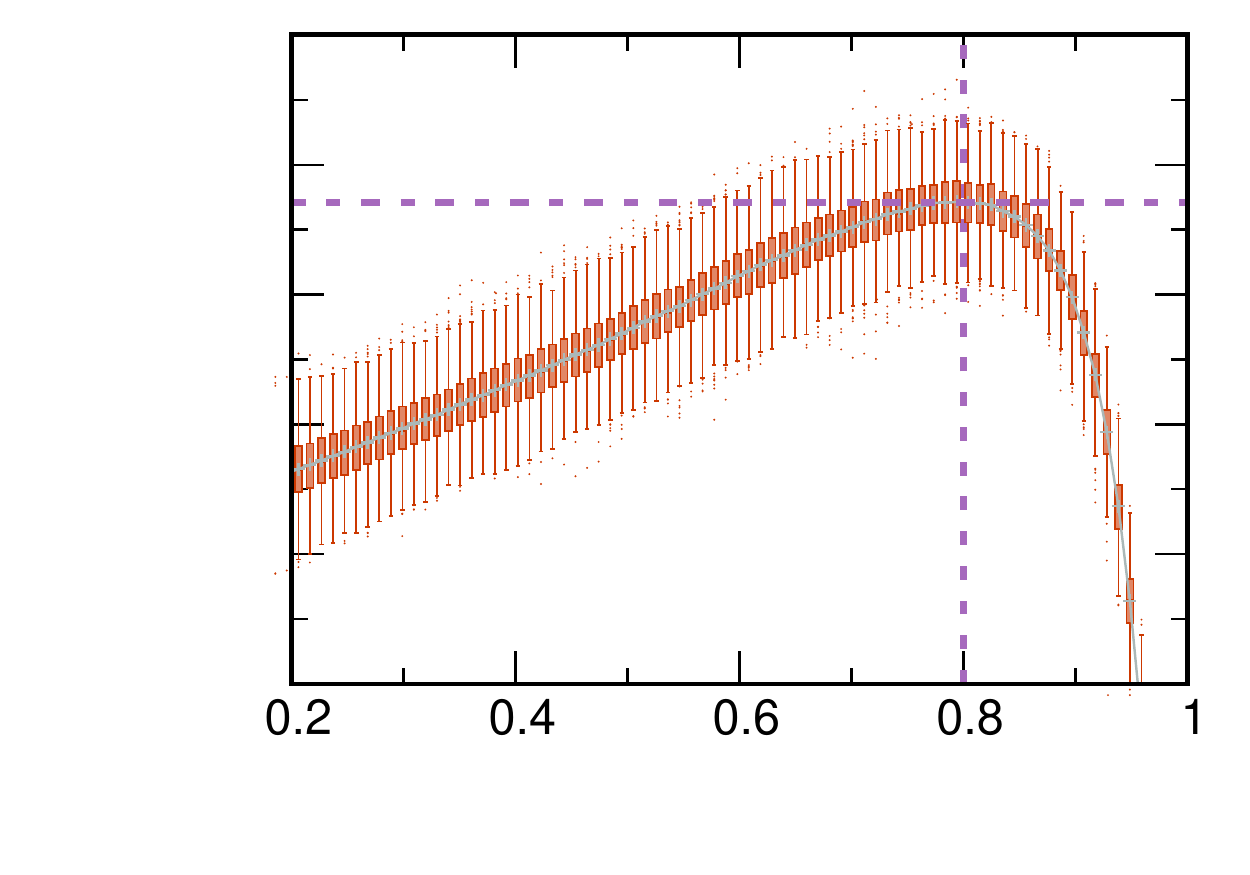}\hspace*{-0.4cm}
\includegraphics[width=0.26\textwidth]{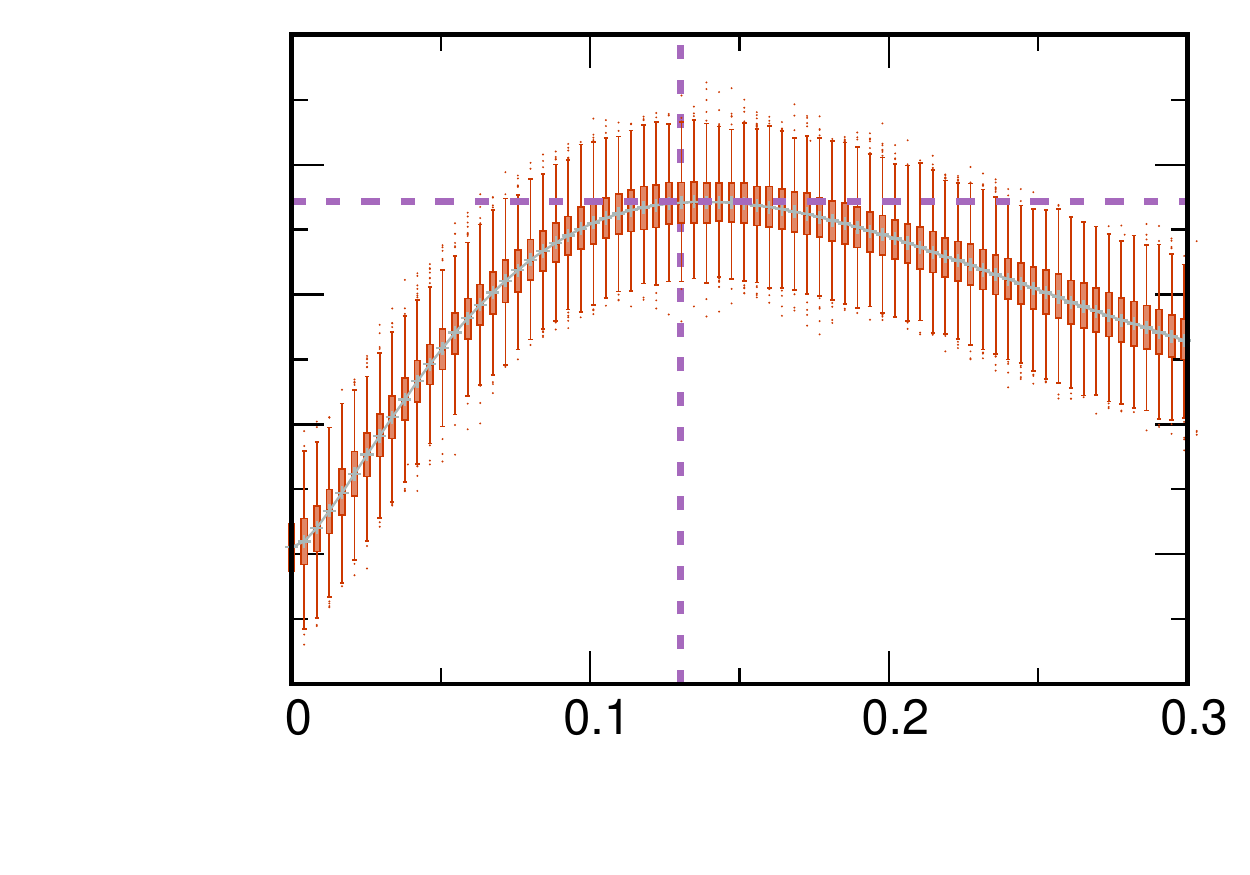}\hspace*{-0.4cm}
\includegraphics[width=0.26\textwidth]{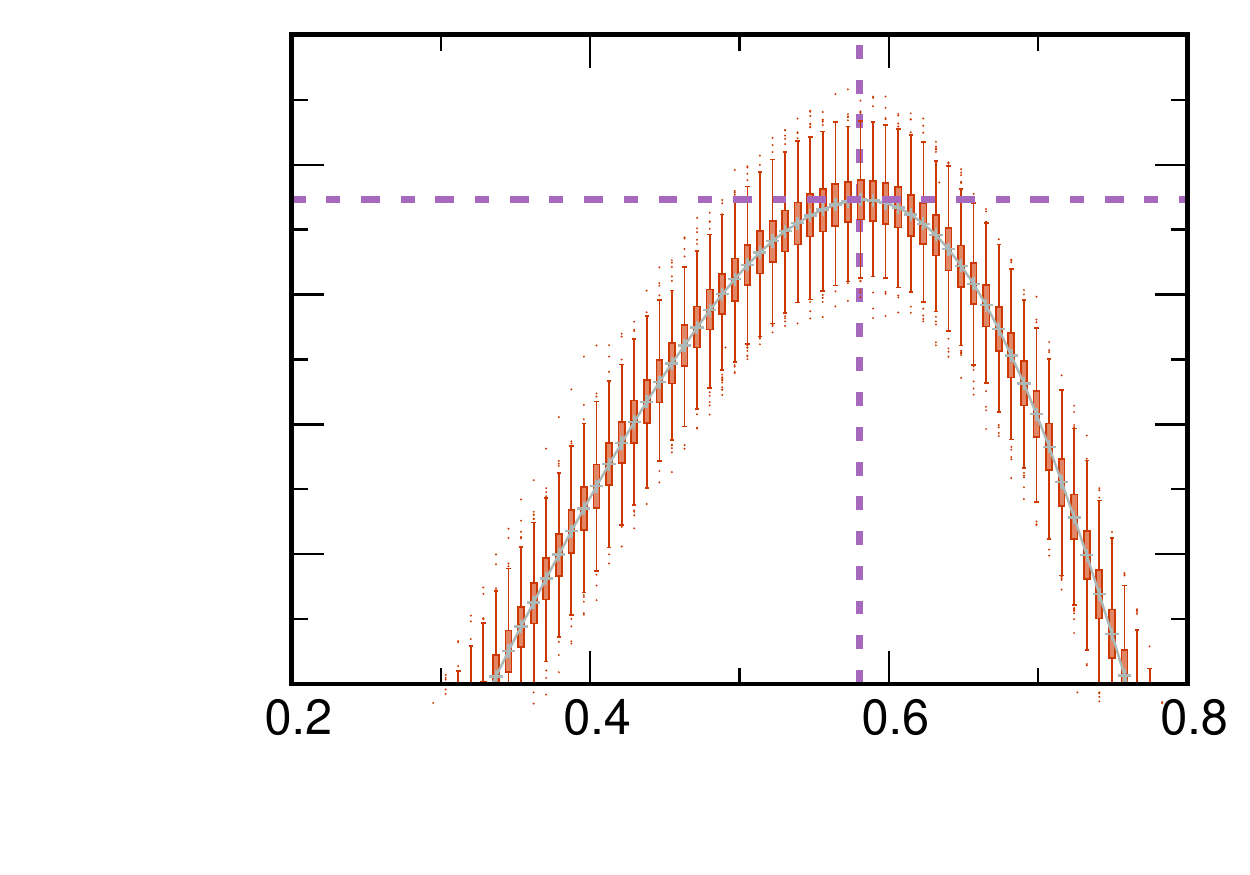}

\vspace*{-0.30cm}
\includegraphics[width=0.26\textwidth]{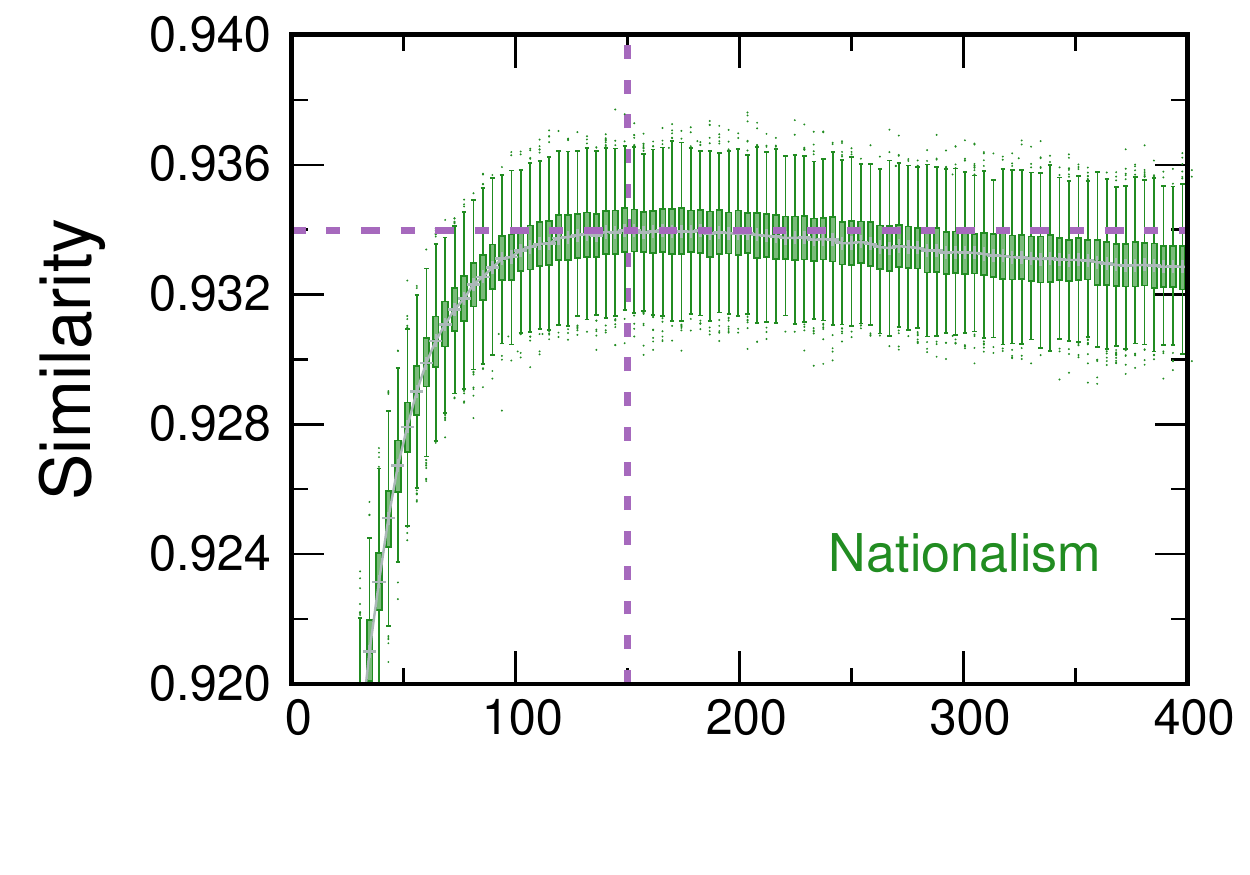}\hspace*{-0.4cm}
\includegraphics[width=0.26\textwidth]{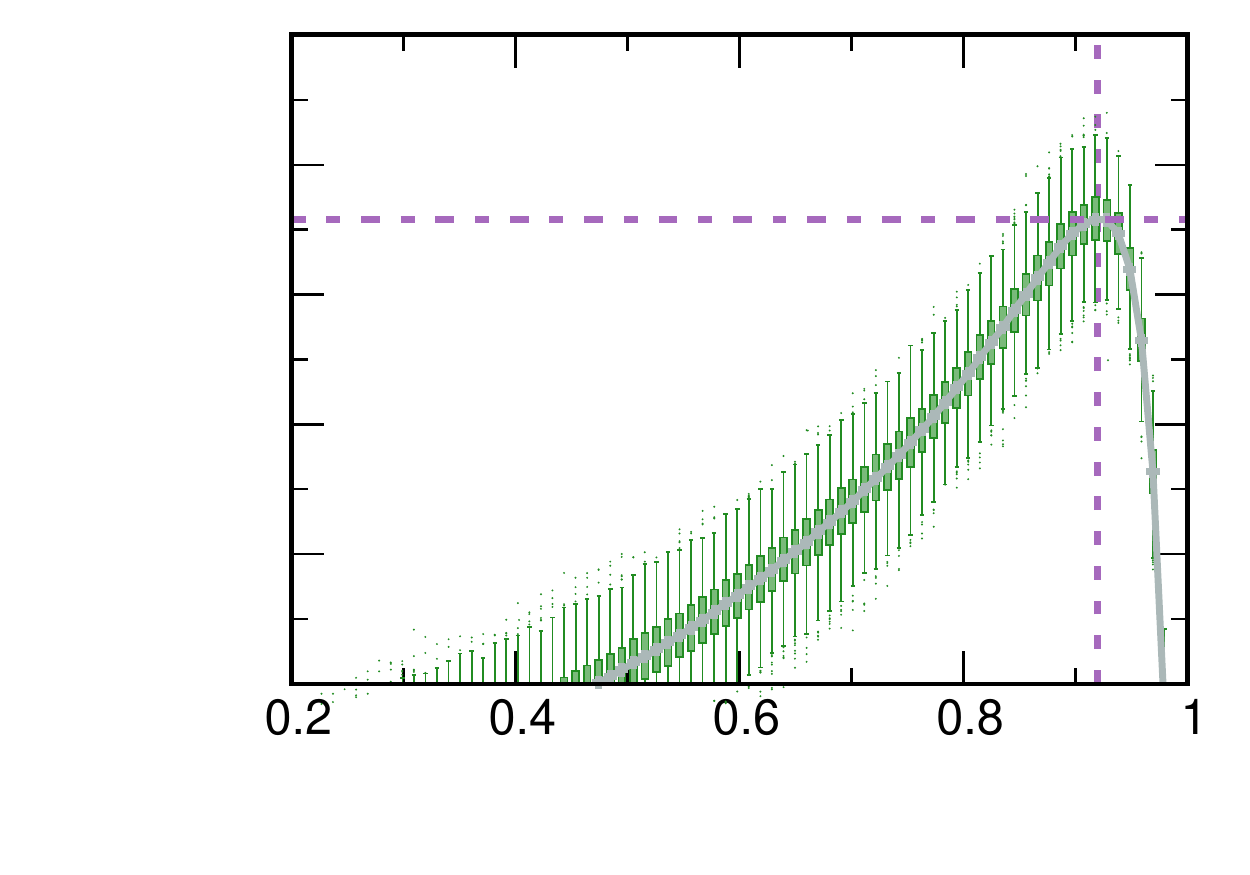}\hspace*{-0.4cm}
\includegraphics[width=0.26\textwidth]{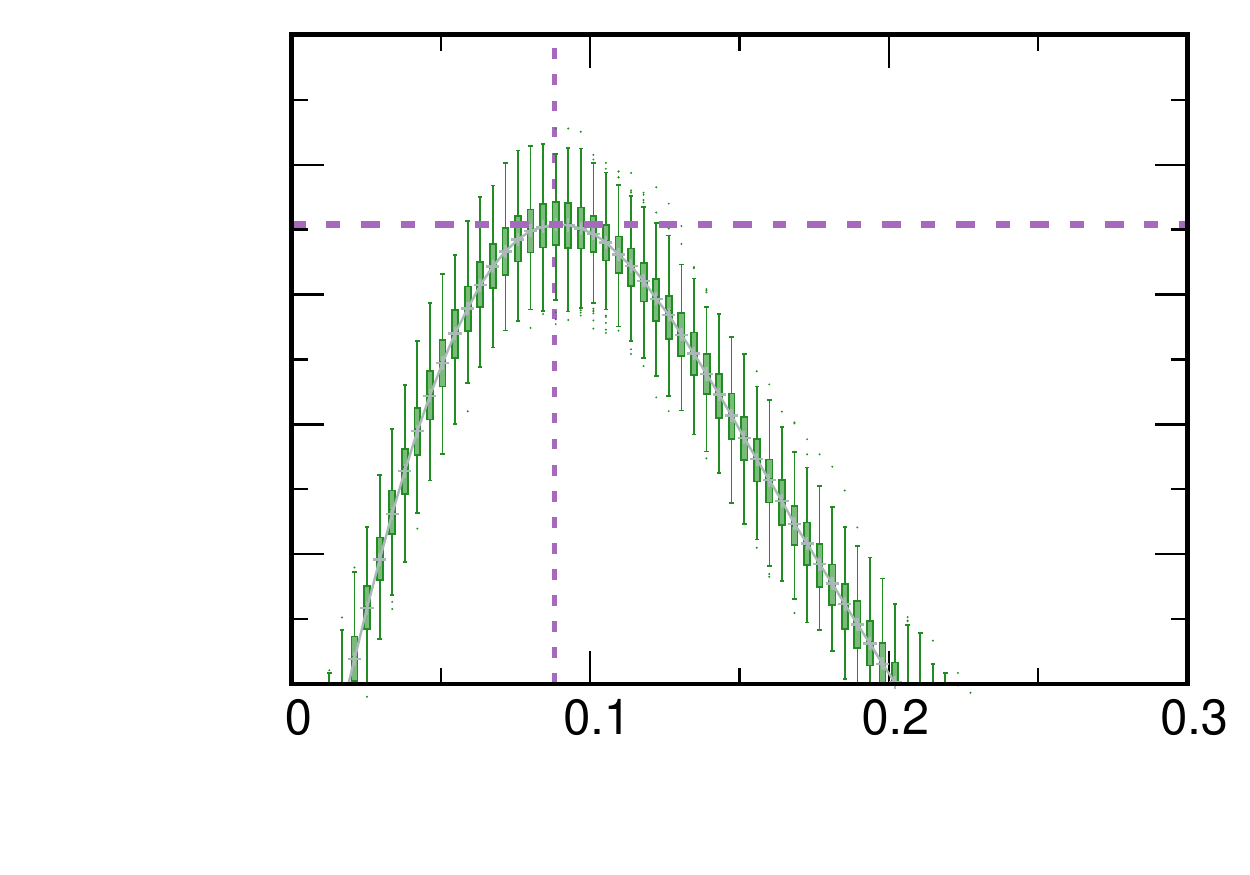}\hspace*{-0.4cm}
\includegraphics[width=0.26\textwidth]{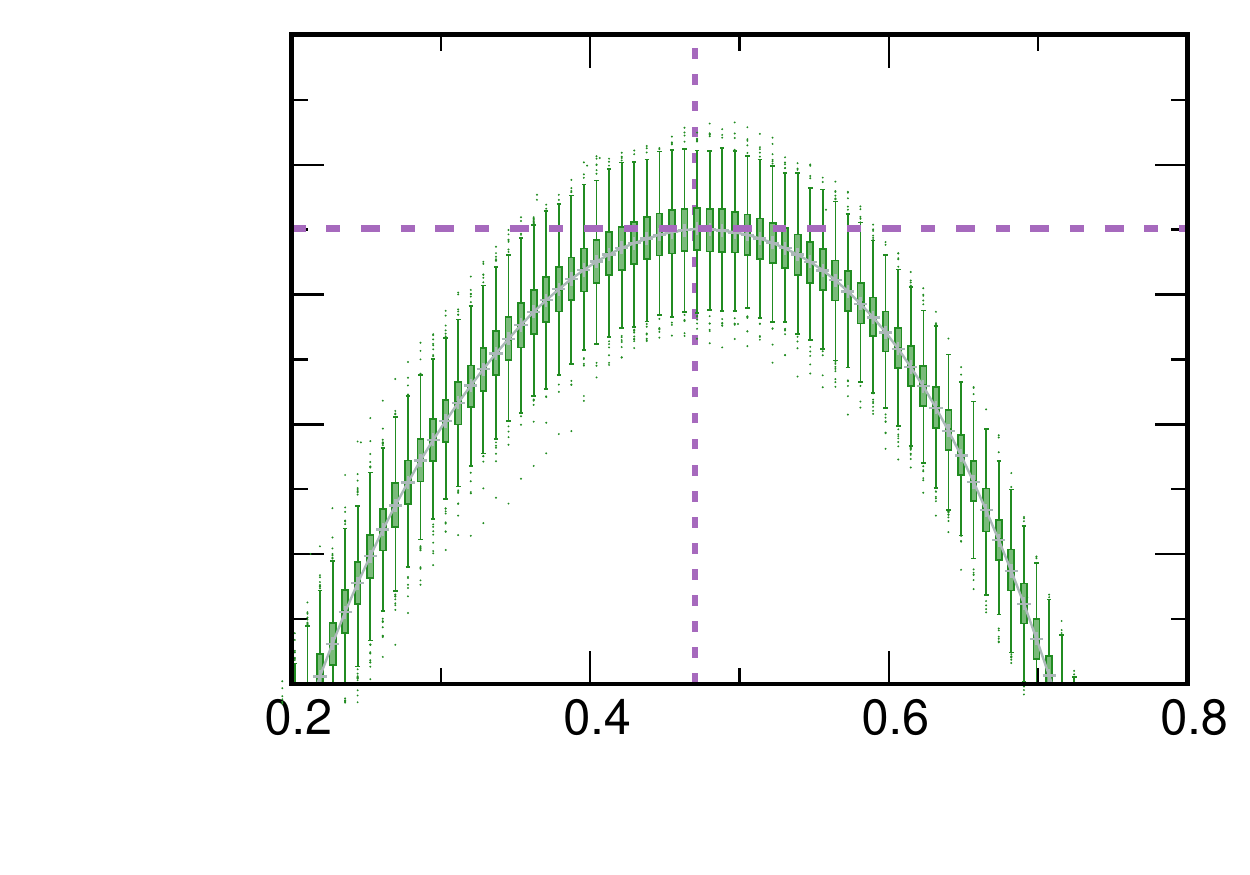}

\vspace*{-0.30cm}
\includegraphics[width=0.26\textwidth]{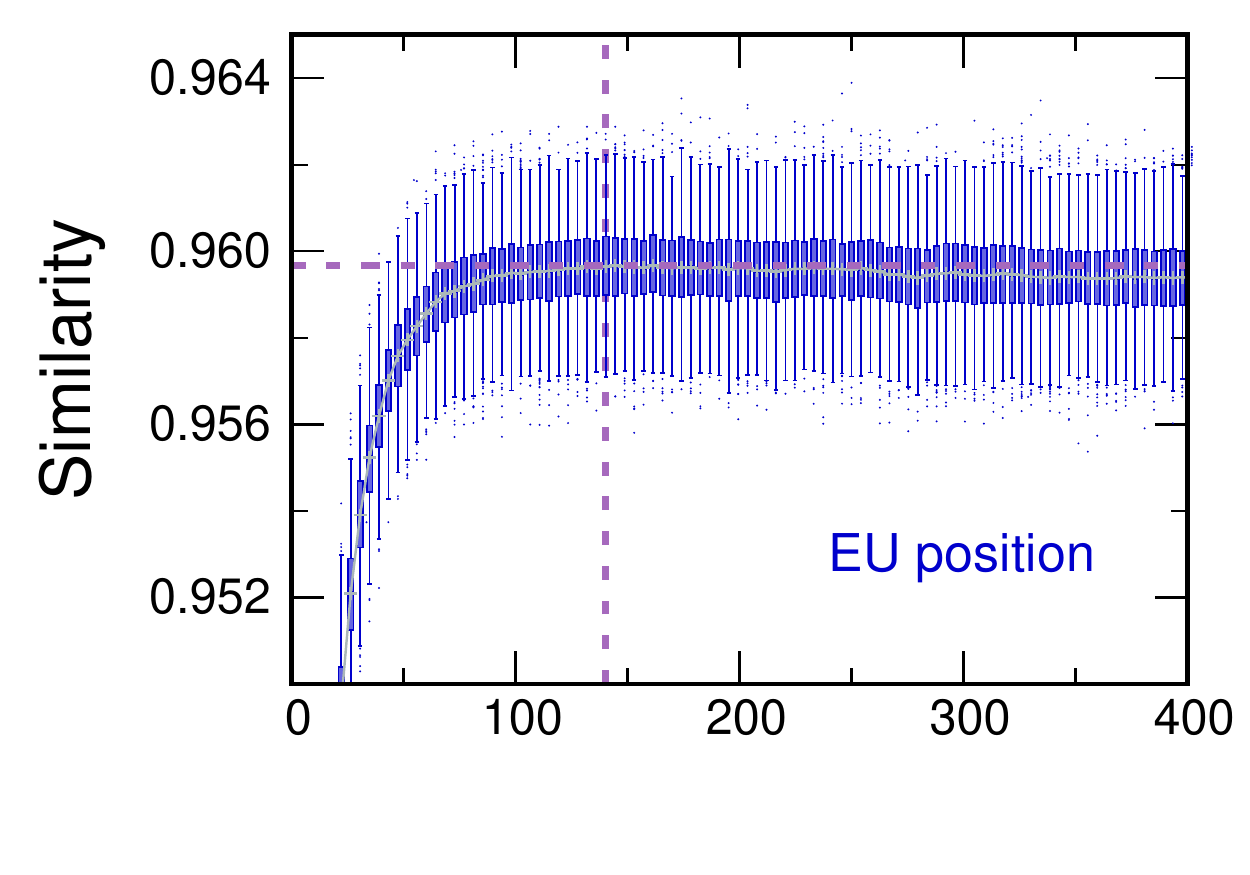}\hspace*{-0.4cm}
\includegraphics[width=0.26\textwidth]{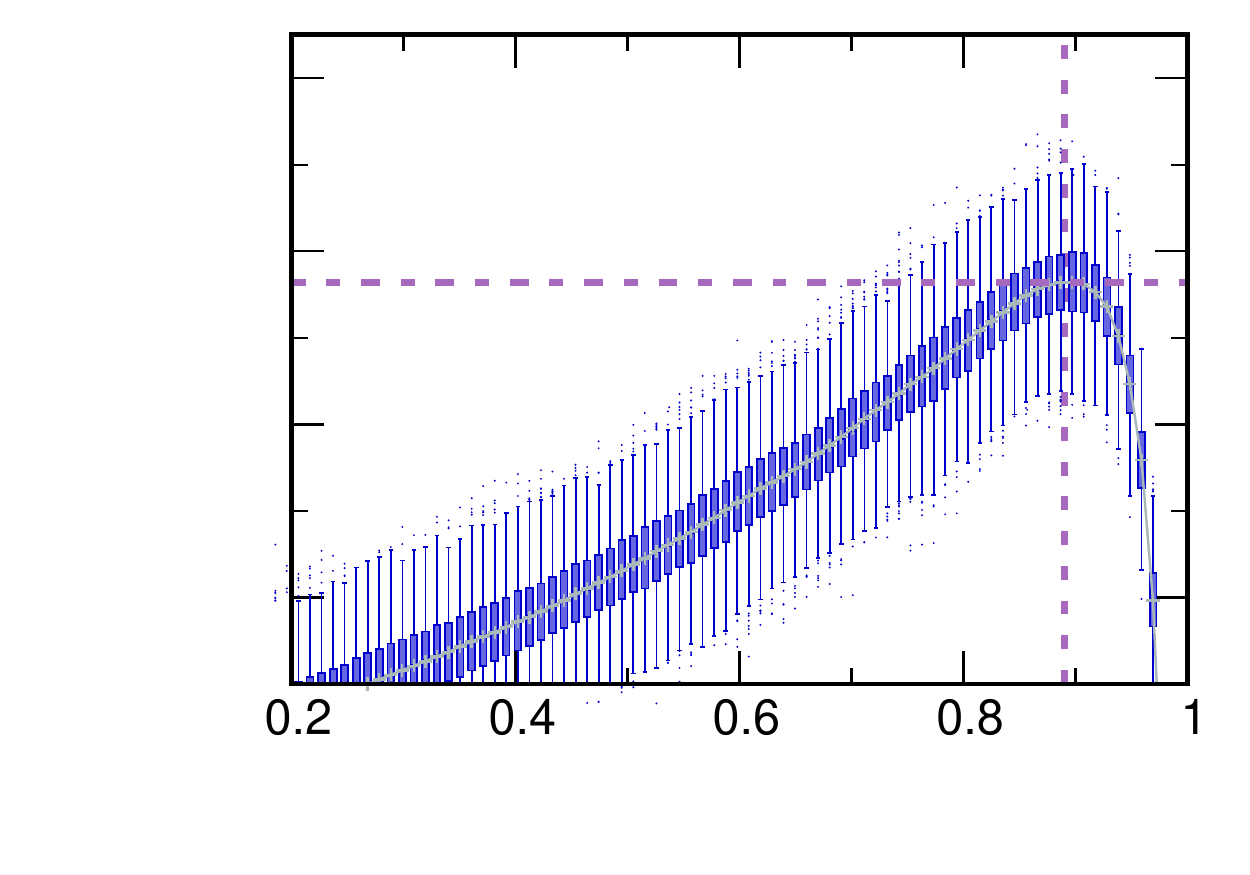}\hspace*{-0.4cm}
\includegraphics[width=0.26\textwidth]{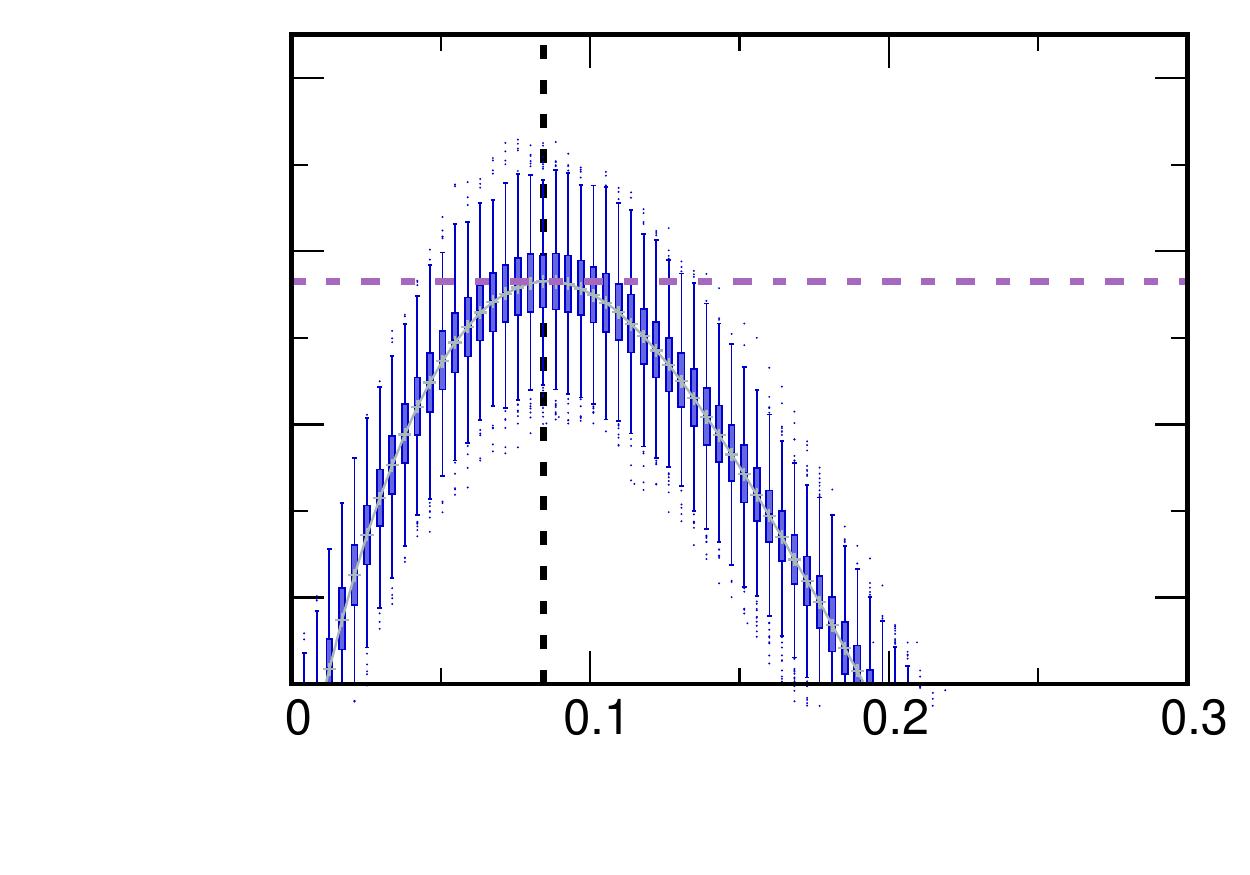}\hspace*{-0.4cm}
\includegraphics[width=0.26\textwidth]{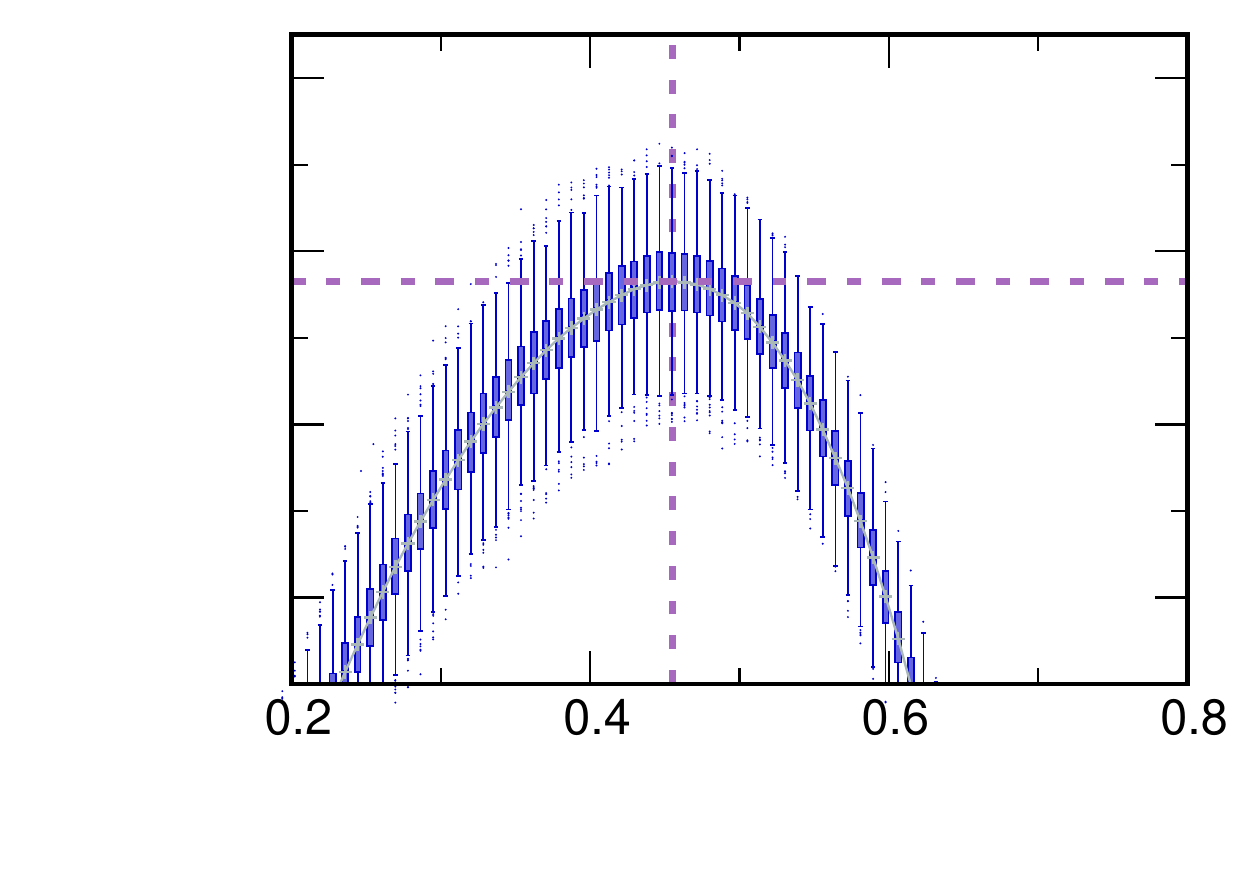}

\vspace*{-0.30cm}
\includegraphics[width=0.26\textwidth]{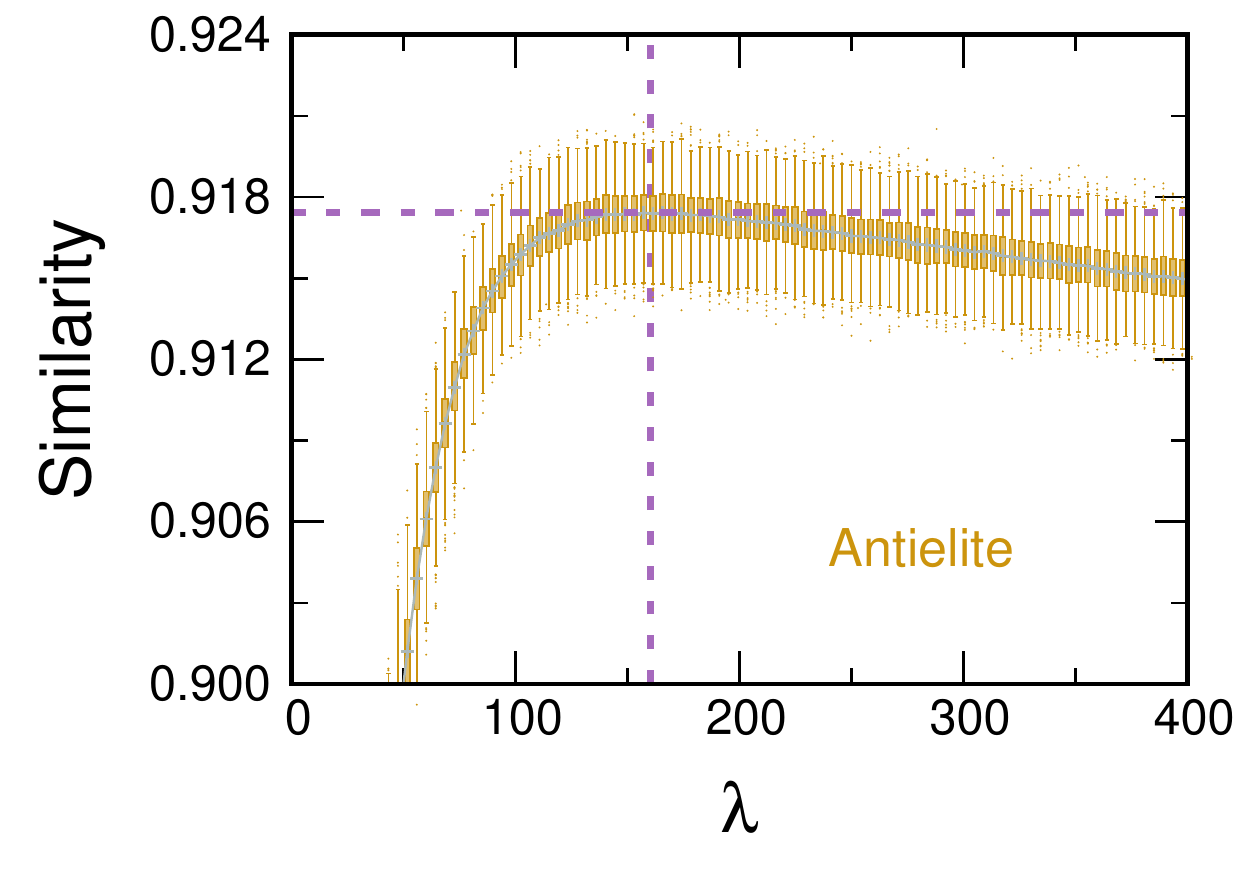}\hspace*{-0.4cm}
\includegraphics[width=0.26\textwidth]{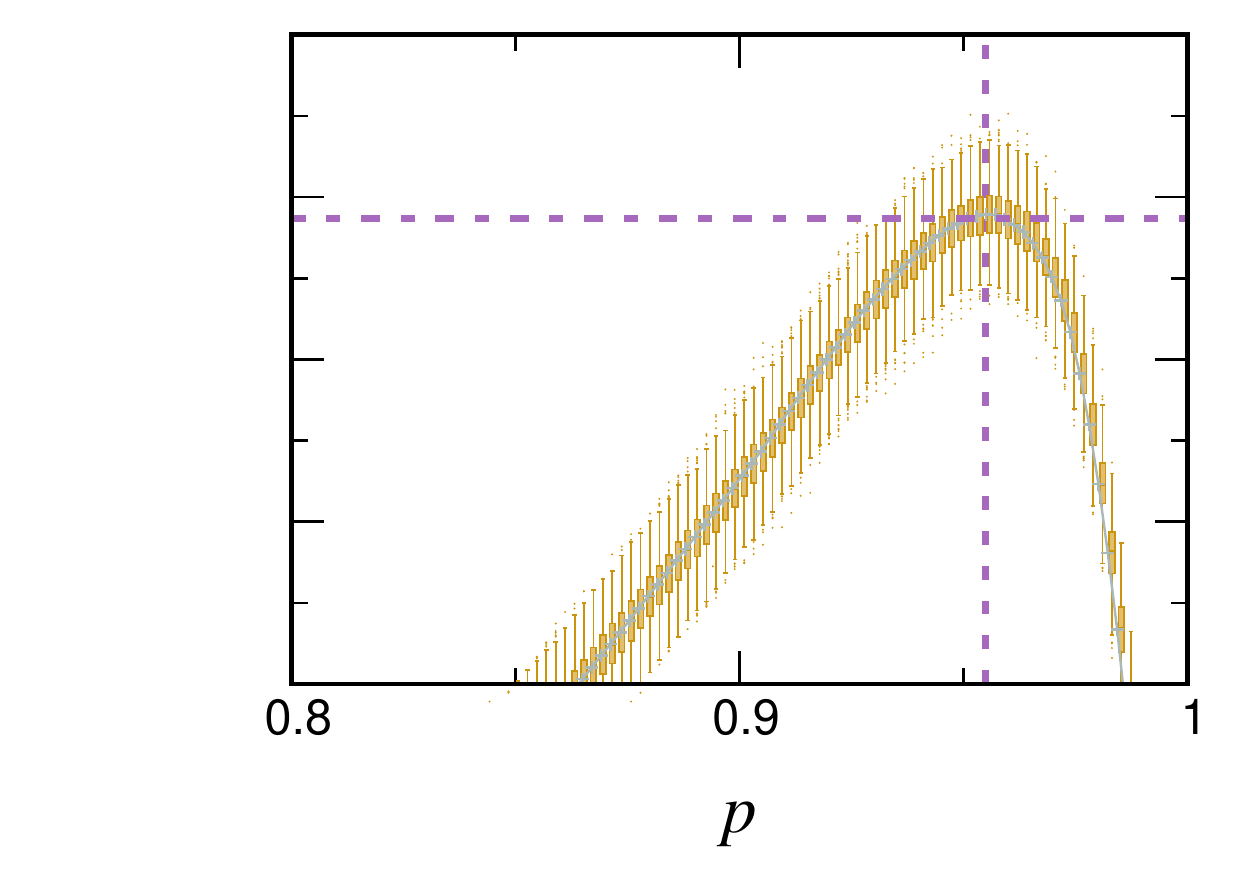}\hspace*{-0.4cm}
\includegraphics[width=0.26\textwidth]{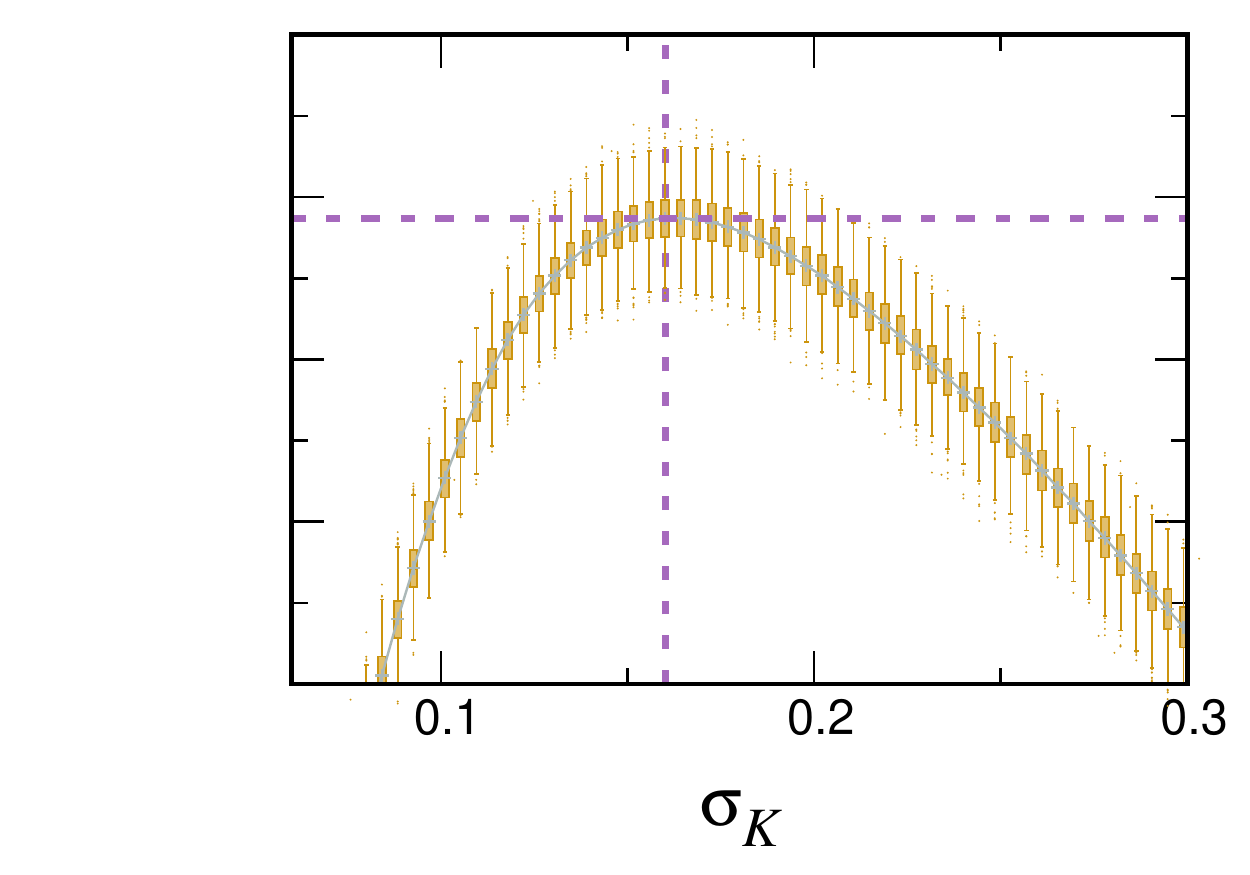}\hspace*{-0.4cm}
\includegraphics[width=0.26\textwidth]{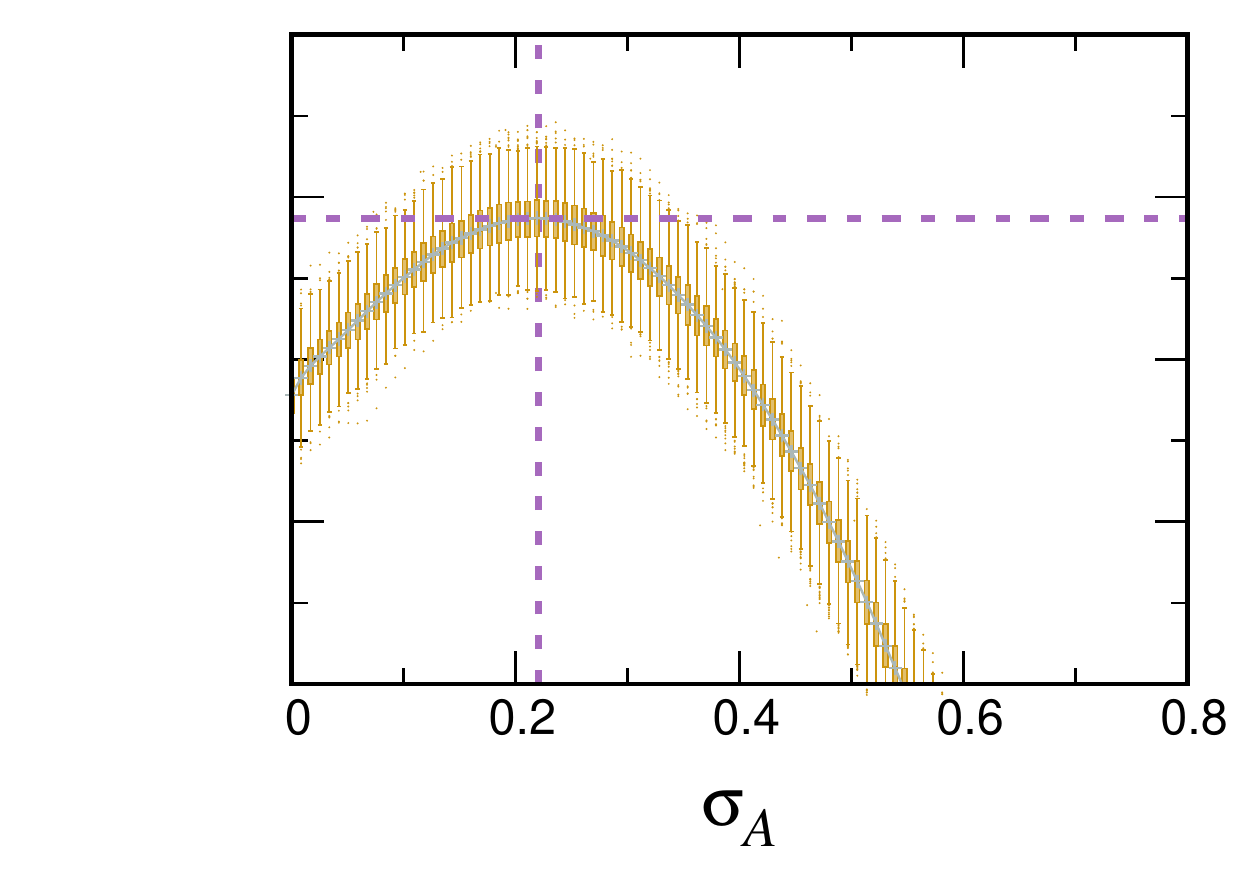}

\caption{Similarity, Eq. (\ref{sim_def1}), between the opinion distribution of the data and model as a function of the parameters, i.e., $\lambda$, $p$, $\sigma_{K}$, $\sigma_{A}$. The color of the errorbars indicate the opinion dimension (LR, NA, EU, AE) as specified in the subplots. The errorbars are calculated by means of a (thermalized) sample of time steps (1000 steps, separated each 10 Monte Carlo steps and after $t_{0}=1000$ thermalizing steps) taken from the simulations (stochastic fluctuations): the boxes correspond to the first and third quartiles, centered around the median (second quartile); while the whiskers extend to a value that lies within
1.5 times the interquartile range. The solid light gray lines are the average similarity of the step sample Eq. (\ref{sim_def2}), while the dashed (dark purple) lines cross at the maximum (average) similarity.}
\label{fitting_curves}
\end{center}
\end{figure}

The shape of the Similarity function Eq. (\ref{sim_def2}) is plotted in Fig. S\ref{fitting_curves} for each opinion dimension (LR, NA, EU, AE) as a function of the parameters $(\lambda, p, \sigma_{K},\sigma_{A})$. The maximum of the similarity function is shown in the same figure and the parameter values at the maximum are specified in Table S\ref{Table_best_parameters}.

\begin{table}[ht]
\centering
\caption{Parameters of the best fit of the model.}
\begin{tabular}{l|cccc|ccc}
 Dim. & $\lambda$ & $p$ & $\sigma_{K}$ & $\sigma_{A}$ & $\alpha_0$ & $\langle I \rangle$  & $\langle I^2 \rangle$ \\
\midrule
 LR & $140$ & $0.80$ & $0.13$ & $0.58$ &  $0.12$ & $0.20$ & $0.28$  \\
 NA & $150$ & $0.92$ & $0.088$ & $0.47$ &  $0.11$ & $0.08$ & $0.10$  \\
 EU & $140$ & $0.89$ & $0.084$ & $0.46$ & $0.12$ & $0.11$ & $0.14$   \\
 AE & $160$ & $0.96$ & $0.16$ & $0.22$ &  $0.11$ & $0.04$ & $0.07$  \\
\bottomrule
\end{tabular}
\label{Table_best_parameters}
\end{table}
Note that we do not specify the errors of the parameters of the model, but they can be directly inferred from the errorbars in Fig. S\ref{fitting_curves} which come from the stochastic fluctuations of the model. More than one (or two) significant figures in the parameters would not correspond to a realistic estimation of the errors. Note also the asymmetry in Fig. S\ref{fitting_curves} of the errors of some parameters, e.g., $\lambda$.

\subsubsection{Importance of the opinion borders}\label{sec_boundary}

\begin{figure}[h!]
\begin{center}
\hspace*{-0.2cm}
\subfloat[]{\label{opinions_borders:a}\includegraphics[width=0.25\textwidth]{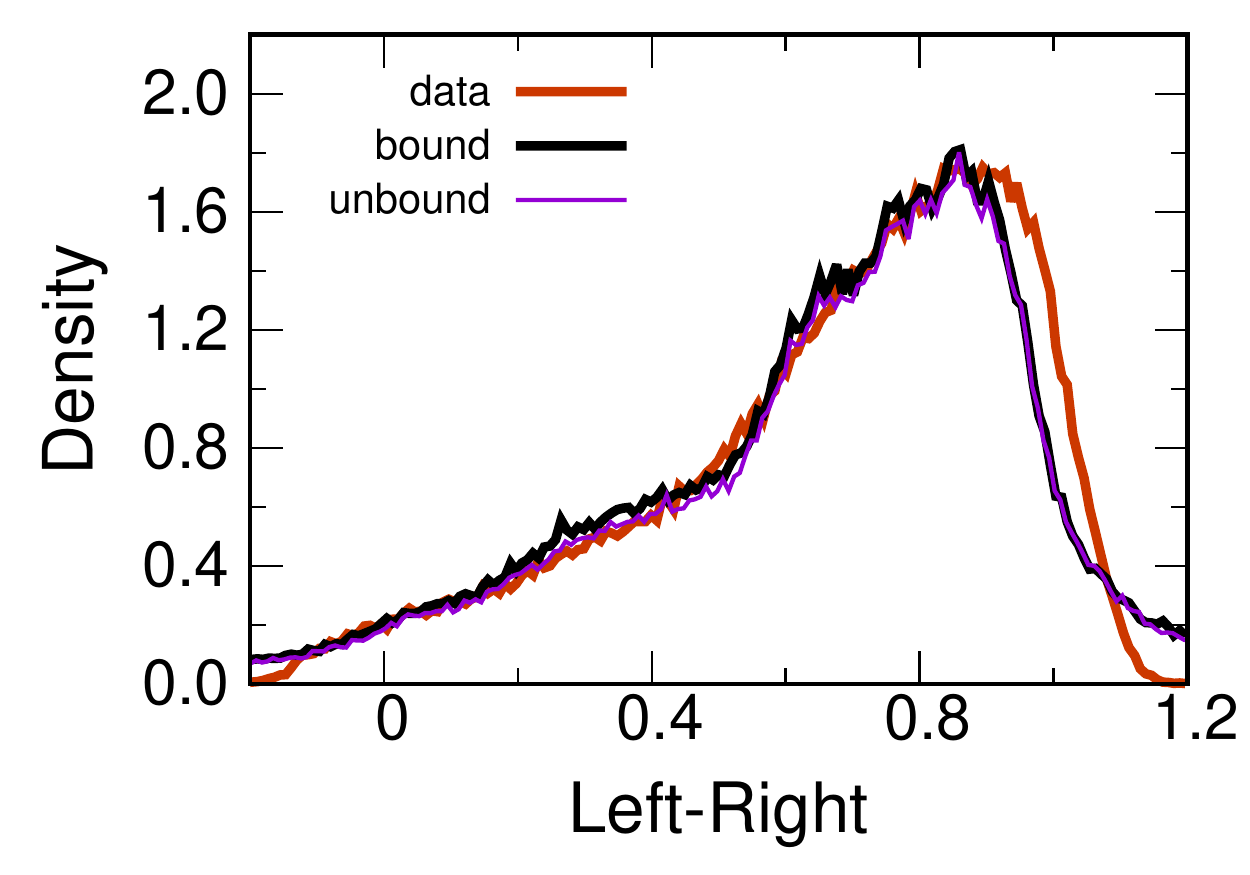}}
\subfloat[]{\label{opinions_borders:b}\includegraphics[width=0.25\textwidth]{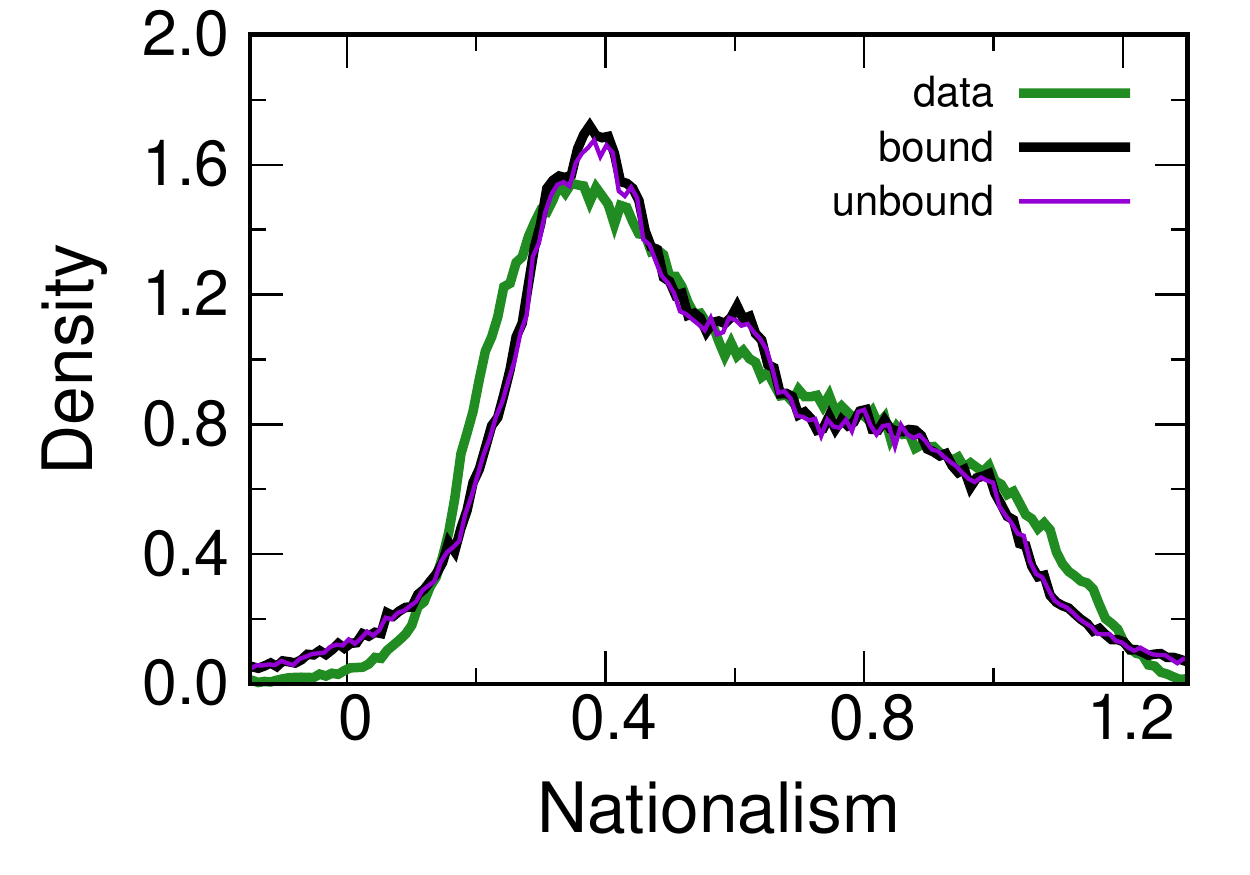}}
\subfloat[]{\label{opinions_borders:c}\includegraphics[width=0.25\textwidth]{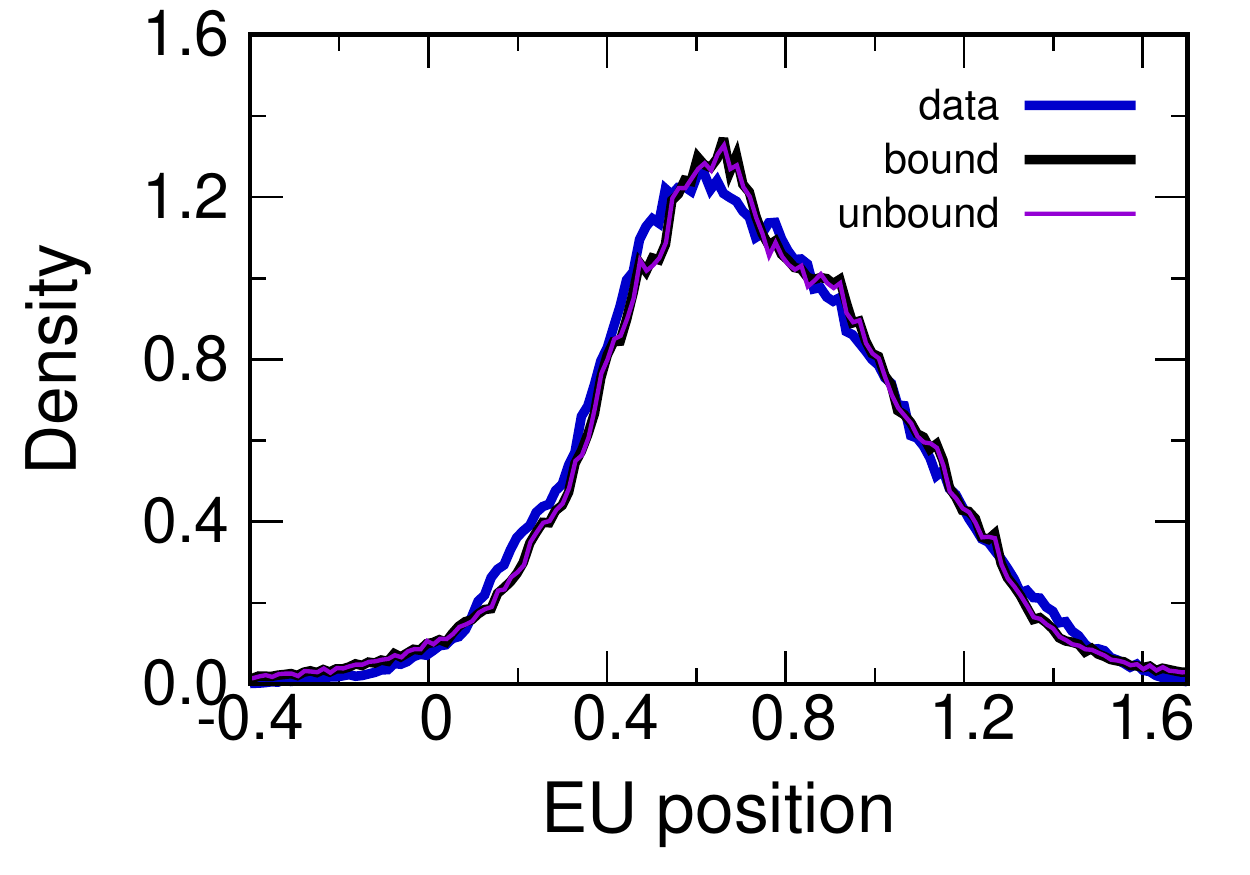}}
\subfloat[]{\label{opinions_borders:d}\includegraphics[width=0.25\textwidth]{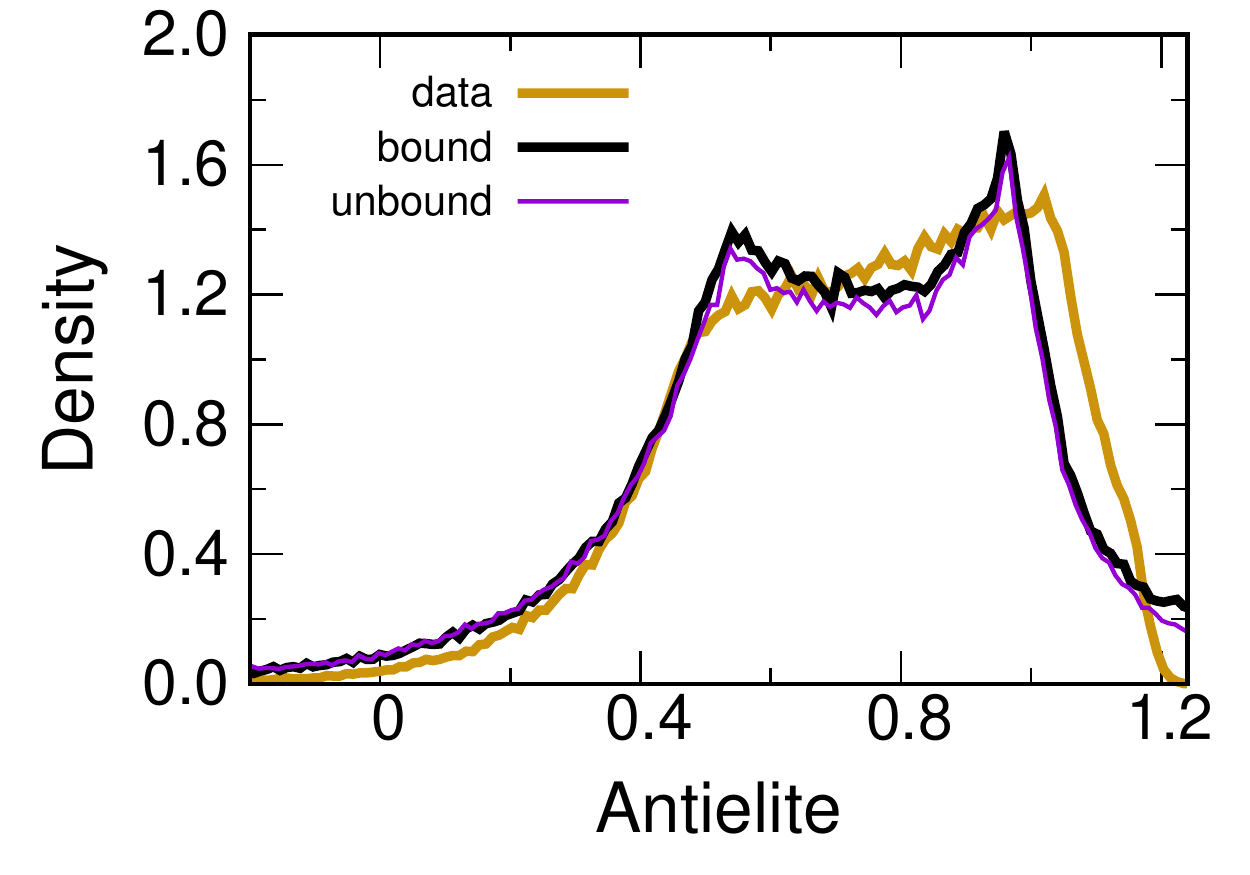}}

\hspace*{-0.2cm}
\subfloat[]{\label{opinions_borders:e}\includegraphics[width=0.25\textwidth]{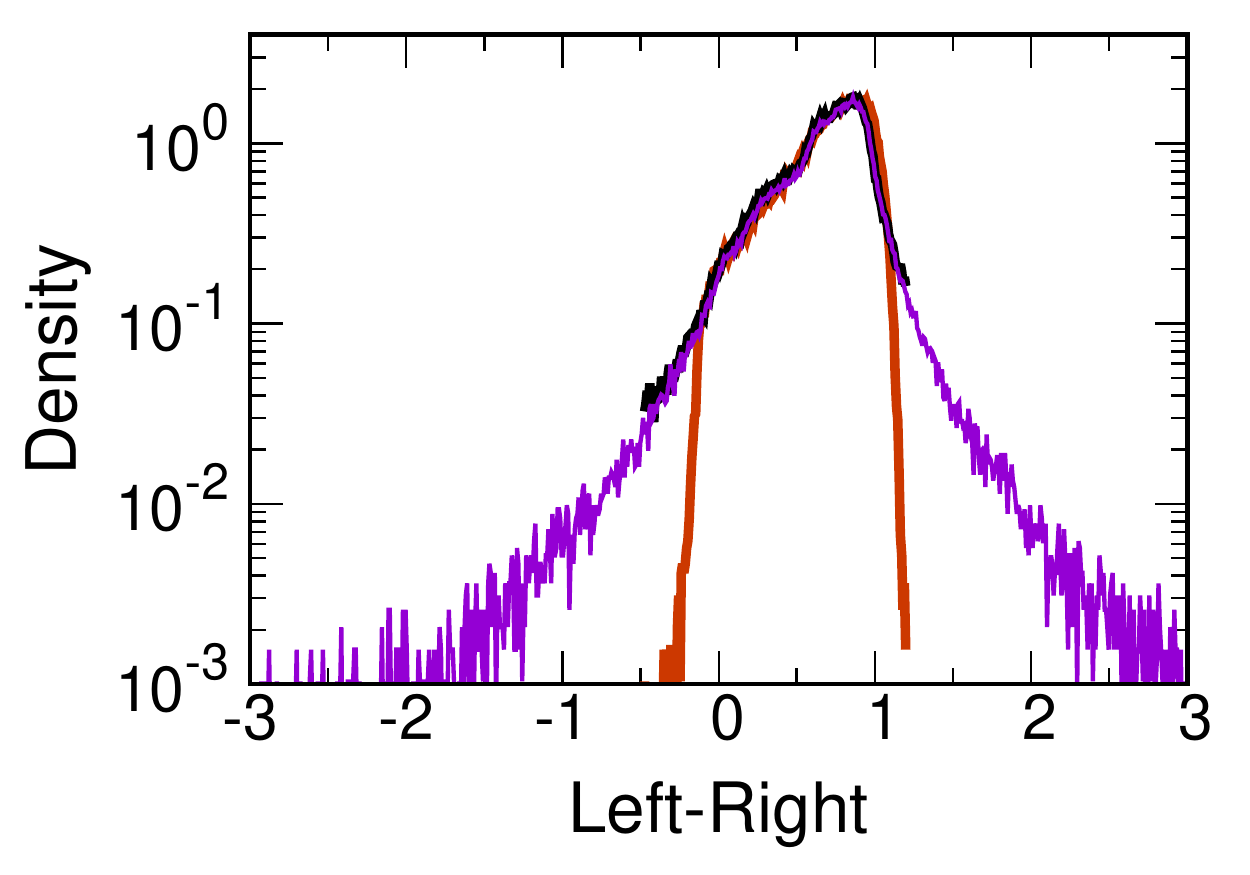}}
\subfloat[]{\label{opinions_borders:f}\includegraphics[width=0.25\textwidth]{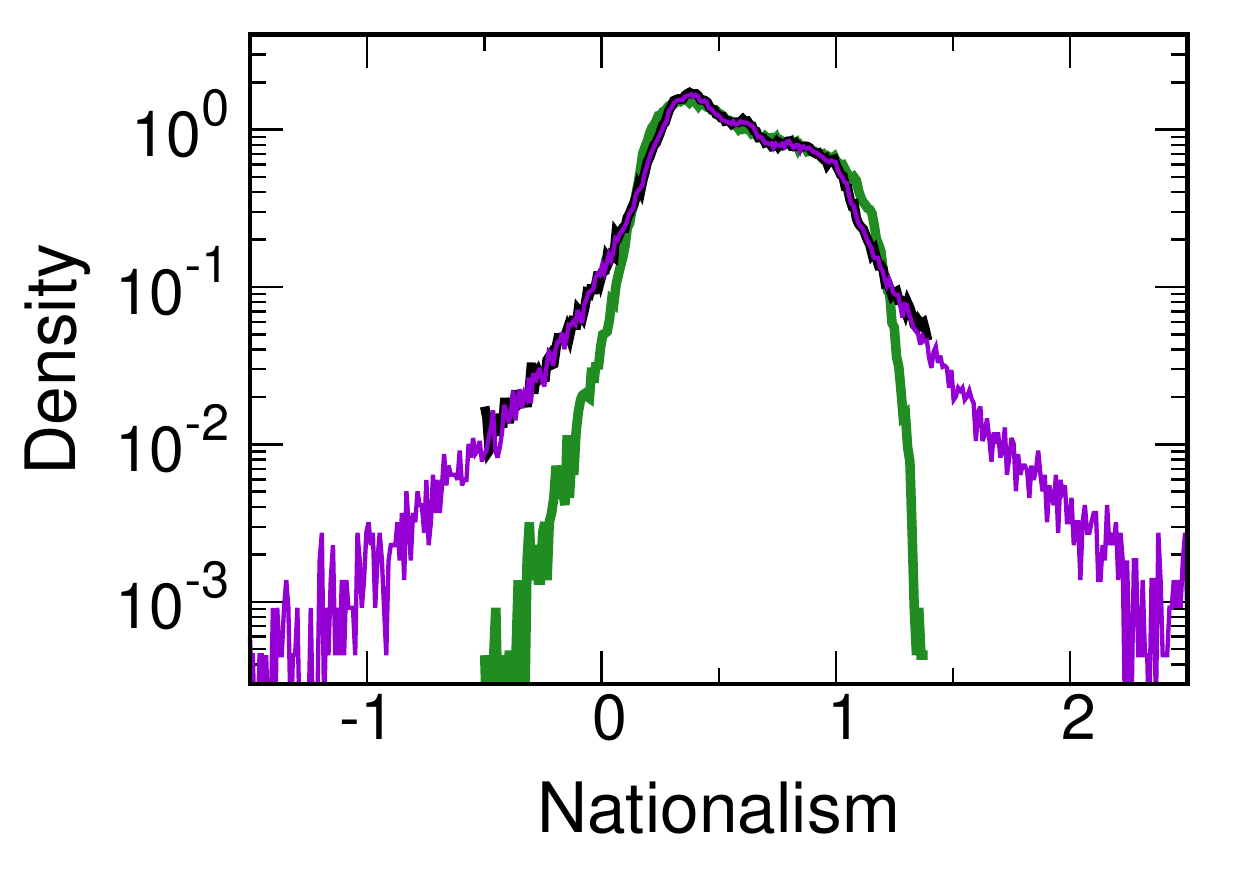}}
\subfloat[]{\label{opinions_borders:g}\includegraphics[width=0.25\textwidth]{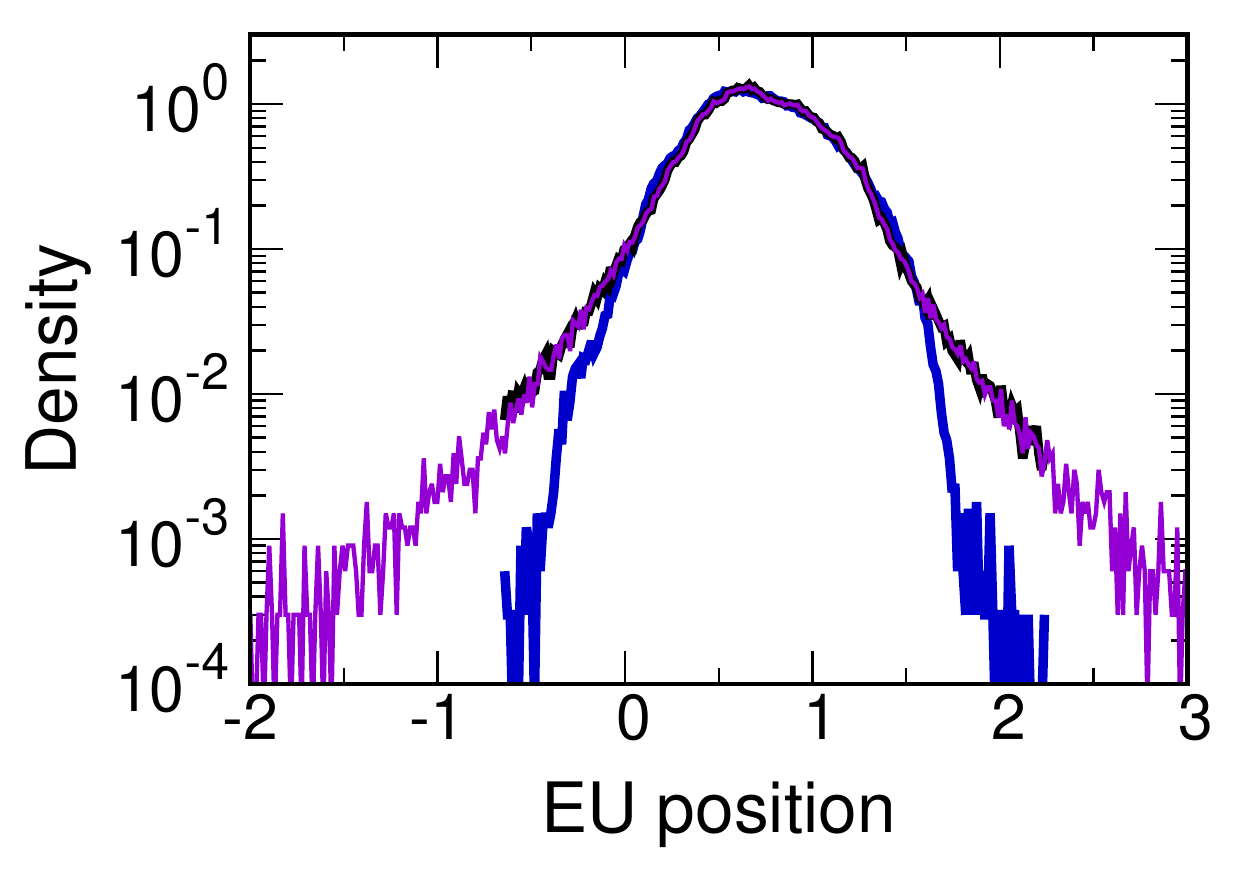}}
\subfloat[]{\label{opinions_borders:h}\includegraphics[width=0.25\textwidth]{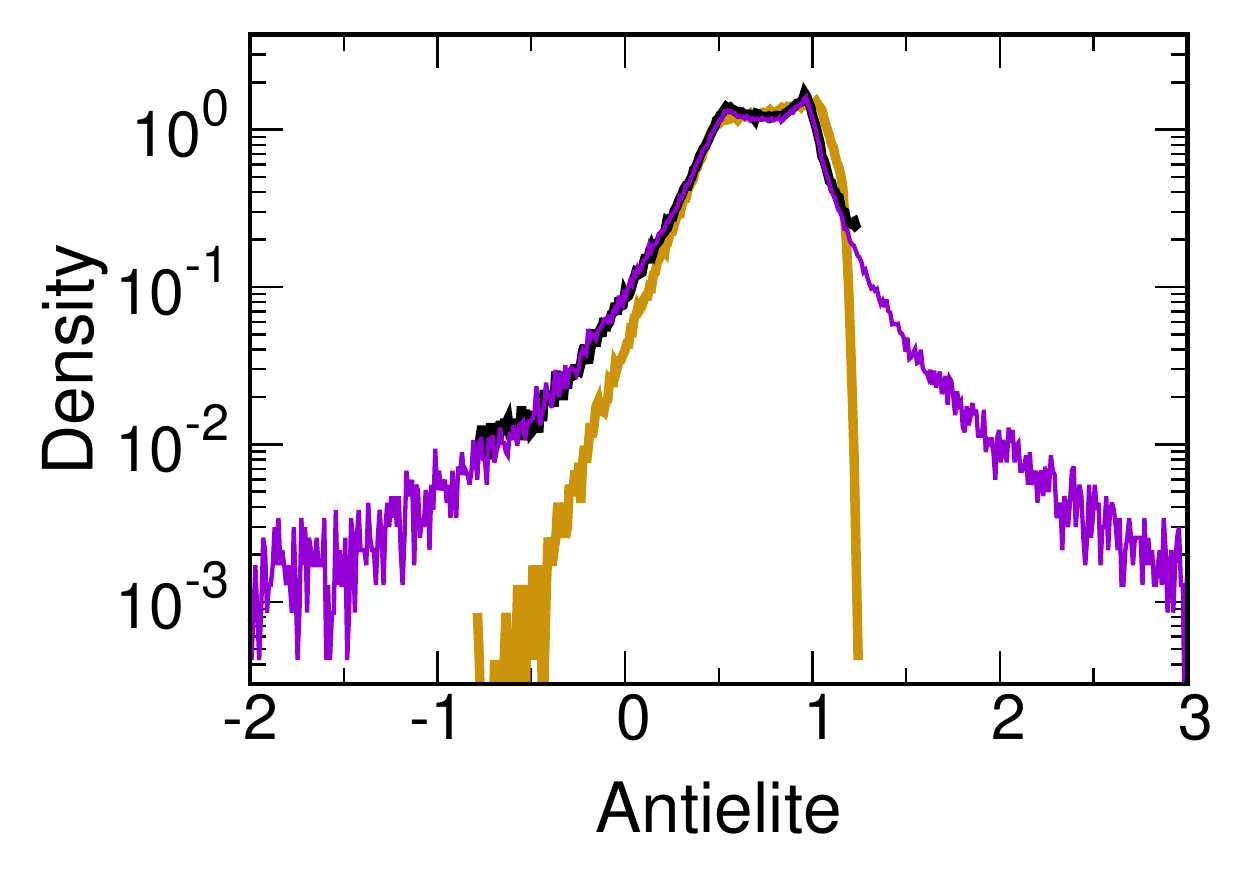}}
\caption{Results of the best fit of the model, Table S\ref{Table_best_parameters}, with and without boundary conditions (bound/unbound) and its comparison with the data. Probability density function of the different opinion variables of the users (LR, NA, EU and AE) in panels (a/e, b/f, c/g, d/h) in linear/logarithmic scale for the vertical axis. The colored solid lines correspond to the values from the data, while the stationary results of the model with boundary conditions are represented by black solid lines and without boundary conditions by purple solid lines.}
\label{opinions_borders}
\end{center}
\end{figure}

During the fitting process, Section \ref{sec_fitting}, we used a boundary condition in the model which amounts to neglecting those influence processes that would result in the opinion of users going beyond some predefined limits $(x_{\text{min}}, x_{\text{max}})$. The reason for this boundary condition is: first in the definition of Similarity Eq. (\ref{sim_def1}), which assumes the existence of these limits, and in the fact that comparing distributions with different limits would result in a worse fitting; and second that the boundary limits improve the model and makes it more realistic. Note that we included in the model influence processes with $I<0$ and $I>1$ which have some limitations in reality, especially for interactions of individuals with extreme opinions. For example, an individual would hardly adopts the opinion of the most extreme individual in the population, and even less overreact to it with $I>1$, especially if their initial difference in opinions is large \cite{Moussaid:2013, Chacoma:2015}. Thus, the boundary condition offers a convenient solution of this limitation in the formulation of the model. In Fig. S\ref{opinions_borders} we compare the stationary results of the best fit of the model, Table S\ref{Table_best_parameters}, with and without boundary conditions altogether with the data. We observe that there is no qualitative difference in the comparison between the model with and without boundary conditions and the data. Note however how the model without boundary condition shows an exponential tail in the opinion distribution that goes beyond the limits of the data.

\subsubsection{Dynamics and stationary of the model}\label{model_dynamics}

\begin{figure}[h!]
 \begin{center}
 \hspace*{-0.2cm}
 \subfloat[]{\label{evolution:a}\includegraphics[width=0.25\textwidth]{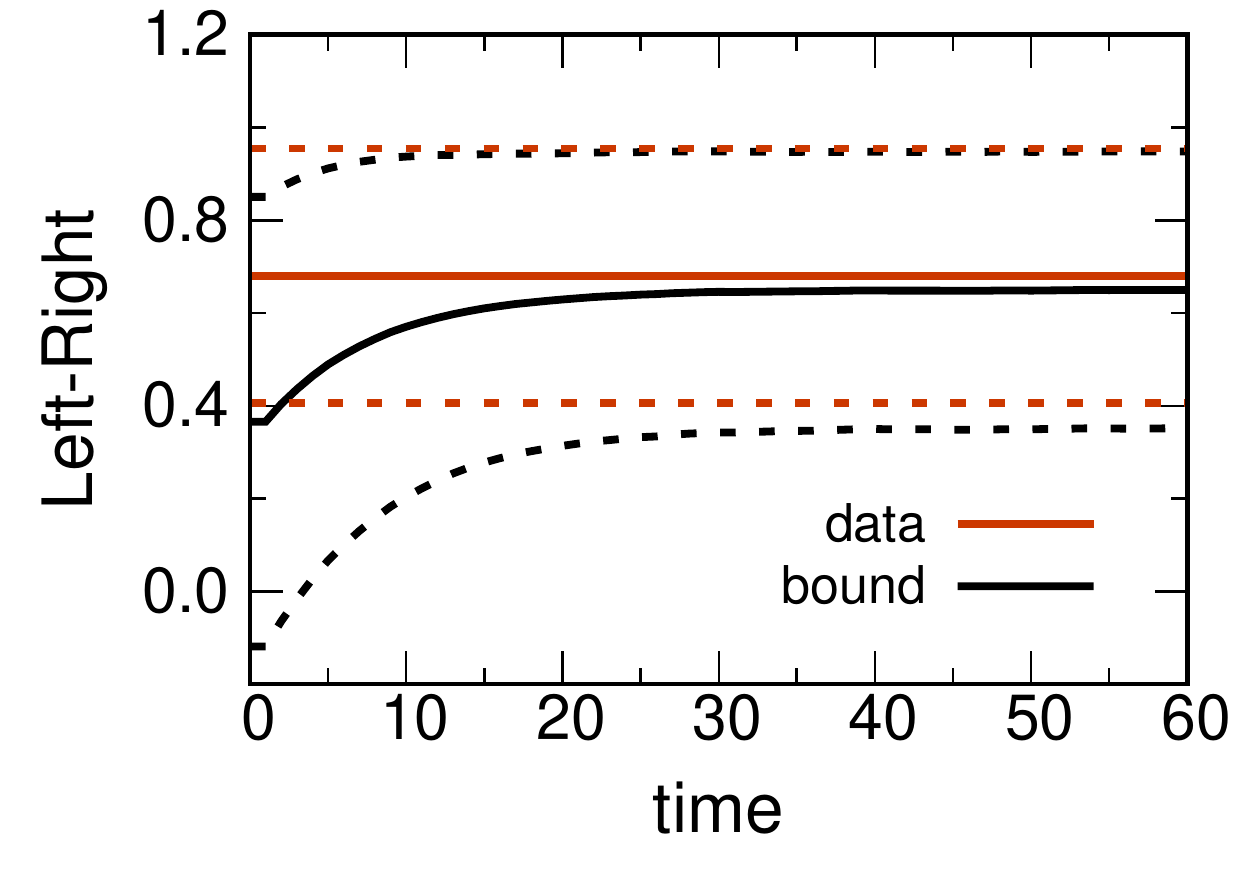}}
 \subfloat[]{\label{evolution:b}\includegraphics[width=0.25\textwidth]{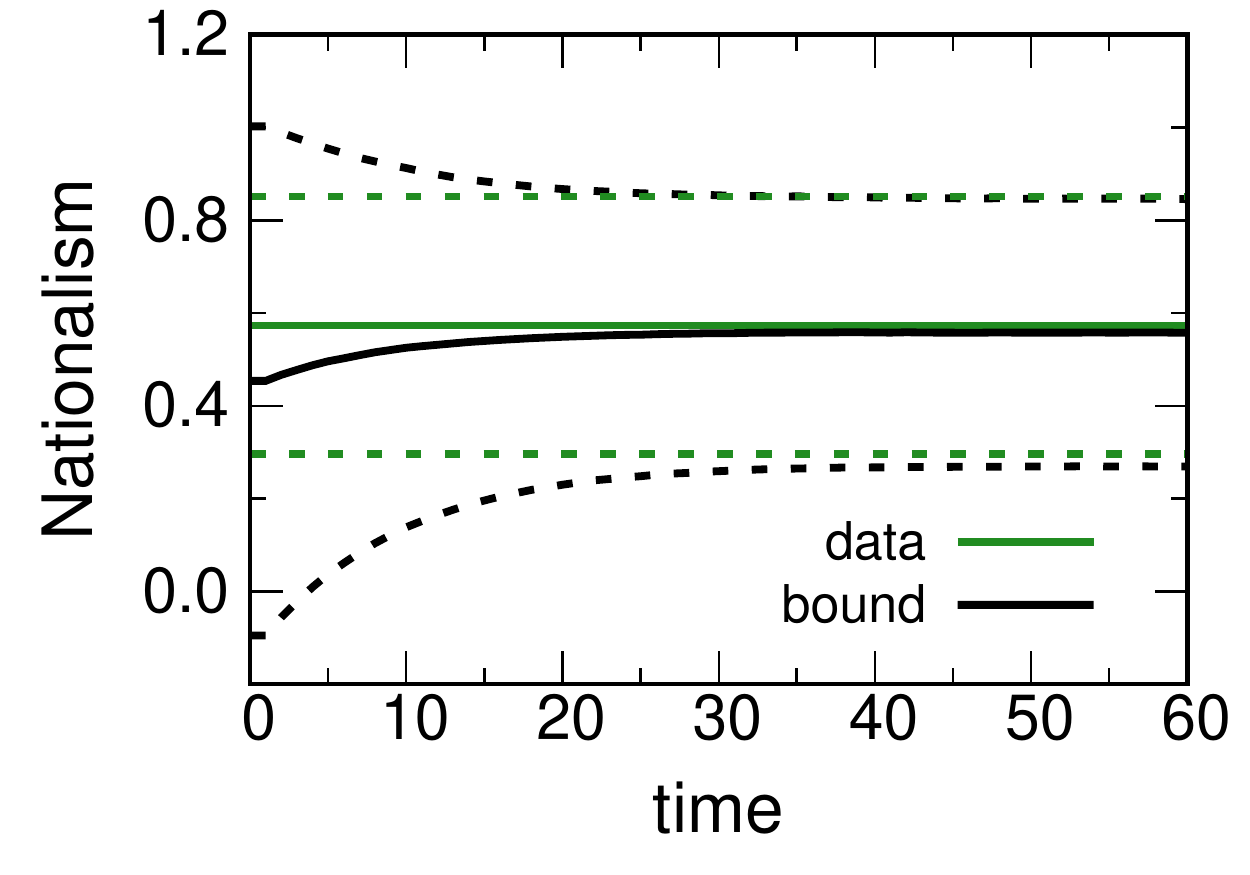}}
 \subfloat[]{\label{evolution:c}\includegraphics[width=0.25\textwidth]{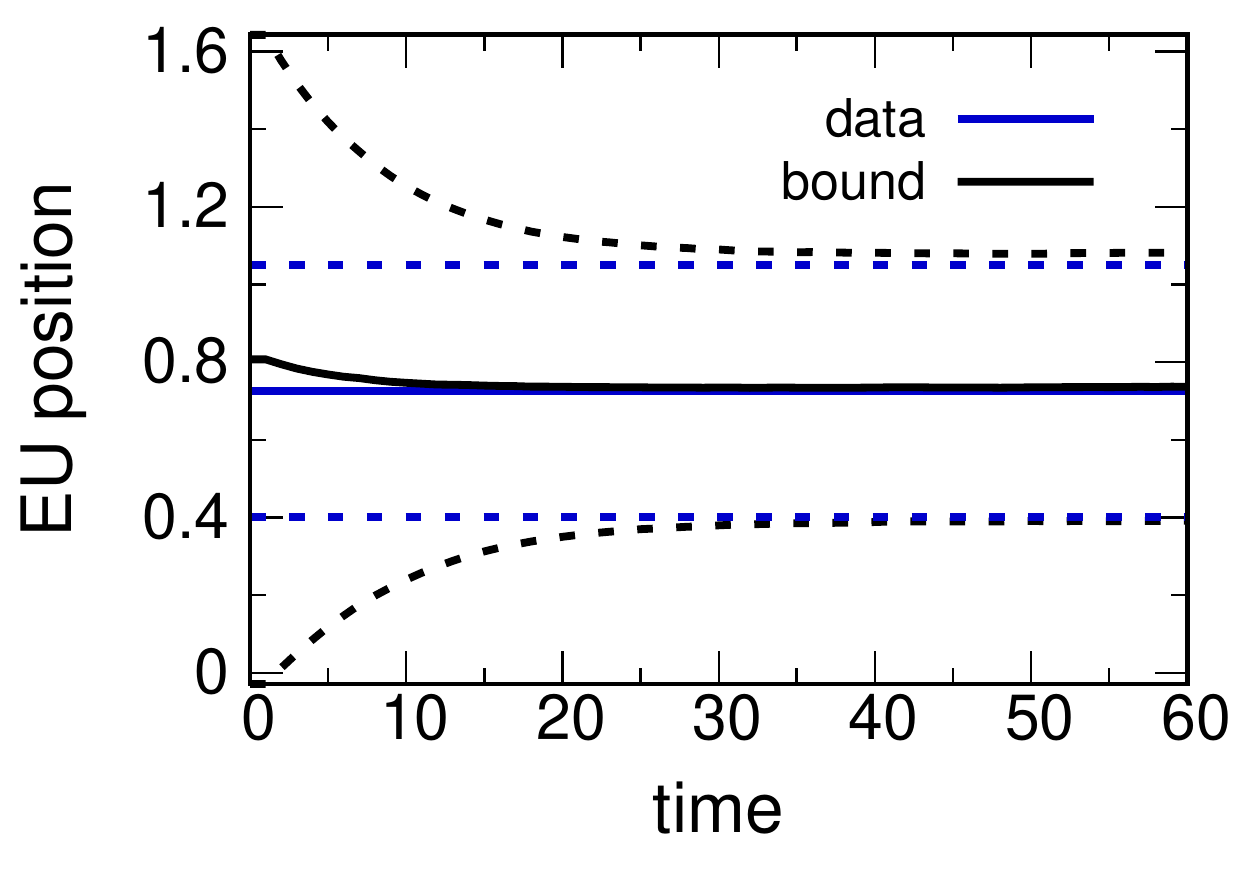}}
 \subfloat[]{\label{evolution:d}\includegraphics[width=0.25\textwidth]{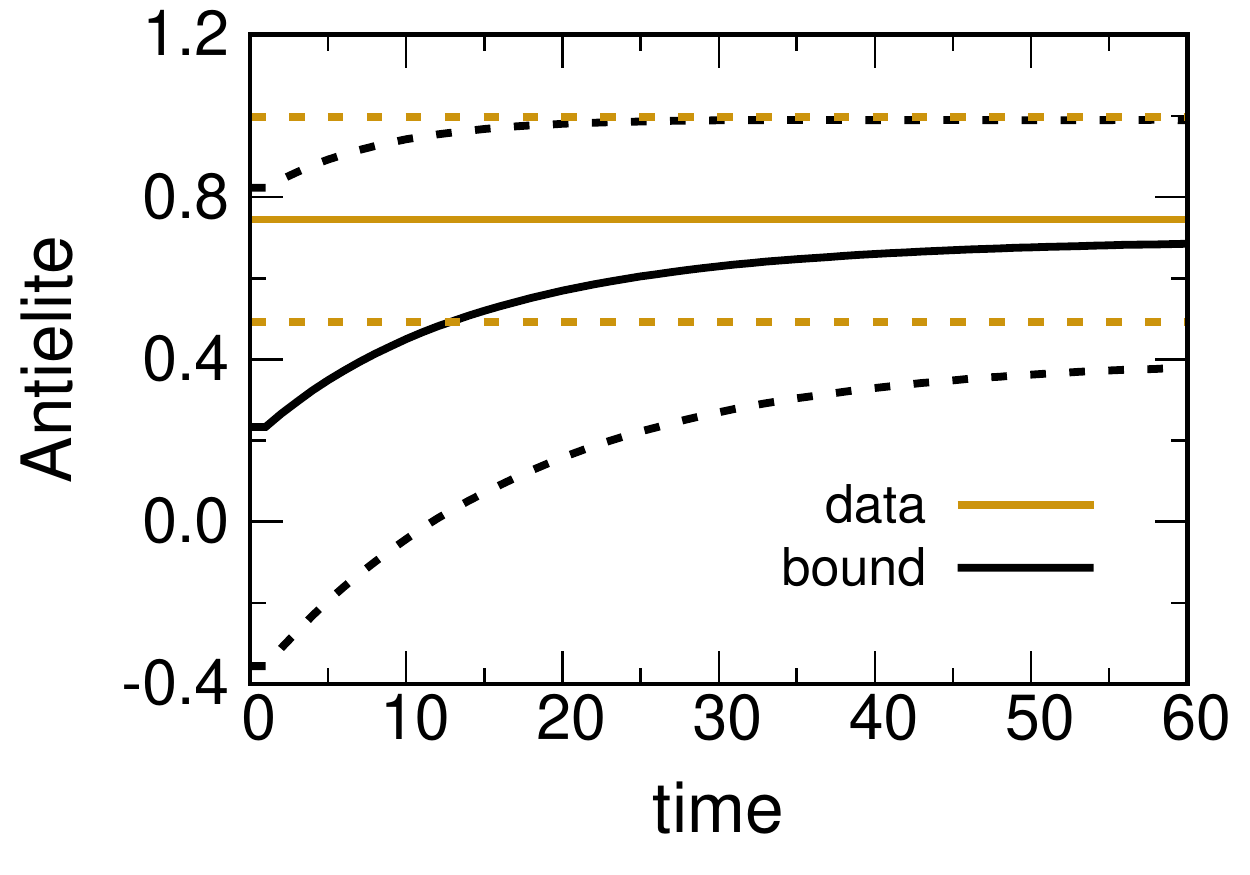}}
 \caption{Time evolution of the results of simulations of the model with boundary conditions and its comparison with data for the different opinion dimension (LR, NA, EU and AE). The black solid lines are the average value of the model $\langle x \rangle(t)$ and the black dashed lines are the average plus/minus the standard deviation $\langle x \rangle(t) \pm \sigma (t)$. The colored solid and dashed lines are the corresponding values coming from the data.}
 \label{evolution}
 \end{center}
 \end{figure}

According to the mean field solution of the model Eqs. (\ref{dynamic_average}, \ref{dynamic_variance}) the average and variance opinion of the users evolves exponentially in time and reaches a steady state value. In Fig. S\ref{evolution} we check that this is also the case in the simulations of the model with boundary conditions on top of the real network. Additionally, note that in the same figure we present a comparison between the stationary average and variance of the model and the results of the data, showing that the model is close to the data.

\subsubsection{Results for communities}\label{sec_results_comm} 
\begin{figure}[h!]
\begin{center}
 \hspace*{-0.2cm}
\subfloat[]{\label{opinions_community:a}\includegraphics[width=0.25\textwidth]{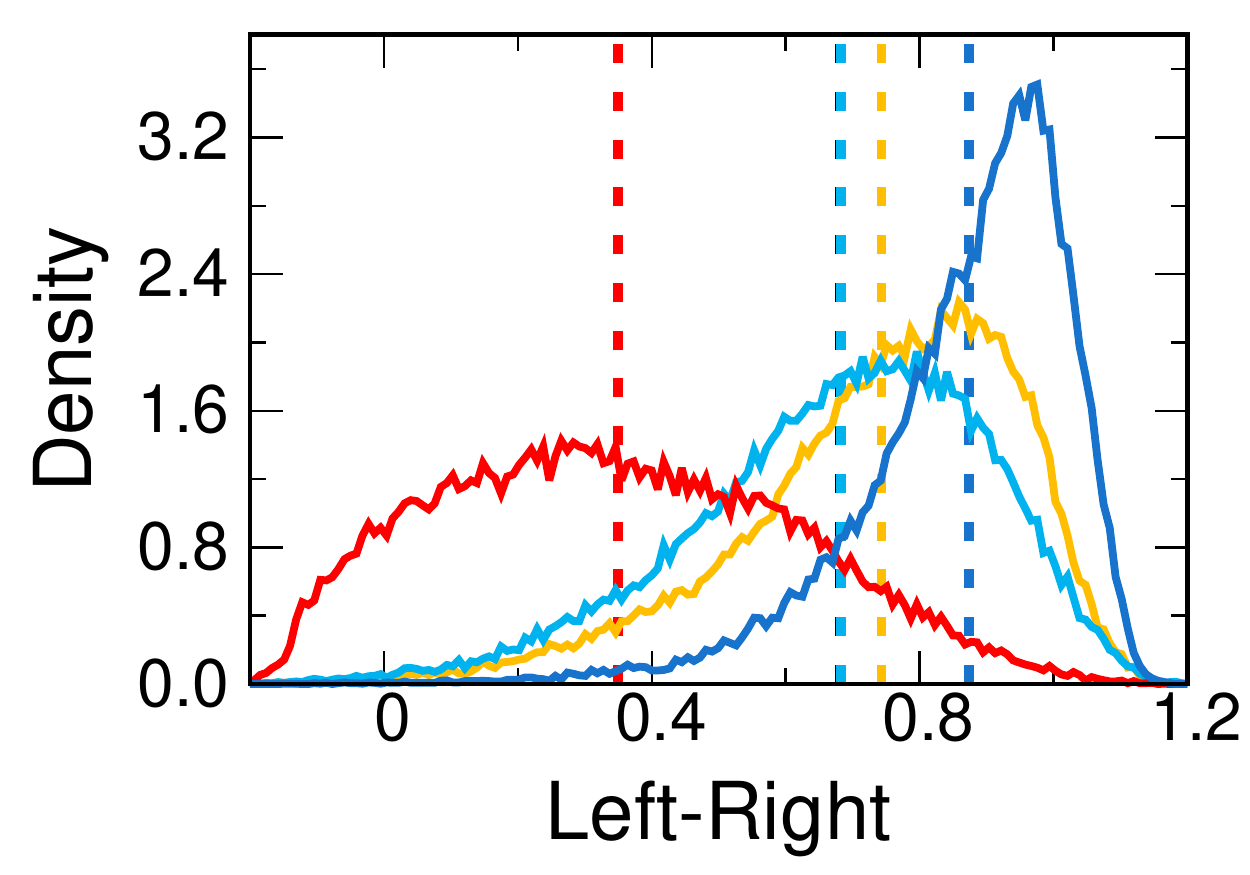}}
\subfloat[]{\label{opinions_community:b}\includegraphics[width=0.25\textwidth]{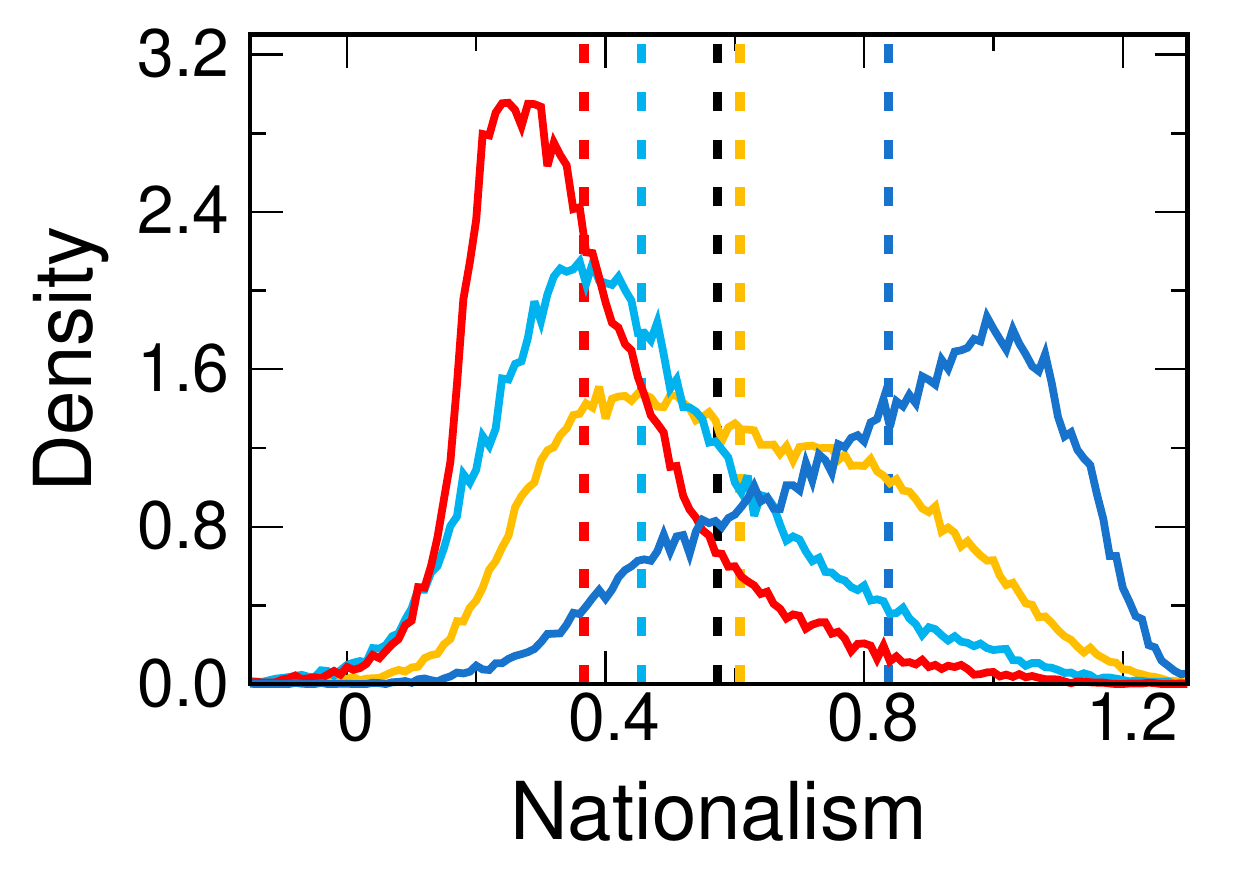}}
\subfloat[]{\label{opinions_community:c}\includegraphics[width=0.25\textwidth]{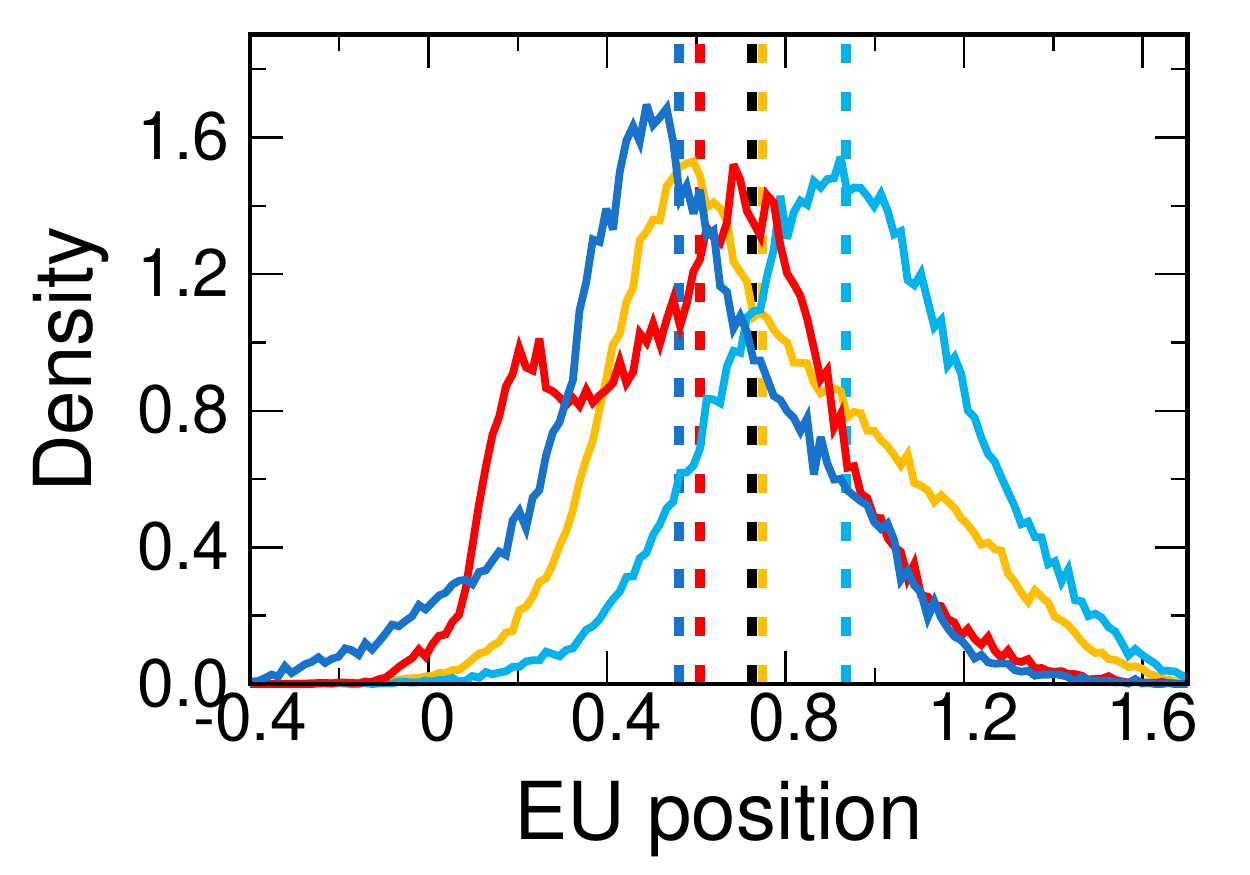}}
\subfloat[]{\label{opinions_community:d}\includegraphics[width=0.25\textwidth]{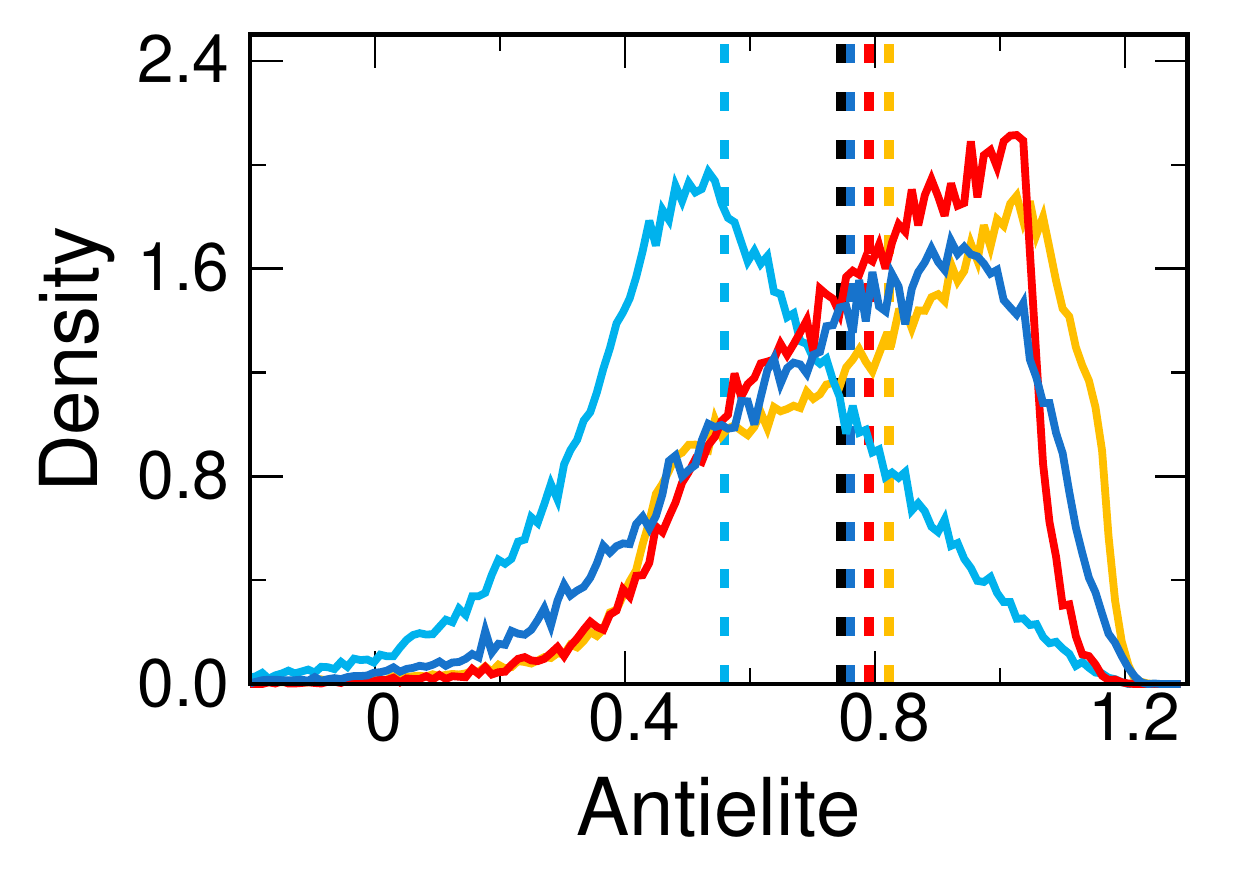}}

 \hspace*{-0.2cm}
\subfloat[]{\label{opinions_community:e}\includegraphics[width=0.25\textwidth]{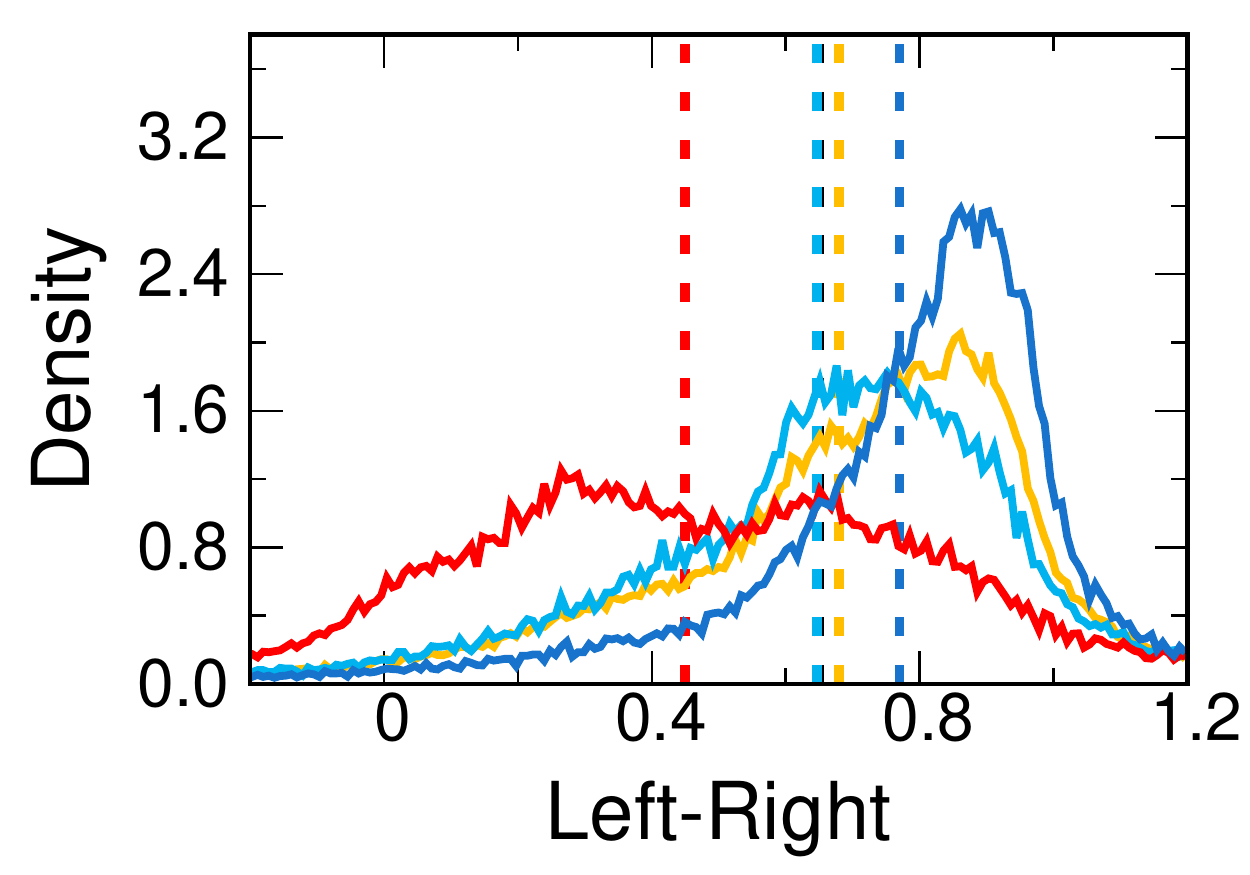}}
\subfloat[]{\label{opinions_community:f}\includegraphics[width=0.25\textwidth]{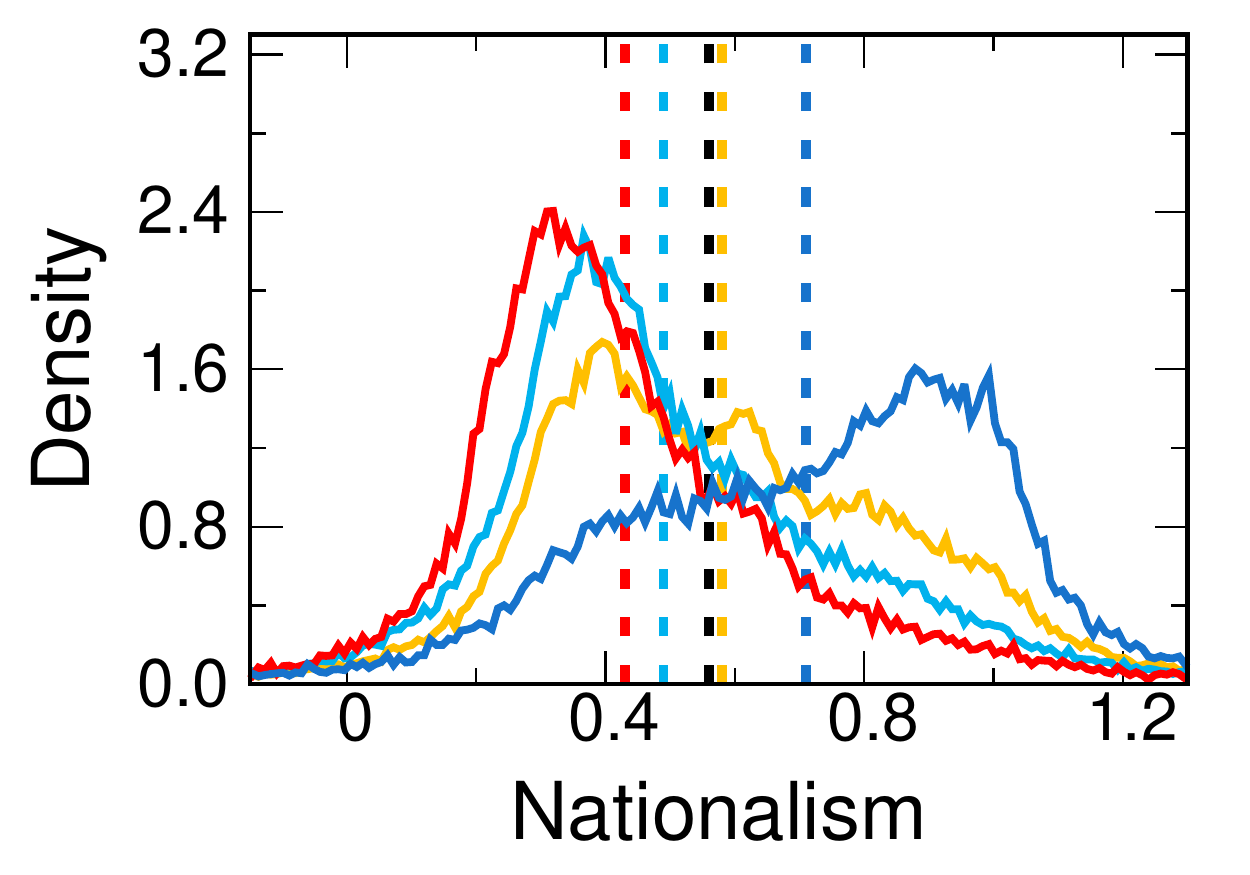}}
\subfloat[]{\label{opinions_community:g}\includegraphics[width=0.25\textwidth]{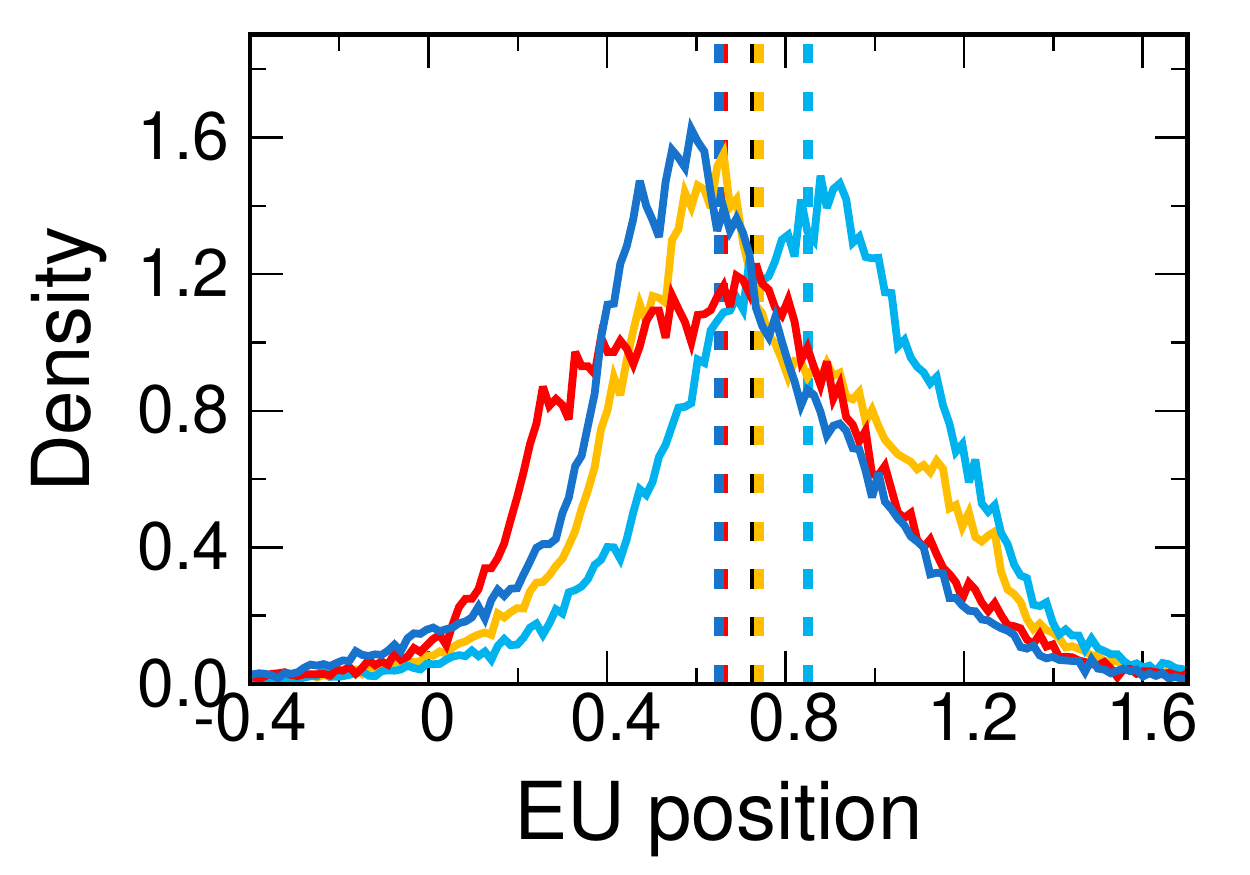}}
\subfloat[]{\label{opinions_community:h}\includegraphics[width=0.25\textwidth]{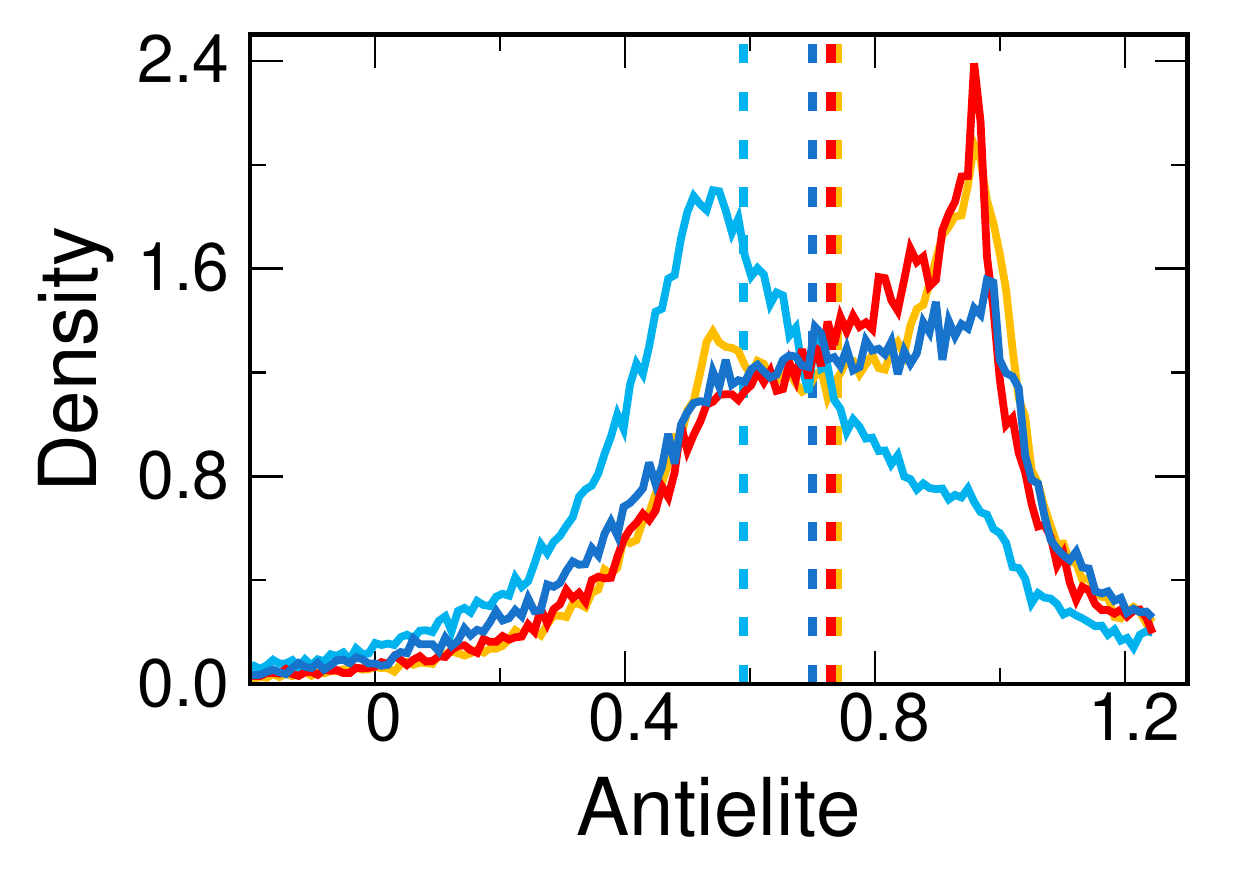}}
\caption{Probability density function of the different opinion variables (LR, NA, EU and AE) of the users belonging to a same community indicated by the color of the lines, i.e., $\alpha$ is red, $\beta$ is light blue, $\gamma$ is yellow and $\delta$ is dark blue. The top panels (a, b, c, d) correspond to the data while the bottom panels are the results of the model (e, f, g, h). The colored dashed lines represent the average opinion of each community, while the black dashed line is the global average.}
\label{opinions_community}
\end{center}
\end{figure}

\begin{figure}[h!]
\begin{center}
\subfloat[]{\label{opinions_distance:a}\includegraphics[width=0.25\textwidth]{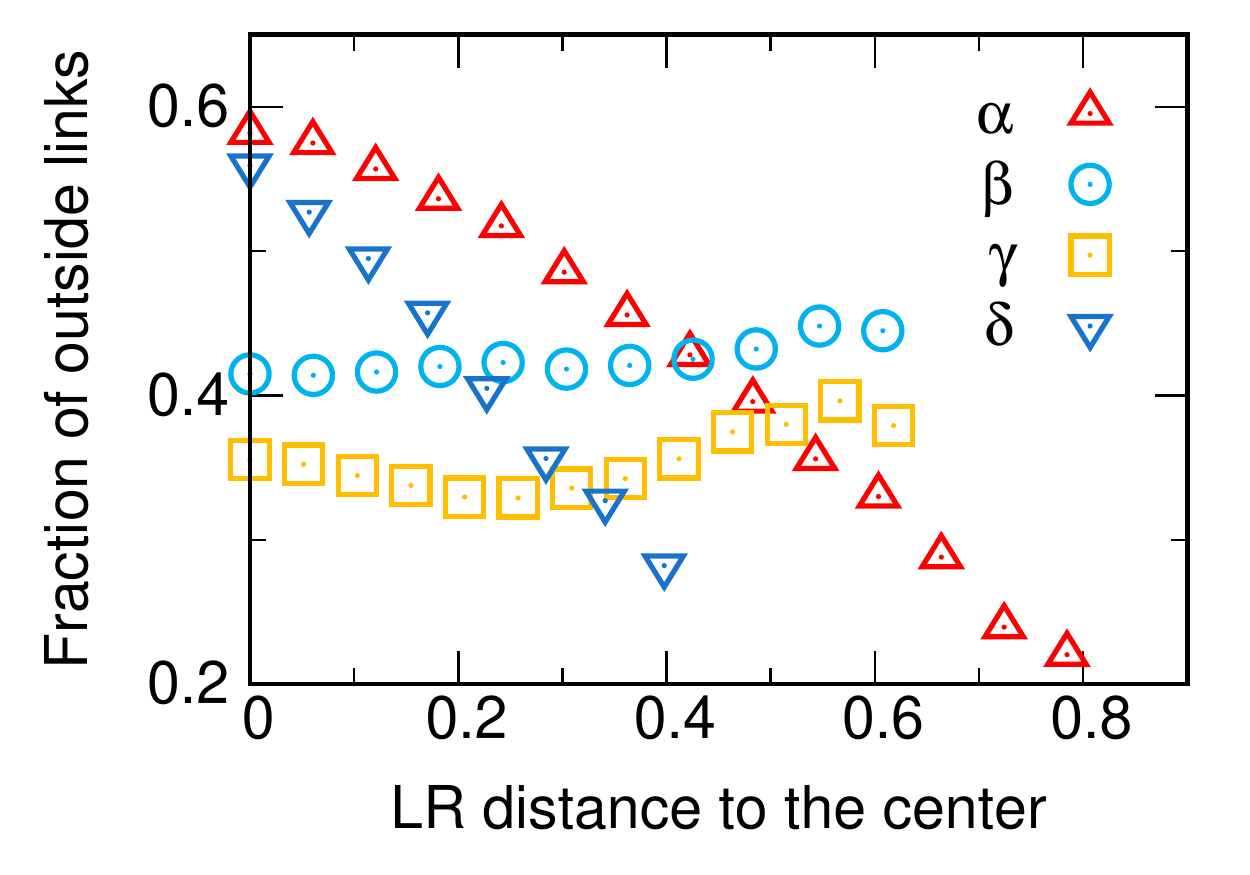}}
\subfloat[]{\label{opinions_distance:b}\includegraphics[width=0.25\textwidth]{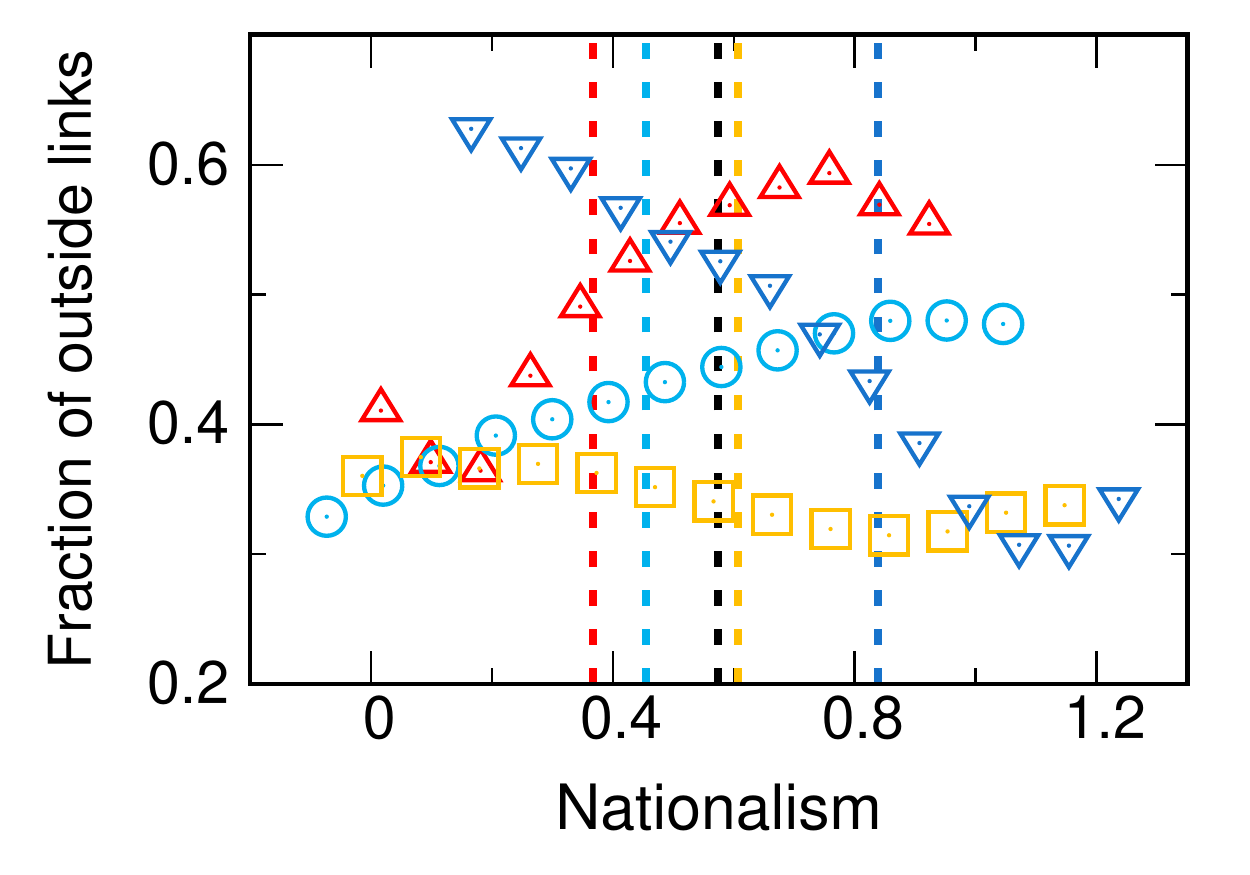}}
\subfloat[]{\label{opinions_distance:c}\includegraphics[width=0.25\textwidth]{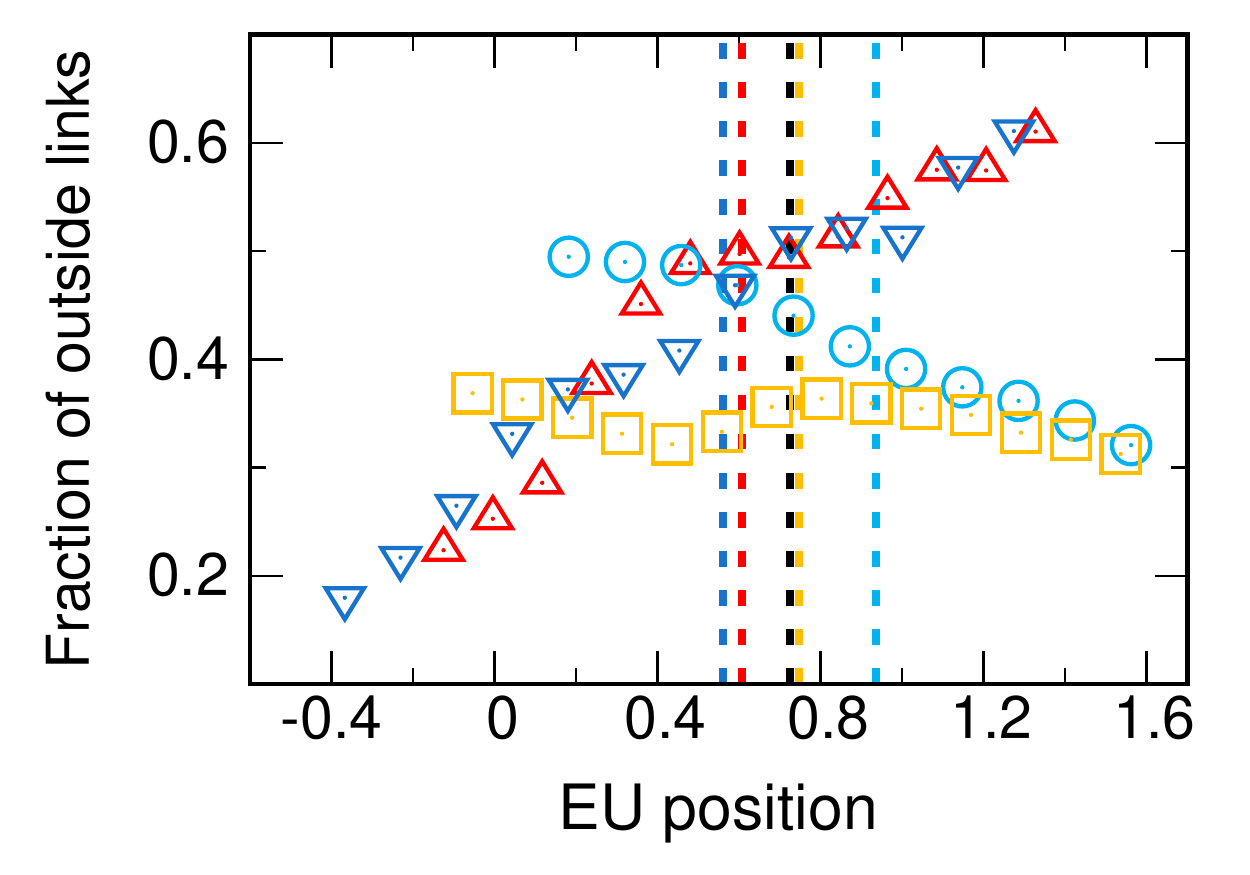}}
\subfloat[]{\label{opinions_distance:d}\includegraphics[width=0.25\textwidth]{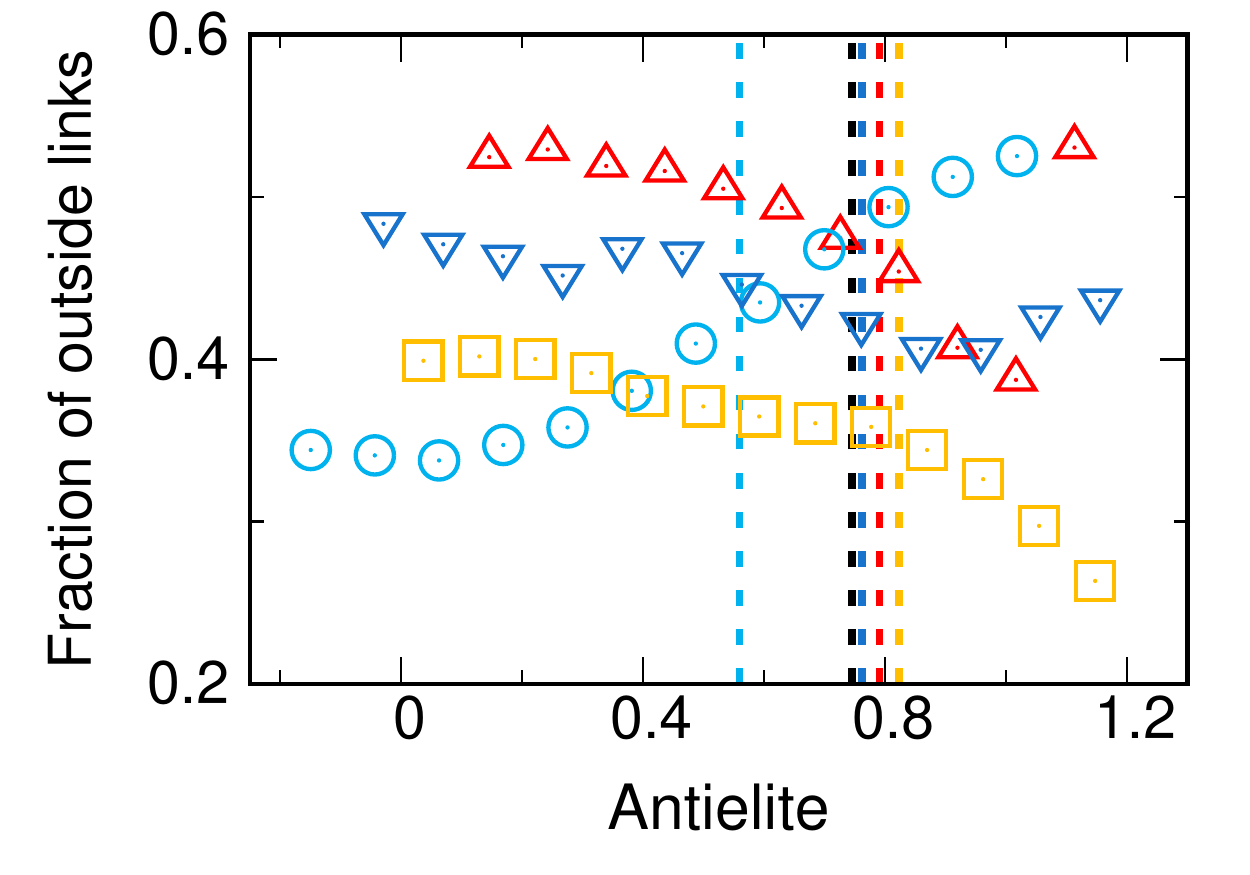}}

\subfloat[]{\label{opinions_distance:e}\includegraphics[width=0.25\textwidth]{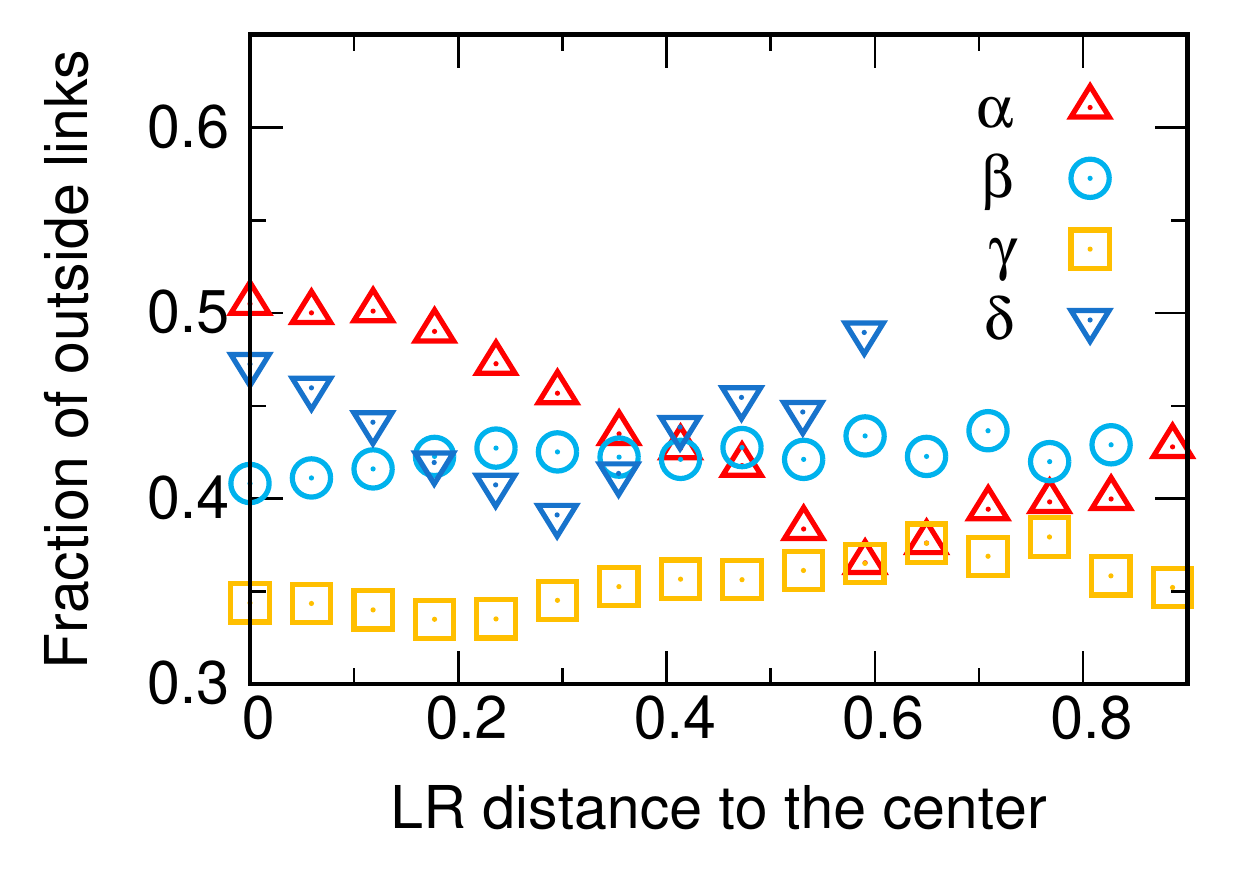}}
\subfloat[]{\label{opinions_distance:f}\includegraphics[width=0.25\textwidth]{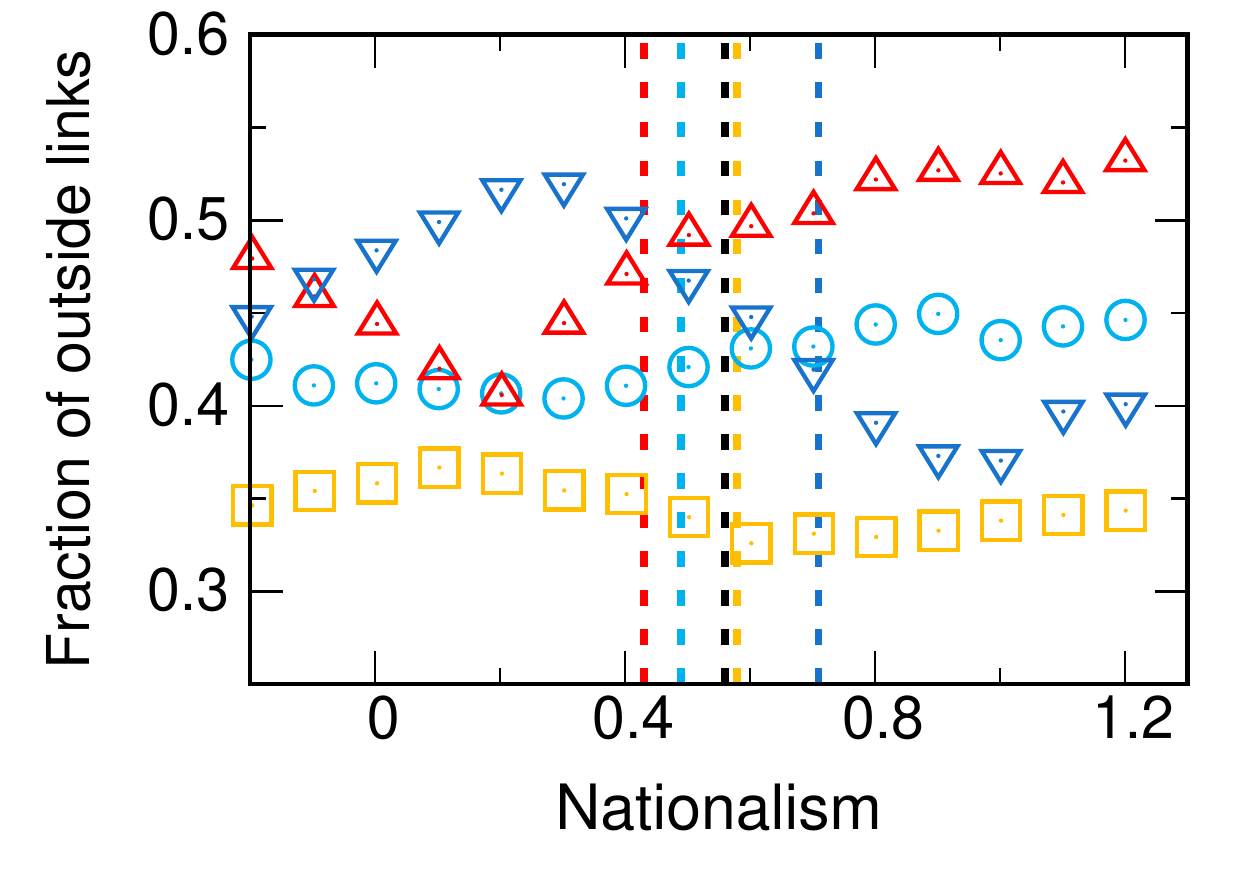}}
\subfloat[]{\label{opinions_distance:g}\includegraphics[width=0.25\textwidth]{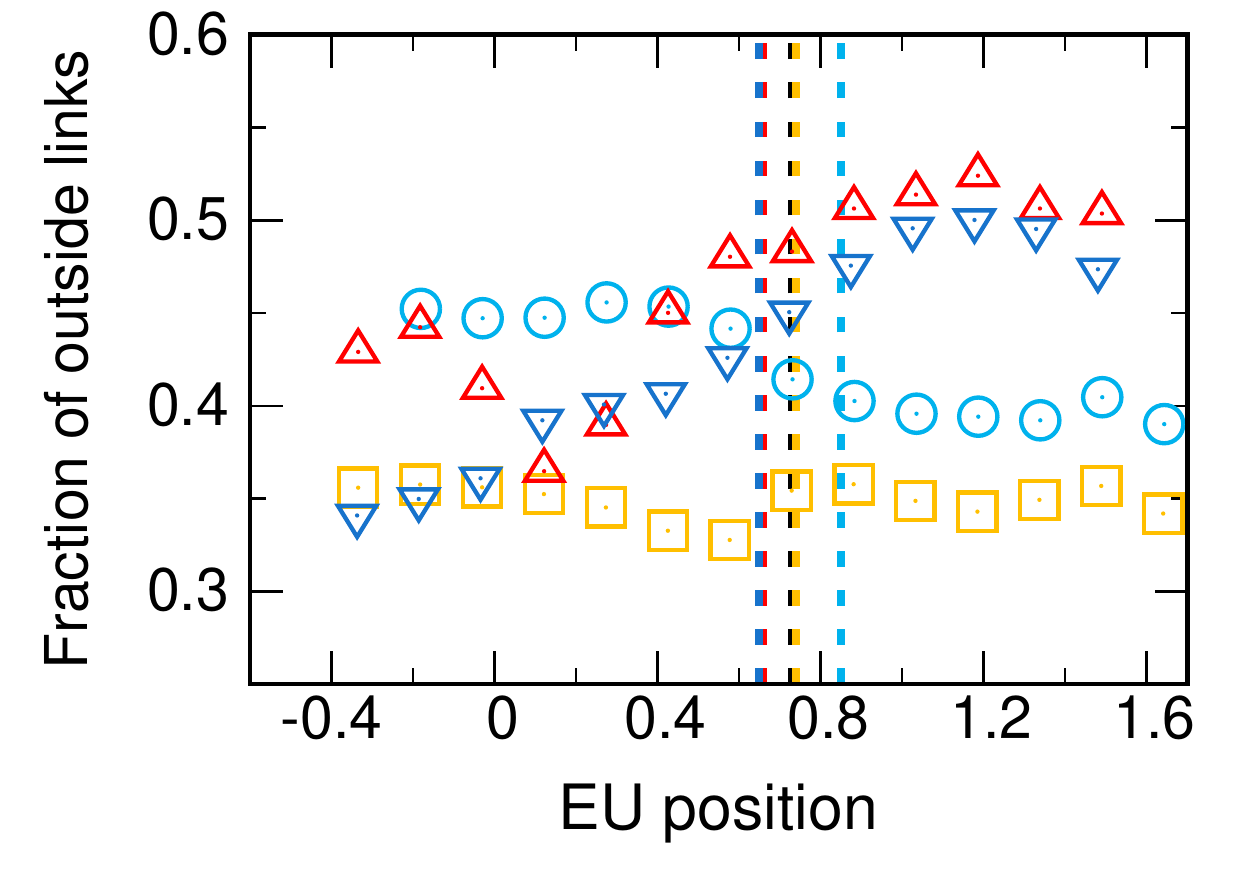}}
\subfloat[]{\label{opinions_distance:h}\includegraphics[width=0.25\textwidth]{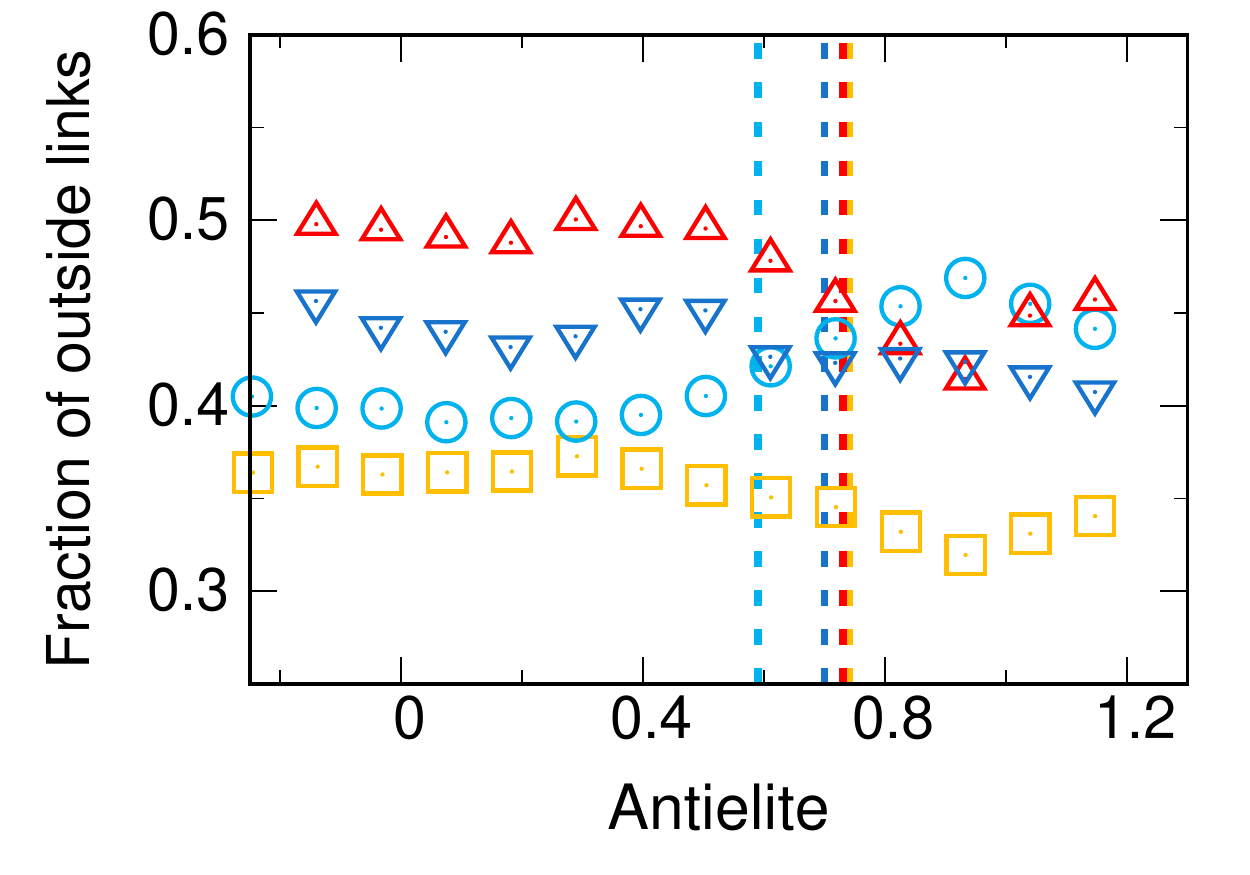}}
\caption{Average link ratio (vertical axis) in the User $\leftrightarrows$ User network between: the links whose target node belongs to a different community than the source node and the total number of links that depart from the source node. The horizontal axis corresponds to the opinion distance of the source nodes to the center for panels (a/e) and the state variables for (b/f, c/g, d/h). The average link ratio is calculated over all users whose opinion distance to the center or state variable lies in the same interval, and the color of the points indicate the community of the source nodes.}
\label{opinions_distance}
\end{center}
\end{figure}

Up to this point we have compared the statistical properties of the opinions of users coming from the model with the data at the global scale, i.e., the whole population. In this section we use the best partition of the User $\leftrightarrows$ User network in four groups $(\alpha, \beta, \gamma, \delta)$, as shown in the main text, to compare the results of the model with the data. In Fig. S\ref{opinions_community} we show the opinion distributions of the different groups computed using the stationary results of the model and the data. In general a good match between model and data is displayed, with some small deviation in the shape of the density function and also in the average values. The average group opinions in the data are farther away from each other than in the model. This is a consequence of the limitations of the model in reproducing the behavior of extreme individuals, as we discussed in detail in the main paper. These limitations and deviations of the model are also observed in Fig. S\ref{opinions_distance} when we evaluate the connectivity between groups as a function of the opinion value of their individuals. Both in data and model the more extreme the opinion of an individual is the more segregated it is in its group interconnectivity. However, this tendency is less pronounced in the model, i.e., the fraction of outside links in Fig. S\ref{opinions_community} decays more smoothly as a function of the opinion values of the individuals. Note also that the over-representation of extreme individuals in the model, see Fig. S\ref{opinions_borders}, produces also some differences in the group interconnectivities of the model as compared to the data.

\bibliographystyle{unsrt}
\bibliography{references}